\documentclass[review]{elsarticle}

\usepackage[]{changes} 

\usepackage[colorlinks,citecolor=blue,linktoc=all,linkcolor=cyan]{hyperref}
\usepackage{graphicx}
\biboptions{sort&compress}

\usepackage[T1]{fontenc}
\usepackage{dsfont}               
\usepackage{mathrsfs}             
\usepackage{slashed}              
\usepackage{amsmath}
\usepackage{amssymb}
\usepackage{amsbsy}
\usepackage{amsfonts}
\usepackage{bbold}
\usepackage{xcolor}
\usepackage{mathtools}
\usepackage{subcaption}
\numberwithin{equation}{section}
\numberwithin{table}{section}
\numberwithin{figure}{section}
\usepackage{soul}
\journal{Progress in Particle and Nuclear Physics}

\usepackage{titlesec}
\usepackage{sectsty}
\titleformat{\section}{\normalfont\Large\bfseries}{\thesection}{1em}{}
\titleformat{\subsection}{\normalfont\large\bfseries}{\thesubsection}{1em}{}
\titleformat{\subsubsection}{\normalfont\normalsize\bfseries}{\thesubsubsection}{1em}{}

\newcommand{\Dmq}{\Delta m^2}
\newcommand{\dCP}{\delta_\mathrm{CP}}

\newcommand{\nova}{NO$\nu$A~}

\newcommand{\ket}[1]{|{#1}\rangle}
\newcommand{\N}[1]{\widetilde \nu}
\newcommand{\Nm}[1]{\ket{\widetilde \nu^{(m)}_{#1}}}
\newcommand{\Nf}[1]{\ket{\widetilde\nu^{(f)}_{#1}}}
\newcommand{\nuf}[1]{\ket{\nu^{(f)}_{#1}}}
\newcommand{\nuem}[1]{\ket{\nu^{(m)}_{#1}}}

\newtheorem{theorem}{Theorem}[section]

\begin{document}
	
	\begin{frontmatter}
		

  \title{Phenomenology of Lepton Masses and Mixing with Discrete Flavor Symmetries}

  
		\author[mymainaddress]{Garv Chauhan}
        \author[mysecondaryaddress]{P. S. Bhupal Dev}
        \author[mythirdaddress]{Ievgen Dubovyk}
        \author[mythirdaddress]{Bartosz Dziewit}
        \author[myfourthaddress]{Wojciech Flieger}
        \author[mythirdaddress]{Krzysztof Grzanka}
        \author[mythirdaddress]{Janusz Gluza\corref{mycorrespondingauthor}}
		\cortext[mycorrespondingauthor]{Janusz Gluza}
		\ead{janusz.gluza@us.edu.pl}
        \author[mythirdaddress]{Biswajit Karmakar}
        \author[mythirdaddress]{Szymon Zięba}

\address[mymainaddress]{Center for Neutrino Physics, Department of Physics, Virginia Tech, Blacksburg, 
USA}

\address[mysecondaryaddress]{Department of Physics and McDonnell Center for the Space Sciences,  Washington University, St. Louis, 
  USA}
\address[mythirdaddress]{Institute of Physics, University of Silesia,  Katowice, Poland}
  
\address[myfourthaddress]{Max-Planck-Institut f\"ur Physik, Werner-Heisenberg-Institut, 
M\"unchen, Germany
}

 \date{\today}

		\begin{abstract}
			The observed pattern of fermion masses and mixing is an outstanding puzzle in particle physics, generally known as the {\it flavor problem}. Over the years, guided by precision neutrino oscillation data, discrete flavor symmetries have often been used to explain the neutrino mixing parameters, which look very different from the quark sector.  In this review, we discuss the application of non-Abelian finite groups to the theory of neutrino masses and mixing in the light of current and future neutrino oscillation data.  We start with an overview of the neutrino mixing parameters, comparing different global fit results and limits on normal and inverted neutrino mass ordering schemes.
   Then, we discuss a  general framework for implementing discrete family symmetries to explain neutrino masses and mixing. We discuss CP violation effects, giving an update of  CP predictions for trimaximal models  with nonzero reactor mixing angle and models with partial $\mu-\tau$ reflection symmetry, and constraining models with neutrino mass sum rules. The connection between texture zeroes and discrete symmetries is also discussed. We summarize viable higher-order groups, which can explain the observed pattern of  lepton   mixing  where the non-zero $\theta_{13}$ plays  an important role. We also review the prospects of embedding finite discrete symmetries in the Grand Unified Theories and with extended Higgs fields. Models based on modular symmetry are also briefly discussed. A major part of the review is dedicated to the phenomenology of flavor symmetries and possible signatures in the current and future experiments at the intensity, energy,  and cosmic frontiers. In this context, we discuss flavor symmetry implications for neutrinoless double beta decay, collider signals, leptogenesis, dark matter, as well as gravitational waves. 
		\end{abstract}
		
		\begin{keyword}
			Discrete Symmetries \sep Flavor mixing \sep 
   CP Violation \sep Neutrino Oscillation \sep Phenomenology 
			
		\end{keyword}

	\end{frontmatter}

	\newpage
	
	\thispagestyle{empty}
	\tableofcontents
	

	 

\section{Introduction}\label{chapter:introduction}

 Over the past few decades, we have seen spectacular progress in understanding neutrinos.
The neutrino quantum oscillation phenomenon
established that at least two
neutrino quantum states are massive, although the masses are tiny, at the sub-electronvolt level,
$m_{\nu}\lesssim {\cal{O}}(0.1)$ eV~\cite{Barger:1999na,Czakon:1999cd}.
A tremendous effort led to this result, as earlier experimental studies of neutrino physics faced the challenge of low event statistics for a  scarce set of observables. It started around half a century ago with the pioneering Homestake experiment~\cite{Davis:1968cp} and the so-called solar neutrino problem ~\cite{Bahcall:1976zz} and culminated with the 2015 Nobel Prize in Physics for the discovery of neutrino oscillations by the Super-Kamiokande and SNO collaborations~\cite{Fukuda:1998mi,SNO:2001kpb, Ahmad:2002jz}, which showed that neutrinos have mass.
     
The simplest neutrino mass theory is based on the three-neutrino ($3\nu$) paradigm with the assumption that flavor states ($\nu_e,\nu_\mu,\nu_\tau$) are mixed with massive states ($\nu_1,\nu_2,\nu_3$) with definite masses ($m_1,m_2,m_3$), at least two of which are non-zero. The standard parametrization of the Pontecorvo–Maki–Nakagawa–Sakata (\texttt{PMNS}) unitary mixing matrix\footnote{{In an equivalent   parametrization~\cite{Schechter:1980gr,Schechter:1980gk,Rodejohann:2011vc}, the lepton mixing matrix can be written in a `symmetrical' form where all three CP violating phases are `physical'. }} reads~\cite{Pontecorvo:1957qd,Maki:1962mu, Kobayashi:1973fv} 
	\begin{align}
  U &=U(\theta_{23})U(\theta_{13}, \delta_{\rm CP})U(\theta_{12})U_M(\alpha_1,\alpha_2) \nonumber \\
  &=\begin{pmatrix}
    1 & 0 & 0 \\
    0 & c_{23}  & {s_{23}} \\
    0 & -s_{23} & {c_{23}}
  \end{pmatrix}
  \begin{pmatrix}
 c_{13} & 0 & s_{13} e^{-i\delta_\text{\rm CP}} \\
    0 & 1 & 0 \\
    -s_{13} e^{i\delta_\text{\rm CP}} & 0 & c_{13}
  \end{pmatrix}
  \begin{pmatrix}
    c_{12} & s_{12} & 0 \\
    -s_{12} & c_{12} & 0 \\
    0 & 0 & 1
  \end{pmatrix}  \begin{pmatrix}
    e^{i \alpha_1} & 0 & 0 \\
    0 & e^{i \alpha_2} & 0 \\
    0 & 0 & 1
  \end{pmatrix},  
\label{upmns1}
\end{align}
where $M$ stands for Majorana neutrinos, $c_{ij} \equiv \cos( \theta_{ij})$, $s_{ij} \equiv \sin(\theta_{ij})$,
and the Euler rotation angles $\theta_{ij}$ can be taken without loss
of generality from the first quadrant, $\theta_{ij} \in [0, \pi/2]$, and the Dirac CP phase $\delta_{\text{\rm CP}}$ and Majorana phases $\alpha_1,\alpha_2$ are in the range $ [0, 2\pi]$ \cite{10.1093/ptep/ptac097c14}. This choice of parameter regions is independent of matter effects~\cite{Gluza:2001de}.  

In the $3\nu$ paradigm, there are two non-equivalent mass orderings: normal mass ordering (NO) with $m_1 < m_2 < m_3$ and
inverted mass ordering (IO) with  $m_3 < m_1 < m_2$. The neutrino masses can be further classified into normal hierarchical mass spectrum (NH) with $m_1 \ll m_2 < m_3$, inverted hierarchical mass spectrum  (IH) with  $m_3 \ll m_1 < m_2$ and quasi-degenerate mass spectrum (QD) with  $m_1 \simeq m_2 \simeq  m_3$~\cite{10.1093/ptep/ptac097c14}.
Within QD, quasi-degenerate NH (QDNH) mass spectrum with  $m_1 \lesssim m_2 \lesssim  m_3$ and quasi-degenerate IH (QDIH) mass spectrum  with  $m_3 \lesssim m_1 \lesssim  m_2$~\cite{Roy:2013yka} can be distinguished. NO/IO notation is sometimes used interchangeably with NH/IH in the literature. However, NO/IO notation is more general since both NH and QDNH are NO and IH and QDIH are IO. For more discussion, see section 14.7 in the PDG review~\cite{10.1093/ptep/ptac097c14}.  In what follows, we will use the NO/IO notation. 
 Neutrino masses can be expressed by the smallest of the three neutrino masses ($m_0$) and experimentally determined mass-squared
differences ($\Delta m_{21}^2, \Delta m_{31}^2, \Delta m_{32}^2$) 
\begin{equation}
\begin{array}{ccc}
\mbox{\underline{Normal mass ordering (NO)}} & \mbox{\underline{Inverted mass ordering (IO)}}
& \\
\begin{array}{l}
m_{1} = m_{0}, \\
m_{2} = \sqrt{m_{0}^2 + \Delta m_{21}^2}, \\
m_{3} = \sqrt{m_{0}^2 + \Delta m_{31}^2}, \\
\end{array}
&
\begin{array}{l}
m_{1} = \sqrt{m_{0}^2 - \Delta m_{21}^2 - \Delta m_{32}^2}, \\
m_{2} = \sqrt{m_{0}^2 - \Delta m_{32}^2}, \\
m_{3} = m_{0}, 
\end{array}
\end{array}
\label{e:NO_and_IO}
\end{equation}
where $\Delta m_{ij}^2 \equiv m_i^2-m_j^2$. The observation of matter effects in the Sun constrains the product $\Delta m^2_{21}\cos{2\theta_{12}}$ to be positive~\cite{Fogli:2002pt}. By definition\footnote{{Considerations of the mass schemes with some negative $\Delta m_{ij}^2$ are not necessary from the point of view of neutrino oscillation parametrization (in vacuum and matter) and may cause double counting only, see Ref.~\cite{Gluza:2001de}.}}, we choose $\Delta m^2_{21}>0$ and $\theta_{12}$ in the first octant, although in presence of non-standard matter effects, a high-octant $\theta_{12}$ is possible~\cite{Miranda:2004nb}.

Depending on their origin, the neutrino oscillation parameters $\theta_{12}$, $\theta_{23}$ and $\theta_{13}$ are often called solar, atmospheric and reactor angles, respectively, while $\Delta m_{21}^2$ and $\Delta m_{31}^2$
are called solar and atmospheric mass-squared differences, respectively.
This naming has a historical background and is related to the neutrino source associated with measured oscillation parameters in the initial neutrino experiments. 

\begin{figure}[h!]
     \centering
     \begin{subfigure}[b]{0.32\textwidth}
         \centering
         \includegraphics[width=\textwidth]{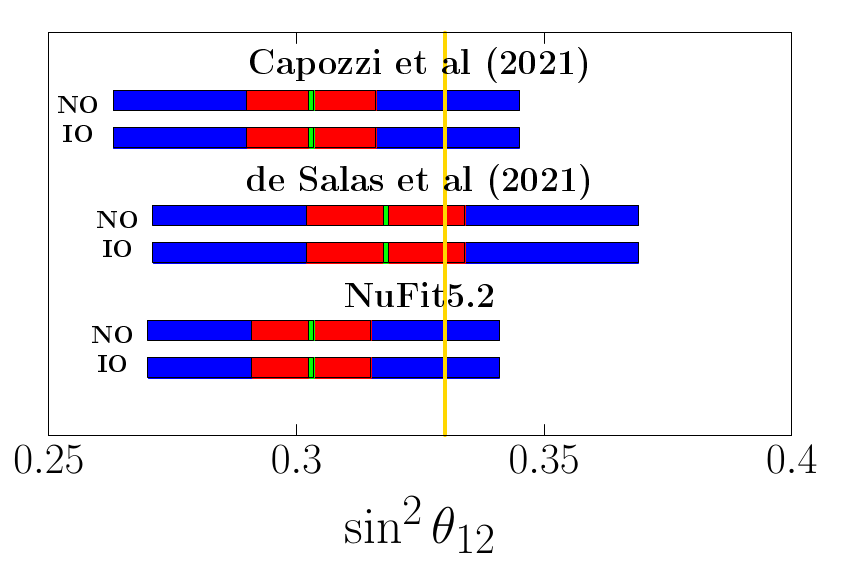}
     \end{subfigure}
     \ 
     \begin{subfigure}[b]{0.32\textwidth}
         \centering
         \includegraphics[width=\textwidth]{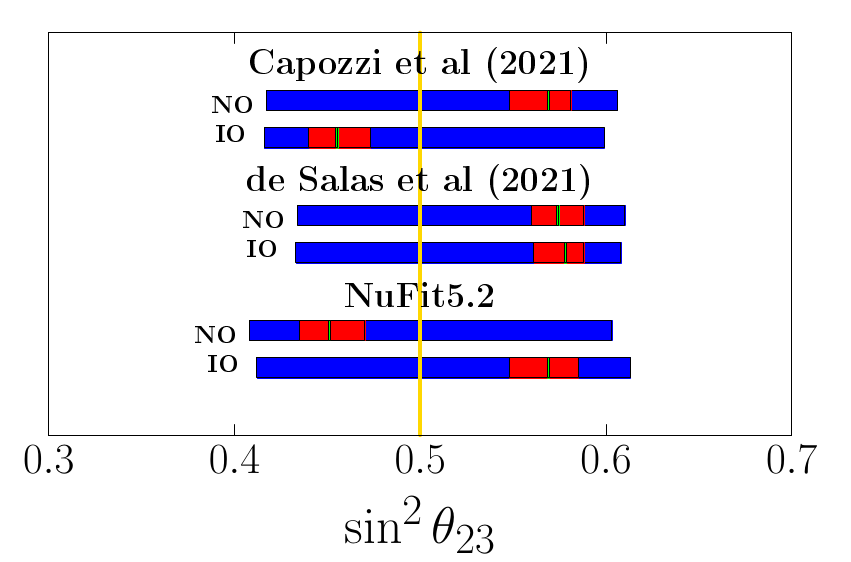}
     \end{subfigure}
     \ 
     \begin{subfigure}[b]{0.32\textwidth}
         \centering
         \includegraphics[width=\textwidth]{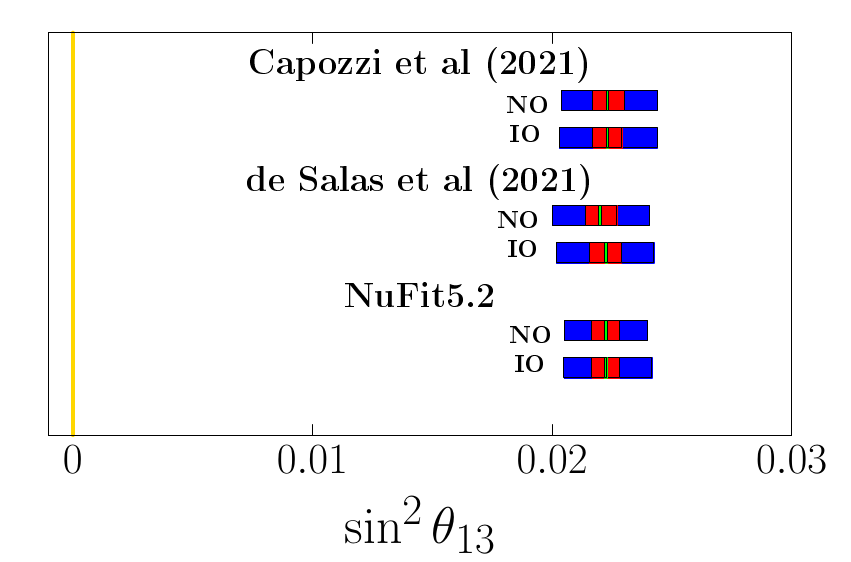}
     \end{subfigure}
\medskip
     \begin{subfigure}[b]{0.32\textwidth}
        \centering
       \includegraphics[width=\textwidth]{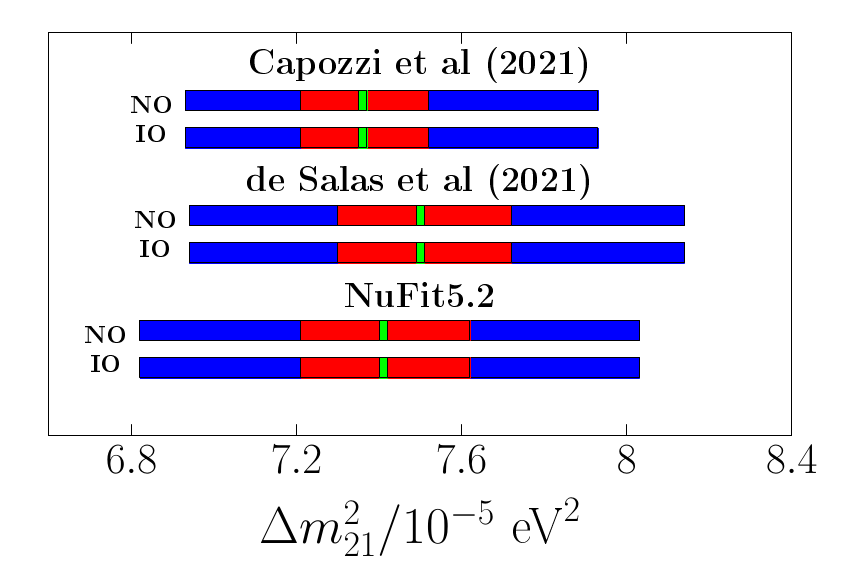}
    \end{subfigure}
    \ 
    \begin{subfigure}[b]{0.32\textwidth}
        \centering
        \includegraphics[width=\textwidth]{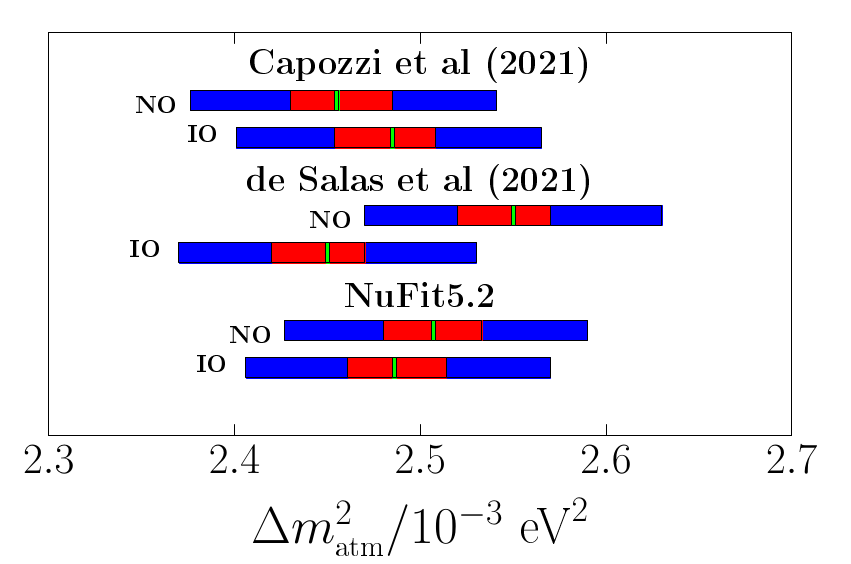}
     \end{subfigure}
     \ 
     \begin{subfigure}[b]{0.32\textwidth}
         \centering
         \includegraphics[width=\textwidth]{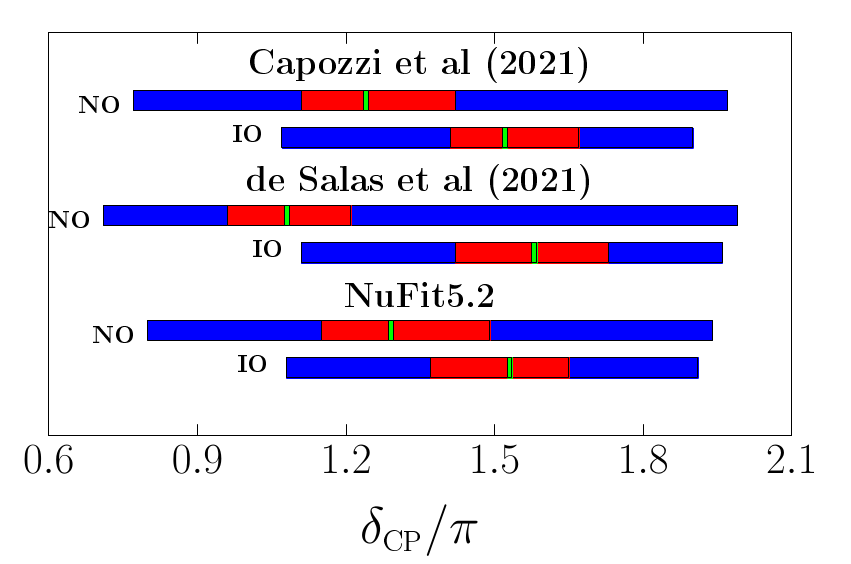}
     \end{subfigure}
        \caption{Schematics of neutrino oscillation data from Table~\ref{tab:neutrino_data}. On the plots, green represents best fits, red 1$\sigma$ ranges, and blue 3$\sigma$ ranges.  The golden vertical lines symbolize the historic tribimaximal (TBM) pattern values, which will be explained in Chapter~\ref{chapter:theory}. The plot is an update of plots presented in  Refs.~\cite{King:2013eh, Agarwalla:2021bzs}.  }
        \label{fig:schematics_new}
\end{figure}

Table~\ref{tab:neutrino_data} summarizes recent global fits for the neutrino oscillation parameters~\cite{NuFIT5.2,Esteban:2020cvm,deSalas:2020pgw,10.5281/zenodo.4726908,Capozzi:2021fjo}, which are used in the present analysis for NO/IO. The data shows a hierarchy between the mass splittings, $\Delta m_{21}^2 \ll \Delta m_{31}^2 \simeq \Delta m_{32}^2$ and preference of NO at an overall level of $\sim2.5 \sigma$ corresponding to $\Delta \chi^{2} \sim 6.4-6.5$.  Lower octant atmospheric angle $\theta_{23}$ best-fit is favored for NO in NuFIT 5.2 \cite{NuFIT5.2,Esteban:2020cvm} and Capozzi et al~\cite{Capozzi:2021fjo} results. These analysis include the most recent Superkamiokande (SK) data (with  $\sin^2 \theta_{23} < 0.5$ preference)~\cite{yasuhiro_nakajima_2020_4134680,Takhistov:2020qhw}. 
The $\theta_{23}$ best-fit octant preference discussed above is illustrated in Fig.~\ref{fig:schematics_new} together with other  neutrino  parameters as given in Table~~\ref{tab:neutrino_data}.
 
 \begin{table}[h!]
 \centering
    \scriptsize
    \begin{tabular}{c c c c c c c c}
    \hline \hline
        Parameter &  Ordering 
        & \multicolumn{2}{c}{NuFIT 5.2 (2022) ~\cite{NuFIT5.2,Esteban:2020cvm}}
        & \multicolumn{2}{c}{
        de Salas et al. (2021) ~\cite{deSalas:2020pgw,10.5281/zenodo.4726908}
        }
        & \multicolumn{2}{c}{
        Capozzi et al. (2021)~\cite{Capozzi:2021fjo}
        }
        \\
        & & 
        bf$\pm 1 \sigma$ & $3\sigma$ range &
        bf$\pm 1 \sigma$ & $3\sigma$ range &
        bf$\pm 1 \sigma$ & $3\sigma$ range 
        \\
         \hline 
        $\sin^2 \theta_{12} / 10^{-1}$
        & NO, IO
        & $3.03^{+0.12}_{-0.12}$&$2.70-3.41$
        & $3.18^{+0.16}_{-0.16}$&$2.71-3.69$
        & $3.03^{+0.13}_{-0.13}$&$2.63-3.45$
        \\
        \hline
         $\sin^2 \theta_{23} / 10^{-1}$
        & NO
        & $4.51^{+0.19}_{-0.16}$&$4.08-6.03$
        & $5.74^{+0.14}_{-0.14}$&$4.34-6.10$
        & $4.55^{+0.18}_{-0.15}$&$4.16-5.99$
        \\
        & IO
        & $5.69^{+0.16}_{-0.21}$&$4.12-6.13$
        & $5.78^{+0.10}_{-0.17}$&$4.33-6.08$
        & $5.69^{+0.12}_{-0.21}$&$4.17-6.06$
        \\
        \hline
         $\sin^2 \theta_{13} / 10^{-2}$
        & NO
        & $2.225^{+0.056}_{-0.059}$&$2.052-2.398$
        & $2.200^{+0.069}_{-0.062}$&$2.000-2.405$
        & $2.23^{+0.07}_{-0.06}$&$2.04-2.44$
         \\
        & IO
        & $2.223^{+0.058}_{-0.058}$&$2.048-2.416$
        & $2.225^{+0.064}_{-0.070}$&$2.018-2.424$
        & $2.23^{+0.06}_{-0.06}$&$2.03-2.45$
        \\
        \hline
         $\delta_{\rm CP} / \pi$
        & NO
        & $1.29^{+0.20}_{-0.14}$&$0.80-1.94$
        & $1.08^{+0.13}_{-0.12}$&$0.71-1.99$
        & $1.24^{+0.18}_{-0.13}$&$0.77-1.97$
         \\
        & IO
        & $1.53^{+0.12}_{-0.16}$&$1.08-1.91$
        & $1.58^{+0.15}_{-0.16}$&$1.11-1.96$
        & $1.52^{+0.15}_{-0.11}$&$1.07-1.90$
        \\
        \hline
         $\Delta m_{21}^2 / 10^{-5} \mbox{eV}^2 $
        & NO, IO
        & $7.41^{+0.21}_{-0.20}$&$6.82-8.03$
        & $7.50^{+0.22}_{-0.20}$&$6.94-8.14$
        & $7.36^{+0.16}_{-0.15}$&$6.93-7.93$
        \\
        \hline
        $\left|{\Delta m_{\rm atm}^2}\right|  / 10^{-3}  \mbox{eV}^2 $
        & NO
        & $2.507^{+0.026}_{-0.027}$&$2.427-2.590$
        & $2.55^{+0.02}_{-0.03}$&$2.47-2.63$
        & $2.485^{+0.023}_{-0.031}$&$2.401-2.565$
         \\
        & IO
        & $2.486^{+0.028}_{-0.025}$&$2.406-2.570$
        & $2.45^{+0.02}_{-0.03}$&$2.37-2.53$
        & $2.455^{+0.030}_{-0.025}$&$2.376-2.541$
         \\
         \hline
        $\Delta \chi^{2}$  & IO - NO  
        & \multicolumn{2}{c}{$6.4$}
        & \multicolumn{2}{c}{$6.4$}
        & \multicolumn{2}{c}{$6.5$}
        \\
        \hline \hline
    \end{tabular}
\caption{
Selection of neutrino oscillation data from NuFIT 5.2 with SK atmospheric data~\cite{NuFIT5.2,Esteban:2020cvm}, 
de Salas et al. neutrino global fit~\cite{deSalas:2020pgw,10.5281/zenodo.4726908} 
and Capozzi et al. global fit~\cite{Capozzi:2021fjo}. Notation used by different groups is unified; namely, the   atmospheric neutrino  mass-squared difference   parameter $\Delta m_{\rm atm}^2$ is defined for NuFIT as 
$\Delta m_{31}^2$ for NO and $\Delta m_{32}^2$ for IO, for de Salas et al. as  $\Delta m_{31}^2$ for NO/IO and for Capozzi et al. as $\Delta m_{31}^2 - \Delta m_{21}^2/2$ for NO/IO.  The lower octant atmospheric angle $\theta_{23}$ best-fit is favored for NO in NuFIT and Capozzi et al. results, which can be related to more recent SK atmospheric data used in analyses. $\Delta \chi^{2}$ represents NO preference over IO at an overall level of $\sim2.5 \sigma$.}
\label{tab:neutrino_data}
\end{table}

\begin{figure}[t!]
    \centering
       \includegraphics[width=0.6\textwidth]{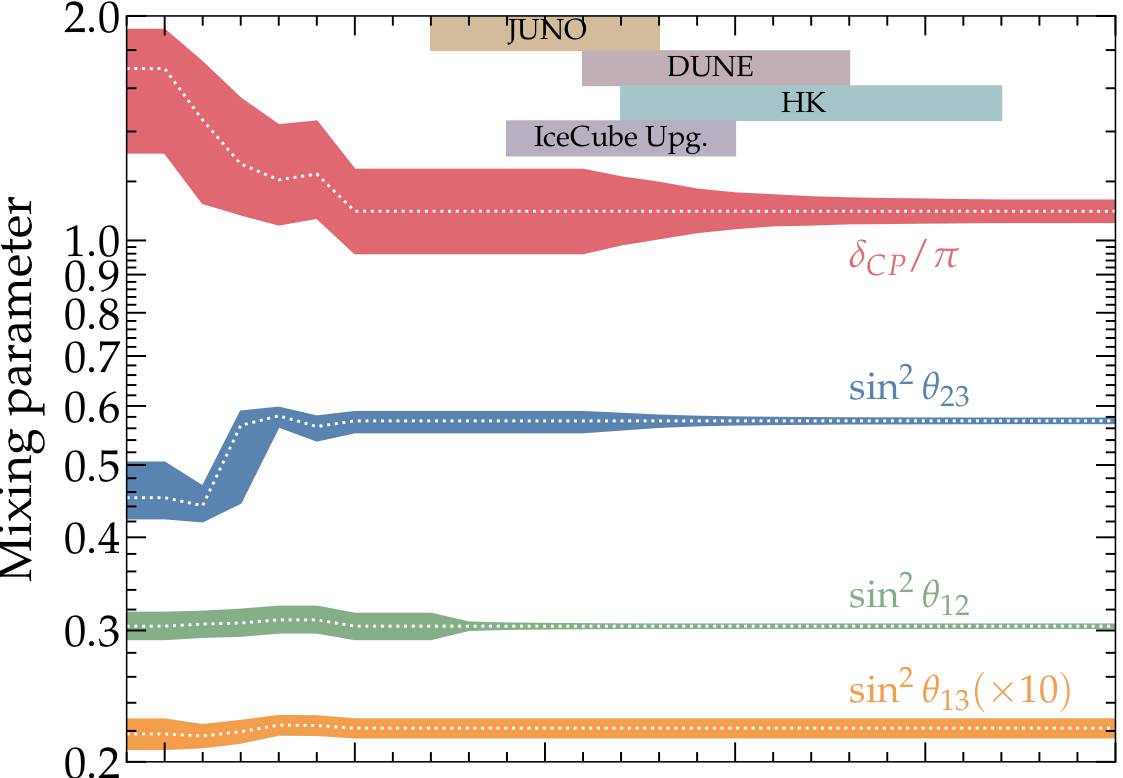}
       
       \hspace{1.5cm}
        \includegraphics[width=0.55\textwidth]{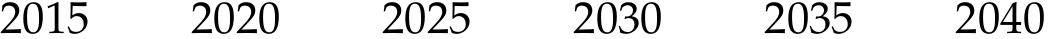}
    \caption{
{Time evolution of the oscillation parameters.  Figure taken from the arXiv version of Ref.~\cite{Song:2020nfh}.}}
    \label{fig:fig1_2012.12893}
\end{figure}

The initial   results by the T2K collaboration~\cite{Abe:2019vii} 
indicated 
CP violation in the lepton sector,  
a preference for NO at the 3$\sigma$ level and also for $\theta_{23}$ in the second octant.  
These data are confirmed with an improved analysis~\cite{T2K:2021xwb, T2K:2023smv}; they continue to prefer normal mass ordering and upper octant of $\theta_{23}$ with a nearly maximal $\delta_{\rm CP}$. 
Also, the \nova collaboration~\cite{NOvA:2021nfi} reports on nonvanishing $\delta_{\rm CP}$ effects, though the best-fit values of CP phase differ between the two groups. Both T2K and \nova prefer NO over IO, but T2K prefers $\delta_{\rm CP}=-90^o$ whereas \nova prefers a region $\delta_{\rm CP} \sim 180^o$.\footnote{The mild tension between T2K and \nova could in principle be resolved by invoking new physics~\cite{Rahaman:2022rfp}.} 
Joint fits between \nova+T2K and Super-K+T2K are ongoing, with the aim to obtain improved oscillation parameter constraints due to resolved degeneracies, and to understand potentially non-trivial systematic correlations~\cite{Denton:2022een}. The next-generation oscillation experiments, such as JUNO~\cite{JUNO:2015sjr}, Hyper-K~\cite{Hyper-Kamiokande:2018ofw}, DUNE~\cite{DUNE:2020lwj} and IceCube upgrade~\cite{IceCube-Gen2:2019fet}, will significantly improve the prospects of measuring $\delta_{\rm CP}$ and determining the mass ordering and the octant of $\theta_{23}$~\cite{Denton:2022een}. 
{Fig.~\ref{fig:fig1_2012.12893} summarizes expected improvements in the precision of determination of oscillation parameters in future experiments in perspective of the next two decades. For each parameter, the dotted white line in Fig.~\ref{fig:fig1_2012.12893} shows the best-fit value and the shaded region around it the $1\sigma$ uncertainty.  See Ref.~\cite{Song:2020nfh} for a complete discussion.} 

Oscillation experiments do not put a limit on the Majorana phases $\alpha_1$, $\alpha_2$. 
However, predictions for the Majorana phases can be obtained using the neutrinoless double beta decay in conjunction with information on the neutrino masses~\cite{Bilenky:1980cx,Minakata:2014jba, Girardi:2016zwz}.  
Also, the oscillation experiments are only sensitivity to the squared mass differences, and not to the individual masses of neutrinos. Therefore, {\it the lightest neutrino mass $m_{0}$ is a free parameter and the other two masses are determined through} Eq.~(\ref{e:NO_and_IO}). However, there are limits on the absolute neutrino mass scale from other experiments, namely, from tritium beta decay~\cite{Formaggio:2021nfz}, neutrinoless double beta decay~\cite{Dolinski:2019nrj}, and precision measurements of the cosmic microwave background (CMB) and large-scale structure (LSS)~\cite{Archidiacono:2020dvx}. We discuss them now.

\begin{itemize}
\item[(i)] A direct and model-independent laboratory constraint on the neutrino mass can be derived from the kinematics of beta decay or electron capture~\cite{Formaggio:2021nfz}. These experiments measure an effective electron neutrino mass 
\begin{align}
    m^2_{\beta}
    = \frac{\sum_i m^2_i |U_{ei}|^2}{\sum_i |U_{ei}|^2}
    =\sum_i m^2_i |U_{ei}|^2 \, ,
\end{align}
{\it assuming $U$ is unitary}. 
This can be expressed through oscillation parameters as ~\cite{Esteban:2018azc}
\begin{equation}
  \label{eq:mbeta}
  \begin{aligned}
    m^2_{\beta}
    &=  
     c_{13}^2 c_{12}^2 m_1^2
    + c_{13}^2 s_{12}^2 m_2^2+s_{13}^2 m_3^2
    = \begin{cases}
      \text{NO: }
      & m^2_0 + \Dmq_{21} c_{13}^2 s_{12}^2+\Dmq_{3\ell} s_{13}^2 \,,
      \\
      \text{IO: }
      & m^2_0 - \Dmq_{21} c_{13}^2 c_{12}^2-\Dmq_{3\ell} c_{13}^2.
    \end{cases}
  \end{aligned}
\end{equation}
Here $\ell=1~(2)$ for NO (IO) in $\Delta m^2_{3\ell}$. 
The current oscillation data impose an ultimate lower bound of $m_\beta > 0.008~(0.047)$ eV for NO (IO). At present, the best direct limit on $m_\beta$ comes from the tritium beta decay experiment KATRIN: $m_{\beta} < 0.8$~eV at 90\%
CL~\cite{KATRIN:2021uub}, with projected sensitivity down to $m_{\beta} < 0.2$~eV at 90\%
CL~\cite{KATRIN:2021dfa}. The future Project~8 experiment using the Cyclotron Radiation Emission Spectroscopy (CRES) technique is expected to reach a sensitivity for $m_{\beta}$ down to 0.04 eV~\cite{Project8:2022wqh}. Some recent advances in the CRES technique were reported in Ref.~\cite{Project8:2022hun}.   Fig.~\ref{fig:memee} (bottom panel) summarizes present and future experimental bounds with corresponding projections to $m_0$ axis in NO scenario. As we can see, IO is completely within the future Project 8 sensitivity.

\begin{figure}[h!]
    \centering
    \includegraphics[width=0.65\textwidth]{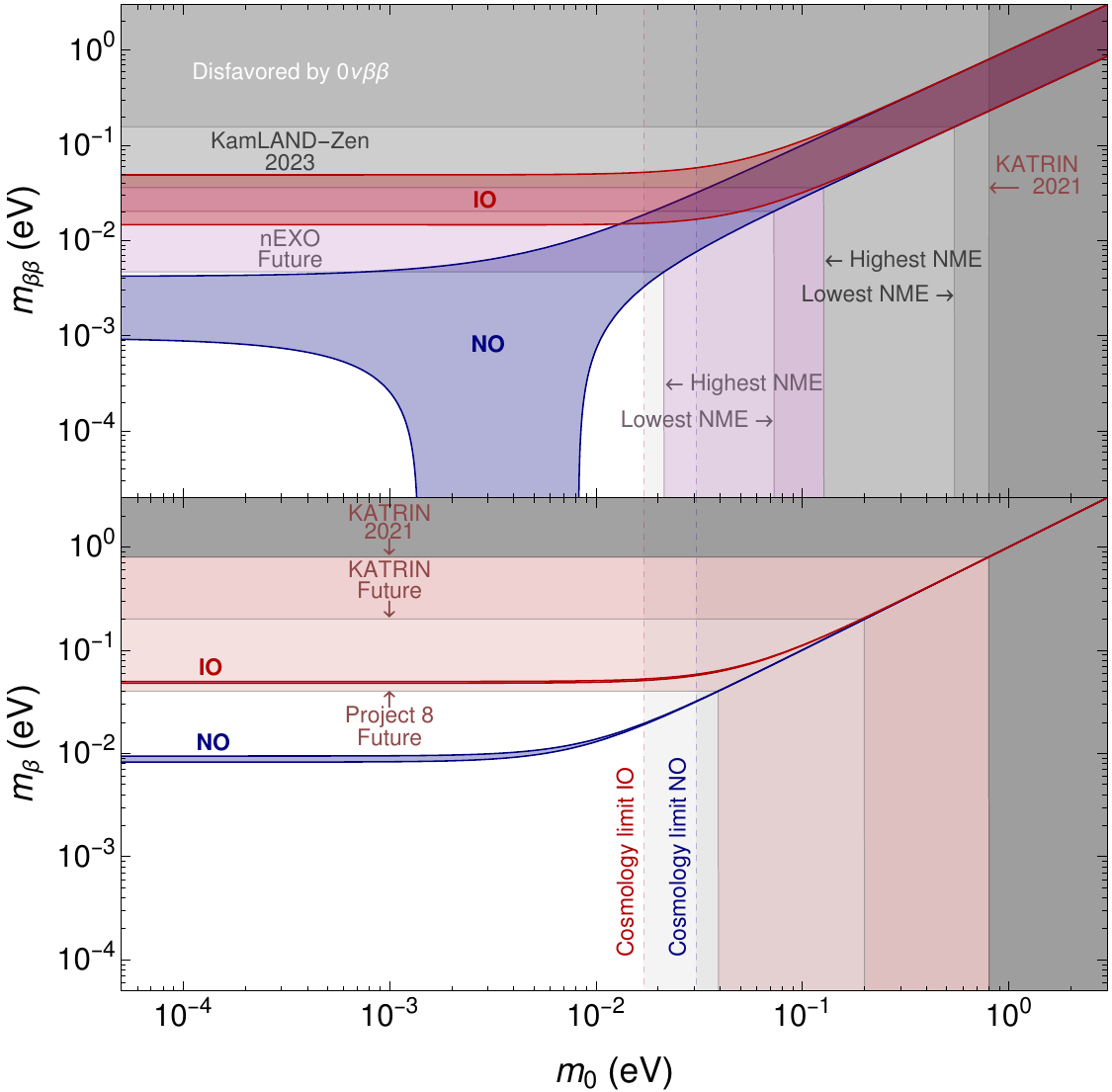}
     \caption{The effective electron neutrino mass $m_{\beta}$ (Eq.~(\ref{eq:mbeta})) and the effective Majorana neutrino mass $m_{\beta\beta}$ (Eq.~(\ref{2nu0})) plotted against the lightest neutrino mass $m_0$. The gray, reddish and light reddish shaded regions represent the KATRIN upper bound ($m_{\beta} < 0.8 \; \mbox{eV}$  at 90\%
CL)~\cite{KATRIN:2021uub}, KATRIN future bound ($m_{\beta} < 0.2 \; \mbox{eV}$ at 90\%
CL)~\cite{KATRIN:2021dfa} and Project 8 future bound ($m_{\beta} < 0.04 \; \mbox{eV}$)~\cite{Project8:2022wqh} respectively. The light gray and light magenta shaded regions represent the current upper limit range from KamLAND-Zen~\cite{KamLAND-Zen:2022tow} ($36-156$ meV at 90\% CL) and future sensitivity range from nEXO~\cite{nEXO:2021ujk} ($4.7-20.3$ meV at 90\% CL) respectively. All regions are presented with NO projections to $m_0$ axis.  The cosmology NO/IO upper limits for $m_0$ correspond to the Planck data~\cite{Planck:2018vyg} (see the discussion in item (iii)).
     }
    \label{fig:memee}
\end{figure}

\item[(ii)] If neutrinos are Majorana particles, neutrinoless double beta decay ($0\nu\beta\beta$) experiments ~\cite{Dolinski:2019nrj} can also provide direct information on neutrino masses via the effective Majorana mass\footnote{According to the Black Box Theorem \cite{Schechter:1980gr} if neutrinoless double beta decay is observed, neutrinos must be Majorana particles, irrespective of whether the light neutrino exchange is dominant or not. For simplicity, we assume the light neutrino exchange.} ~\cite{Esteban:2018azc}
\begin{multline}
  m_{\beta\beta} = \Big| \sum_i m_i U_{ei}^2 \Big|  
  = \Big| m_1 c_{13}^2 c_{12}^2 e^{i 2\alpha_1} +
  m_2 c_{13}^2 s_{12}^2 e^{i 2\alpha_2} +
  m_3 s_{13}^2 e^{-i 2\dCP} \Big|
  \\
  = \begin{cases}
    \text{NO:}
    & m_0\, \Big|c_{13}^2 c_{12}^2 e^{i 2(\alpha_1-\dCP)}
    + \sqrt{1 + \frac{\Dmq_{21}}{m_0^2}} \, c_{13}^2 s_{12}^2 e^{i 2(\alpha_2-\dCP)}
    + \sqrt{1 + \frac{\Dmq_{3\ell}}{m_0^2}}\, s_{13}^2 \Big|
    \\
    \text{IO:}
    & m_0\, \Big| \sqrt{1-\frac{\Dmq_{3\ell} +\Dmq_{21}}{m_0^2}} \,
    c_{13}^2 c_{12}^2 e^{i 2(\alpha_1 - \dCP)}
    + \sqrt{1-\frac{\Dmq_{3\ell}}{m_0^2}}\,
    c_{13}^2 s_{12}^2 e^{i 2(\alpha_2-\dCP)} + s_{13}^2 \Big|.
  \end{cases} \label{2nu0}
\end{multline}

The present best upper limit on $m_{\beta\beta}$ comes from the KamLAND-Zen experiment using $^{136}{\rm Xe}$: $m_{\beta\beta}<0.036 - 0.156$ eV at 90\% CL~\cite{KamLAND-Zen:2022tow}, where the range is due to the nuclear matrix element (NME) uncertainties. Several next-generation experiments are planned with different isotopes~\cite{Adams:2022jwx}, with ultimate discovery sensitivities to $m_{\beta\beta}$ down to 0.005 eV.  Based on Table~\ref{tab:neutrino_data}, 
 the update for $m_{\beta\beta}$ predictions from Eq.~\eqref{2nu0} as a function of the lightest neutrino mass is plotted in Fig.~\ref{fig:memee}. 
The light gray shaded region shows the current upper limit range for $m_{\beta\beta}$ ($36-156$ meV at 90\% CL) from KamLAND-Zen~\cite{KamLAND-Zen:2022tow} (comparable limits were obtained from GERDA~\cite{GERDA:2020xhi}), whereas the light magenta shaded region gives the future upper limit range for $m_{\beta\beta}$ ($4.7-20.3$ meV at 90\% CL) from nEXO~\cite{nEXO:2021ujk}, with the shaded area in each case arising from NME uncertainties and corresponding projections to the $m_0$ axis in the NO scenario. Comparable future sensitivities are discussed for other experiments, such as LEGEND-1000~\cite{LEGEND:2021bnm} and THEIA~\cite{Theia:2019non}, not shown in this plot. The dark gray shaded region is disfavored by KATRIN~\cite{KATRIN:2021uub}. 
  The vertical dashed lines are the cosmological upper limits (for NO and IO) on the sum of neutrino masses ($\sum m_i < 0.12 \; \mbox{eV}$ at 95\% CL) from Planck~\cite{Planck:2018vyg}; see the next item and Fig.~\ref{fig:mass}.

\item[(iii)] Massive neutrinos impact CMB and LSS. Thus, precision cosmological data restrict neutrino masses~\cite{Lesgourgues:2006nd}.   
Here we use the most stringent limit from Planck~\cite{Planck:2018vyg}: $\sum m_i < 0.12 \; \mbox{eV}$ at 95\% CL (Planck TT,TE,EE+lowE+lensing+BAO). A slightly stronger limit of $\sum m_i < 0.09 \; \mbox{eV}$ at 95\% CL has been obtained in Refs.~\cite{Palanque-Delabrouille:2019iyz,DiValentino:2021hoh}, while  Ref.~\cite{DiValentino:2021imh} has argued in favor of a weaker limit of $\sum m_i < 0.26 \; \mbox{eV}$. {In general, the cosmological limits on the neutrino mass highly depend on the cosmological datasets and priors used, and therefore, are less robust than direct laboratory measurements.} Sum of light neutrino masses  $\sum m_i$ is plotted in Fig.~\ref{fig:mass} against the lightest neutrino mass $m_0$ (left), effective electron neutrino mass $m_\beta$ (middle) and effective Majorana mass $m_{\beta\beta}$ (right) with the NuFIT $3\sigma$ oscillation parameters from Table~\ref{tab:neutrino_data} for both NO and IO scenarios. The horizontal gray-shaded region represents the current Planck upper limit~\cite{Planck:2018vyg}. Future cosmology sensitivity forecast is represented by the brown-green shaded area, with an uncertainty of the order of 15 meV \cite{Lee:2013bxd} (see also ~\cite{Chang:2022tzj}). The dashed lines represent the lowest allowed values of $\sum m_i=58$ meV (NO) and 97 meV (IO) by current oscillation data.

\begin{figure}[t!]
    \centering
    \includegraphics[width=1\textwidth]{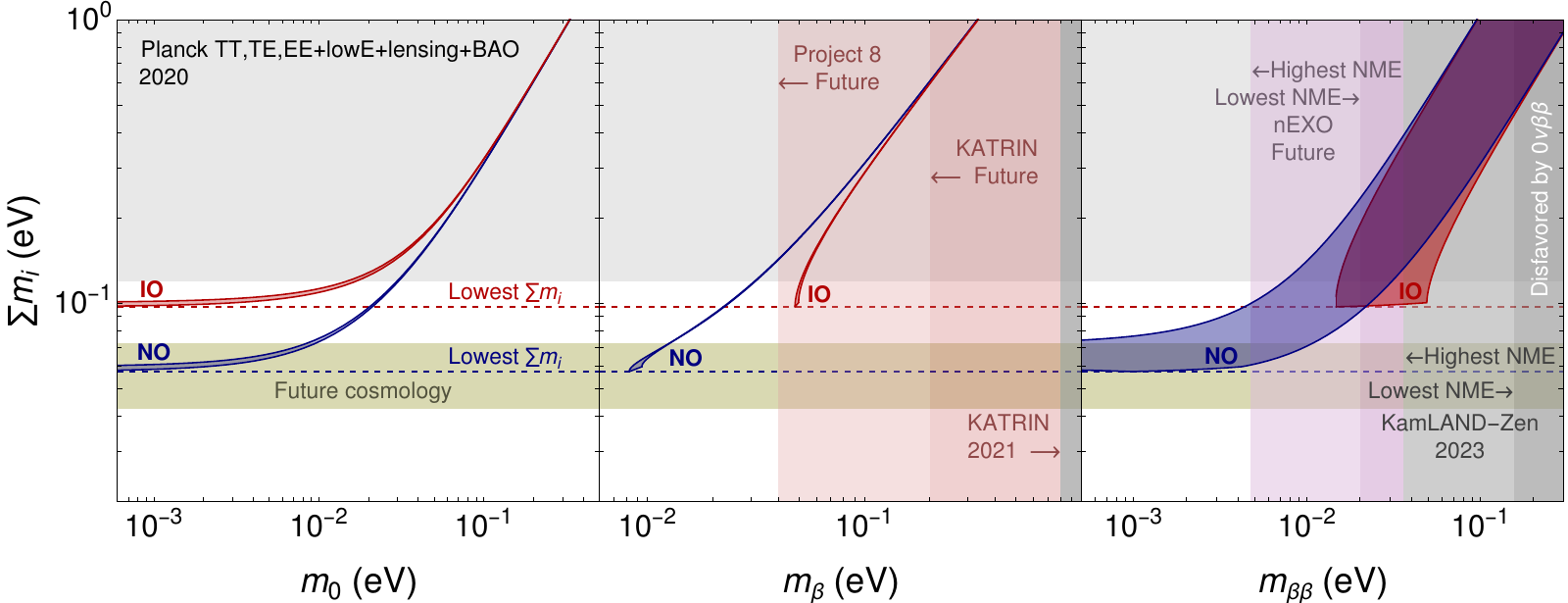}
    \caption{{Sum of light neutrino masses  $\sum m_i$ plotted against the lightest neutrino mass $m_0$ (left), effective electron neutrino mass $m_\beta$ (middle) and effective Majorana mass $m_{\beta\beta}$ (right) with the NuFIT $3\sigma$  oscillation parameters from Table~\ref{tab:neutrino_data}. The horizontal gray-shaded region represents the current Planck upper limit~\cite{Planck:2018vyg}. Future cosmology sensitivity forecast is represented by a brown-green shaded area, with an uncertainty of the order of 15 meV \cite{Lee:2013bxd} (see also Ref.~\cite{Chang:2022tzj}). The lowest allowed values of $\sum m_i$ for NO and IO from NuFIT data are also shown by the dashed lines. Other current and future exclusion regions shown here are described in Fig.~\ref{fig:memee}.
    The plot is an updated variation of the plot presented in Ref.~\cite{DiValentino:2021hoh}.}
    }
    \label{fig:mass}
\end{figure}
\end{itemize}

\begin{figure}[t!]
    \centering
    \includegraphics[width=0.65\textwidth]{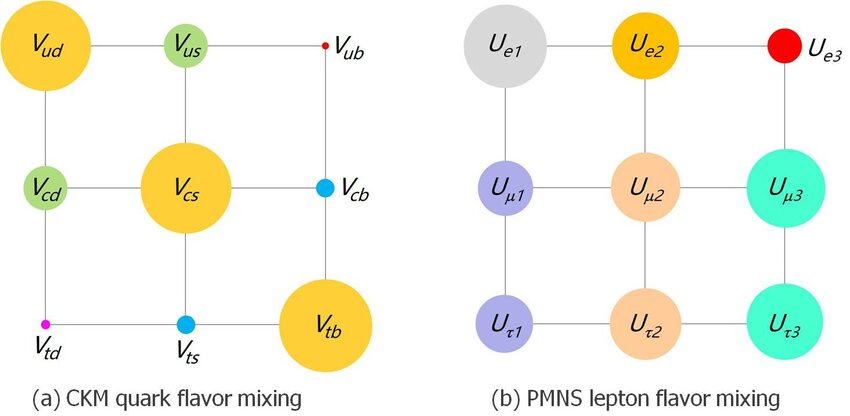}
    \caption{Flavor puzzle. Sizes of circles represent magnitudes of mixing among quarks (on the left) and neutrinos (on the right). Figure taken from the arXiv version of Ref.~\cite{Xing:2022uax}.}
    \label{fig:puzzle}
\end{figure}
  
Understanding the pattern of neutrino mixing is crucial because it is part of the long-standing flavor puzzle. As seen with the naked eye in Fig.~\ref{fig:puzzle}, the mixing patterns for quarks and neutrinos are intrinsically very different. The neutrino case is more "democratic" except for the $U_{e3}$ parameter, which is proportional to the small $s_{13}$ in Eq.~(\ref{upmns1}).  
In fact, before the non-zero $\theta_{13}$ discovery by Daya Bay~\cite{An:2012eh} and RENO~\cite{Ahn:2012nd}  in 2012, the reactor mixing angle $\theta_{13}$ was thought to be vanishingly small. Guided by the $\theta_{13} \sim 0^{\circ}$ assumption and to be consistent with the observed solar and atmospheric mixing angles, several flavor mixing schemes were postulated.  By substituting $\theta_{13}=0^{\circ}$ and $\theta_{23}=45^{\circ}$ in the general lepton mixing matrix given in Eq.~(\ref{upmns1}), up to the phase matrix $U_M$, most of the popular mixing schemes such as bi-maximal (BM)~\cite{Vissani:1997pa,Fukugita:1998vn, Barger:1998ta, Baltz:1998ey}, tribimaximal (TBM)~\cite{Harrison:2002er, Harrison:2002kp},  hexagonal (HG)~\cite{Albright:2010ap}, and golden ratio (GR)~\cite{Datta:2003qg, Kajiyama:2007gx, Everett:2008et, Rodejohann:2008ir, Adulpravitchai:2009bg, Ding:2011cm} mixing schemes can be altogether written as 
\begin{eqnarray}
				U_0=	\left(
					\begin{array}{ccc}
						c_{12} &s_{12} &0\\
						-\frac{s_{12}}{\sqrt{2}} &\frac{c_{12}}{\sqrt{2}} &-\frac{1}{\sqrt{2}}\\
						-\frac{s_{12}}{\sqrt{2}} &\frac{c_{12}}{\sqrt{2}} &\frac{1}{\sqrt{2}}
					\end{array}
					\right).\label{ugen0}
				\end{eqnarray}
Substituting $s_{12}=1/\sqrt{2},~1/\sqrt{3},~1/2$, and $\tan\theta_{12}=1/\varphi$ (with $\varphi=(1+\sqrt{5})/2$ being the golden ratio),  one can explicitly obtain the fixed mixing schemes BM, TBM, HG and GR\footnote{There exists an alternate version of GR mixing where $\cos\theta_{12}=\varphi/2$~\cite{Rodejohann:2008ir, Adulpravitchai:2009bg}.} respectively. 
In each case the Dirac CP phase $\delta_{\rm CP}$ is undefined as $\theta_{13}=0$ and can be easily extended to include the non-vanishing Majorana phases defined in Eq. (\ref{upmns1}) via the definition $U_{0}\rightarrow U_{0} U_M$.  
Using the diagonalization relation 
\begin{align}
    M_\nu=U_0^\star {\rm diag} (m_1,m_2,m_3)U_0^{\dagger}, 
\end{align}
such a mixing matrix  can  easily diagonalize a $\mu-\tau$ symmetric (transformations $\nu_{e} \rightarrow\nu_{e}$, $\nu_{\mu}\rightarrow\nu_{\tau}$, $\nu_{\tau}\rightarrow\nu_{\mu}$ under which the neutrino mass term remains unchanged) neutrino mass matrix of the form~\cite{Xing:2022uax}  \begin{eqnarray}
			M_\nu=	\left(
					\begin{array}{ccc}
						A &B &B\\
					B & C & D\\
						B &D  &C
					\end{array}\label{mmutau}
					\right), 
\end{eqnarray}
where the elements $A,B,C$ and $D$   are in general complex. With $A+B=C+D$ this matrix yields the TBM mixing pattern 
\begin{align}
U_{\rm TBM}=\left(
 					\begin{array}{ccc}
 						\sqrt{\frac{2}{3}} &\frac{1}{\sqrt{3}} &0\\
 						-\frac{1}{\sqrt{6}} &\frac{1}{\sqrt{3}} &-\frac{1}{\sqrt{2}}\\
 						-\frac{1}{\sqrt{6}} &\frac{1}{\sqrt{3}} &\frac{1}{\sqrt{2}}
 					\end{array}
 					\right) \, .
      \label{eq:TBM}
      \end{align}
Such first-order approximations of the neutrino oscillation data  motivated  theorists to find other symmetry-based aesthetic  frameworks which can  lead  towards these fixed mixing matrices. 

In this regard, non-Abelian discrete groups turned out to be popular as appropriate flavor symmetries for the lepton sector. Discrete groups have always played a key role in physics starting from crystallographic groups in solid-state physics, to discrete symmetries such as $C$, $P$, and $T$, which have shaped our understanding of nature. In neutrino physics, for a long time, various discrete groups such as $S_3$, $A_4$, $S_4$, $A_5$, $T'$, $\Delta(27)$, $D_n$, $T_{7}$, $\Delta(6n^2)$~\cite{Ishimori:2010au,Altarelli:2010gt,King:2013eh}  etc. have been extensively used  to explain fermion mixing. 
Among the various discrete groups used for this purpose, $A_4$ emerged as the most widely adopted choice 
initially proposed as an underlying family symmetry for quark sector~\cite{Wyler:1979fe, Branco:1980ax}.  
Interestingly,  in the last decade, thanks to the reactor neutrino experiments Double Chooz~\cite{Abe:2011fz}, Daya Bay~\cite{An:2012eh}, and  RENO~\cite{Ahn:2012nd} (also T2K~\cite{T2K:2013ppw}, MINOS~\cite{MINOS:2013utc}, and others~\cite{wwwubound}), the reactor neutrino mixing angle is conclusively measured to be `large' (see Table~\ref{tab:neutrino_data}).  In addition to this, as mentioned earlier, a non-zero value of the Dirac CP phase  $\delta_{\rm CP}$ is favored by the oscillation experiments. Such an observation has ruled out the possibility of simple neutrino mixing schemes like in Eq.~\eqref{ugen0}. Therefore, it is consequential to find modifications, corrections or the successors of the above mixing schemes which are still viable; this will be discussed in the next Chapter. Models based on non-Abelian discrete flavor symmetries often yield interesting predictions and correlations among the neutrino masses, mixing angles and CP phases.
~Involvement of such studies may have broader applications in various aspects of cosmology (matter-antimatter asymmetry of the Universe, DM, gravitational waves), collider physics and other aspects of particle physics, which will be discussed in subsequent Chapters. 

Moreover, the light neutrino sector and masses connected with three families of neutrinos and charged leptons can be a reflection of a more general theory where weakly interacting (sterile) neutrinos exist with much higher masses (GeV, TeV, or higher up to the GUT scale). A related problem is the symmetry of the {\it full} neutrino mass matrix, including the sterile sector. 
In this framework, the known neutrino mass and flavor states can be denoted by ${ \nuem i }$ and 
${ \nuf \alpha }$, respectively, where $i=1,2,3$ and $\alpha=e,\mu,\tau$. Any extra, beyond SM (BSM) sterile mass and flavor states (typically much heavier than the active ones) can be denoted by $\Nm j$ and $\Nf \beta$, respectively for $j,\beta=1,\ldots, n_R$.
In this general scenario mixing between an extended set of neutrino mass states $\{\nuem i,\Nm j\}$ with flavor states $\{\nuf \alpha,\Nf \beta\}$ is described by
\begin{align}
\begin{pmatrix} { \nuf \alpha} \\ 
 \Nf \beta\end{pmatrix}  &=
 \begin{pmatrix} {{ U }} & V_{lh} \\  V_{hl} & V_{hh}
  \end{pmatrix} 
\begin{pmatrix} {\nuem i } \\ 
\Nm j\end{pmatrix} 
\equiv {\cal U}  \begin{pmatrix} {{ \nuem i }} \\ \Nm j\end{pmatrix}\;.
\label{ugen1}
\end{align}
The SM flavor states $\nuf \alpha$ are then given by 
\begin{equation}
\nuf \alpha = 
\sum_{i=1}^3\underbrace{\left( U  \right)_{\alpha i} {\nuem i}}_{\rm SM \;part}
+ 
\sum_{j=1}^{n_R}\underbrace{
{ \left(V_{lh}  \right)}_{\alpha j} \Nm j}
_{\rm BSM \; part}\;. \label{gen}
\end{equation}  

The mixing matrix $\cal{U}$ in (\ref{ugen1})  diagonalizes a general neutrino mass matrix 
\begin{equation}\label{ssmm}
M_{\nu}=
\left(
\begin{array}{cc}
M_L & M_{D} \\
M_{D}^T & M_{R} 
\end{array}
\right) , 
\end{equation}
\noindent
using a congruence transformation
\begin{equation}
{\cal U}^{T}M_{\nu}{\cal U} \simeq {\rm diag}(m_i, M_j).
\label{congr}
\end{equation}
The structure and symmetry of the heavy neutrino sector $M_R$  in Eq.~(\ref{ssmm}), altogether with  $M_D$ variants, influence the masses and mixing of the light sector, beginning with the seesaw type of models ~\cite{Flieger:2020lbg}. 
{The extended {\it unitary} mixing matrix ${\cal U}$ in (\ref{ugen1}), with nonzero submatrices $V$, makes $U$ nonunitary. In fact, oscillation experiments do not exclude such cases, giving the following ranges of elements~\cite{Ellis:2020hus} (present analysis and future projections): 
\begin{align}
\left\lvert U^{\rm Current} \right\rvert^2_{3\sigma} &= \left( \begin{array}{c c c} \left[0.606, 0.742\right] & \left[0.265, 0.337\right] & \left[0.020, 0.024\right] \\ \left[0.051, 0.270\right] & \left[0.198, 0.484\right] & \left[0.392, 0.620\right] \\ \left[0.028, 0.469\right] & \left[0.098, 0.685\right] & \left[0.140, 0.929\right] \end{array}\right), \label{eq:Upres} \\
\left\lvert U^{\rm Future} \right\rvert^2_{3\sigma} &= \left( \begin{array}{c c c} \left[0.653, 0.699\right] & \left[0.291, 0.311\right] & \left[0.020, 0.024\right] \\ \left[0.074, 0.108\right] & \left[0.355, 0.454\right] & \left[0.447, 0.561\right] \\ \left[0.129, 0.359\right] & \left[0.212, 0.423\right] & \left[0.349, 0.595\right] \end{array}\right). \label{eq:Ufut}
\end{align}
As can be seen from these numbers, the current (and future) precision on the nonunitarity of the neutrino mixing matrix is still far away from the ultra-high precision achieved in the quark sector~\cite{UTfit:2022hsi}.   The interval matrices (\ref{eq:Upres}) and (\ref{eq:Ufut}) include nonunitary cases. This information can be used to derive bounds between known three neutrino flavors and additional neutrino states~\cite{Bielas:2017lok,Flieger:2019eor}. Neutrino mixing constructions based on discrete flavor symmetries discussed in the next chapters are based on unitary $3 \times 3$ mixing matrices and variants of non-unitary distortions.  
}

The number of model-building options available with discrete flavor symmetries is vast, and many have already been thoroughly reviewed in the literature. In Ref.~\cite{Ishimori:2010au}, the authors have presented a pedagogical review of various non-Abelian discrete groups, including their characters, conjugacy classes,  representation, and tensor products, which are essential for particle physics phenomenology. In Ref.~\cite{Altarelli:2010gt}, the authors discussed the application of non-Abelian finite groups to the theory of neutrino masses and mixing with $\theta_{13}\sim 0$ to reproduce fixed mixing schemes like TBM  and  BM, based on finite groups like $A_4, S_4$, etc. After measurement of the reactor mixing angle, in Ref.~\cite{King:2013eh}, the authors reviewed various discrete family symmetries and their (in)direct model-building approaches. They also discussed combining grand unified theories with discrete family symmetry to describe all quark and lepton masses and mixing. In Ref.~\cite{Petcov:2017ggy}, the author reviewed the scenarios for flavor symmetry combined with generalized CP symmetry to understand the observed pattern of neutrino mixing and the related predictions for neutrino mixing angles and leptonic Dirac CP violation. Finally, along with conventional flavor symmetric approaches, in Ref.~\cite{Feruglio:2019ybq}, the authors also reviewed the modular invariance approach to the lepton sector. Many other excellent reviews also partially cover issues discussed here~\cite{Frampton:1994rk,Xing:2020ijf,deAdelhartToorop:2011re, Tanimoto:2015nfa, Xing:2015fdg, King:2020ldi,SajjadAthar:2021prg}; see also the  Snowmass contributions~\cite{Gehrlein:2022nss, Altmannshofer:2022aml, Chauhan:2022gkz,  Almumin:2022rml}. Apart from the update to the mentioned reviews, our main focus in this review is a discussion on the phenomenology and testability of discrete flavor symmetries at the energy, intensity, and cosmic frontier experiments.

The rest of the review is organized as follows. In Chapter \ref{chapter:theory}, we present a general framework for understanding neutrino masses and mixing with non-Abelain discrete flavor symmetries,  discuss the compatibility of a few surviving mixing schemes with present neutrino oscillation data, and elaborate on explicit flavor models. We also mention various neutrino generation mechanisms and possible consequences once we augment them with discrete flavor symmetries. Then, we discuss the implications of combining flavor symmetries with CP, higher order discrete groups,  Grand Unified Theories, extended Higgs sector, and finally, allude to the recently revived modular invariance approach to address the flavor problem. In  Chapter \ref{sec:intensity},  we discuss the impact of discrete flavor symmetries in intensity frontiers such as neutrino oscillation experiments, neutrinoless double beta decay, lepton flavor and universality violation. Then in  Chapter \ref{sec:collider}, with some specific examples, we elaborate on the role of flavor symmetry at colliders, which includes a discussion on prospects of right-handed neutrino detection at colliders, lepton flavor violation and constraints on the $h \to \gamma \gamma $ decay width. In Chapter \ref{sec:cosmic}, we elaborate on the consequences of flavor symmetry at cosmic frontier, including studies on DM, leptogenesis and gravitational waves.  Finally, in Chapter \ref{sec:sum}, we summarize and conclude.   
\section{Flavor Symmetry and Lepton Masses and Mixing: Theory}\label{chapter:theory}

From Eq.~(\ref{upmns1}) we find that the neutrino mixing matrix is expressed in terms of mixing angles and CP violating phases, and we are yet to understand the experimentally observed mixing pattern~\cite{Esteban:2020cvm}. The masses and mixing of the leptons (as well as of the quarks) are obtained from the Yukawa couplings related to the families. Therefore, {\it{it is natural to ask whether any fundamental principle governs such mixing patterns.}}

\subsection{General Framework}\label{sec:genfw}

The primary approaches which try to address the issue of the neutrino mixing pattern include (i) random analysis without imposing prior theories or symmetries on the mass and mixing matrices \cite{Hall:1999sn,deGouvea:2012ac, Haba:2000be}; (ii) more specific studies with imposed mass or mixing textures for which models with underlying symmetries can be sought \cite{Xing:2002ta, Zhou:2004wz,Dziewit:2006cg, Ludl:2014axa,Fukugita:2016hzf}, and finally, (iii) theoretical studies where some explicit symmetries at the Yukawa Lagrangian level are assumed and corresponding extended particle sector is defined.
In the anarchy hypothesis (i),
the leptonic mixing matrix manifests as a random draw from an unbiased distribution of unitary  $3 \times 3$ matrices and does not point towards any principle or its origin. This hypothesis does not make any correlation between the neutrino masses and mixing parameters. However, it predicts probability distribution for the parameters which parameterize the mixing matrix.
{Though random matrices cannot solve
fundamental problems in neutrino physics, they generate
intriguing hints on the nature of neutrino mass matrices. For instance, in Ref.~\cite{Gluza:2011nm}, preference has been observed towards random models of neutrino masses with sterile neutrinos}.
In the intermediate approach (ii), some texture zeros of neutrino mass matrices can be eliminated.
For instance, in Ref.~\cite{Ludl:2014axa} 570 (298)
inequivalent classes of texture zeros in the Dirac (Majorana) case were found.  
For both cases, about 75\% of the classes are compatible with the data. In the case of maximal texture zeros in the neutrino and charged lepton mass matrices, there are only
about 30 classes of texture zeros for each of the four categories defined by Dirac/Majorana
nature and normal/inverted ordering of the neutrino mass spectrum.  
Strict texture neutrino mass matrices can also be discussed in phenomenological studies. For more, see section \ref{sec:massmodels}.

In what follows, we will discuss the symmetry-based approach (iii) to explain the non-trivial mixing in the lepton sector known as family symmetry or horizontal symmetry~\cite{Berezhiani:1990jj, Grimus:2011fk}. Such fundamental symmetry in the lepton sector can easily explain the origin {or the guiding principle} of neutrino mixing, which is considerably different from quark mixing. {Hence the observed pattern of the lepton mixing, which is part of  SM flavor puzzle, may have a deeper theoretical foundation.}  
Incidentally, both  Abelian and non-Abelian family symmetries have the potential to shed light on the Yukawa couplings. The Abelian symmetries (such as Froggatt-Nielsen symmetry \cite{Froggatt:1978nt}) only point towards a hierarchical structure of the Yukawa couplings, whereas non-Abelian symmetries are more equipped to explain the non-hierarchical structures of the observed lepton mixing as observed by the oscillation experiments.  
 
 If we consider a family symmetry $G_f$, the three generations of leptons and quarks can be assigned to irreducible representations or multiplets, hence unifying the flavor of the generations. If $G_f$ contains a triplet representation ($3$), all three fermion families can follow {the} same transformation properties. For example, let us consider that non-zero neutrino mass is generated through the Weinberg operator $H L^T c LH $ where the lepton and Higgs doublets transform as a triplet ($\mathbf{\bar{3}}$) and singlet under a family symmetry, say,   $SU(3)$.   To construct $SU(3)$ invariant operator, an additional scalar field $\Phi$  (also known as {\it{flavon}}~\cite{Froggatt:1978nt}) is introduced, and the effective operator takes the form $H L^T \Phi^T \Phi LH $. {A} suitable vacuum alignment ($\langle \Phi \rangle \propto (u_1, u_2, u_3)^T$) for the flavon is inserted in such a way that the obtained mass matrix is capable of appropriate mixing pattern.
As a result, $G_f$ is spontaneously broken once flavons acquire non-zero vacuum expectation values (VEV). Continuous family symmetry such as $U(3)$, $O(3)$ (and their subgroups $SU(3)$ and $SO(3)$) can in principle be used for this purpose
to understand the neutrino mixing.
However, the non-Abelian discrete flavor symmetric approach is much more convenient as in such a framework, obtaining the desired vacuum alignment (which produces correct mixing) of the flavon can be obtained easily~\cite{Altarelli:2005yp, Altarelli:2005yx}.
At this point, it is worth mentioning that these non-Abelian discrete symmetries can also originate from a continuous symmetry~\cite{Koide:2007sr,Merle:2011vy, Adulpravitchai:2009kd, Luhn:2011ip, Wu:2012ria, King:2018fke, Ludl:2010bj, Joshipura:2016quv, Joshipura:2015dsa}. For example widely used discrete groups such as $A_4, S_4, A_5, \Delta(27), T_7$  can originate from the  {continuous} group $SU(3)$ \cite{King:2013eh}. In another example~\cite{King:2018fke}, the authors showed that continuous $SO(3)$  can also give rise to $A_4$, further broken into smaller $Z_3$ and $Z_2$ symmetries. A few years back, it was proposed that various non-Abelian discrete symmetries can also originate from superstring theory through compactification of extra dimensions and known as {the} modular invariance approach~\cite{Altarelli:2005yx, deAdelhartToorop:2011re, Feruglio:2017spp, deMedeirosVarzielas:2023crv}.

In this report, we concentrate on all these aspects of discrete family symmetries discussed above and their implications for understanding lepton mixing and its extensions. The model building with flavor symmetries is not  trivial  since the underlying flavor symmetry group $G_f$ must be broken. {Due to the imposition of the flavor symmetry   $G_f$, fields contained in the theory must transform under irreducible representations of the same $G_f$. For instance, let us consider  $\psi(x)$ to be a matter field in a general irreducible representation $r$ of $G_f$. If $\rho_r(g)$ is the unitary representation matrix for the element $g$ in the irreducible representation $r$,  under the  action of $G_f$,  $\psi(x)$ transforms as 
\begin{eqnarray}
\psi(x)\xrightarrow[]{G_f}\rho_r(g)\psi(x),\quad g\in G_f. 
\end{eqnarray}
Suppose the left-handed lepton fields are assumed to transform under the 3-dimensional irreducible representation of $G_f$. In that case, there is an opportunity to unify the three generations of lepton families at some high scale.
 Therefore, to distinguish the different flavors of the lepton families, at low energies, considered flavor symmetry $G_f$ has to be broken. Often, $G_f$ is considered to be spontaneously broken after flavon fields ($\Phi$, mentioned earlier) acquire VEV in a suitable direction, discussed in Section~\ref{sec:flavon-VEV}. These flavon fields transform trivially under the SM gauge group but transform non-trivially in specific irreducible
representations of $G_f$.} 
 Usually, this symmetry $G_f$ is considered to exist at some large scale (sometimes with proximity to GUT scale~\cite{Altarelli:2010gt}) and to be broken at lower energies with residual symmetries of the charged lepton and neutrino sectors, represented by the subgroups $G_e$ and $G_\nu$, respectively. Therefore, to obtain definite predictions and correlations of the mixing, the choice of the non-Abelian discrete group $G_f$ and its breaking pattern to yield remnant subgroups $G_e$ and $G_\nu$ shapes the model building significantly. 
{In the basis where the charged lepton mass matrix is not diagonal and $G_e \in G_f$ is the residual symmetry in the charged lepton sector, it demands invariance of $M_\ell^{\dagger}M_{\ell}$, where $M_\ell$ is the charged lepton mass matrix. This implies
\begin{eqnarray}
\rho_r(g_e)^{\dagger}M_\ell^{\dagger}M_{\ell}\rho_r(g_e)=M_\ell^{\dagger}M_{\ell} \, ,
\end{eqnarray}
where $G_e$ transforms the  lepton doublets  ($\tilde{\ell_{\alpha}}=\tilde{\ell_{e}},\tilde{\ell_{\mu}},\tilde{\ell_{\tau}}$) as 
\begin{eqnarray}
  \tilde{\ell_{\alpha}}\xrightarrow[]{G_e}(\rho_r(g_e))_{\ell\ell'}  \tilde{\ell'_{\alpha}} \, . 
\end{eqnarray}
$\rho_r(g_e)$ gives the action of $G_e$ on $\tilde{\ell_{\alpha}}$ and $g_e$ is an element of $G_e$. Similarly, if $G_{\nu}$ is the residual symmetry in the Majorana neutrino mass matrix $M_{\nu}$, one can write \begin{eqnarray}
\rho_r(g_{\nu})^{T}M_{{\nu}}\rho_r(g_{\nu})=M_{{\nu}} \, , 
\end{eqnarray}
where $g_{\nu}$ is the element of $G_{\nu}$ and the action of  $G_{\nu}$ on the left-handed neutrino fields can be written as 
\begin{eqnarray}
\nu_{\tilde{\ell} L} \xrightarrow[]{G_\nu}(\rho_r(g_{\nu}))_{\ell\ell'} \nu_{\tilde{\ell'} L}. 
\end{eqnarray}
Hence, from the above equation one can write 
\begin{eqnarray}
\rho_r(g_{\nu})^{\dagger}M_{\nu}^{\dagger}M_{{\nu}}\rho_r(g_{\nu})=M_{\nu}^{\dagger}M_{{\nu}}.
\end{eqnarray}
From the transformation of the charged lepton and Majorana neutrino mass matrices, it implies that $M_\ell^{\dagger}M_{\ell}$ and  $M_{\nu}^{\dagger}M_{{\nu}}$ commutes with $\rho_r(g_e)$ and $\rho_r(g_{\nu})$, respectively. It is therefore evident that  $M_\ell^{\dagger}M_{\ell}$ and $\rho_r(g_e)$  can be diagonalized by the same matrix, say, $U_{e}$. On the other hand,  $M_{\nu}^{\dagger}M_{{\nu}}$ and $\rho_r(g_{\nu})$ can also be diagonalized by the same matrix, say, $U_{\nu}$. Hence one can write the diagonalization relation as 
\begin{eqnarray}
    U^{\dagger}_e\rho_{r}(g_e)U_e=\rho^{(\rm diag)}_{r}(g_e),\\
     U^{\dagger}_{\nu}\rho_{r}(g_{\nu})U_{\nu}=\rho^{(\rm diag)}_{r}(g_{\nu}).
\end{eqnarray}
Thus with known $G_f$ and $\rho_r(g_f)$ one can easily determine/constrain both \{$G_e$, $\rho_r(g_e)$\} and \{$G_{\nu}$,  $\rho_r(g_{\nu})$\} easily. In the basis where the charged lepton mass matrix is not diagonal, the PMNS matrix given Eq.~(\ref{upmns1})  receives contributions from both $U_e$ and $U_{\nu}$, $i.e.$, $U=U_e^{\dagger}U_{\nu}$. Therefore with consistent choice of  $G_f, \rho_r(g_f), G_e$ and $G_{\nu}$ one can determine $U$. For this reason models with discrete flavor symmetries are very predictive in nature. 
}
Without any residual symmetry, the flavor $G_f$ loses its predictivity markedly. For a detailed discussion on the choice of various discrete symmetries and their generic predictions, see~Refs.~\cite{King:2013eh, Petcov:2017ggy, Tanimoto:2015nfa}. 
\begin{table}[]
    \centering
    \begin{tabular}{|c |c |c |c| c|}
			\hline
			 Group & Order  & Irreducible Representations  & Generators \\
			\hline
			$A_4$ & 12  & $1, 1', 1'', 3$  &  $S,T$ \\ 
			$S_4$ &  24 & $1,1',2,3,3'$  &   $S,T(U)$ \\
			$T'$ & 24 &  $1,1',1'',2,2',2'',3$      & $S,T(R)$  \\
			$\Delta(27)$ & 27 & $1_{r,s} (r,s=0,1,2), 3_{01, 02}$ & $C,D$ \\
			$A_5$ & 60 & $1,3,3',4,5$& $\Tilde{S}$, $\Tilde{T}$\\
			\hline
    \end{tabular}
    \caption{Basic characteristics of {a} few small groups with triplet irreducible representations. For details, see Ref.~\cite{Ishimori:2010au}. For instance, a possible representation for generators of the $S_4$ group is defined in the text, see Eqs.~(\ref{STUgen}) and (\ref{STU}).}
    \label{tab:sg}
\end{table}
    In Table~\ref{tab:sg}, we mention the basic details such as order or number of elements (first and second columns), irreducible representations (third column) and generators (fourth column) of small groups (which contain at least one triplet) such as $A_4,  S_4, T'$  $\Delta(27)$ and $ A_5$. A pedagogical review, including catalogues of the generators and multiplication rules of these widely used non-Abelian discrete groups, can be found in~Ref.~\cite{Ishimori:2010au}.

Now, for model building {purposes}, there {exist} various approaches based on the breaking pattern of $G_f$ into its residual symmetries, also known as  direct, semi-direct and indirect approaches ~\cite{King:2013eh, Tanimoto:2015nfa}. After breaking of $G_f$,  different residual symmetries {exist} for charged lepton (typically $G_{e}=Z_3$) and neutrino sector (typically  $G_{\nu}=Z_2 \times Z_2$, also known as the Kline symmetry). It is known as {the} direct approach. In {a} semi-direct approach, one of the generators of the residual symmetry is assumed to be broken. {On the} contrary, in the indirect approach, no residual symmetry of flavor groups remains intact, and the flavons acquire special vacuum alignments whose alignment is guided by the flavor symmetry. Usually, different flavons take part in the charged lepton and neutrino {sectors}. To show how the family symmetry {shapes} the flavor model building, let us consider $G_f=S_4$ as {a} guiding symmetry. Geometrically, this group can be seen as the symmetry group of a rigid cube, a group of permutation four objects. Therefore, the order of the group is $4!=24$ and  the  elements can be conveniently generated  by the generators $S,T$ and $U$ satisfying the relation 
\begin{equation}
    S^2=T^3=U^2=1~~{\rm and}~~ST^3=(SU)^2=(TU)^2=1. \label{STUgen}
\end{equation}
In their irreducible triplet representations, these three generators  can be written as~\cite{Ishimori:2010au,Altarelli:2010gt} 
\begin{eqnarray}
				S=	\frac{1}{3}\left(
					\begin{array}{ccc}
						-1 &2 &2\\
						2 &-1 &2\\
						2 &2 &-1
					\end{array}
					\right);   \quad 
					T=	\left(
					\begin{array}{ccc}
						1 &0 &0\\
						0 &\omega^2 &0\\
						0 &0 &\omega
					\end{array}
					\right) ~~{\rm and }~~
					U=	\mp \left(
					\begin{array}{ccc}
						1 &0 &0\\
						0 & 0 &1\\
						0 &1 & 0
					\end{array}
					\right). \label{STU}
				\end{eqnarray}
where $\omega=e^{2i\pi/3}$. These generators can also be expressed as 3-dimensional irreducible (faithful) real representation matrices
\begin{eqnarray}
				S= \left(
					\begin{array}{ccc}
						-1 &0 &0\\
						0 & 1 &0\\
						0 &0 & -1
					\end{array}
					\right);   \quad 
					T=	\frac{1}{2}\left(
					\begin{array}{ccc}
						1 &\sqrt{2} &1\\
						\sqrt{2} &0 &-\sqrt{2}\\
						-1 &\sqrt{2} &-1
					\end{array}
					\right) ~~{\rm and }~~
					U=	\mp \left(
					\begin{array}{ccc}
						1 &0 &0\\
						0 & 1 &0\\
						0 &0 & -1
					\end{array}
					\right). \label{eq:STUreal}
				\end{eqnarray} 
In the direct approach the charged lepton mass matrix ($M_\ell$) respects the generator $T$ whereas the neutrino mass matrix ($M_{\nu}$)  respects the generators $S,U$ satisfying the conditions  
\begin{equation}
    T^{\dagger}M_{\ell}^{\dagger}M_{\ell}T=M_{\ell}^{\dagger}M_{\ell},~ S^T M_{\nu} S = M_{\nu}~{\rm and }~U^T M_{\nu} U = M_{\nu}, \label{eq:s4tbm1}
\end{equation}
which leads to~\cite{Tanimoto:2015nfa}
\begin{equation}
    [T, M_{\ell}^{\dagger}M_{\ell}]=[S,M_{\nu}]=[U, M_{\nu}]=0\label{eq:s4tbm2}. 
\end{equation}
The non-diagonal matrices $S, U$ can be  diagonalized by the TBM mixing matrix given in Eq. (\ref{eq:TBM}). Therefore, the TBM mixing scheme can be elegantly derived from the direct approach of the $S_4$ group. For generic features of semi-direct and indirect approaches to the  flavor model building, we refer the readers to~\cite{King:2013eh, King:2020ldi, Tanimoto:2015nfa}.   The TBM mixing pattern explained here can be generated using various discrete groups. For detailed models and groups see $A_4$~\cite{Ma:2001dn,Ma:2004zv, Altarelli:2005yp,Altarelli:2005yx}, $S_4$~\cite{Bazzocchi:2012st, Bazzocchi:2008ej}, $\Delta(27)$~\cite{deMedeirosVarzielas:2006fc}, $T'$~\cite{Frampton:2008bz}. In addition,  explicit  models with discrete flavor symmetry for BM~\cite{Girardi:2015rwa,Lam:2008rs,Frampton:2004ud, Merlo:2011vc}, GR~\cite{Everett:2008et, Feruglio:2011qq, Ding:2011cm}, HG~\cite{Kim:2010zub} mixing can easily be constructed.

\subsection{Flavor Symmetry, Nonzero $\theta_{13}$ and Nonzero $\delta_{\rm CP}$ \label{sec:th13CP}}

 After precise measurement of the non-zero value of the reactor mixing angle $\theta_{13}$~\cite{Abe:2011fz, An:2012eh, Ahn:2012nd, MINOS:2013utc, T2K:2013ppw} the era of fixed patterns (such as BM, TBM, GR, HG mixing) of the lepton mixing matrix is over. Also, as mentioned earlier,  long baseline neutrino oscillation experiments such as T2K~\cite{T2K:2019bcf} and NO$\nu$A \cite{NOvA:2019cyt} both hint at CP violation in the lepton sector. Therefore, each of the fixed patterns needs some modification to be consistent with the global fit of the neutrino oscillation data \cite{NuFIT5.2,Esteban:2020cvm,deSalas:2020pgw,10.5281/zenodo.4726908,Capozzi:2021fjo}. There are two distinct ways of generating a mixing pattern that appropriately deviates from fixed mixing schemes such as BM, TBM, GR, and HG. The first approach is based on symmetry assertion, which demands considering larger symmetry groups that contain a larger residual symmetry group compared to the fixed mixing schemes such us TBM~\cite{deAdelhartToorop:2011re,deAdelhartToorop:2011nfg,Hernandez:2012sk,Ding:2012xx,Ge:2011ih,Hernandez:2012ra}. On the other hand, in the second approach, the setups for the BM, TBM, GR, and HG mixing schemes are supplemented by an additional ingredient which breaks these structures in a well-defined and controlled way~\cite{Luhn:2013lkn}. This can be achieved in various ways. An apparent source for such corrections can be introduced through the charged lepton sector~\cite{Antusch:2005kw, Altarelli:2009gn, Roy:2012ib, Roy:2015cka, Dev:2011bd}. Thus, in models where in the neutrino sector the  mass matrix solely reproduces the mixing scheme, a  non-diagonal charged lepton sector will contribute to the PMNS matrix $U=U_{\ell}^{\dagger}U_{\nu}$ where $U_{\ell}$ and $U_{\nu}$ respectively are the diagonalizing matrices of the charged lepton and neutrino mass matries.  In addition, one can also consider small perturbations around the BM/TBM/GR/HG vacuum-alignment conditions~\cite{Branco:2012vs,Feruglio:2007uu,King:2010bk,Barry:2010zk},  which can originate from higher dimensional operators in the flavon potential yielding desired deviation. The minimal flavon field content can also be extended to incorporate additional contributions to the neutrino mass matrix to achieve correct deviation from fixed mixing schemes~\cite{Borah:2013jia, Karmakar:2014dva,Karmakar:2015jza,Karmakar:2016cvb,Bhattacharya:2016lts,Bhattacharya:2016rqj}.  
{To summarize, the fixed mixing schemes can still be regarded as a first approximation, necessitating specific corrections to include non-zero $\theta_{13}$ and
 $\delta_{\rm CP}$.}
 For example,  even if the TBM mixing is obsolete, two successors are still compatible with data. These are called TM$_1$ and TM$_2$ mixing and {are} given by
\begin{eqnarray}
	|U_{{\rm TM}_1}|=	\left(
					\begin{array}{ccc}
						\frac{2}{\sqrt{6}} &* &*\\
						\frac{1}{\sqrt{6}} &* &*\\
						\frac{1}{\sqrt{6}} &* &*
					\end{array}
					\right)~{\rm and}~|U_{{\rm TM}_2}|=	\left(
					\begin{array}{ccc}
						* &\frac{1}{\sqrt{3}} &*\\
						* &\frac{1}{\sqrt{3}} &*\\
						* &\frac{1}{\sqrt{3}} &*
					\end{array}
					\right), \label{utm}
				\end{eqnarray}
respectively. Clearly,  Eq. (\ref{utm}) shows that TM$_1$ and TM$_2$ mixings preserve the first and the second column of the TBM mixing matrix given in Eq. (\ref{eq:TBM}). Here, the reactor mixing angle becomes a free parameter, and the solar mixing angle can stick close to its TBM prediction.

To illustrate this, let us again consider the discrete flavor symmetry $G_f=S_4$. In contrast to the breaking pattern mentioned in Eqs.~(\ref{eq:s4tbm1}), (\ref{eq:s4tbm2}),  $S_4$ is considered to be broken spontaneously into $Z_3=\{1, T, T^2\}$ (for the charged lepton sector) {and} $Z_{2}=\{ 1, SU\}$ (for the neutrino sector) such that it satisfies  
\begin{equation}
    [T, M_{\ell}^{\dagger}M_{\ell}]=[SU,M_{\nu}]=0\label{eq:s4tbm3}. 
\end{equation}
Following the above prescription, the matrix that diagonalizes $SU$ (see Eq.~(\ref{STU})) can be written as $U_{\rm TBM}U_{23}(\theta,\gamma)$ where the `23' rotation matrix is given by ($c_{\theta}=\cos\theta$, $s_{\theta}=\sin\theta$ and $\gamma$ is  the associated  phase factor)
\begin{eqnarray}
				U_{{23}}=	\left(
					\begin{array}{ccc}
						1 &0 & 0\\
						0 &c_{\theta} &s_{\theta}e^{-i\gamma}\\
						0 &-s_{\theta}e^{i\gamma}&c_{\theta}
					\end{array}
					\right). \label{u23}
\end{eqnarray}
 The obtained effective mixing matrix is called $U_{{\rm TM}_1}$ and can be written as 
\begin{eqnarray}
				U_{{\rm TM}_1}=	\left(
					\begin{array}{ccc}
						\frac{2}{\sqrt{6}} &\frac{c_{\theta}}{\sqrt{3}} &\frac{s_{\theta}}{\sqrt{3}}e^{-i\gamma}\\
						-\frac{1}{\sqrt{6}} &\frac{c_{\theta}}{\sqrt{3}}-\frac{s_{\theta}}{\sqrt{2}}e^{i\gamma} &-\frac{s_{\theta}}{\sqrt{3}}e^{-i\gamma}-\frac{c_{\theta}}{\sqrt{2}}\\
						-\frac{1}{\sqrt{6}} &\frac{c_{\theta}}{\sqrt{3}}-\frac{s_{\theta}}{\sqrt{2}}e^{i\gamma} &-\frac{s_{\theta}}{\sqrt{3}}e^{-i\gamma}+\frac{c_{\theta}}{\sqrt{2}}
					\end{array}
					\right),\label{ums4tm1}
				\end{eqnarray}
 The above matrix has the TM$_1$ mixing structure  mentioned in Eq. (\ref{utm}). This is also an example of the method of a semi-direct approach to the flavor model building. Similarly, the generic structure for the structure for TM$_2$ mixing matrix can be written as \begin{eqnarray}
				U_{{\rm TM}_2}=	\left(
					\begin{array}{ccc}
						\frac{2c_{\theta}}{\sqrt{6}} &\frac{1}{\sqrt{3}} &\frac{2s_{\theta}}{\sqrt{6}}e^{-i\gamma}\\
			-\frac{c_{\theta}}{\sqrt{6}}+\frac{s_{\theta}}{\sqrt{2}}e^{i\gamma} &\frac{1}{\sqrt{3}}  &-\frac{s_{\theta}}{\sqrt{3}}e^{-i\gamma}-\frac{c_{\theta}}{\sqrt{2}}\\
				-\frac{c_{\theta}}{\sqrt{6}}+\frac{s_{\theta}}{\sqrt{2}}e^{i\gamma} &\frac{1}{\sqrt{3}} &-\frac{s_{\theta}}{\sqrt{3}}e^{-i\gamma}+\frac{c_{\theta}}{\sqrt{2}}
					\end{array}
					\right)\label{ums4tm2}. 
				\end{eqnarray}

\noindent
The above discussion shows special cases of the TBM mixing, which can still be relevant for models with discrete flavor symmetries. Now, imposing sufficient corrections to the other fixed mixing schemes like BM, GR, and HG we can make them consistent with observed data~\cite{Feruglio:2011qq,CarcamoHernandez:2022bka,Cooper:2012bd,deMedeirosVarzielas:2013tbq, Gehrlein:2014wda,Wilina:2022pzy,Duarah:2012bd}. 
The modified mixing matrix can be obtained by lowering the residual symmetry $G_{\nu}$ for the neutrino sector. This generates a correction matrix for these fixed mixing patterns. The general form of these corrections can be summarized as~\cite{Frampton:2004ud,Petcov:2017ggy,Hernandez:2013vya} \begin{equation}
    U= U_{e}^{\dagger} U_{0} U_{p},
\end{equation}
where $U_0$ is the general form of the relevant fixed pattern mixing scheme mentioned in Eq. (\ref{ugen0}), $U_{e}$ is the generic correction matrix  and  $U_{p}$ is additional phase matrix contributing in the Dirac and Majorana phases mentioned in Eq.~(\ref{upmns1}). {\emph{These corrections help us to obtain interesting correlations among $\sin\theta_{12}, \sin\theta_{23}, \sin\theta_{13}$ and $\delta_{\rm CP}$  of PMNS mixing matrix}~\cite{Petcov:2014laa}. 
\begin{table}[h!]
    \centering
    \begin{tabular}{|c|c|c|}
    \hline
          & TM$_1$ & TM$_2$ \\
    \hline      
        $|U_{e2}|$ & $\left|\frac{\cos\theta}{\sqrt{3}}\right|$ & $\frac{1}{\sqrt{3}}$\\
    \hline     
    $|U_{e3}|$ & $\left|\frac{\sin\theta}{\sqrt{3}} \right|$ & $\left|\frac{2\sin\theta}{\sqrt{6}}\right|$\\
    \hline    
    $|U_{\mu 3}|$ & $\left|\frac{\cos\theta}{\sqrt{2}} + \frac{\sin\theta}{\sqrt{3}} e^{-i\gamma}\right|$ &  $\left|-\frac{\cos\theta}{\sqrt{2}} - \frac{\sin\theta}{\sqrt{6}} e^{-i\gamma}\right|$\\
    \hline     
    $\sin^2\theta_{12}$ & $1-\frac{2}{3-\sin^2\theta}$ & $\frac{1}{3-2\sin^2\theta}$\\
    \hline     
    $\sin^2\theta_{13}$ & $\frac{1}{3}\sin^2\theta$ & $\frac{2}{3}\sin^2\theta$\\
    \hline   
    $\sin^2\theta_{12}$ & $\frac{1}{2} \left( 1- \frac{\sqrt{6}\sin 2 \theta \cos\gamma}{3-\sin^2\theta} \right)$ & $\frac{1}{2} \left( 1+ \frac{\sqrt{3}\sin 2 \theta \cos\gamma}{3-\sin^2\theta} \right)$\\
    \hline    
    $J_{CP}$ & $-\frac{1}{6\sqrt{6}}\sin2\theta \sin\gamma$
 &  $-\frac{1}{6\sqrt{3}}\sin2\theta \sin\gamma$
 \\
    \hline 
$\sin\delta_{\rm CP}$ & $-\frac{(5+\cos2\theta)\sin\gamma}{\sqrt{(5+\cos2\theta)^2-24 \sin^22\theta\cos^2\gamma}}$ & $-\frac{(2+\cos2\theta)\sin\gamma}{\sqrt{(2+\cos2\theta)^2-3 \sin^22\theta\cos^2\gamma}}$\\
\hline
    \end{tabular}
    \caption{Mixing parameters in the TM$_1$ and TM$_2$ scenarios. {For more details, see ~Ref.~\cite{Petcov:2014laa}. The elements $U_{e2}, U_{e3}, U_{\mu 3}$ are defined in Fig.~\ref{fig:puzzle}.}
    \label{tab:tm12prediction}}
\end{table}

 In Table~\ref{tab:tm12prediction}, we mention the typical predictions for TM$_1$ and TM$_2$ mixing matrices, including the Jarlskog invariant   $J_{CP}=c_{12} s_{12} c_{23} s_{23} c_{13}^2 s_{13} \sin\delta_{\rm CP}$ \cite{Jarlskog:1985ht}.

 The correlations among the neutrino mixing 
angles ($\theta_{23}, \theta_{12}, \theta_{13}$) and phase ($\delta_{\rm CP}$) for TM$_1$ and TM$_2$ respectively  can be written as~\cite{deMedeirosVarzielas:2012apl}
 \begin{eqnarray}
     \label{eq:tm1formulas}
     \mbox{TM}_1 &:&  s_{12}^2 = \frac{1-3 s_{13}^2}{3 - 3 s_{13}^2}, 
    ~~~\cos \delta_{\rm CP} = \frac{
    (1-5s_{13}^2)(2s_{23}^2-1)
    }{
    4 s_{13}s_{23}\sqrt{2(1-3s_{13}^2)(1-s_{23}^2)}
    },
    \\
    \label{eq:tm2formulas}
     \mbox{TM}_2 &:& s_{12}^2 = \frac{1}{3 - 3 s_{13}^2}, 
    ~~~\cos \delta_{\rm CP} = -\frac{
    (2-4s_{13}^2)(2s_{23}^2-1)
    }{
    4 s_{13}s_{23}\sqrt{(2-3s_{13}^2)(1-s_{23}^2)}
    }.   
\end{eqnarray}

\begin{figure}[h!]
    \centering
\includegraphics[width=0.49\textwidth]{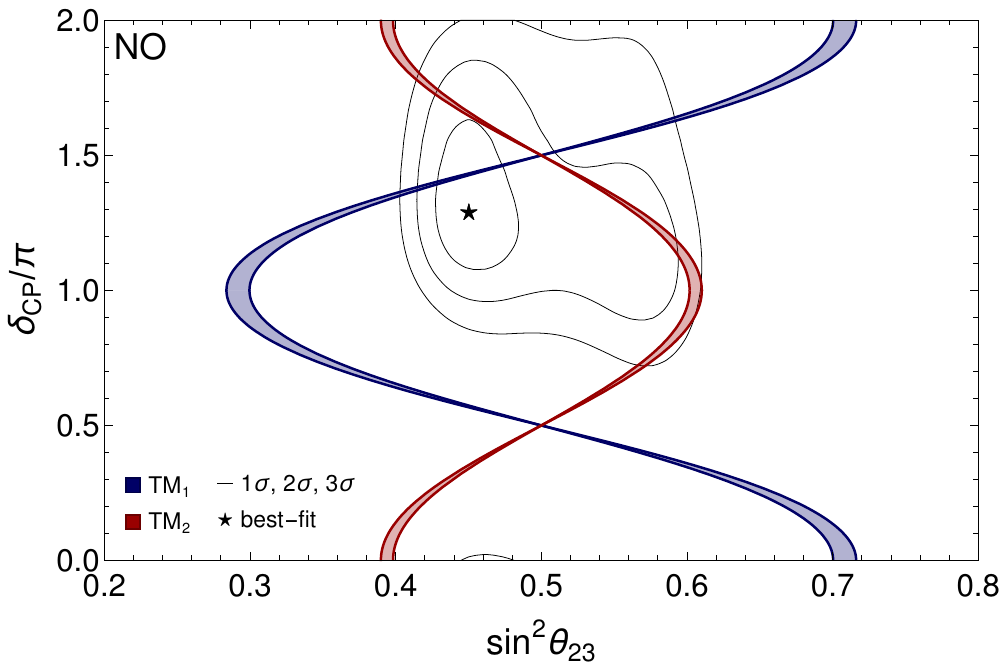}
\includegraphics[width=0.49\textwidth]{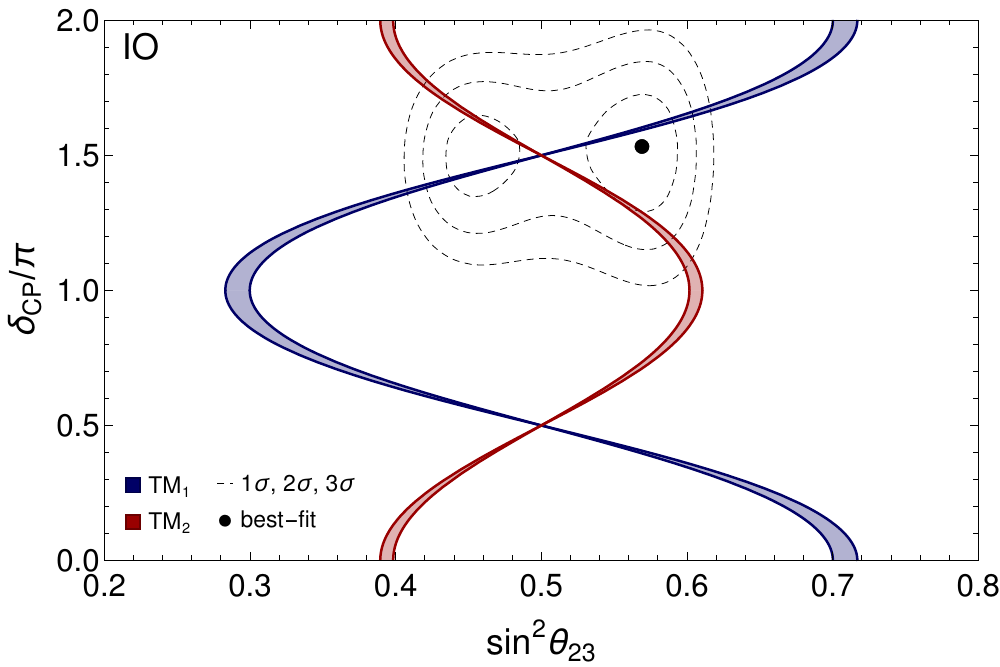}
    \caption{$\delta_{\rm CP}$ plotted against $\sin^2\theta_{23}$ within TM$_1$ and TM$_2$ models using
    Eqs.~(\ref{eq:tm1formulas}) and (\ref{eq:tm2formulas}) and with the NuFIT 5.2   oscillation data \cite{NuFIT5.2,Esteban:2020cvm} from Table~\ref{tab:neutrino_data}. The $1\sigma$, $2\sigma$, $3\sigma$    regions (also in next figures) were derived from $\chi^2$ tables (NuFIT 5.2 data files \cite{NuFIT5.2data} with SK atmospheric data) for corresponding two-dimensional projections of the global analysis, minimized for a given  mass ordering.
    The blue (red) shaded region represents TM$_1$ (TM$_2$) model predictions for $\sin^2\theta_{13}$ in the $3\sigma$ range. 
    }
    \label{fig:deltatm1tm2nhih}
\end{figure}

\begin{figure}[h!]
    \centering
    \includegraphics[width=0.49\textwidth]{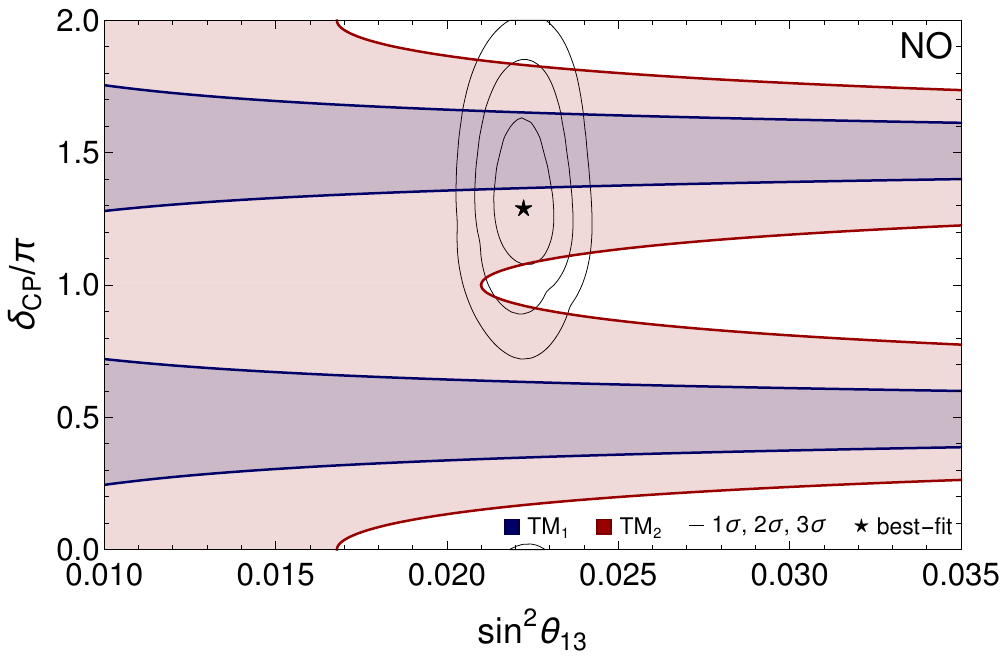}
    \includegraphics[width=0.49\textwidth]{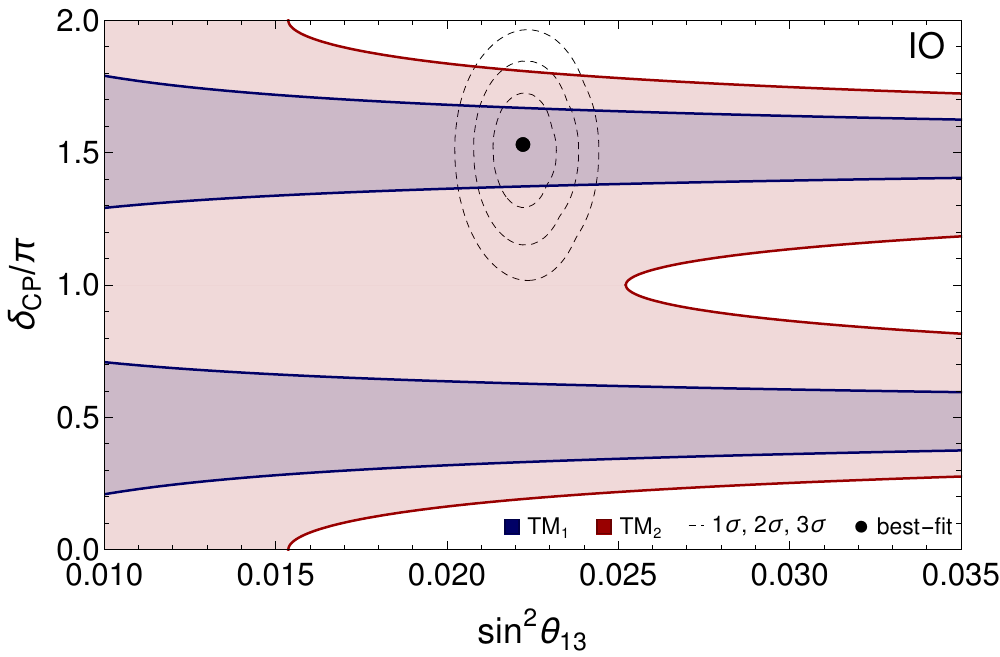}
    \caption{$\delta_{\rm CP}$ plotted against $\sin^2\theta_{13}$ within TM$_1$ and TM$_2$ models using
    Eqs.~(\ref{eq:tm1formulas}) and (\ref{eq:tm2formulas}) and with the NuFIT 5.2   oscillation data \cite{NuFIT5.2,Esteban:2020cvm} from Table~\ref{tab:neutrino_data}. 
    The blue (red) shaded region represents TM$_1$ (TM$_2$) model predictions for $\sin^2\theta_{23}$ in the $3\sigma$ range. 
    }
    \label{fig:deltas13tm1tm2nhih}
\end{figure}

\begin{figure}[h!]
    \centering
    \includegraphics[width=0.49\textwidth]{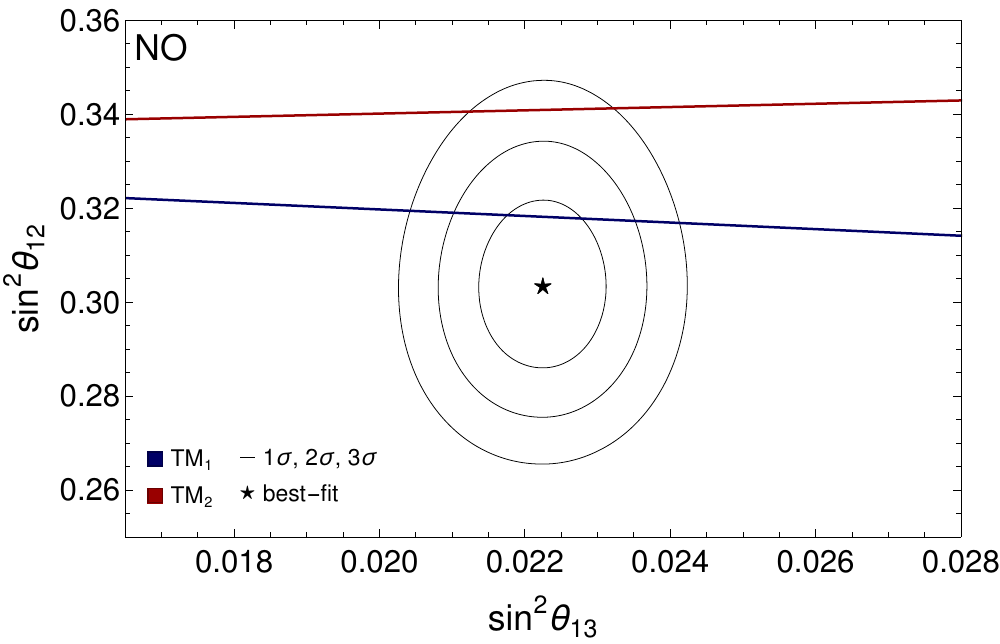}
    \includegraphics[width=0.49\textwidth]{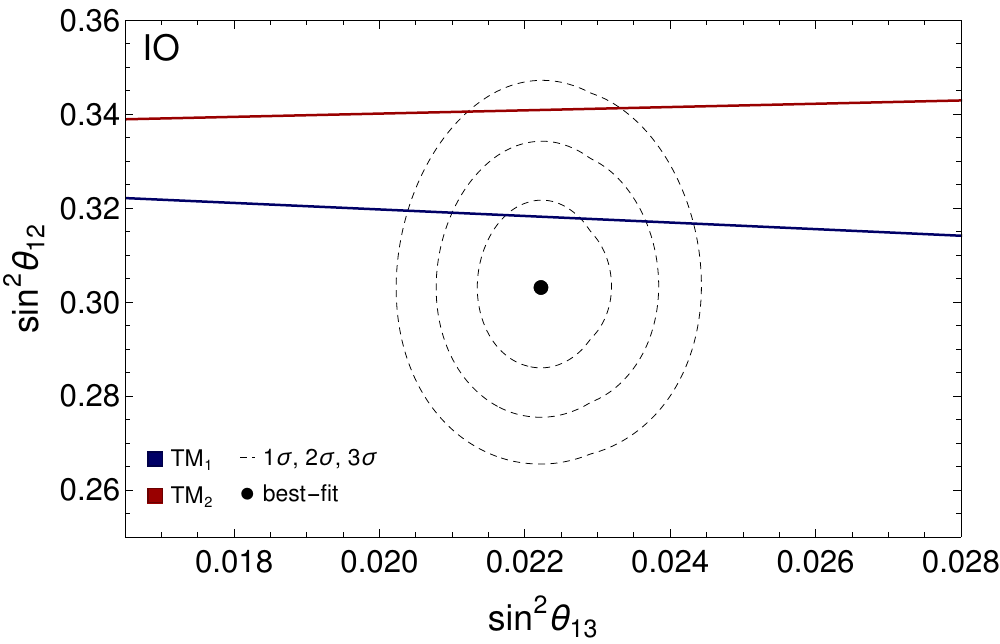}
     \caption{$\sin^2\theta_{12}$ plotted against $\sin^2\theta_{13}$ for TM$_1$ and TM$_2$ mixing [cf.~Eqs.~(\ref{eq:tm1formulas}) and (\ref{eq:tm2formulas})] with the NuFIT 5.2 oscillation data \cite{NuFIT5.2,Esteban:2020cvm}. 
    }
    \label{fig:s12s13tm1tm2nh}
\end{figure}

This helps us illuminate the feasibility of these models in the context of present neutrino oscillation data. In this regard, following Eqs. (\ref{eq:tm1formulas}) and (\ref{eq:tm2formulas}), 
we have plotted  correlations in $\sin^2\theta_{23}-\delta_{\rm CP}$, $\sin^2\theta_{13}-\delta_{\rm CP}$ and $\sin^2\theta_{12}-\sin^2\theta_{13}$ planes for both NO and IO in Figs.~\ref{fig:deltatm1tm2nhih}, \ref{fig:deltas13tm1tm2nhih} and  \ref{fig:s12s13tm1tm2nh}, respectively. 
In these plots  
we have also shown the 1$\sigma$, 2$\sigma$, 3$\sigma$ allowed regions of neutrino oscillation data, based on the two degrees of freedom (2 dof)
tabularized data given in Ref.~\cite{ NuFIT5.2data}.  
The best-fit values are denoted by  $\star$ ($\bullet$) for NO (IO).
The $\sin^2\theta_{23}-\delta_{\rm CP}$ correlation  
plotted in Fig.~\ref{fig:deltatm1tm2nhih} is {important} because of the existing ambiguities on octant of $\theta_{23}$ (i.e., whether it is $\theta_{23}>45^{\circ}$ or $\theta_{23}<45^{\circ}$ ) and precise value of $\delta_{\rm CP}$. Here the blue and red shaded regions represent the predicted correlation of $\theta_{23}$ and $\delta_{\rm CP}$ for TM$_1$ and TM$_2$  
for 3$\sigma$ allowed range of $\sin^2\theta_{13}$.  For correlations $\sin^2\theta_{13}-\delta_{\rm CP}$  in Fig.~\ref{fig:deltas13tm1tm2nhih},  
the spreading of the shaded region depends on the 3$\sigma$ allowed range of $\theta_{23}$.  In Fig.~\ref{fig:s12s13tm1tm2nh},  the $\sin^2\theta_{12}-\sin^2\theta_{13}$ correlations  
for TM$_1$ and TM$_2$ mixing schemes 
yield tight constraints on the allowed ranges of $\sin^2\theta_{12}$:  NO (IO) $0.3168-0.3195~ (0.3167-0.3195)$ for TM$_1$ and $0.3405-0.3413 ~(0.3405-0.3413)$ for TM$_2$. Furthermore, the $\sin^2\theta_{12}$ prediction for the TM$_2$ mixing lies at the edge of the 3$\sigma$ allowed region. 
Thus, a {\it precise measurement of $\sin^2\theta_{12}$  can potentially rule out the \rm{TM$_2$} mixing scheme and corresponding flavor symmetric models}.   
{More on allowed ranges  among TM$_1$ and TM$_2$ parameters coming from correlations of oscillation data will be given in Section \ref{sec:oscexp}.}
{Explicit models to obtain TM$_1$ and TM$_2$ mixing can be found on various occasions in the literature~\cite{Shimizu:2011xg, King:2011zj, Luhn:2013lkn, deMedeirosVarzielas:2012apl}. In Chapter~\ref{sec:cosmic}, we will discuss models for TM$_1$ and TM$_2$ mixing with $A_4$ non-Abelian discrete flavor symmetry having implications in dark matter and leptogenesis and their connection with neutrino masses and mixings.  }

As mentioned earlier, the mixing matrix in Eq.~(\ref{eq:TBM}) and the corresponding mass matrix   Eq.~(\ref{mmutau}) obeys the underlying $\mu-\tau$ symmetry (also known as the $\mu-\tau$ permutation symmetry). As this feature is outdated now for obvious reasons, there is another class of flavor CP model known as $\mu-\tau$ reflection symmetry~\cite{Xing:2015fdg}. This symmetry can be expressed  as the transformation: 
\begin{equation}\label{eq:reflection}
    \nu_{e} \rightarrow \nu_{e}^C, ~\nu_{\mu} \rightarrow \nu_{\tau}^C,~ \nu_{\tau} \rightarrow 
\nu_{\mu}^C
\end{equation}
where `C' stands for the charge conjugation of the corresponding neutrino field under which the neutrino mass term remains unchanged.  The scheme leads to the predictions $\theta_{23}= 45^\circ$, $\delta_{\rm CP}= 90^\circ ~{\rm or}~ 270^{\circ}$. This mixing scheme is still experimentally viable~\cite{Esteban:2020cvm}.  Under the discussed $\mu-\tau$ symmetry, the elements of the lepton mixing matrix satisfy
\begin{equation}\label{umutau}
 |U_{\mu i}|=|U_{\tau i}|~~~~~{\rm where}~~~{i=1,2,3}. 
\end{equation}

Such a mixing scheme is also known as cobimaximal (CBM) mixing scheme~\cite{Grimus:2017itg}. Eq. (\ref{umutau}) indicates that the moduli of $\mu$ and $\tau$  flavor elements of the $3\times 3$ neutrino mixing matrix are equal. 
With these constraints, the neutrino mixing matrix can be parametrized as~\cite{Harrison:2002et,Grimus:2003yn}
\begin{eqnarray}\label{u3b3}
U_{0}&=&\left(
\begin{array}{ccc}
 u_1  & u_2   & u_3 \\
v_1  & v_2   & v_3 \\
v^*_1  & v^*_2   & v^*_3 
\end{array}
\right), 
\end{eqnarray}
where the entries in the first row, $u_i$'s are real (and non-negative) with trivial (vanishing) values of the Majorana phases.  Here $v_i$  satisfies the orthogonality condition 
$2{\rm Re}(v_jv_k^*)=\delta_{jk}-u_k u_k$.  In~Ref.~\cite{Grimus:2003yn} it was argued that the mass matrix leading to  the mixing matrix given in Eq.~(\ref{u3b3}) 
can be written as 
\begin{eqnarray}
{M}_0&=&\left( \label{m3b3}
\begin{array}{ccc}
 a  & d   & d^* \\
d  & c   & b \\
d^* & b   & c^*
\end{array}
\right), 
\end{eqnarray}
where $a,b$ are real and $d,c$ are complex parameters. As a consequence of the symmetry given in Eqs.(\ref{umutau})-(\ref{m3b3}),
we obtain the  predictions for maximal $\theta_{23}=45^\circ$ and $\delta_{\rm CP}= 90^\circ ~{\rm or}~ 270^{\circ}$ in the basis where the 
charged leptons are considered to be diagonal. {Based on the choice of diagonalizing  the phase factor of $M_0$ in Eq.~(\ref{m3b3}), the Majorana phases for $\mu-\tau$ reflection symmetry are predicted to be $0^{\circ}$ or $90^{\circ}$}. This scheme, interestingly, still leaves room for nonzero $\theta_{13}$. Realization of such a mixing pattern is possible with various discrete flavor symmetries ($A_4, \Delta(27)$, etc.), for example, see Refs.~\cite{Chakraborty:2019rjc,CarcamoHernandez:2020udg,Vien:2020dlk,Vien:2021bsv, Vien:2022cxr,Karmakar:2022cbm}. 

Earlier, we mentioned that fixed mixing schemes such as BM, TBM, GR, HG are ruled out and require specific corrections to accommodate non-zero $\theta_{13}$ and $\delta_{\rm CP}$. These corrections can also provide a possible deviation from $\theta_{23}=45^{\circ}$. For example, to achieve experimentally viable TM$_1$ and TM$_2$ mixing, we have shown instances where additional contribution to the neutrino sector over TBM mixing generates necessary corrections. However, this can be achieved in various ways. A correction in the charged lepton sector is one such possibility. This can be achieved by additional non-trivial contribution in the charged lepton sector
~\cite{Branco:2012vs}. 

{
From Eq.~(\ref{umutau}), constraints on $\theta_{23}$, $\theta_{13}$ and $\delta_{\rm CP}$ are obtained as
\begin{eqnarray}
    \theta_{23}=\frac{\pi}{4}, \quad s_{13}\cos\delta_{\rm CP}=0, 
\end{eqnarray}
so $\theta_{23}$ is found to be exactly $45^{\circ}$, and for non-zero $\theta_{23}$,  $\delta_{\rm CP}$ is predicted to be $90^{\circ}$ or $270^{\circ}$. Although these predictions are within the allowed range by the global fits of neutrino oscillation data summarized in Table~\ref{tab:neutrino_data}, precise future measurements may depart from these exact values. In fact, all the best-fit values by the three global fits mentioned in Table~~\ref{tab:neutrino_data} deviate from  $\theta_{23}=45^{\circ}$ for either mass hierarchy. All these arguments can be considered as an argument for a possible departure from the $\mu-\tau$ reflection symmetry. This motivates us to assume a `partial $\mu-\tau$' reflection symmetry\footnote{For a review on $\mu-\tau$ flavor symmetry and its possible breaking, see Ref.~\cite{Xing:2015fdg}.}~\cite{Xing:2014zka} where the condition given in Eq.~(\ref{umutau}) holds only for a single column of $U_0$ in Eq.~(\ref{u3b3}).  
}
Given this, it seems well motivated to introduce partial $\mu-\tau$ reflection symmetry, for which the $\delta_{\rm CP}$ and $\theta_{23}$ are not fixed but correlated \cite{Chakraborty:2018dew}
\begin{eqnarray}
     \label{eq:mutau1}
     |U_{\mu 1}|=|U_{\tau 1}| &:& 
    ~~~\cos \delta_{\rm CP} = \frac{
    (c_{23}^2-s_{23}^2)(c_{12}^2 s_{13}^2-s_{12}^2)
    }{
    4 c_{12}s_{12}c_{23}s_{23}s_{13}
    },
    \\
    \label{eq:mutau2}
     |U_{\mu 2}|=|U_{\tau 2}| &:& 
     ~~~\cos \delta_{\rm CP} = \frac{
    (c_{23}^2-s_{23}^2)(c_{12}^2- s_{12}^2s_{13}^2)
    }{
    4 c_{12}s_{12}c_{23}s_{23}s_{13}
    }.
\end{eqnarray}
{Such a framework can also have its origin  in flavor symmetry~\cite{Joshipura:2016quv}}. A similar investigation of partial $\mu-\tau$ reflection symmetry can also be performed for the 3+1 neutrino mixing scheme, denoted as (3+1)$\nu$, leading to the $\delta_{\rm CP} / \theta_{23}$ correlations as given below~\cite{Chakraborty:2019rjc}
\begin{eqnarray}
|U_{\mu 1}|=|U_{\tau 1}| &:& ~~ \cos\delta_{\rm CP}=\frac{
                  (a_1^2+b_1^2)-(c_1^2+d_1^2)
                  }
                  {
                  2(c_1 d_1-a_1 b_1)
                  }\label{eq:11}\\
|U_{\mu 2}|=|U_{\tau 2}| &:& ~~ \cos\delta_{\rm CP}=\frac{
                  (a_2^2+b_2^2)-(c_2^2+d_2^2)
                  }
                  {
                  2(a_2 b_2-c_2 d_2)
                  }\label{eq:22}        \\
|U_{\mu 3}|=|U_{\tau 3}| &:&  ~~\cos\delta_{\rm CP}=\frac{
                  (a_3^2+b_3^2)-(c_3^2+d_3^2)
                  }
                  {
                  2(a_3 b_3-c_3 d_3)
                  }\label{eq:33}  
                   \\
{|U_{\mu 4}|=|U_{\tau 4}|} &:&  ~~{\tan^2\theta_{24}=\sin^2\theta_{34}}
                  \label{eq:44}  
\end{eqnarray}
where 
\begin{equation}
\begin{aligned}
 a_1&=c_{12}s_{13}s_{23}s_{24}s_{34}-c_{12}c_{23}c_{34}s_{13}, &
 b_1&=c_{34}s_{12}s_{23}-c_{12}c_{13}c_{24}s_{14}s_{34}+c_{23}s_{12}s_{24}s_{34}, \\
 c_1&=c_{23}c_{24}s_{12}+c_{12}c_{13}s_{14}s_{24}, &
 d_1&=c_{12}c_{24}s_{13}s_{23},\\
 a_2&=c_{12}(c_{34}s_{23}+c_{23}s_{24}s_{34})+s_{12}c_{13}c_{24}s_{14}s_{34}, &
 b_2&=s_{12}(s_{13}s_{23}s_{24}s_{34}-c_{23}c_{34}s_{13}), \\
 c_2&=c_{12}c_{23}c_{24}-s_{12}c_{13}s_{14}s_{24},&
 d_2&=s_{12}c_{24}s_{13}s_{23}, \\
 a_3&=c_{13}(c_{23}c_{34}-s_{23}s_{24}s_{34}), &
 b_3&=c_{24}s_{13}s_{14}s_{34}, \\
 c_3&=s_{13}s_{14}s_{24}, &
 d_3&=c_{13}c_{24}s_{23}. 
 \label{corpara}
\end{aligned}
\end{equation}

The $\delta_{\rm CP} - \theta_{23}$ correlations in Eqs. (\ref{eq:mutau1}) and (\ref{eq:mutau2}) for  the partial $\mu-\tau$ reflection symmetry ($|U_{\mu 1}|=|U_{\tau 1}|$ , $|U_{\mu 2}|=|U_{\tau 2}|$)  are given in Fig.~\ref{fig:deltatmutau}. These correlations partially overlap the $1\sigma$, $2\sigma$, $3\sigma$  regions  but do not include the best-fit values.

\begin{figure}[h!]
    \centering
    \includegraphics[width=0.49\textwidth]{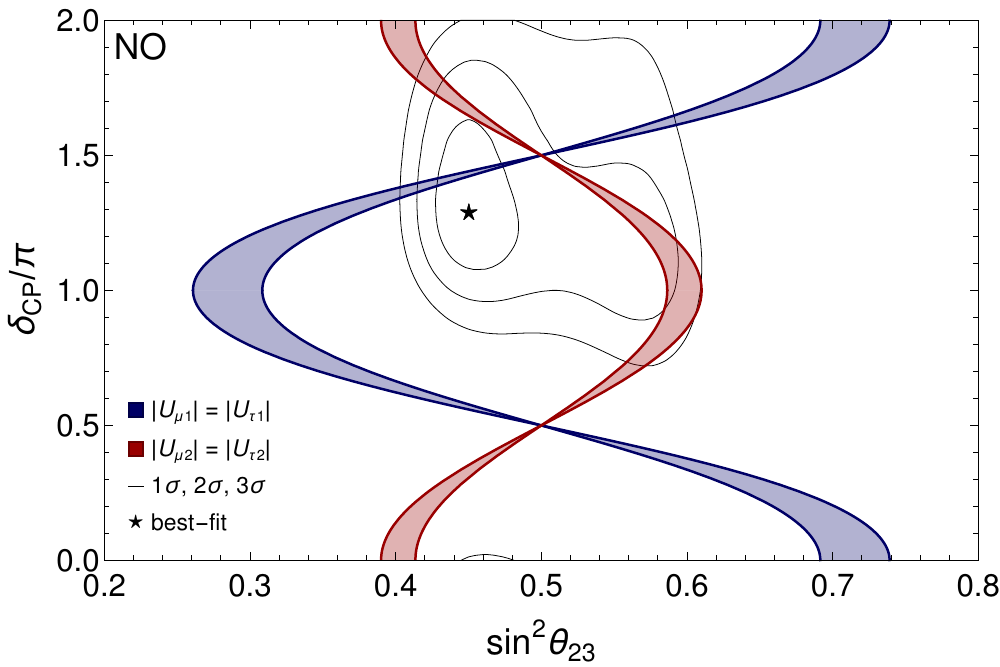}
    \includegraphics[width=0.49\textwidth]{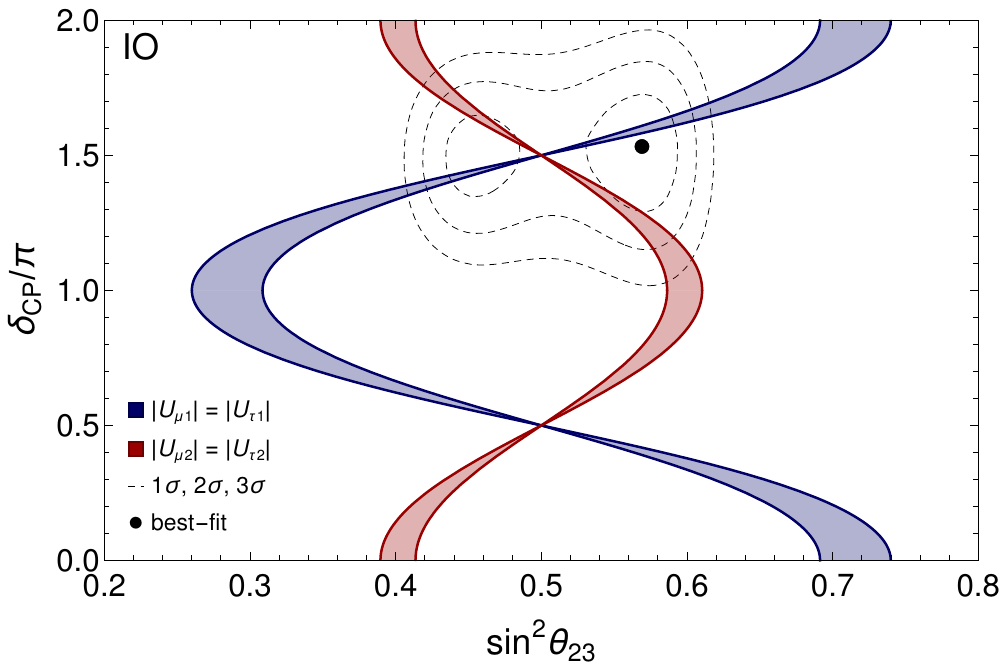}
    \caption{$\delta_{\rm CP}$ plotted against $\sin^2\theta_{23}$ within partial 
    $\mu-\tau$ reflection symmetry predictions 
    (Eqs.~(\ref{eq:mutau1}) and (\ref{eq:mutau2}))
    and the NuFIT 5.2  oscillation data \cite{NuFIT5.2,Esteban:2020cvm}. 
   The blue (red) shaded region represents $|U_{\mu 1}|=|U_{\tau 1}|$ ($|U_{\mu 2}|=|U_{\tau 2}|$) model predictions with $\sin^2\theta_{12}$ and $\sin^2\theta_{13}$ in the $3\sigma$  range.   
    }
    \label{fig:deltatmutau}
\end{figure}

\begin{figure}[h!]
    \centering
    \includegraphics[width=0.49\textwidth]{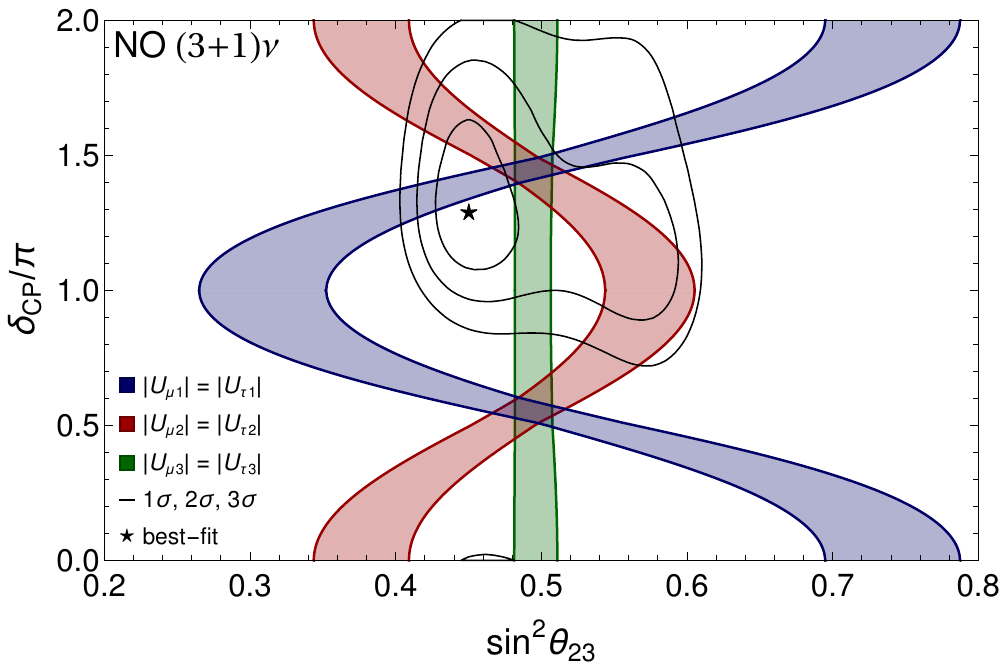}
    \includegraphics[width=0.49\textwidth]{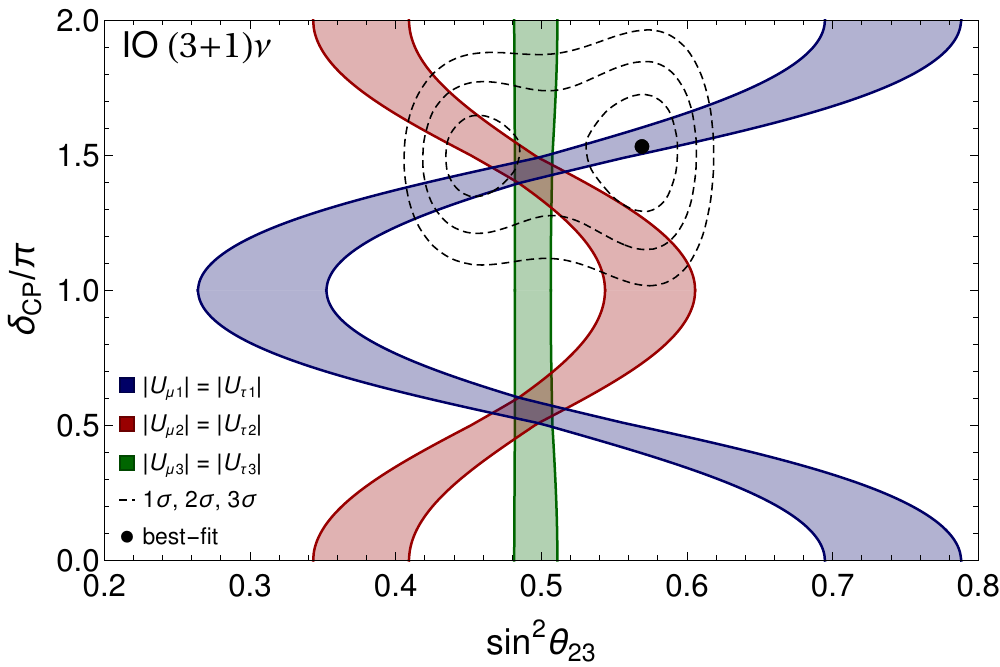}
    \caption{$\delta_{\rm CP}$ plotted against $\sin^2\theta_{23}$ in the (3+1)$\nu$ case within partial 
    $\mu-\tau$ reflection symmetry scenario \cite{Chakraborty:2019rjc} and NuFIT 5.2 (with SK atmospheric) oscillation data \cite{NuFIT5.2,Esteban:2020cvm}.  
    The blue, red, green regions represent $|U_{\mu 1}|=|U_{\tau 1}|$ , $|U_{\mu 2}|=|U_{\tau 2}|$, $|U_{\mu 3}|=|U_{\tau 3}|$ symmetry predictions respectively. The $\sin^2\theta_{12}$ and $\sin^2\theta_{13}$ vary in $3\sigma$   range (Table~\ref{tab:neutrino_data}) while the (3+1)$\nu$ oscillation parameters ($\theta_{14},\theta_{24},\theta_{34}$) vary in  representative $3\sigma$ ranges given in Table~I of Ref.~\cite{Chakraborty:2019rjc} 
    ($\theta_{14}: 4^\circ-10^\circ$, 
    $\theta_{24}: 5^\circ-10^\circ$, 
    $\theta_{34}: 0^\circ-11^\circ$, 
    $\delta_{14}=\delta_{24}=0^\circ$). 
    }
    \label{fig:deltatmutau3p1}
\end{figure}

The results for the (3+1)$\nu$ scenario are gathered in Fig.~\ref{fig:deltatmutau3p1}. From the overlapped regions in Fig. \ref{fig:deltatmutau3p1}, it is clear that if we demand a total $\mu-\tau$ reflection symmetry, i.e.,  $|U_{\mu i}|=|U_{\tau i}|$ for all four columns  then the atmospheric mixing angle $\theta_{23}$ is restricted within  a narrow region around
$45^{\circ}$ and the Dirac CP phase is also restricted around the maximal CP
violating values. Since the present best fit values for $\theta_{23}$  clearly favors a deviation from $45^{\circ}$ (see Fig.~\ref{fig:schematics_new}), a partial  $\mu-\tau$ reflection symmetry in the (3+1)$\nu$ scenario may accommodate appropriate deviation. From Fig.~\ref{fig:deltatmutau3p1}, we find that  $\delta_{\rm CP} - \theta_{23}$ correlations are partly overlapping the ($1\sigma$, $2\sigma$ and $3\sigma$)  $\delta_{\rm CP} - \theta_{23}$  regions for all three
 partial $\mu-\tau$ reflection symmetries but only for the
$|U_{\mu 1}|=|U_{\tau 1}|$ symmetry, the $\delta_{\rm CP} -\theta_{23}$ correlation region includes the best-fit value for IO. In addition to this it is worth mentioning that the partial $\mu-\tau$ reflection symmetry in the fourth column ($|U_{\mu 4}|=|U_{\tau 4}|$) restricts $\theta_{34}$ within the range $4.9^{\circ}-9.8^{\circ}$~\cite{Chakraborty:2019rjc}.
{In Section \ref{sec:oscexp}, we will further stress that the allowed parameter space within these mixing schemes can be further constrained by employing the correlation among the oscillation parameters obtained from global analysis of neutrino oscillation data. Furthermore,  the implications of fixed values of the Majorana phases in case of $\mu-\tau$ reflection symmetries will be discussed in Section \ref{sec:bb0nu}.}

\subsection{Flavor Symmetry and Neutrino Mass Models} \label{sec:massmodels}

Apart from the observed pattern of neutrino mixing, the origin of tiny neutrino mass is still unknown to us.
Over decades, the exclusive evidence for non-zero neutrino masses stems solely from neutrino oscillation experiments, which are sensitive to the mass-squared differences and not to the absolute scale of neutrino masses (see Table~\ref{tab:neutrino_data}). On the other hand,  bounds on absolute neutrino masses come from cosmological surveys~\cite{Planck:2018nkj,Planck:2018vyg} and the end-point spectrum of tritium beta decay. Combining all these results, we come to the neutrino mass ranges discussed in the Introduction, which are at the mili-electronvolt scale at the most. Based on the established neutrino mass spectrum, different models try to explain it. {For a detailed discussion of many mechanisms to generate light neutrino masses, the readers are referred to~Refs.~\cite{Akhmedov:1999uz,King:2003jb,Mohapatra:2006gs,Valle:2006vb,deGouvea:2016qpx, Cai:2017jrq} and references therein. Here, we briefly mention a few of them that are frequently used in realistic flavor model building.   } 

{\underline{Seesaw models}}	

One of the most popular ways to generate tiny neutrino mass is to start with the high scale suppressed lepton number violating Weinberg operator $H L^T LH/\Lambda $ mentioned earlier, which is non-renormalizable. This can give rise to various 
seesaw mechanisms such as type-I, type-II, type-III, inverse and linear seesaw ~\cite{Minkowski:1977sc, Mohapatra:1979ia, Yanagida:1979as, GellMann:1980vs, Glashow:1979nm, Schechter:1980gr, deGouvea:2016qpx}. They can be embedded in the form of Eq.~(\ref{ssmm}), see Ref.~\cite{Flieger:2022ekj}. {However, to elucidate the observed pattern of neutrino mixings, one must include additional ingredients such as discrete flavor symmetries.} 
Examples of discrete flavor symmetric models for type-I, type-II, and inverse seesaw which are very efficient in explaining tiny mass as well as correct mixing as observed by the neutrino oscillation experiments can be found in~Refs.~\cite{Karmakar:2014dva,Karmakar:2015jza,Karmakar:2016cvb,Datta:2021zzf,Yang:2011fh,Yamanaka:1981pa,Brown:1984dk,Ma:2005pd,Hagedorn:2006ug,Zhang:2006fv,Ishimori:2008fi,Grimus:2009pg,Dutta:2009bj,Meloni:2009cz,Ge:2010js,Dong:2010zu,Ishimori:2010fs,Park:2011zt}. 
Based on their scale, these flavor symmetric seesaw mechanisms may include a wide range of phenomenological implications in lepton flavor violation, collider phenomenology or leptogenesis~\cite{Karmakar:2017eit}.

{\underline{Radiative mass models}}	

Another class of models that can be connected with flavor symmetries are radiative neutrino mass models in which masses of neutrinos are absent at the tree-level and are generated at 1- or higher-loop orders. {These models explain the lightness of neutrino masses with sizable Yukawa couplings and suppression provided by the loop factor. A broad review of various radiative neutrino models can be found in Ref.~\cite{Cai:2017jrq}. The key feature of these models is that they can be verified experimentally because the masses of exotic particles that take part in the neutrino mass generation are in the  TeV range, which the current colliders' experiments can probe. Furthermore, these models may contribute to electric dipole moments, anomalous magnetic moments and meson decays, matter-antimatter asymmetry~\cite{Cai:2017jrq,Hagedorn:2018spx,Baumholzer:2019twf}. Most interestingly, some radiative models naturally incorporate potential DM candidates~\cite{Ma:2006km,Restrepo:2013aga,Cacciapaglia:2020psm}. Additional symmetries that explain tiny neutrino masses also stabilize the DM.   On top of that, radiative neutrino mass models with discrete flavor symmetries can also explain the observed lepton mixing for obvious reasons, for instance, modular $S_3$ and $A_4$ \cite{Okada:2021aoi,Nomura:2022hxs, Nomura:2021pld,Otsuka:2022rak,Okada:2019xqk}.  
}

\underline{Neutrino mass sum rules}

Neutrino mass mechanisms augmented with discrete flavor symmetries predict a range of neutrino masses and mixings and can yield interesting correlations
between several observables, such as leptonic mixing angles,  phases, and neutrino masses. Models that have these features enhance the testability at neutrino experiments. Discrete flavor symmetric models can give rise to sum rules for neutrino mixing angles and masses. The mixing sum rules relate the leptonic mixing angles to
the Dirac CP-violating phase $\delta_{\rm CP}$~\cite{King:2005bj,Masina:2005hf,Antusch:2005kw,Antusch:2007rk,Girardi:2015vha}. For example, considering appropriate deviations, the approximate mixing sum rules can be written as
~\cite{Marzocca:2011dh,Antusch:2011qg,Spinrath:2016cwn}
\begin{eqnarray}
    \sin^2\theta_{12}&\simeq& \frac{1}{2}+\sin{\theta_{13}}\cos\delta ,\\
    \sin^2{\theta_{12}}&\simeq& \frac{1}{3}+\frac{2\sqrt{2}}{3}\sin{\theta_{13}}\cos\delta, 
\end{eqnarray}
for    BM and TBM mixing, respectively. For the implication of the mixing sum rules at the neutrino oscillation experiments, see~Ref.~\cite{Antusch:2007rk}.
On the other hand,  the mass sum rules~\cite{Altarelli:2008bg,Hirsch:2008rp,  Chen:2009um, Altarelli:2009kr,Barry:2010yk,Dorame:2012zv}, which describe the interrelation between the three complex neutrino eigenvalues are particularly important because of their substantial implication~\cite{Dorame:2011eb,Morisi:2012fg,King:2013psa, Cirigliano:2022oqy} in the prediction of the effective mass parameter ($m_{\beta\beta}$) appearing in the neutrinoless double beta decay described in Eq. (\ref{2nu0}). Theoretically, it is natural to wonder if these sum rules are related to residual or accidental symmetry.  The neutrino mass sum rules can originate from discrete flavor symmetric models in which neutrino masses can arise from the aforementioned mass generation mechanisms.  However, the most general mass sum rule can be written as~\cite{King:2013psa}
\begin{equation}
    A_1  \tilde{m}_1^{p}e^{i\chi_1}+A_2  \tilde{m}_2^{p}e^{i\chi_2}+A_3  \tilde{m}_3^{p}e^{i\chi_3}=0,\label{eq:sumrulegen} 
\end{equation}
where $\tilde{m_i}$ are three complex mass eigen values, $p\neq 0, \chi_1 \in [0, 2\pi], A_i > 0$. Here $A_i$ and $\chi_i$ stand for appropriate complex coefficients and phase factors (without the Majorana phases). The power $q$ of the complex mass eigenvalues characterizes the sum rule. For example, a simple sum rule can be obtained when a type-I seesaw mass mechanism is augmented by $A_5$ non-Abelian discrete flavor symmetry leading to ~\cite{Cooper:2012bd} 
\begin{eqnarray}
    \frac{1}{\tilde{m}_1}+\frac{1}{\tilde{m}_2}=\frac{1}{\tilde{m}_3}. \label{eq:suminv}
\end{eqnarray}
Comparing Eq. (\ref{eq:sumrulegen}) and  Eq. (\ref{eq:suminv}) we find that $A_i=1, \chi_{1,2}=0, \chi_3=\pi$ and $p=-1$. Such mass sum rules (i.e., when $p=-1$) are called {\it inverse sum rules}, obtained from a diverse combination of neutrino mass mechanisms and discrete flavor symmetries~\cite{Altarelli:2005yx, Altarelli:2008bg,Hagedorn:2009jy,Chen:2009gy,Karmakar:2014dva}. 

\begin{table}[h!]
\centering 
\begin{small}
\begin{tabular}{c|c|c}
\hline
Sum Rule & Group & Seesaw Type\\ \hline
$\tilde{m}_1+\tilde{m}_2=\tilde{m}_3$ & $A_4$\cite{Barry:2010zk,Ding:2010pc,Ma:2005sha,Ma:2006wm,Honda:2008rs,Brahmachari:2008fn}; $S_4$\cite{Bazzocchi:2009pv}; $A_5$\cite{Everett:2008et}  &Weinberg \\
$\tilde{m}_1+\tilde{m}_2=\tilde{m}_3$  &$\Delta(54)$\cite{Boucenna:2012qb}; $S_4$\cite{Bazzocchi:2009da} &Type II \\\hline
$\tilde{m}_1+2\tilde{m}_2 = \tilde{m}_3$ & $S_4$\cite{Mohapatra:2012tb} &Type II \\\hline
$2\tilde{m}_2+\tilde{m}_3=\tilde{m}_1$ & $A_4$
\cite{Barry:2010zk,Altarelli:2005yp,Altarelli:2006kg,Ma:2006vq,Bazzocchi:2007na,Bazzocchi:2007au,Lin:2008aj,Ma:2009wi,Ciafaloni:2009qs,Ma:2005sha,Ma:2006wm,Honda:2008rs,Brahmachari:2008fn,Altarelli:2005yx,Chen:2009um}  &Weinberg \\
  
  & $S_4$\cite{Bazzocchi:2008ej,Feruglio:2013hia}; $T'$\cite{Chen:2007afa,Ding:2008rj,Chen:2009gf,Feruglio:2007uu,Merlo:2011hw,Chen:2009gy}; $T_7$\cite{Luhn:2012bc} & \\
$2\tilde{m}_2+\tilde{m}_3=\tilde{m}_1$ & $A_4$\cite{Fukuyama:2010mz} & Type II \\\hline
$\tilde{m}_1+\tilde{m}_2=2\tilde{m}_3$ & $S_4$\cite{Ding:2013eca}& Dirac  \\
$\tilde{m}_1+\tilde{m}_2=2\tilde{m}_3$ & $L_e - L_\mu - L_\tau$\cite{Lindner:2010wr} &Type II \\\hline
$\tilde{m}_1 + \frac{\sqrt{3}+1}{2} \tilde{m}_3 = \frac{\sqrt{3}-1}{2} \tilde{m}_2$ & $A_5'$\cite{Hashimoto:2011tn} &Weinberg \\\hline
$\tilde{m}_1^{-1}+\tilde{m}_2^{-1}=\tilde{m}_3^{-1}$ & $A_4$\cite{Barry:2010zk}; $S_4$\cite{Bazzocchi:2009da,Ding:2010pc}; $A_5$\cite{Ding:2011cm,Cooper:2012bd}&Type I \\
$\tilde{m}_1^{-1}+\tilde{m}_2^{-1}=\tilde{m}_3^{-1}$&$S_4$\cite{Bazzocchi:2009da}&Type III\\\hline
$2\tilde{m}_2^{-1}+\tilde{m}_3^{-1}=\tilde{m}_1^{-1}$ &$A_4$\cite{Barry:2010zk,Morisi:2007ft,Altarelli:2008bg,Adhikary:2008au,Lin:2009bw,Csaki:2008qq,Altarelli:2009kr,Hagedorn:2009jy,Burrows:2009pi,Ding:2009gh,Mitra:2009jj,delAguila:2010vg,Burrows:2010wz,Altarelli:2005yx,Chen:2009um}; $T'$\cite{Chen:2009gy}&Type I \\
$\tilde{m}_1^{-1}+\tilde{m}_3^{-1}=2\tilde{m}_2^{-1}$ & $A_4$\cite{He:2006dk,Berger:2009tt,Kadosh:2010rm}; $T'$\cite{Lavoura:2012cv} &Type I \\\hline
$\tilde{m}_3^{-1} \pm 2i\tilde{m}_2^{-1}=\tilde{m}_1^{-1}$ & $\Delta(96)$\cite{King:2012in}& Type I \\\hline
$\tilde{m}_1^{1/2}- \tilde{m}_3^{1/2}=2\tilde{m}_2^{1/2}$ & $A_4$\cite{Hirsch:2008rp} &Type I \\
$\tilde{m}_1^{1/2}+\tilde{m}_3^{1/2}=2\tilde{m}_2^{1/2}$ & $A_4$\cite{Adulpravitchai:2009gi} &Scotogenic \\\hline
$\tilde{m}_1^{-1/2}+\tilde{m}_2^{-1/2}=2\tilde{m}_3^{-1/2}$ & $S_4$\cite{Dorame:2012zv}&Inverse\\\hline 
\end{tabular}
\end{small}
\caption{ Sum rules for the complex light neutrino mass eigenvalues $\tilde{m}_i$ defined in Eq.~(\ref{eq:sumrulegen}) obtained with various combinations of neutrino mass generation mechanisms and discrete flavor symmetries. {The table is adopted from  Ref.~\cite{King:2013psa}, see also Refs.~\cite{Barry:2010yk,Cirigliano:2022oqy}.}} \label{tab:sr}
\end{table}

In Table~\ref{tab:sr}, we have mentioned various simple sum rules for the complex light neutrino mass eigenvalues ($\tilde{m}_i$) obtained with combinations of neutrino mass generation mechanism and discrete flavor symmetries~\cite{Barry:2010yk,King:2013psa,Cirigliano:2022oqy}. Similar but less simple mass sum rules can also be obtained for models with modular symmetry. The authors reported in Ref. \cite{Gehrlein:2020jnr} four different mass sum rules for models based on modular symmetries where a
residual symmetry in the lepton sector is preserved. The reported sum rules (SR) within these modular invariance
approaches  (which also follow the most general form given in Eq.~(\ref{eq:sumrulegen}))
are called  SR 1 (Case I and Case II), SR 2, SR 3, and SR 4~\cite{Gehrlein:2020jnr}. Studies show that {these sum rules can not be conclusively connected to a particular mass generation mechanism, discrete symmetry, or any remnant symmetry in the lepton sector}~\cite{Gehrlein:2017ryu}. Rather, they are more connected to minimal breaking of the symmetries, which introduces a minimal number of parameters related to nonzero neutrino mass eigenvalues. However, the Majorana phases (connected with the complex mass eigenvalues $\tilde{m}_i$) also appear in the mass sum rules, making them ideal observable to test them in the neutrinoless double beta decay experiments. The occurrence of such sum rules severely constrains the prediction for $m_{\beta\beta}$, see Fig.~\ref{fig:sumrule} in Section \ref{sec:bb0nu}. A detailed discussion of the role of individual sum rules can be found in Refs.~\cite{Dorame:2011eb,Barry:2010yk,Morisi:2012fg,King:2013psa, Cirigliano:2022oqy,Gehrlein:2017ryu}. 

{\underline{{Flavor} Models: Majorana vs  Dirac neutrinos}}	

While neutrino oscillation experiments are insensitive to the nature of neutrinos, experiments looking for lepton number violating signatures can probe the Majorana nature of neutrinos. Neutrinoless double beta decay is one such lepton number violating process which has been searched for at several experiments without any positive result so far but giving stricter bounds on the effective neutrino mass, as discussed in Chapter \ref{chapter:introduction}. Although negative results at neutrinoless double beta decay experiments do not prove that the light neutrinos are of Dirac nature, it is nevertheless suggestive enough to come up with scenarios predicting Dirac neutrinos with correct mass and mixing. There have been several
proposals already that can generate tiny Dirac neutrino masses~\cite{Babu:1988yq,Peltoniemi:1992ss,Aranda:2013gga, Ma:2015mjd, CentellesChulia:2016rms, Abbas:2016qbl}.  In Refs.~\cite{Borah:2017dmk,Borah:2018nvu,CentellesChulia:2018gwr,CentellesChulia:2018bkz}, the authors showed that it is possible to propose various seesaw mechanisms (type-I, inverse and linear seesaw) for Dirac neutrinos with $A_4$ discrete flavor symmetry. Here the symmetry is chosen in such a way that it naturally explains the hierarchy among different terms in the neutrino mass matrix, contrary to the conventional seesaws where this hierarchy is ad-hoc. 

{\underline{Texture zeroes}}	

When a flavor neutrino mass matrix contains zero elements, this is called a texture zeros mass matrix \cite{Frampton:2002yf,Xing:2002ta,Hagedorn:2005kz,Merle:2006du}. Models with texture zeros can originate from the Froggatt–Nielsen mechanism \cite{Froggatt:1978nt} or certain flavor symmetries, see e.g. \cite{Grimus:2004hf,Ding:2022aoe}.
Texture zeroes make the neutrino mass and mixing models simpler. Such constructions are interesting as they lead to a reduction of independent mass parameters in theory\footnote{ The number of free parameters in neutrino models can also be reduced with the requirement of zero mass determinant
\cite{Branco:2002ie,Chauhan:2006uf} or the zero trace ("zero-sum" $m_{\nu_1}+m_{\nu_2}+m_{\nu_3}=0$) condition \cite{He:2003nt,Chamoun:2023bjc}.}. There is a vast literature on the subject, e.g. see the list of references in the recent works \cite{Minamizawa:2022fch,Ding:2022aoe} and discussed literature below. 

In the three-generation scenario, the low energy Majorana neutrino mass matrix $M_\nu$ is a $3 \times 3$ complex symmetric matrix having six independent elements given by
\begin{eqnarray}
M_{\nu}&=&
       \begin{pmatrix}
        m_{ee} & m_{e\mu} & m_{e\tau} \\
m_{e \mu}& m_{ \mu \mu} & m_{\mu \tau} \\
m_{e \tau }& m_{\mu \tau} & m_{\tau \tau}\\
       \end{pmatrix}.
\end{eqnarray}
\\
For three generations and one-zero textures, all six one-zero textures $G_1-G_6$ in the three-generation neutrino flavor scenario
 {  
\begin{eqnarray}
&& {\rm G}_1:
\left( 
\begin{array}{*{20}{c}}
0 & \times & \times \\
- & \times & \times \\
- & - & \times \\
\end{array}
\right),
\quad
 {\rm G}_2:
\left( 
\begin{array}{*{20}{c}}
\times & 0 & \times \\
- & \times & \times \\
- & - & \times \\
\end{array}
\right), \quad 
{\rm G}_3:
\left( 
\begin{array}{*{20}{c}}
\times & \times &0 \\
- & \times & \times \\
- & - & \times \\
\end{array}
\right),
\quad \nonumber \\ &&
{\rm G}_4:
\left( 
\begin{array}{*{20}{c}}
\times & \times & \times \\
- & 0 & \times \\
- & - & \times \\
\end{array}
\right), \quad 
 {\rm G}_5:
\left( 
\begin{array}{*{20}{c}}
\times & \times & \times \\
- & \times & 0 \\
- & - & \times \\
\end{array}
\right),
\quad 
{\rm G}_6:
\left( 
\begin{array}{*{20}{c}}
\times & \times & \times \\
- & \times & \times \\
- & - &0 \\
\end{array}
\right)
\label{eq:onezero}
\end{eqnarray}
}
can accommodate the experimental data \cite{Bora:2016ygl}, see also Refs.~\cite{Kitabayashi:2020ajn,Merle:2006du,Lashin:2011dn}. 
The crosses ``$\times$'' stand for the non-zero entries and "$-$" represent symmetric elements (the matrices are assumed to be symmetric). 
There are 15 possible two-zero textures categorised in different classes, as shown in Table~\ref{tab:two-zero}. {{The constraints can be released for non-Hermitian textures (Dirac neutrino cases). For instance, for light Dirac neutrinos (less CP phases) and non-Hermitian textures, even textures with four zeros are allowed. And most of the one-zero, two-zero and three-zero textures are permitted. Four-zero textures are tightly constrained, with only six allowed out of 126 possibilities \cite{Borgohain:2020csn}. For more investigations on Dirac neutrino mass textures, see \cite{Ahuja:2009jj,Benavides:2020pjx,Lenis:2023lgq,Zhao:2023vkb}. }}
\begin{table}[h!]
\begin{center}
\begin{small}
\begin{scriptsize}
\begin{tabular}{|c|c|c|c||c|c|c|}
\hline $ A_1$& $A_2$ &  &  & $D_1$ & $D_2$ & \\
\hline $\left(
\begin{array}{ccc}
0 & 0 & \times \\
0 & \times & \times \\ 
- & - & \times
\end{array}
\right)$ & $\left(
\begin{array}{ccc}
0 & \times & 0\\ 
- & \times & \times \\
0 & - & \times
\end{array}
\right)$  & & & $\left(
\begin{array}{ccc}
\times &\times & \times  \\  - & 0 & 0 \\ - & 0 & \times
\end{array}
\right)$
& 
$\left(
\begin{array}{ccc}
\times & \times & \times \\  - & \times &0 \\ - & 0 &0 
\end{array}
\right)$

& \\

\hline $ B_1$ & $B_2$ & $B_3$ & $B_4$ & $E_1$ & $E_2$ & $E_3$ \\
\hline
 $\left(
\begin{array}{ccc}
\times & \times & 0 \\  - &0 & \times \\ 0 & - & \times
\end{array}
\right)$  & $\left(
\begin{array}{ccc}
\times& 0 & \times \\  0 &\times & \times \\ - & - & 0 
\end{array}
\right)$& $\left(
\begin{array}{ccc}
\times & 0 & \times \\  0 & 0 & \times \\ - & - & \times
\end{array}
\right)$  & $\left(
\begin{array}{ccc}
\times & \times& 0 \\  - & \times & \times \\0 & - & 0
\end{array}
\right)$ & 
$\left(
\begin{array}{ccc}
0 & \times & \times \\ - & 0 & \times \\ - & - &\times
\end{array}
\right)$
& $\left(
\begin{array}{ccc}
0 & \times & \times \\  - & \times & \times \\ - & - & 0
\end{array}
\right)$
& $\left(
\begin{array}{ccc}
0 &\times & \times \\ - & \times & 0 \\ - & 0 & \times
\end{array}
\right)$
\\
\hline $C$& 
& & & $F_1$ & $F_2$ & $F_3$ \\
\hline
 $\left(
\begin{array}{ccc}
\times & \times & \times \\ - & 0 & \times \\ - & - & 0
\end{array}
\right)$ & & & & $\left(
\begin{array}{ccc}
\times & 0 & 0 \\  0 & \times & \times \\0 & - & \times
\end{array}
\right)$
& $\left(
\begin{array}{ccc}
\times& 0 & \times \\  0 & \times & 0 \\ - & 0 & \times
\end{array}
\right)$ & 
$\left(
\begin{array}{ccc}
\times &\times & 0 \\ - & \times &0 \\ 0 & 0 & \times
\end{array}
\right)$ \\
\hline
\end{tabular}
\end{scriptsize}
\caption{Possible two-zero textures in the three-generation neutrino flavor scenario.}
\label{tab:two-zero}
\end{small}
\end{center}
\end{table} 
The sum of crosses in each matrix in Eq.~(\ref{eq:onezero}) and Table~\ref{tab:two-zero} gives the number of independent parameters of 5(4) for textures with one (two) zeroes. Textures with more than two independent zeroes appear to be excluded by the experiments\footnote{Extending the flavor space to the 3+1 neutrino framework, 15 out of 210 textures with four-zero textures are allowed \cite{Borah:2016xkc}.} (three independent parameters are not enough to accommodate neutrino data).

In the $3 \times 3$ scenario, among the 15 possible textures, only 7 are phenomenologically allowed \cite{Frampton:2002yf, Dev:2006qe,Xing:2002ta}.
In Ref.~\cite{Guo:2002ei}, nine patterns were compatible with data (two of them only marginally for considered experimental data).  
The results, in general, depend strongly on available data. After measurement of non-zero $\theta_{13}$, the updated analysis can be found in~Refs.~\cite{Fritzsch:2011qv,Liao:2013saa,Grimus:2012zm} and the number of viable textures have been reduced significantly, namely: 

\begin{enumerate}
    \item Class $A$ is allowed only for NH.
    \item Class $B$ is allowed for both NH and IH. $B_1$ and $B_4$ predict negative values of $\cos\delta_{\rm CP}$ whereas the classes $B2$ and $B3$ predict positive values of $\cos\delta_{\rm CP}$.  The textures $B1$ and $B3$ predict $\theta_{23}$ in the lower octant and
    the textures $B2$ and $B4$ predict $\theta_{23}$ in the upper octant for NH. The predictions are opposite for the IH.
    \item Class $C$ class is allowed mainly in the IH.  This class is marginally allowed in the NH when $\theta_{23}$ is close to $45^\circ$. In this class when $\theta_{23} < 45^\circ$, one must have $-90^\circ < \delta_{\rm CP} < 90^\circ$ and when
   $\theta_{23} > 45^\circ$, one must have $90^\circ < \delta_{\rm CP} < 270^\circ$.
   \item The textures in the classes $D$, $E$ and $F$ are forbidden by the data.
\end{enumerate}

 In Ref.~\cite{Dziewit:2006cg} numerical scan over neutrino parameters using adaptive Monte Carlo generator confirmed seven allowed patterns with two zeroes (the CP conserving case) and identified additional cases for the non-degenerated neutrino masses (mass ordered scenarios). 
In Ref.~\cite{Hagedorn:2004ba}, the stability of phenomenological consequences of texture zeros under radiative corrections in the type-I see-saw scenario is discussed.
It has been shown that additional patterns are allowed under certain conditions due to these effects.
Comparing these results with the classification of two-zero textures, three of the six forbidden textures turn out to agree with experimental data due to the renormalization group evolution of the Yukawa couplings  
\[
        \begin{pmatrix}
        \times & 0 & 0 \\ 0 & \times & \times \\ 0 & - & \times
        \end{pmatrix} ,
        \begin{pmatrix}
        \times & 0 & \times \\ 0 & \times & 0 \\ -  & 0 & \times
        \end{pmatrix} ,
        \begin{pmatrix}
        \times & \times & 0 \\ - & \times & 0 \\ 0 & 0 & \times
        \end{pmatrix} .
\]
\noindent
The matrices of the form
\[
        \begin{pmatrix}
        0 & \times & \times \\ - & 0 & \times \\ - & - &\times
        \end{pmatrix} ,
        \begin{pmatrix}
        0 & \times & \times \\ - & \times & \times \\ - & - &0
        \end{pmatrix} ,
        \begin{pmatrix}
        0 & \times & \times \\ - & \times & 0 \\ - & 0 & \times
        \end{pmatrix} ,
\]
remain forbidden. Table~\ref{tab:RadGenSummary}, extracted from Ref.~\cite{Hagedorn:2004ba}, summarizes the situation. In this work the quasi-degenerate cases are also considered, however, they are already excluded by cosmological and $(\beta \beta)_{0\nu}$ data, see Figs.~\ref{fig:mass} and \ref{fig:memee} (so not repeated in Table~\ref{tab:RadGenSummary}).

\begin{table}[h!]
\centering
\renewcommand{\arraystretch}{1.8}
\begin{tabular}{llc}
\hline \\[-5.5ex]
Neutrino masses & Majorana phases & \\
\hline
Normal ordering, $m_1 \approx 0$ & arbitrary &
 \Large $\left(\begin{smallmatrix}
 \makebox[.8ex]{\rule[.1ex]{0ex}{.6ex}$\scriptstyle\cdot$} &\cdot &\cdot
 \\
 \cdot &\makebox[.8ex]{\rule[.1ex]{0ex}{.6ex}$\scriptstyle\cdot$} &\cdot
 \\
 \cdot &\cdot &\makebox[.8ex]{\rule[.1ex]{0ex}{.6ex}$\scriptstyle\cdot$}
 \end{smallmatrix}\right)$
\\
\hline
Inverted ordering, $m_3 \approx 0$ & $\varphi_1\approx\varphi_2$ &
 \Large $\left(\begin{smallmatrix}
 \cdot & \circ & \circ \\
 \circ & \cdot & \cdot \\
 \circ & \cdot & \cdot
 \end{smallmatrix}\right)$
\\
& $\varphi_1\not\approx\varphi_2$ &
 \Large $\left(\begin{smallmatrix}
 \makebox[.8ex]{\rule[.1ex]{0ex}{.6ex}$\scriptstyle\cdot$} &\cdot &\cdot
 \\
 \cdot &\makebox[.8ex]{\rule[.1ex]{0ex}{.6ex}$\scriptstyle\cdot$} &\cdot
 \\
 \cdot &\cdot &\makebox[.8ex]{\rule[.1ex]{0ex}{.6ex}$\scriptstyle\cdot$}
 \end{smallmatrix}\right)$
\\
\hline
\end{tabular}
\caption{Possible positions of radiatively generated texture zeros in
 the neutrino mass matrix, marked by a ``{\Large$\scriptstyle\circ$}''.
 For $\varphi_1\approx\varphi_2\approx\pi$, at most, 3 of the four zeros can
 be produced simultaneously. Table taken from Ref.~\cite{Hagedorn:2004ba} where a convention for Majorana phases in the PMNS matrix is $\phi_{1,2} = -\alpha/2$, see Eq.~(\ref{upmns1}).
}
\label{tab:RadGenSummary}
\end{table}

The question is if models with zeroes in neutrino mass matrix constructions (either in the effective mass matrix $M_\nu$ or in $M_D, M_R$) have anything to do with discrete symmetries. \emph{The answer is positive.}  
There are examples in the literature where texture zeros are related to discrete symmetries. For instance, in Ref.~\cite{Verma:2021koo}, the $A_4$-based texture one-zero neutrino mass model within the inverse seesaw mechanism for DM is discussed.
The obtained effective neutrino mass matrix is in the form
  \begin{equation}
  M_{\nu}= 
  \begin{pmatrix}
    X+X^{'}& 0 & \Delta + \Delta^{'}  \\
     0 & \Delta^{'}   & X^{'}  \\
    \Delta + \Delta^{'} & X^{'} & \Delta^{''}\
   \end{pmatrix}.
 \end{equation}
Not-primed and doubly-primed elements gather the light neutrino mass matrix contributions while primed elements come from the inverse seesaw mass mechanism construction. For more on a connection between mass matrix textures and discrete symmetries and further references, see Ref.~\cite{Zhao:2020cjm} (the inverse neutrino mass matrix with one texture zero and TM mixing) or Ref.~\cite{Chamoun:2023vnn} (textures of neutrino mass matrix with $S_4$ symmetry in which some pairs of mass matrix elements are equal, up to the sign).
{There are also Abelian realizations of phenomenological texture zeros, e.g. for two-zero neutrino textures see Ref.~\cite{GonzalezFelipe:2014zjk} and for inverse seesaw see Ref.~\cite{Camara:2020efq}.}

{Finally, we give some remarks on applying more formal notions coming from matrix theory to neutrino physics, indicating already applied or potential applications of the theory to flavor discrete symmetries, in particular texture zeroes. This is a relatively new subject discussed in the literature. We will consider an inverse eigenvalue
(singular value) problem (IEP), mixing parametrization using Frobenius norm, eigenvector-eigenvalue relations and `magic' mass sum rules. In all cases, we point to possible or explicit (non-Abelian discrete group) connections with texture zeroes. 

We start with general remarks on matrix theory and neutrino physics.  The matrix theory is used in neutrino theory in \cite{Bielas:2017lok}  where theoretical background is settled for mixing matrices linked to singular values and matrix dilations. In \cite{Flieger:2019eor}  new limits on light-heavy mixings using the concept of dilation and sine-cosine decomposition were derived and in [85] - relations between neutrino mass spectrum and mixing (relations among eigenvalues and eigenfunctions) in the context of all kind of seesaw mechanism has been discussed.

Recently an important relation between eigenvalues and eigenvectors has been rediscovered in the context of neutrino physics \cite{Denton:2019ovn, Denton:2019pka}. This relation complements nicely the mentioned above constructive methods, providing a tool for simultaneous studies of mass and mixing matrices. It gives a simple expression for the elements of eigenvectors in terms of eigenvalues
 \begin{equation}
 \vert v_{ij} \vert^{2} \prod_{k=1, k\neq i}^{n}( \lambda_{i}(A) - \lambda_{k}(A)) = 
 \prod_{k=1}^{n-1}( \lambda_{i}(A) - \lambda_{k}(M_{j}))
 \label{val_vec_iden}
 \end{equation}
 where $v_{ij}$ is the $j^{th}$ component of the eigenvector $v_{i}$, $\lambda_{i}(A)$ $i=1,...,n$ are increasingly ordered eigenvalues of the $n\times n$ Hermitian matrix $A$ and $\lambda_{k}(M_{j})$ $k=1,...,n-1$ are eigenvalues of the submatrix $M_{j}$ obtained by deleting $j^{th}$ row and column from $A$. 

 The relation in Eq.~(\ref{val_vec_iden}) provides a powerful tool for studying neutrino physics, allowing for the simultaneous modelling of masses and mixing, i.e. having mass spectrum coming from discrete flavor symmetries, phenomenological predictions for neutrino mixing effects can be studied, and vice versa.

The  above formula is valid only for Hermitian matrices, reducing its applicability to specific models. However, the eigenvector-eigenvalue identity has already been extended to singular values \cite{Xu:2020abu}, allowing us to study neutrino mass and mixing matrices more generally.
In case of Majorana neutrinos, the phase determination of the eigenvectors in \eqref{val_vec_iden}
 is based on the Frobenius covariants \cite{Krivoruchenko:2023tih} defined as
 \begin{align}
F_{i} = \prod_{j=1,j\neq i}^{k} \frac{1}{\lambda_{i} - \lambda_{j}} \left( M_{\pm} M_{\mp} - \lambda_{j} I \right) \,,   
 \end{align}
 where $\lambda_{j}$, for $j=1,\hdots, k$ are eigenvalues of the matrix $M_{\pm} M_{\mp}$ with $M_{\pm}$ being a complex symmetric matrix. In general elements of the mixing matrix can be written as
 \begin{align}
 U_{\alpha j} = e^{i \phi_{\alpha j}} \vert U_{\alpha j} \vert \,,   
 \end{align}
 where $\alpha, j = 1, \hdots, n$.
 The phases can be expressed as
 \begin{align}
 e^{2i\phi_{\gamma j}} = \frac{m_{j} \langle \gamma \vert F_{j+} \vert \gamma \rangle}{\langle \gamma \vert M_{L} F_{j+} \vert \gamma \rangle} \,,    
 \end{align}
 where $F_{j+}$ are Frobenius covariants of the neutrino mass matrix.
 Thus the complete mixing matrix element is given by
 \begin{align}
 U_{\alpha j} = \sqrt{\frac{\langle \gamma \vert M_{L} F_{j+} \vert \gamma \rangle}{m_{j}}} \frac{\langle \alpha\vert F_{j+} \vert \gamma \rangle}{\langle \gamma \vert F_{j+} \vert \gamma \rangle} \,.  
 \end{align}
 In \cite{Chiu:2022qrk} a different issue in neutrino physics has been addressed by using the eigenvector-eigenvalue identity \eqref{val_vec_iden}. Namely, a derivation of the general mixing parameters, given by the rephasing invariant quantities, has been done in terms of eigenvalues of the mass matrix and its minors.}
 
The textures of neutrino matrices can also be studied using IEP.   
It is a method that reconstructs a matrix from a given spectrum \cite{Chu_1998,chu_golub_2002}. This is an important field on its own with many applications. One application which seems natural from the neutrino physics point of view is the reconstruction of the neutrino mass matrix from experimental constraints. Especially important would be one class of IEP, namely an inverse eigenvalue problem with prescribed entries whose goal can be stated as follows: given a set $\mathcal{L} = \lbrace i_{\nu}, j_{\nu} \rbrace_{\nu=1}^{l}$, $1 \leq i_{\nu}, j_{\nu} \leq n$, a set of $l$ values $\lbrace a_{1}, \hdots , a_{l} \rbrace$ and a set of n values $\lbrace \lambda_{1}, \hdots , \lambda_{n} \rbrace$ find a matrix $A \in \mathbb{M}^{n \times n}$ such that
\begin{align}
&\sigma(A) = \lbrace \lambda_{1}, \hdots , \lambda_{n} \rbrace \,, \\
&A_{i_{\nu}j_{\nu}} = a_{\nu} \text{ for } \nu =1 \hdots , l \,,
\end{align}
where $\sigma(A)$ is a spectrum of the matrix $A$. Some of the classical results of IEP are Schur-Horn theorem \cite{schur_1923, horn_1954}, Mirsky theorem \cite{mirsky_1958}, Sing-Thompson theorem \cite{sing_1976, thompson_1977}. For example, the Mirsky theorem says
\begin{theorem}
A square matrix with eigenvalues $\lambda_{1} , \hdots , \lambda_{n}$ and main diagonal elements $a_{1}, \hdots , a_{n}$ exists if and only if
\begin{align}
\sum_{i=1}^{n} a_{i} = \sum_{i=1}^{n} \lambda_{i} \,.    
\end{align}
\end{theorem}
There are many extensions of these classical results, allowing arbitrary location of the prescribed elements, see for example Ref.~\cite{CHU200485}. 
This method can be applied to studies of discrete symmetries in the neutrino sector. As an example one can systematically examine the structure of the mass matrix in the texture-zeros approach, where one would like to reconstruct a matrix with a prescribed spectrum agreeing with the experimental observations and with zero matrix elements in specific locations. However, it is also well suited for studies of more general structures of the neutrino mass and mixing matrices.
Some applications of the IEP in the neutrino sector have already been done, e.g., in Ref.~\cite{Flieger:2020lbg} the discussion of the mass spectrum in the seesaw scenario is given. The application of the inverse singular value problem in the study of neutrino mixing matrices was considered in Ref.~\cite{Flieger:2019eor}.  
 
A different approach, also using tools from the matrix theory, to the modelling of the structure of the neutrino mass matrix has been proposed in Ref.~\cite{Hollik:2017get}, where it has been applied to the Altarelli–Feruglio model \cite{Altarelli:2005yp, Altarelli:2005yx} and its perturbation from the TBM regime. This approach is based on the observation that one of the invariants of the $n \times n$ complex matrix $A$ is a square of the Frobenius norm
\begin{align}
R^{2}=\Vert A \Vert_{F}^{2} = Tr\left( A A^{\dag} \right) = \sum_{i,j}^{n} \vert a_{ij} \vert^{2} \,, 
\label{sphere_frob}
\end{align}
which can be interpreted as an equation of the $n^{2}$-dimensional hyper-sphere. Thus, it makes it natural to express elements of the matrix $A$ in terms of spherical coordinates. For the physically interesting case of $n=3$, elements of the neutrino mass matrix $M_\nu$ (assumed to be real) can be parametrized as
\begin{align}
&M_{11}= R \sin(\chi) \left( \prod_{i=1}^{6} \sin(\phi_{i})\sin(\phi_{7}) \right), \; M_{12}= R \sin(\chi) \left( \prod_{i=1}^{6} \sin(\phi_{i})\cos(\phi_{7}) \right) \,, \nonumber \\
&M_{13}= R \sin(\chi) \left( \prod_{i=1}^{5} \sin(\phi_{i})\cos(\phi_{6}) \right), \;
M_{21}= R \sin(\chi) \left( \prod_{i=1}^{4} \sin(\phi_{i})\cos(\phi_{5}) \right) \,, \nonumber \\
&M_{22}= R \sin(\chi) \left( \prod_{i=1}^{3} \sin(\phi_{i})\cos(\phi_{4}) \right), \;
M_{23}= R \sin(\chi) \left( \prod_{i=1}^{2} \sin(\phi_{i})\cos(\phi_{3}) \right) \,, \nonumber \\
&M_{31}= R \sin(\chi)\sin(\phi_{1})\cos(\phi_{2}) \,, \;
M_{32}= R \sin(\chi)\cos(\phi_{1}) \,, \; 
M_{33}= R \cos(\chi) \,.
\end{align}
As a consequence of this parametrization, we get interrelations between different elements, and also, it is straightforward to produce texture-zeros by a particular choice of angles.
Moreover, the Frobenius norm can also be expressed in terms of singular values
\begin{align}
\Vert A \Vert_{F} = \sqrt{ \sum_{i=1}^{q} \sigma_{i}^{2} } \,,   
\end{align}
where $q$ is the rank of $A$ and similarly to \eqref{sphere_frob} can be interpreted as a $q$-dimensional sphere. 
For the normalized matrix in the case $n=3$,
$\bar{A} = \frac{A}{\Vert A \Vert_{F}}$\,,   
the normalized singular values can then be defined as
$\bar{\sigma}_{1} = \sin \alpha \sin \beta$\,,  $\bar{\sigma}_{2} = \sin \alpha \cos \beta$,  $\bar{\sigma}_{3} = \cos \alpha$,
where $\alpha, \beta \in [0, \frac{\pi}{2}]$. Thus, one can see that only two angles $\alpha$ and $\beta$ are necessary to describe normalized singular values of $\bar{A}$ reflecting the fact that only two independent mass ratios are relevant. As we have seen the $9$-dimensional sphere \eqref{sphere_frob} carries more information than the one given in the singular value space, requiring in total $8$ angles to describe it fully. The additional 6 angles are related to the unitary matrices of the singular value decomposition of $\bar{A}$
\begin{align}
\bar{A} = L^{\dag} \Sigma R \,.    
\end{align} 
Finally, we can express angles $\alpha$ and $\beta$ in terms of singular values as
\begin{align}
&\sin \alpha = \sqrt{\frac{\bar{\sigma}_{1}^{2} + \bar{\sigma}_{2}^{2}}{\bar{\sigma}_{1}^{2} + \bar{\sigma}_{2}^{2} + \bar{\sigma}_{3}^{2}}} \,, \; \sin \beta =  \sqrt{\frac{\bar{\sigma}_{1}^{2}}{\bar{\sigma}_{1}^{2} + \bar{\sigma}_{2}^{2}}} \,.
\end{align}
 
Another interesting realization of discrete flavor symmetries with specific mixing matrices can be realized in the framework of so-called magic matrices. An $n\times n$ matrix $A$ is {\it magic} if the row sums and the column sums are all equal to a common number 
$\alpha$:
\begin{equation}
\sum_{i=1}^nA_{ij}=\sum_{j=1}^nA_{ij}=\alpha.
\end{equation}

Neutrino mass matrix \cite{Bjorken:2005rm} is magic, in the sense that the sum of each column and the sum of each row are all identical, see also   Ref.~\cite{Lam:2006wy}.
Zeros in the magic neutrino mass matrix are discussed in Ref.~\cite{Gautam:2016qyw}. In  Ref.~\cite{Verma:2019uiu}, the magic neutrino mass model within the type-I
and II seesaw mechanism paradigm are investigated. It is based on the neutrino mass model triggered by the $A_4$ discrete flavor symmetry where
a minimal scenario with two right-handed neutrinos and with broken $\mu-\tau$ symmetry  leads to leptogenesis.  

{\underline{Neutrino and quark models \label{sec:neuquarks}}}	

Neutrino and quark sectors are qualitatively different regarding their mixing and mass patterns. However, there are ambitious trials to consider them together. For instances, in Ref.~\cite{Ahn:2011yj}  the quark-lepton complementarity (QLC) relations \cite{Raidal:2004iw,Minakata:2004xt,Hochmuth:2007wq,deAdelhartToorop:2010vtu,Barranco:2010we} are considered with $\theta_{12} +\theta_{12}^q \simeq 45^o$ and $\theta_{23} +\theta_{23}^q \simeq 45^o$. The QLC relations indicate that there could be a quark-lepton symmetry based on a flavor
symmetry.  In  Ref.~\cite{Ahn:2011yj} a discrete symmetry $A_4 \times Z_2$ is considered in the context of charged leptons and quarks, and tribimaximal neutrino mixing, see also Ref.~\cite{Feruglio:2007uu}.  We should mention that there is also an intriguing, the so-called King-Mohapatra-Smirnov (KMS) relation \cite{Minakata:2004xt,Antusch:2005ca,King:2012vj,Farzan:2006vj}  $U = U^{\star}_{CKM} U^{\star}_{TBM}$, which gives $\vert U^{13} \vert \simeq \vert \sin{\theta_C}/\sqrt{2} \vert \simeq 0.156$. For a recent discussion on the possible origin of this relation, see  Ref.~\cite{Albergaria:2023vck}. The KMS relation will also be discussed in section \ref{sec:gut} on the discrete symmetries and the GUT scale. For the application of dihedral groups to the lepton and quark sectors, see also Ref.~\cite{Lu:2019gqp}.

\subsection{Flavor and Generalized CP Symmetries} \label{sec:flavorgcp} 

As discussed earlier (see Eq.~(\ref{upmns1}), the PMNS mixing matrix is characterized by three mixing angles which are well measured as well as three CP phases, namely the Dirac CP phase ($\delta_{\rm CP}$), and the two Majorana phases, which are largely unconstrained. The flavor symmetry approach can predict the three mixing angles, making exploring how CP phases can be predicted similarly attractive. A hint of how this can be achieved can be seen in the transformations given in Eq. (\ref{eq:reflection}) for the $\mu$-$\tau$ reflection symmetry. This overall symmetry operation can be seen as a canonical CP transformation augmented with the $\mu$-$\tau$ exchange symmetry, which will later be a case of the ``generalized" CP transformation. This example predicts both atmospheric mixing angle and $\delta_{\rm CP}$ to be maximal. Therefore, such transformations represent an interesting extension of the discrete symmetry framework discussed earlier with an additional invariance under a CP symmetry.
We will first discuss the basic properties of CP transformations. We then discuss combining CP and flavor symmetry group $G_f$ in a consistent framework. 

The action of a generalized CP transformation $X$ on a field operator is given by 
\begin{equation}
    \varphi'(x_0,\vec{x})\, = \,  X \, \varphi^*(x_0,-\vec{x}) \, ,
\end{equation}
where $X$ is a constant unitary matrix i.e. $XX^\dagger=\mathbb{1}$. The requirement of CP invariance on the neutrino mass matrix $M_\nu$ leads to the condition
\begin{equation}
    X^T M_\nu X \, = \, M_\nu^* \, .
    \label{eq:mnuCP}
\end{equation}
For example, in the case of the $\mu$-$\tau$ reflection symmetry, the mass matrix given in Eq. (\ref{m3b3}) is invariant under  
\begin{equation}
    \mathcal{S}^T {M}_0 \mathcal{S} = {M}^*_0, 
    \label{eq:mnuCPMuTau}
\end{equation}
where the transformation matrix is given by 
\begin{eqnarray}
 \mathcal{S}&=&\left( \label{S}
\begin{array}{ccc}
 1  & 0   & 0 \\
0  & 0   & 1 \\
0 & 1   & 0
\end{array}
\right).
\end{eqnarray}
On comparing Eq.~\eqref{eq:mnuCPMuTau} with Eq.~\eqref{eq:mnuCP}, we can easily identify $\mathcal{S}$ as the CP transformation $X$ in this case. Note that if neutrinos are Dirac particles, condition in Eq.~\eqref{eq:mnuCP} is replaced by $X^\dagger M_\nu^\dagger M_\nu X \, = \, (M_\nu^\dagger M_\nu)^*$. It can be shown explicitly that only a generalized transformation leads to vanishing CP invariants, thus leading to vanishing CP phases \cite{Branco:2011zb,Feruglio:2012cw}. {For general CP transformations, in particular,
in the context of finite groups,
see \cite{Chen:2014tpa,Ratz:2019zak}}.

Now consider the case with both flavor and CP symmetry. The existence of both discrete flavor and generalized CP symmetries determines the possible structure of the generalized CP symmetry matrices. For example, if we consider a discrete flavor symmetry $G_f$ in the lepton sector, the transformation matrix given in Eq. (\ref{S}) satisfy the {\it consistency condition} given by~\cite{Feruglio:2012cw, Holthausen:2012dk}
\begin{equation}
     \mathcal{S} \rho(g)^*  \mathcal{S}^{-1} =\rho (u(g)),
     \label{eq:cons}
\end{equation}
(or equivalently in the general case $S\rightarrow X$) where $u$ is an automorphism of a group of $G_f$ which maps an element $g \in G_f$ into $g'=u(g)\in G_f$ where the latter belong to the conjugacy class of $g^{-1}$. Since this automorphism is class-inverting, it is an outer automorphism\footnote{In an outer automorphism, the automorphism cannot be represented by a conjugation with a group element i.e. $g\rightarrow hgh^{-1}, \, h \notin G_f$} of $G_f$. One can derive this condition by applying a generalized CP symmetry transformation, subsequently, a flavor transformation associated with the group element $g=G_{f}$ and an inverse generalized CP symmetry transformation in order. As the Lagrangian remains unchanged, the resulting transformation must correspond to an element of $G_f$.  This can lead to interesting predictions for the leptonic CP phases (Dirac and Majorana) and the mixing angles.

\begin{figure}[h!]
\centering
\includegraphics[width=0.6\textwidth]{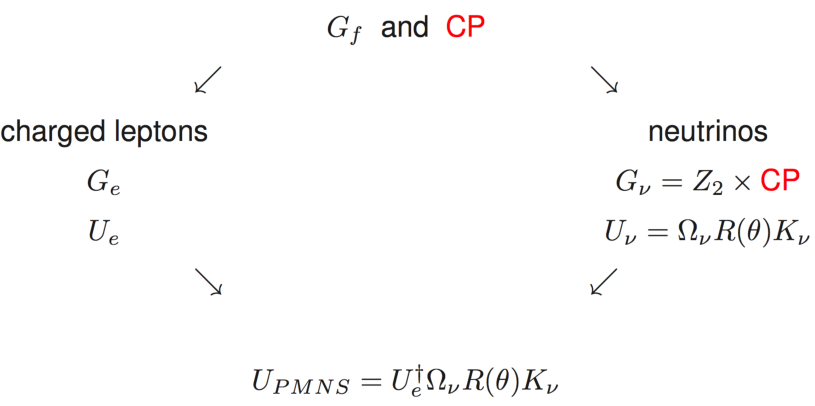}
\caption{General scheme  considered in literature for ways of merging discrete flavor groups with CP effects; figure taken from Ref.~\cite{claudiaMTTD23}, {see also \cite{Hagedorn:2024skt}.} For a concrete example, see Section \ref{sec:delta6pheno} and associated phenomenological studies in the subsequent sections.}
\label{fig:claudia_sketch} 
\end{figure}

Following Ref.~\cite{claudiaMTTD23}, in Fig.~\ref{fig:claudia_sketch}, it is indicated that typically CP symmetry is considered in the neutrino sector.
As an example, we consider $G_f= S_4$ and CP ~\cite{Feruglio:2013hia,Penedo:2018gpg,  Li:2013jya, Li:2014eia,Penedo:2017vtf}. 
Our discussion follows  specifically the case $G_f= S_4$, $G_\nu= Z_2 \times \, CP$ and $G_e= Z_3$ (see ~Ref.~\cite{Feruglio:2013hia} for details). Given this choice of residual group for $G_e$, the only possible choice for a generator of the group is $Q= T$. In this case, it is also found that all choices of generator $Z$ and $X$ are related through similarity transformations to the following three options: $Z=S$, $Z=SU$ and $Z=U$. Using the consistency relation as given in Eq.~\eqref{eq:cons} along with the unitary symmetric nature of $X$, the following forms of CP transformations are (in the 3-dim real representation of $S_4$, see Eq.~\eqref{eq:STUreal}) : $X_1\propto 1$, $X_2\propto S$, $X_3\propto U$, $X_4\propto SU$, $X_5\propto TST^2 S$, $X_6\propto T^2STS$, where $S,T,U$ are the three generators of the group $S_4$. Also note that $X_i$'s being proportional to the generators of $G_f$ is just a coincidence in this case and does not hold generally. 

Few other such examples are listed here for generalized CP symmetry transformation with various discrete groups such as  $\Delta(27)$~\cite{ Nishi:2013jqa,Ohki:2023zsn}, $A_4$~\cite{Ding:2013bpa}, $\Delta(48)$~\cite{Ding:2013nsa,Ding:2014hva}, $\Delta(6n^2)$~\cite{King:2014rwa,Hagedorn:2014wha}, $\Delta(96)$~\cite{Ding:2014ssa}, $A_5$~\cite{Feruglio:2015jfa, DiIura:2018fnk}, $\Delta(3n^3)$~\cite{Ding:2015rwa}.

\subsection{Higher Order Discrete Groups} 

In the above, we have discussed fixed mixed schemes (BM, TBM, GR, HG), which are already ruled out by data and mixing schemes (TM$_{1}$,TM$_{2}$,CBM)  which are consistent with observations.   We have also seen that smaller discrete groups (such as $A_4, S_4, \Delta(27), T_7$) still can explain the correct mixing with 'appropriate adjustments' in the old flavor symmetric models. We will discuss a few aspects of explaining lepton mixing with larger groups. In this technique, we look for new groups that predict a different leptonic mixing pattern. Therefore,  our new method is to start with much higher order groups ($G_f$), which essentially breaks down to two groups $G_e$ and $G_{\nu}$ for charged leptons and neutrino sector.   For example, when we start  
with a discrete group of the order less than 1536, and the residual symmetries are fixed at $G_e=Z_3$ and $G_{\nu}=Z_2 \times Z_3$, the only surviving group which can correct neutrino mixing are $\Delta(6 n^2)$ with $n=10$,   $(Z_{18} \times Z_6)\times S_3$ and $\Delta(6 n^2)$ with $n=16$~\cite{Holthausen:2012wt}. In a more updated study, considering discrete subgroups of $U(3)$, it has been shown that the smallest group which satisfies 3$\sigma$ allowed range of neutrino oscillation data is $\Delta(6n^2)$ with $n=18$, i.e., the order of the group is 1944~\cite{Joshipura:2016quv}. This study  also proposes an analytical formula   to predict full columns of the lepton mixing matrix. 
For a review of similar studies, the reader is referred to~Ref.~\cite{Yao:2015dwa} and references therein.

\subsection{Flavor Symmetry and Grand Unified Theory \label{sec:gut}} 

To have a complete understanding of the flavor problem in particle physics, propositions are there to combine  family  symmetry and Grand
Unified Theory (GUT)~\cite{Georgi:1974sy, Pati:1974yy, Fritzsch:1974nn}. These special versions of  
Grand Unified Theories of Flavor include a discrete flavor symmetry giving GUT predictions through the associated Clebsch factors~\cite{Georgi:1979df, Antusch:2013rxa} explaining the observed lepton mixing. This construction can provide a novel connection between the smallest lepton mixing angle, e.g., as discussed in Section \ref{sec:neuquarks} in the form of the KMS relation \cite{Minakata:2004xt,Antusch:2005ca,King:2012vj,Farzan:2006vj} between the reactor mixing angle  $\theta_{13}$ and the largest quark mixing angle, the Cabibbo angle $\theta_C$:
\begin{equation}
    \theta_{13}=\theta_C / \sqrt{2} \, .
\end{equation}
See other works on the subject ~\cite{Antusch:2013kna, Antusch:2012fb, Marzocca:2011dh,Lu:2018oxc,Albergaria:2023vck}.
The symmetry choice makes several combinations of discrete flavor symmetry and GUT possible. For classification of such models, the readers are referred to the review~\cite{King:2017guk} and references therein for explicit models. These models, however, greatly depend on the fields' symmetry-breaking pattern and vacuum alignment. In addition, such unified constructions have consequences in leptogenesis~\cite{Bjorkeroth:2016lzs, deAnda:2017yeb,Gehrlein:2015dxa,Gehrlein:2015dza, deAnda:2018oik} and at LHC~\cite{Belyaev:2018vkl}. Recently, GUT frameworks have also been implemented in modular invariance approach~\cite{Charalampous:2021gmf, Ding:2021eva, King:2021fhl,Chen:2021zty}.

\subsection{Flavor Symmetry and the Higgs Sector}\label{sec:flavor-higgs}
{There are many possibilities to extend the SM Higgs sector, e.g. by introducing singlet scalar fields, two- (2HDM) and multi-Higgs doublets (NHDM), and triplet multiplets. These models include charged and neutral Higgs bosons with rich phenomenology, including modifications of the SM-like Higgs couplings, flavor-changing neutral currents (FCNC), CP violation effects, and in cosmology, scalar DM candidates, and modification of the phase transitions in the early Universe.  The 2HDM and NHDM doublets as copies of the SM Higgs doublet, in analogy to fermion generations,  is a natural choice. Also, many BSM models, including gauge unification models, supersymmetry, and even string theory constructions, inherently lead to several Higgs doublets at the electroweak scale \cite{Ivanov:2017dad}.}
 However, SM scalar sector extensions lead to many free parameters. For example, the free parameters exceed one hundred in the three-Higgs-doublet model (3HDM). One can impose additional flavor symmetry to reduce the number of free parameters.
To discuss BSM Higgs sectors in the context of flavor symmetries, the starting point is the Yukawa Lagrangian. The imposition of a flavor symmetry on the leptonic part of the Yukawa Lagrangian has been discussed in  Refs. \cite{Lam:2007ev,King:2017guk, Altarelli:2010gt}.
In the SM, the application of family symmetry is limited due to Schur's lemma \cite{Chaber:2018cbi}, which implies that for three-dimensional mass matrices of charged leptons and neutrinos, their diagonalization matrices are proportional to identity. Thus, the PMNS matrix becomes trivial. This drawback can be overcome in two ways. One approach is to break the family symmetry group by a scalar singlet, flavon \cite{King:2013eh}. Despite many attempts, it has failed to reconstruct the PMNS matrix. In the second approach, a non-trivial mixing can be achieved by extending the Higgs sector by additional multiplets \cite{Machado:2010uc,deMedeirosVarzielas:2021zqs, Varzielas:2022lye}. 
The general classification of which symmetry groups can be implemented in the scalar sector of the 2HDM is studied in Refs.  \cite{Nishi:2006tg, Ferreira:2010yh, Ivanov:2005hg}. Analogous analysis of an overall possible set of finite reparametrization symmetry groups in the 3HDM is presented in Ref.~\cite{Ivanov:2012fp}. For the four-Higgs-doublet model, a systematic study of finite non-Abelian symmetry groups that can be imposed on the scalar sector is given in Ref.~\cite{Shao:2023oxt}.

Recent activities in the field of flavor symmetry and the Higgs sector are mostly related to either to new analytical and model building studies (see e.g. Refs.~\cite{Darvishi:2019dbh,Bento:2022vsb,Kalinowski:2021lvw}) or to numerical verification of existing ones. For the second approach, multi-Higgs doublet models were examined, which include two Higgs doublets  (2HDM) \cite{Chaber:2018cbi} and three Higgs doublets (3HDM) \cite{vergeest:2022mqm}.\footnote{{Discrete flavor groups are also considered in composite Higgs models \cite{Goertz:2021ndf,delAguila:2010vg}.}}  Finite, non-Abelian, discrete subgroups of the $U(3)$ group up to order 1035 were investigated to search for specific groups that could explain the masses and mixing matrix elements of leptons. For both 2HDM and 3HDM models, both Dirac and Majorana neutrinos were examined. From the model building point of view, it was also assumed that the total Lagrangian has a full flavor symmetry, but the Higgs potential is only form invariant~\cite{Branco:2011iw,Davidson:2005cw}, meaning that only the scalar potential coefficients may change while the terms in the potential do not vary.  
The scan was performed with the help of the computer algebra system GAP \cite{PhysRevD.92.096010}. A limitation to subgroups with irreducible three-dimensional faithful representations has been performed to reduce an enormous number of evaluated subgroups significantly. {The results were utterly negative for 2HDM \cite{Chaber:2018cbi} { - for such an extension of the SM, up to considered highest dimensional discrete groups, there is no discrete symmetry that would fully match the masses and parameters of the mixing matrix for leptons.}} However, 3HDM provides nontrivial relations among the lepton masses and mixing angles, leading to nontrivial results \cite{vergeest:2022mqm}. Namely, {\it some of the scanned groups provide either the correct neutrino masses and mixing angles
or the correct masses of the charged leptons}. The group $\Delta(96)$ is the smallest group compatible with the
experimental data of neutrino masses and PMNS mixing whereas $S_4$ is an approximate symmetry for Dirac
neutrino mixings, with parameters staying within 3{$\sigma$} of the measured $\theta_{12}, \theta_{23}, \theta_{13}$, and $\delta_{\rm CP}$.
{Thus, phenomenological investigations based on the 3HDM results or further theoretical studies of discrete groups beyond 3HDM are worth future studies.}

 \subsection{Modular Symmetry}

An appealing proposal to understand the flavor pattern of fermions {{is}} the idea of modular symmetry~\cite{Altarelli:2005yp,Feruglio:2017spp}. 
In conventional approaches, a plethora of models exist based on non-Abelian discrete flavor symmetries and finite groups. The spectrum of the models is so large that it is difficult to obtain a clear clue of the underlying flavor symmetry. Additionally, there are a few major disadvantages to using this conventional approach.  
Firstly, the effective Lagrangian of a typical flavor model includes a large set of flavons. Secondly, a {typical} vacuum alignment of flavons essentially determines the flavor structure of quarks and leptons, {severely affecting the model predictions}. Often, auxiliary symmetries are also needed to forbid unwanted operators from contributing to the mass matrix.  The third and most crucial disadvantage of conventional approaches is the breaking sector of flavor symmetry, bringing many unknown parameters, hence compromising the minimality. On the contrary, the primary advantage of models with modular symmetry~\cite{ deAdelhartToorop:2011re, Feruglio:2017spp} is that the flavon fields might not be needed, and the flavor symmetry can be uniquely broken by the vacuum expectation value of the modulus $\tau$. {This modular invariance approach added new avenues in flavor model building, pointing towards a deep theoretical origin such as string compactifications, see references in~\cite{Feruglio:2019ybq, Feruglio:2017spp, Ohki:2020bpo}.  Geometrically, modular symmetry can be considered compact spaces such as torus $T^2$  and orbifolds $T^2/Z_2$.  The modular invariance hypothesis assumes the action is invariant under the modular group and so far in the literature, only supersymmetric modular invariant theories have been proposed.
In this construction, along with the matter fields (such as leptons, quarks, and the Higgs fields), the coupling parameters also transform under the modular group.} 
Here, the Yukawa couplings are written as modular forms, functions of only one complex parameter i.e., the modulus $\tau$, {which} {transforms}  non-trivially under the modular symmetry. Furthermore, all the higher-dimensional operators in the superpotential are completely determined by modular invariance if supersymmetry is exact; hence auxiliary Abelian symmetries are not needed in this case. Like  {in} the conventional model-building approach, models with modular symmetry can also be highly predictive. The fundamental advantage is that the neutrino masses and mixing parameters can be  {expressed by} a few input parameters. {In the following, we briefly review the modular symmetry within the framework of $N=1$ global supersymmetry. The action $\mathcal{S}$ of the chiral superfields can be written as 
\begin{eqnarray}
    \mathcal{S}=\int d^4xd^2\theta d^2 \bar{\theta}K(\chi^{I},\bar\chi^I, \tau,\bar\tau)+\left[\int d^4xd^2\theta W(\chi^{I},\tau)+h.c.\right]\label{eq:modS}
\end{eqnarray}
where $ K(\chi^{I},\bar\chi^I, \tau,\bar\tau)$ is the K\"{a}hler potential, a real gauge invariant function of the chiral superfields $\chi^I$ and $\tau$ is a complex parameter known as modulus and their Hermitian conjugates are $\bar\chi^I, \bar\tau$. On the other hand the superpotential $W(\chi^{I},\tau)$ is a holomorphic function of  $\chi^I$ and $\tau$. The integration goes over both space-time coordinates $x$ and $\theta, \bar\theta$ are Gra{\ss}mann variables. Modular symmetry is the invariance of the action $\mathcal{S}$  under modular transformation and respects the SM (or GUT) gauge symmetry. The modular group $\bar \Gamma$ is the group of a linear transformation $\gamma$ acting on the modulus $\tau$ belonging to the upper-half complex plane as 
\begin{eqnarray}
    \tau\rightarrow \tau'=\gamma \tau = \frac{a \tau +b}{c \tau + d},  
\end{eqnarray}
where $a,b,c,d \in Z$ and $ad-bc=1, {\rm Im}(\tau) >0$. Here the modular group $\bar \Gamma$ is isomorphic the projective special linear group $PSL(2,Z)=SL(2,Z)/Z_2$, where $Z_2=\{I, -I\}$, $I$ is the identity element and $SL(2,Z)$ is the special linear group of $2\times 2$ matrices with integer identity. $PSL(2,Z)$ is also known as the inhomogeneous modular group whereas $SL(2,Z)$ is called {\it the homogeneous modular group} or {\it  the full modular group}.  By two transformations $S$ and $T$ satisfying 
\begin{eqnarray}
  S^2=(ST)^3=I\label{eq:modST},  
\end{eqnarray}
the modular transformation can be generated by these generators represented as 
\begin{eqnarray}
     S=\begin{pmatrix}
    0 & 1  \\
    -1 & 0 
  \end{pmatrix},\quad T=\begin{pmatrix}
    1 & 1  \\
    0 & 1 
  \end{pmatrix} .
\end{eqnarray}
The generators $S$ and $T$ are also referred to as modular inversion and translation, under which we obtain 
\begin{eqnarray}
    \tau \xrightarrow[]{\text{S}}-\frac{1}{\tau}, \quad \tau \xrightarrow[\text{}]{\text{T}}\tau+1. 
\end{eqnarray}
The modular group $\Gamma$ has a series of infinite normal subgroups $\Gamma(N),\; N=1,2,3..$, called principle congruence group and  defined as
\begin{eqnarray}
    \Gamma(N)=\left\{\begin{pmatrix}
    a & b  \\
    c & d 
  \end{pmatrix} \in SL(2,Z),\quad\begin{pmatrix}
    a & b  \\
    c & d 
  \end{pmatrix}=\begin{pmatrix}
    0 & 1  \\
    1 & 0 
  \end{pmatrix}\quad({\rm mod}~~ N)\right\}.
\end{eqnarray}
For $N=2$, we define $\bar\Gamma(2)\equiv\Gamma(2)/\{I,-I\}$. Therefore, for $N>2$,  one can obtain $\bar\Gamma(N) \equiv\Gamma(N)$  as $-I$ do not belong to $\Gamma (N)$. Here the elements of $\bar\Gamma(N)$ are in one-to-one correspondence with the associated linear transformations. The quotient groups, defined as
\begin{eqnarray}
   \Gamma_N\equiv PSL(2,Z)/\bar\Gamma(N)\equiv\bar\Gamma/\bar\Gamma(N)
\end{eqnarray}
}
\{are the inhomogeneous  finite modular groups, isomorphic to $PSL(2,Z)$ where along with the condition in Eq.~(\ref{eq:modST}), it also satisfy $T^N=I$. Interestingly,  the finite modular groups $\Gamma_N$ with $N=2,3,4,5$ are  isomorphic to the permutation groups $S_3$, $A_4$, $S_4$, $A_5$, respectively. As stated earlier, such non-Abelian discrete flavor symmetric groups are widely used in lepton flavor models. This is why just like conventional flavor models,  frameworks based on modular symmetries are also predictive.}

{Now, modular forms of level $N$ are holomorphic functions $f(\tau)$ transforming under the action of $\bar\Gamma(N)$ as 
\begin{eqnarray}
    f(\gamma\tau)=(c\tau+d)^kf(\tau), \quad\gamma =\begin{pmatrix}
    a & b  \\
    c & d 
  \end{pmatrix}\in \bar\Gamma(N)
\end{eqnarray}
where $k(\geq 0)$ is known as modular weight and $(c\tau+d)^k$ is known as automorphy factor. Note that modular forms of level $N$ and weight $k$ form a linear space in finite dimension. Thus, if  $f_{1,2}(\tau)$ are modular forms of level $N$ and  weight $k$ then $f_1(\tau)+f_2(\tau)$ is also a modular form of level $N$ and  weight $k$. Hence, the set of the modular forms of level $N$ and weight $k$ forms a vector space denoted by $Mod^{(N)}_k$. Similarly, if $f(\tau)$ is a  modular form of level $N$ and weight $k$, $f(\gamma \tau)$ is also a modular form of level $N$ and weight $k$. This is true even if $\gamma \not\in \Gamma(N)$. Therefore, it is possible to choose a basis such that a transformation of a set of modular forms $f_i(\tau)$ is
\begin{eqnarray}
    f_i(\gamma\tau)=(c\tau+d)^k\rho(\gamma)_{ij}f_j(\tau),\quad \gamma \in \bar \Gamma, \label{eq:modfi}
\end{eqnarray}
where $\{f_i\}$ is a basis of $Mod^{(N)}_k$, and $\rho$ is a unitary representation of $\Gamma (N)$. This is the basic building block of the modular invariance approach to the flavor problem as formulated in Ref.~\cite{Feruglio:2017spp}. 
Now, to construct modular invariant action, we need modular
transformations for chiral superfields. If the chiral superfields $\chi^I$ in Eq.~(\ref{eq:modS}) are assumed to have modular form at level $N$ with weight $-k$ under certain  representation, following Eq.~(\ref{eq:modfi}) we can write the relevant transformation as~\cite{Feruglio:2017spp} 
\begin{eqnarray}
 \tau\rightarrow\gamma \tau = \frac{a \tau +b}{c \tau + d},\quad \chi^I \rightarrow   (c\tau+d)^{-k} \rho^I(\gamma) \chi^I
\end{eqnarray}}
{Under this transformation, the action $\mathcal{S}$ in Eq.~(\ref{eq:modS}) demands invariance of the superpotential $W(\chi^{I},\tau)$ while the K\"{a}hler potential  $ K(\chi^{I},\bar\chi^I, \tau,\bar\tau)$ allowed to change by a K\"{a}hler transformation, written as~\cite{Wess:1992cp}   
\begin{eqnarray}
    &W(\chi^{I},\tau) \rightarrow W(\chi^{I},\tau), 
    &K(\chi^{I},\bar\chi^I, \tau,\bar\tau) \rightarrow K(\chi^{I},\bar\chi^I, \tau,\bar\tau)  +F(\chi^{I},\tau) + \bar F(\bar\chi^{I},\bar\tau). 
\end{eqnarray}
An example of $ K(\chi^{I},\bar\chi^I, \tau,\bar\tau)$ which satisfies such condition can be written as~\cite{Feruglio:2017spp} 
\begin{eqnarray}
    K(\chi^{I},\bar\chi^I, \tau,\bar\tau) =-h\Lambda^2{\rm log}(-i\tau+i\bar\tau)+\sum_{I}\frac{|\chi_I|^2}{(-i\tau+i\bar\tau)^{k_I}}
\end{eqnarray}
where $h$ is a positive constant. As pointed out in Ref.~\cite{Chen:2019ewa} K\"{a}hler potential is not tightly constrained by the modular symmetry, there may be additional terms consistent with modular symmetry hence reducing the predictivity of the model. However,  this can be addressed by the top-down string-motivated models and eclectic flavor groups~\cite{Nilles:2020nnc,Ohki:2020bpo,Nilles:2020kgo}. }
With this setup, the superpotential $W(\chi^{I},\tau$) is also invariant under the modular transformation and can be expanded in terms of the supermultiplets $\chi^{(I^{}_{i})}_{}$ (for $i = 1, \cdots, n$) as
\begin{eqnarray}
W(\chi^{I},\tau)= \sum_{n}^{}\sum_{\{I^{}_{1},\dots,I^{}_{n}\}}^{} Y^{}_{I^{}_1\dots I^{}_n}(\tau)\chi_{}^{(I^{}_1)}\cdots\chi_{}^{(I^{}_n)} \; ,
\end{eqnarray}
where the modular forms transform as 
\begin{eqnarray}
Y^{}_{I^{}_1\dots I^{}_n}(\tau) \rightarrow (c\tau+d)^{k^{}_Y}_{} \rho^{}_{Y} (\gamma) Y^{}_{I^{}_1 \dots I^{}_n}(\tau). \; 
\end{eqnarray}
Here the coefficients $Y^{}_{I^{}_1\dots I^{}_n}(\tau)$ take the modular forms of weight $k_Y$, level $N$ of the finite modular group, which are the key elements of the modular symmetry approach. In the theory, $k^{}_Y$ and $\rho^{}_{Y}$ must satisfy 
\begin{eqnarray}
    k^{}_{Y} = k^{}_{I^{}_1} +  \cdots + k^{}_{I^{}_N}, \quad \rho_Y \otimes \rho_{I_1} \otimes ...\otimes \rho_{I_n}, 
\end{eqnarray}
which can be used to constrain the charge assignments of superfields and modular forms. As mentioned earlier finite modular groups  of different levels ($N$) are isomorphic to permutation  groups.  
For example,  {for} the modular group $\Gamma^{}_{4} \simeq S^{}_{4}$, the functions $Y (\tau)$ are modular forms of the level $N = 4$ and weight 2  {with} five linearly independent modular forms ($Y^{}_i(\tau)$ for $i = 1, 2, \cdots, 5$).  These five linearly independent forms $Y^{}_i(\tau)$ arrange themselves into two irreducible representations of $ S^{}_{4}$, a doublet $2$ and a triplet $3'$ which can be written as~\cite{Penedo:2018nmg} 
\begin{eqnarray}
Y^{}_{\bf 2}(\tau) \equiv \left(\begin{matrix} Y^{}_{1}(\tau) \\ Y^{}_2 (\tau)  \end{matrix}\right) \; , \quad Y^{}_{3^{\prime}_{}} (\tau) \equiv  \left(\begin{matrix} Y^{}_{3}(\tau) \\ Y^{}_4 (\tau) \\ Y^{}_{5} (\tau) \end{matrix}\right). \; 
\end{eqnarray}
The modular forms $Y^{}_{2}(\tau)$ and $Y^{}_{3^{\prime}_{}} (\tau)$  dictate the flavor structure of the charged lepton and neutrino mass matrices, controlling the neutrino masses and mixing pattern. There are already several activities adopting this approach, for a few examples see Refs.~\cite{Kobayashi:2018vbk, Penedo:2018nmg, Kobayashi:2018wkl, Kobayashi:2021pav,deAnda:2018ecu, Okada:2018yrn, Novichkov:2018ovf, Ding:2019xna, Wang:2020dbp, Wang:2019ovr, Wang:2019xbo, Wang:2020lxk, Wang:2021mkw, Ding:2021zbg, Mishra:2023cjc} and references therein. {Since its inception in 2017~\cite{Feruglio:2017spp},  both bottom-up and top-down modular invariance approaches are being studied to address issues related to quark and lepton flavors. Although a vast majority of literature is based on bottom-up consideration, in recent times top-down approaches are also getting deserved attention. For discussions on eclectic flavor symmetry and modulus stabilization see Refs.~\cite{Nilles:2020nnc,Ohki:2020bpo,Nilles:2020kgo,Ishiguro:2020tmo,Knapp-Perez:2023nty,King:2023snq}. In Section \ref{sec:flavorgcp}, we have discussed combinations of discrete flavor and generalized CP
symmetries leading towards interesting predictions for resudal symmetries and associated phases. Generalized CP can also be extended to the modular symmetry, pointing towards a unified picture of CP~\cite{Baur:2019kwi,Novichkov:2019sqv}. In this case, the consistency conditions invoke the modulus to transform as $\tau\rightarrow -\tau^{*}$ up to modular transformations under the action of CP symmetry.   Such a framework reduces the number of free parameters in modular-invariant models as well as the allowed values for the neutrino oscillation parameters~\cite{Novichkov:2019sqv}. Mathematically consistent theories for modular symmetry involve extra dimensions. Compactifications of these extra dimensions lead to multiple moduli ($\tau_i$) instead of a single modulus field ($\tau$) assumed in bottom-up models. Hence it is essential to build a generalized framework for modular symmetric flavor models. This can be achieved considering factorizable moduli~\cite{deMedeirosVarzielas:2019cyj}, symplectic modular invariance~\cite{Ding:2020zxw} approaches. One of the drawbacks of many modular symmetry models is that the charged lepton mass hierarchy is explained by introducing one free parameter for each family, which is fitted from the observation. This issue can be addressed by introducing additional  singlet~\cite{King:2020qaj} or triplet~\cite{Criado:2019tzk}  flavon fields. Unfortunately, these are the fields that we tried to get rid of in the modular invariance approach. Earlier, in Section~\ref{sec:massmodels}, we discussed the concept of texture zero in neutrino mass matrices which leads to constrained predictions. In the bottom-up approach,  by adjusting the weights of the matter fields ($k$) the texture zeros in the fermion matrices can be achieved~\cite{Lu:2019vgm, Ding:2022aoe, Zhang:2019ngf}. Modular symmetry can also be implemented in various GUT frameworks, e.g.  $SU(5)$~\cite{deAnda:2018ecu}, $SO(10)$~\cite{Ding:2021eva} or flipped $SU(5)$~\cite{Charalampous:2021gmf}.}  As modular symmetric models make it possible to understand fermionic mixing, there is a scope for phenomenological studies of mixings and possible connections with matter-antimatter asymmetry and DM~\cite{Wang:2019ovr, Nomura:2023kwz, Kang:2022psa, Nomura:2022hxs}. However, testing the modular symmetry in the context of current and future neutrino experiments like T2HK JUNO,  NO$\nu$A, and DUNE, makes it very hard to obtain robust correlations among the neutrino oscillation parameters. A discussion in this direction can be found in Section \ref{sec:oscexp}.  Nonetheless, the sum rules obtained in models based on modular symmetries with residual symmetries may shed some light in this regard~\cite{Gehrlein:2020jnr}. {For a comprehensive understanding of modular symmetry and detailed explicit frameworks, the readers are referred to the recent review~\cite{Ding:2023htn} and references therein.} 

\subsection{Flavor Symmetry Breaking and Flavon VEV alignment}\label{sec:flavon-VEV}

{In non-Abelian discrete flavor symmetric models it is also important to explain observed neutrino masses and mixing issues related to the neutrino mass generation mechanism and breaking of the considered flavor symmetry $G_f$. In Section~\ref{sec:massmodels},  we have discussed various mechanisms of neutrino mass generation, which yields an effective light neutrino mass matrix after EW symmetry breaking. Generically, Dirac Yukawa and Majorana couplings are dynamically generated from the operators invariant under $G_f$ with or without the involvement of the flavon fields, discussed earlier. These flavon fields transform as a singlet under SM gauge symmetry. In Section~\ref{sec:genfw}, we have already mentioned the Abelian $U(1)_F$ flavor symmetry, which first illustrated the observed quark mixing and mass hierarchies~\cite{Froggatt:1978nt}. In this mechanism, a flavon field $\Theta$ breaks the $U(1)_F$ symmetry by acquiring a vacuum expectation value. If we consider $\Lambda$ to be the cutoff scale of the considered $U(1)_F$ symmetry, one can define a small dimensionless parameter $\epsilon=\langle \Theta \rangle/\Lambda < 1 $. This $U(1)_F$  symmetry is then assigned in such a way that the different powers of the expansion parameter $\epsilon$ explain the hierarchy of the associated fermion masses. As already pointed out, despite its simplicity, such Abelian flavor symmetries are not equipped enough (compared to their non-Abelain counterparts) to explain non-trivial neutrino mixing as well as various mixing schemes. The flavons involved in non-Abelain discrete flavor symmetries usually have masses larger than the EW scale, introducing a higher energy scale ($\Lambda$) in the theory. The vacuum expectation values of these flavons determine the energy scale at which the symmetry is broken. To avoid introducing additional scales into the theory, an alternative approach has been explored where both the flavor and electroweak symmetries are broken simultaneously. This is achieved by incorporating multiple copies of the SM Higgs doublet that transform non-trivially under the flavor group. The readers are referred to  Section~\ref{sec:flavor-higgs} for more details on the symmetry breaking of multi-Higgs models. In the following, we will mainly discuss the scenario where flavor symmetry $G_f$ is broken at a higher scale compared to the EW scale. }

{As pointed out in Section~\ref{sec:genfw}, the flavor symmetry $G_f$ is spontaneously broken down to different subgroups in the charged lepton $G_e$ and neutrino sector $G_{\nu}$. In many cases, to explain correct neutrino masses and mixing flavon fields are introduced which transforms non-trivially under $G_f$ and acquires vacuum expectation values in different, dictating the flavon VEV alignment. Let us illustrate this mechanism with a simple example~\cite{Altarelli:2005yx} in the context of vastly used $A_4$  non-Abelian discrete symmetry. Here two triplet  (namely, $\varphi_T$ and $\varphi_S$) and a singlet ($\xi$) $A_4$ flavons are considered. In this framework, the lepton doublet transforms as a triplet whereas the charged leptons transform as $1,1''$ and $1'$, respectively. The flavon fields introduced here and in flavor models are considered to be gauge singlet matter fields. With this, the contributions in charged lepton and neutrino mass matrices can be written as~\cite{Bhattacharya:2016lts} 
\begin{eqnarray}
    -\mathcal{L}=y_{\alpha_i}\bar\ell\varphi_TH \alpha_i+\ell H\ell H (y_1\xi-y_2\varphi_S)+h.c.,
\end{eqnarray}
where$\alpha_i$'s are the right-handed charged leptons and $y$'s are respective coupling constants. When the triplet flavon fields developed VEV in the direction 
\begin{eqnarray} \label{eq:flavonVEV}
\langle \varphi_T \rangle \propto (v_T,0,0)^T, \quad \langle \varphi_S \rangle   \propto (v_S,v_S,v_S)^T, 
\end{eqnarray}
the $A_4$ symmetry is broken down to $G_e=Z_3$ and $G_\nu=Z_2$ in the charged lepton and neutrino sectors, respectively. These mismatched residual symmetries account for the associated neutrino mixing.  The obtained mass matrix can be diagonalized by the TBM mixing~\cite{Harrison:2002er} mentioned in Eq.~(\ref{eq:TBM}). Similar vacuum alignment can also be considered where deviations from TBM (to include nonzero $\theta_{13}$\footnote{For a discussion on vacuum alignment with less constrained residual symmetries and hierarchy of the elements of neutrino mass matrix, see~\cite{King:2013eh}.}) is exercised~\cite{Karmakar:2014dva}. Such vacuum alignment must have a dynamical origin, and often auxiliary or shaping symmetries  (which also forbids $\varphi_T \leftrightarrow \varphi_S$ exchange in the Lagrangian) included to achieve this. Finding appropriate vacuum alignment in flavor symmetric constructions is quite challenging.  In non-supersymmetric models,  the most general scalar potential constructed by the two $A_4$ triplet flavon does not allow the desired VEV configuration mentioned in Eq.~(\ref{eq:flavonVEV}).  Such vacuum alignment problem is not exclusive to $A_4$ symmetry, but a general shortcoming of many discrete flavor symmetry groups.  For example, the presence of a $A_4$ invariant contribution $(\varphi^{\dagger}_T\varphi_T)(\varphi^{\dagger}_S\varphi_S)$ in the flavon potential forbids us to achieve the minima.  
By minimizing the flavon scalar potential w.r.t. $\varphi_T$ and $\varphi_S$, one can obtain six minimization conditions that we would like to satisfy in terms of only two unknown $v_T$ and $v_S$ written in Eq.~(\ref{eq:flavonVEV}). Hence, a higher number of algebraically independent minimization conditions only leads to more fine-tuned relations among the potential parameters. Fortunately, there are various ways to obtain the desired vacuum alignment discussed in literature~\cite{King:2013eh}. In models with extra dimensions, this can be achieved by localizing two flavons in different branes~\cite{Altarelli:2005yp} (hence suppressing the cross-couplings) or with an explicit breaking of flavor symmetry via boundary conditions~\cite{Kobayashi:2008ih}. In supersymmetric theories, if flavor symmetry breaking scale is higher than the supersymmetry breaking, the appropriate flavor VEV can be obtained via the $F$-terms of the driving fields~\cite{Altarelli:2005yx, deMedeirosVarzielas:2005qg}. If the $D$-terms can also play an important role here if the flavons transform under a new gauge symmetry~\cite{Ross:2004qn}. Correct vacuum for spontaneously broken flavor symmetries can also be obtained by introducing a soft supersymmetry breaking term~\cite{Feruglio:2009iu}. In non-supersymmetric theories, by embedding the flavor symmetry responsible for neutrino mixing in a higher larger group it is also possible to obtain proper vacuum alignment~\cite{Holthausen:2011vd, Krishnan:2019ftw}. Considering multiple flavons contraction on higher-order operators flavon VEV can also be realized through the mechanism known as effective alignments~\cite{deMedeirosVarzielas:2017hen, DeMedeirosVarzielas:2019xcs}. Recently, in Ref.~\cite{Hagedorn:2023mrg}, the authors have shown that in the case of softly broken supersymmetry appropriate vacuum alignment (up to a rescaling) can be employed in low-energy non-supersymmetric theories.}

\subsection{Continuous Flavor Symmetry Models}

{So far, we have only focused on the discrete flavor symmetries. There is also a vast literature on flavor models based on continuous symmetries. The simplest candidate is an Abelian flavor symmetry $U(1)_F$, with Froggatt-Nielson~\cite{Froggatt:1978nt} and horizontal symmetries~\cite{Leurer:1992wg, Leurer:1993gy} being concrete examples. However, its gauging is challenging because of the anomaly cancellation conditions, which require additional degrees of freedom, like right-handed neutrinos, vector-like fermions, or supersymmetry~\cite{Ibanez:1994ig, Allanach:2018vjg, Costa:2019zzy,Smolkovic:2019jow, Tavartkiladze:2022pzf}. Another possibility is flavor-specific $U(1)$ models, such as $U(1)_{L_\alpha-L_\beta}$~\cite{He:1991qd, Foot:1994vd, Choubey:2004hn, Ota:2006xr, Heeck:2011wj, Araki:2012ip, Araki:2019rmw} and $U(1)_{L_e-L_\mu-L_\tau}$~\cite{Petcov:1982ya, Altarelli:2005pj, Meloni:2011ac, Arcadi:2022ojj}, which are anomaly-free even within the SM.  

The size of the $U(n)$ group depends on the gauge group of the model, e.g. in the framework of the SM, the maximal symmetry of the kinetic terms of the fermions is $U(3)^5$, while it is reduced to $U(3)$ in the case of $SO(10)$ in which all fermions of one generation including the right-handed neutrino are unified into a 16-dimensional representation. The presence of more than one generation can only be explained with a non-Abelian symmetry which has two- and three-dimensional irreducible representations. Therefore, many models tried to go beyond the $U(1)$ to be more realistic. For instance, in Ref.~\cite{Barbieri:1996ww}, supersymmetric models with a $U(2)$ group are discussed in the context of $SU(5)$ and $SO(10)$ GUTs; in these models however, only $\theta_{23}$ is in general large, while $\theta_{12}$ and $\theta_{13}$ are both small. A more realistic non-supersymmetric $U(2)$ model of flavor was recently proposed in Ref.~\cite{Linster:2018avp}. Models with $SO(3)$~\cite{King:2005bj, Reig:2018ocz}, $SU(2)$~\cite{Shaw:1992gk, Darme:2023nsy, Greljo:2023bix} and $SU(3)$~\cite{King:2003rf, deMedeirosVarzielas:2005ax, Grinstein:2010ve} flavor symmetries are more promising, as they allow to unify all three generations, but they must be strongly broken.

 The discrete symmetries have certain advantages over the continuous ones. Firstly, the former in general contain several smaller  representations suitable to host the three fermion generations. Secondly, it is well-known that the spontaneous breaking of a continuous global (local) symmetry leads to the appearance of Goldstone (gauge) bosons. Thirdly,  the vacuum alignment is in general achieved in a simpler way with discrete flavor symmetries. For all these reasons, we mostly focus on the  discrete symmetry groups here.   }

 \section{Flavor Symmetry at Intensity Frontier \label{sec:intensity}}

Flavor symmetry can potentially provide an important link between the outstanding puzzle of neutrino mass-mixing with various other aspects of particle physics, cosmology, and astroparticle physics, such as neutrinoless double-beta decay, lepton flavor violating decays, the nature of DM,  baryon asymmetry of the Universe, nonstandard interactions, etc. If a flavor symmetry connects these seemingly uncorrelated sectors,  the constraints from various cosmological, collider, and neutrino experiments may probe the existence of such symmetry. Such connections and correlations allow us to probe discrete flavor symmetries at many frontiers. In this Chapter we discuss the so-called intensity frontier experiments which include low-energy rare processes like neutrinoless double beta decay, LFV, and long-baseline neutrino experiments. Since these rare LNV/LFV processes are predicted to be either absent or highly suppressed in the SM, even a single unambiguous event detection would signal new physics.  

\subsection{Neutrino Oscillation Experiments}\label{sec:oscexp}

\begin{figure}[h!]
    \centering
    \includegraphics[width=0.4\textwidth]{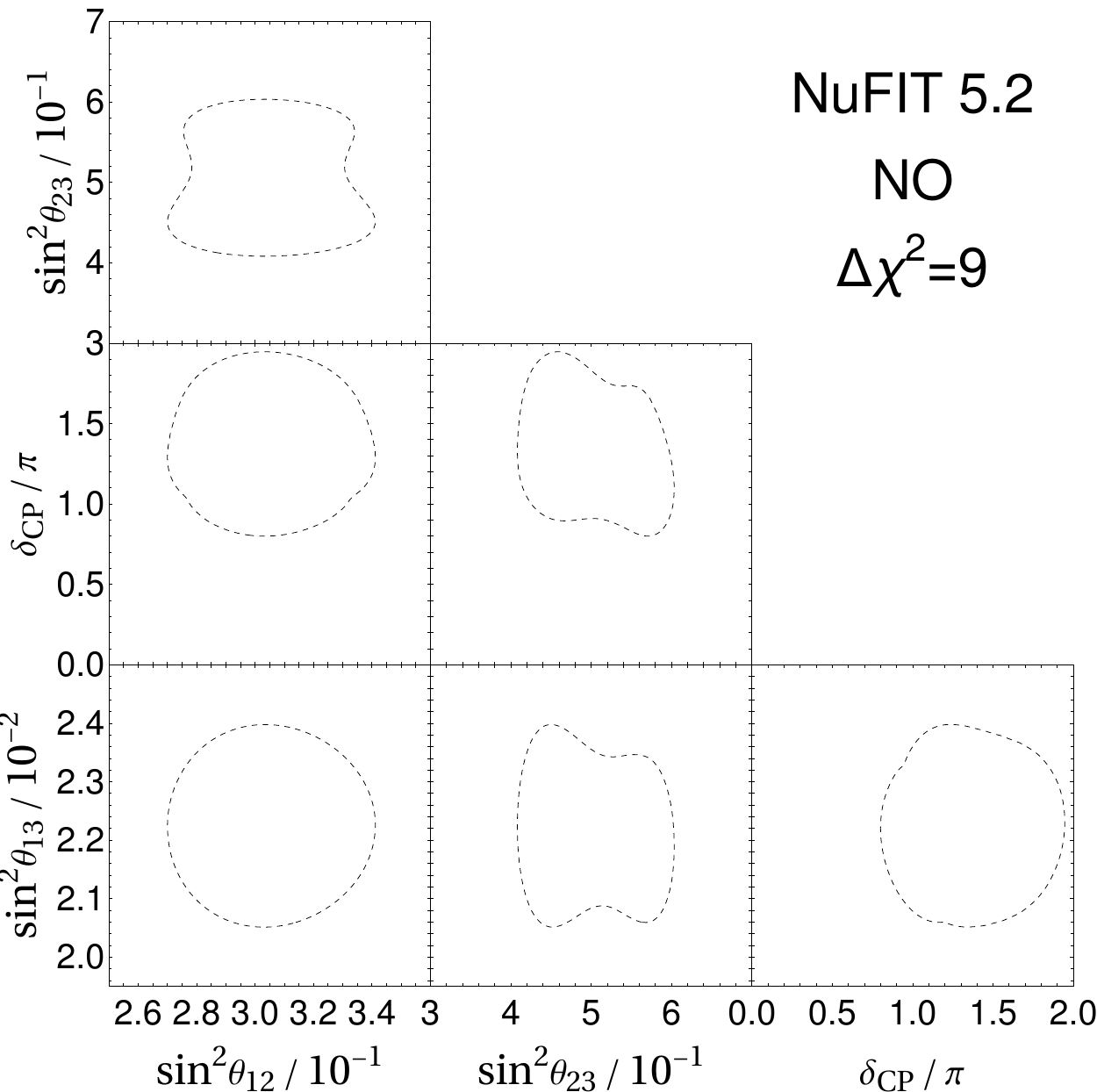}
    \includegraphics[width=0.4\textwidth]{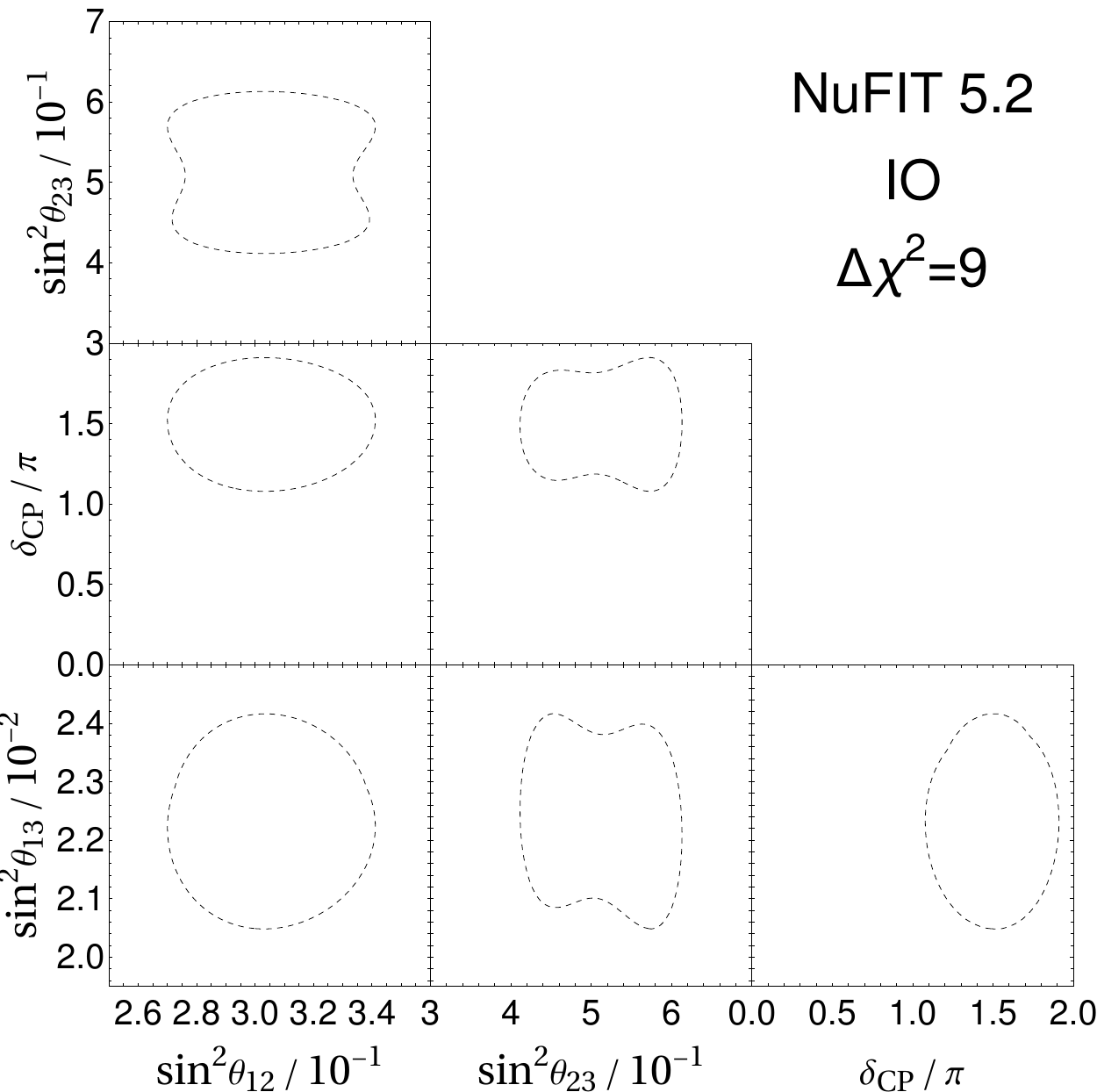}
       \includegraphics[width=0.4\textwidth]{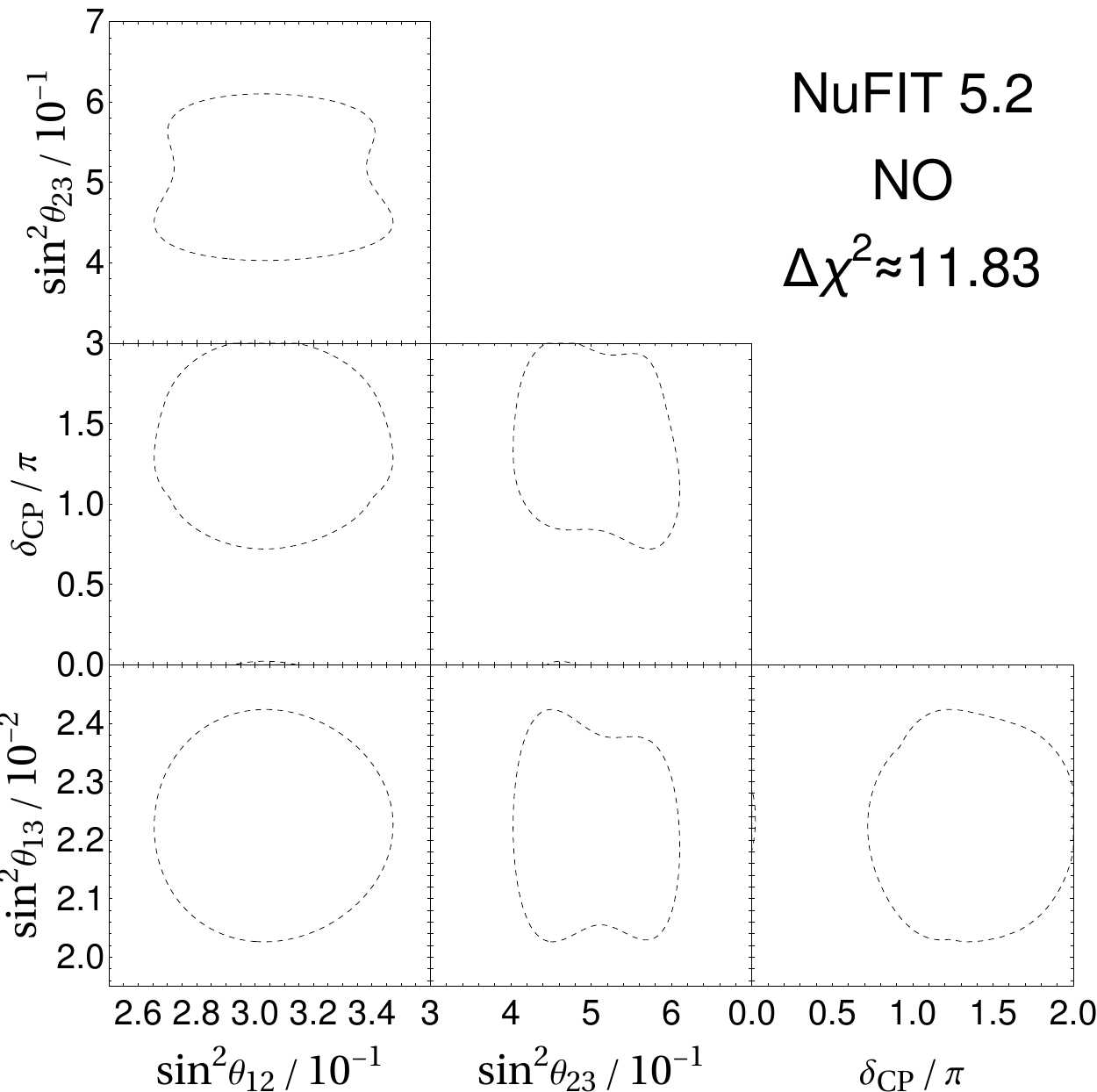}
    \includegraphics[width=0.4\textwidth]{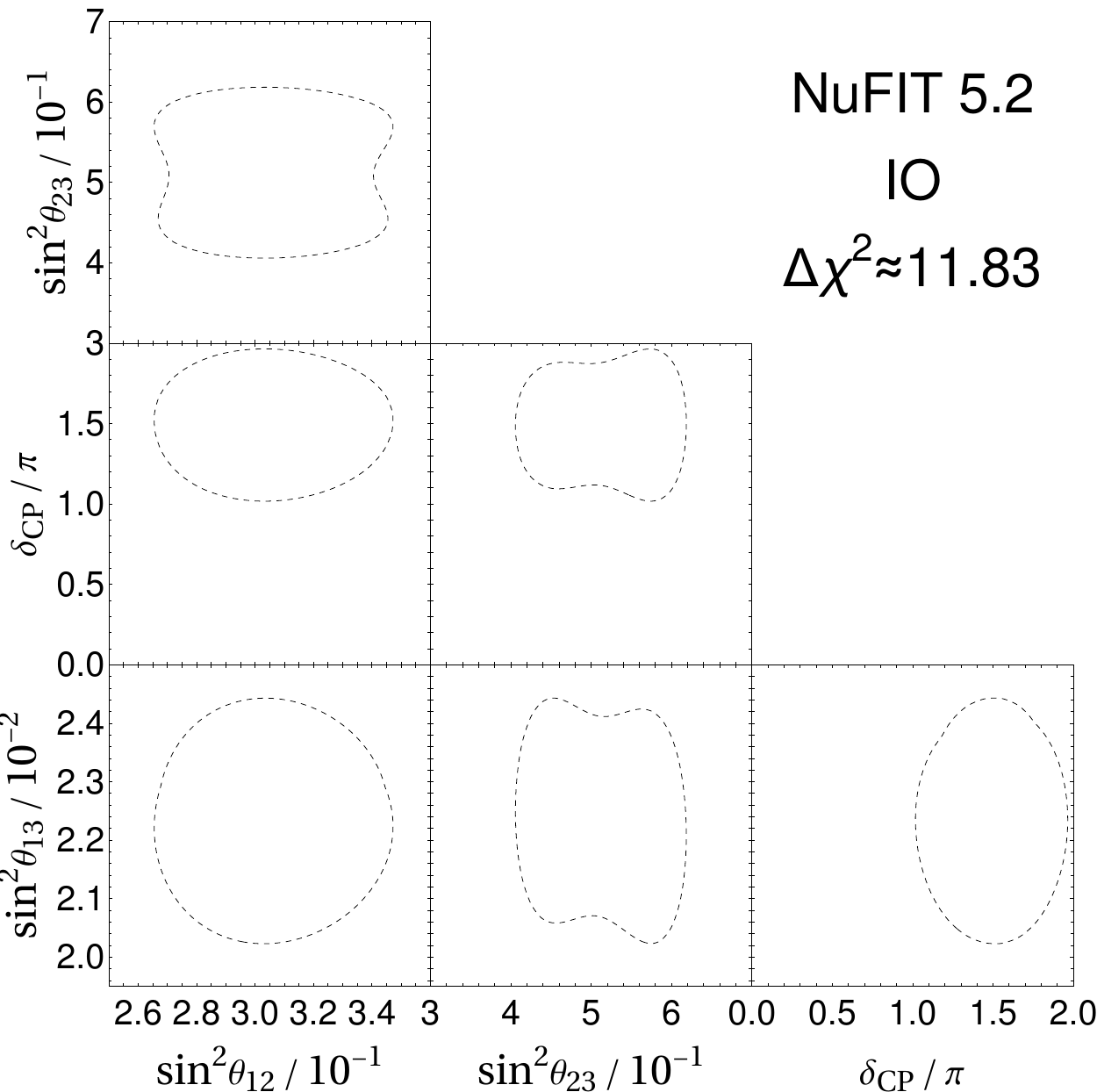}
     \caption{  
     {Sample correlations of oscillation parameters at $\Delta\chi^2=9$ and $\Delta\chi^2=11.83$
     for NO and IO mass ordering. Plots based on NuFIT data~\cite{NuFIT5.2data}.} 
     }
    \label{fig:correlations9chisqnoio}
\end{figure}

{In Chapter \ref{chapter:theory} we have introduced two variants of TBM, namely TM$_1$ and TM$_2$ mixing schemes and discussed bounds on mixing parameters in both models, see Figs.~\ref{fig:deltatm1tm2nhih}-\ref{fig:deltatmutau3p1}. However, we can get stringer bounds on parameters if we take into account correlations among data obtained for global fits in oscillation experiments \cite{Karmakar:2024ejm}. In Fig.~\ref{fig:correlations9chisqnoio} we show some of such correlations among parameters. 
}
 
\begin{figure}[h!]
    \centering
    \includegraphics[width=0.49\textwidth]{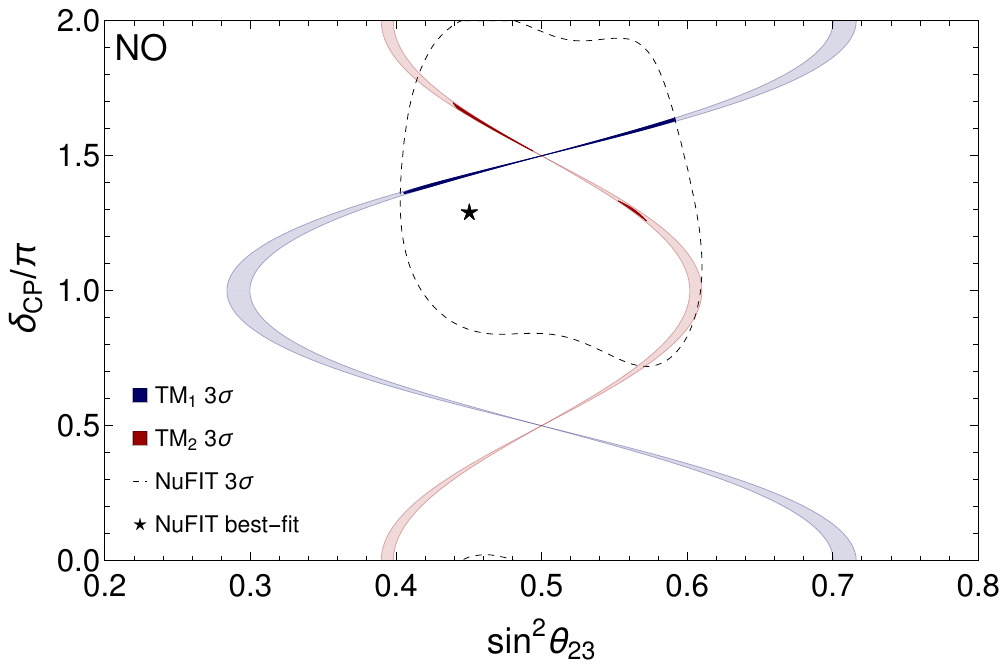}
    \includegraphics[width=0.49\textwidth]{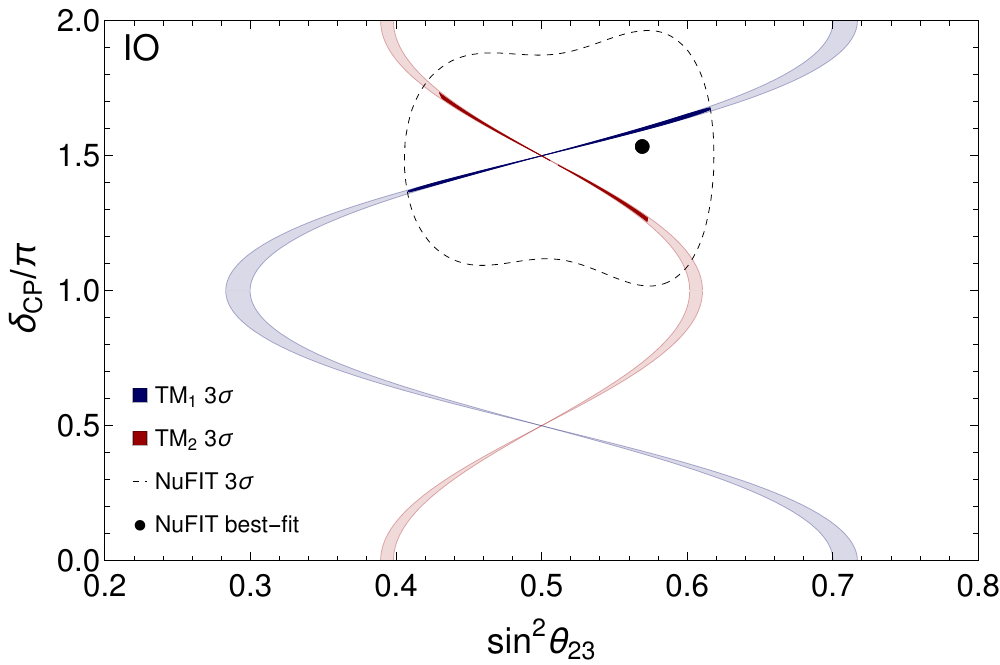}
       \includegraphics[width=0.49\textwidth]{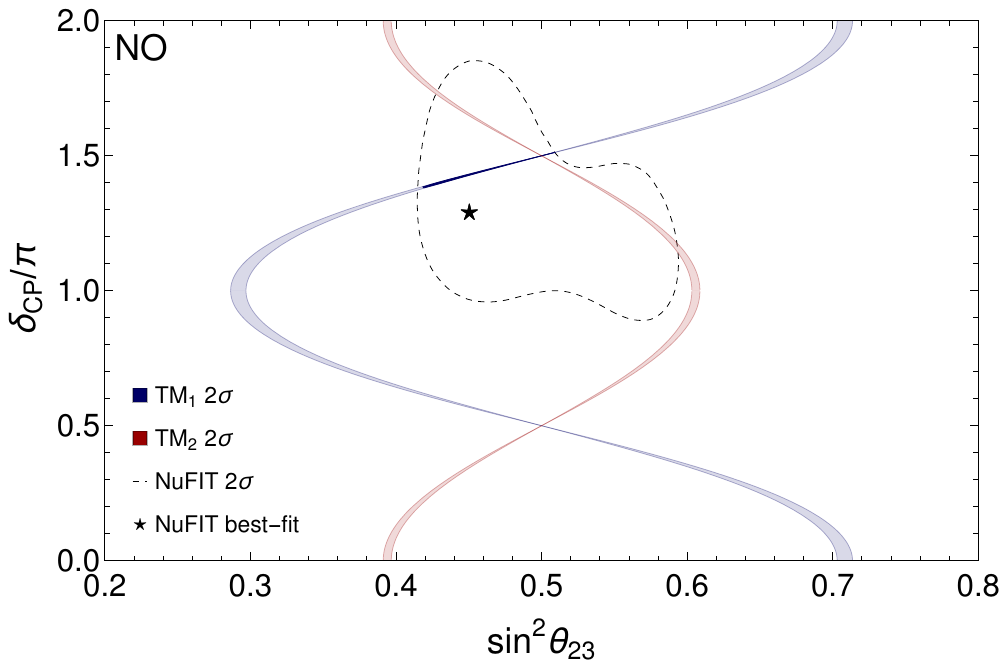}
    \includegraphics[width=0.49\textwidth]{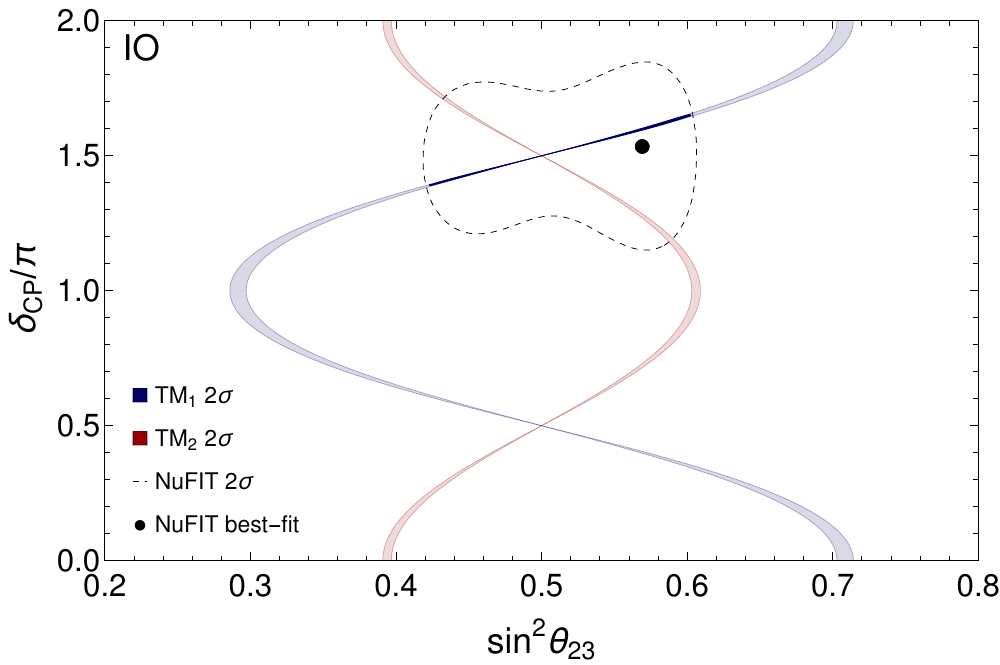}
     \caption{{$\delta_{\rm CP}$ plotted against $\sin^2\theta_{23}$ within TM$_1$ and TM$_2$ models using
     Eqs.~(\ref{eq:tm1formulas}) and (\ref{eq:tm2formulas}) and with the NuFIT 5.2   oscillation data files \cite{NuFIT5.2data} at the 2$\sigma$ and 3$\sigma$  C.L. (2dof). We can note no solutions at the 2$\sigma$ C.L. for the TM$_2$ mixing scheme.}
    }
    \label{fig:dcps23tm1tm2c3sigma}
\end{figure}

{{\underline{Constraints on flavor mixing model parameters coming from global fits with data correlations, an example.}}

Using the ranges given in Section \ref{sec:th13CP} (see e.g. Fig.~\ref{fig:s12s13tm1tm2nh}) and correlations from NuFIT as in Fig.~\ref{fig:correlations9chisqnoio}, we can further constrain the predictions for TM$_{1,2}$ mixing schemes. For example, the darker-shaded regions in Fig.~\ref{fig:dcps23tm1tm2c3sigma} (inside the dashed regions)  represent the allowed areas in the $\delta_{\rm CP} - s_{23}^2$ plane, significantly constraining the  theoretical prediction for the TM$_2$ mixing. Similar constraints can also be imposed on  $\delta_{\rm CP}-s_{13}^2$ and $s_{23}^2-s_{13}^2$ planes.  
In Table~\ref{tab:constr_data}  we have summarized the final prediction on $\theta_{12,23,13}$ and $\delta_{\rm CP}$ for TM$_{1,2}$ mixing schemes, for both mass orderings.
The results in the Table
should be compared with `model independent' results of global fits discussed in Chapter \ref{chapter:introduction} given in Table~\ref{tab:neutrino_data} and Fig.~\ref{fig:schematics_new}, which  at the 3$\sigma$ level (1dof, $\Delta\chi^2=9$) are 
\begin{eqnarray}
\rm{NO:} &&\theta_{13} \in (8.23^{\circ}, 8.91^{\circ}), \theta_{12} \in (31.31^{\circ}, 35.74^{\circ}), \theta_{23} \in (39.7^{\circ}, 51.0^{\circ}),   \delta_{\rm CP} \in (144^{\circ}, 350^{\circ}), \label{eq:NO}\\
\rm{IO:} && \theta_{13} \in (8.23^{\circ}, 8.94^{\circ}), \theta_{12} \in (31.31^{\circ}, 35.74^{\circ}), \theta_{23} \in (39.9^{\circ}, 51.5^{\circ}), \delta_{\rm CP} \in (194^{\circ}, 344^{\circ}). \label{eq:IO}
\end{eqnarray}
}

\begin{table}[h!]
 \centering
    \scriptsize
    \begin{tabular}{c c c c c c c c}
    \hline \hline
        &
        &  \multicolumn{3}{c}{ $\rm{TM}_1$}
            &\multicolumn{3}{c}{ $\rm{TM}_2$} 
            \\
        Parameter &  Ordering 
            & 
        $\Delta\chi^2\approx6.18$
        &  
        $\Delta\chi^2=9$
        &  
        $\Delta\chi^2\approx11.83$
        &  
         $\Delta\chi^2\approx6.18$
        &  
        $\Delta\chi^2=9$
        &  
         $\Delta\chi^2\approx11.83$\\
         \hline 
          $\theta_{13}/^{\circ}$
        & NO
        &$8.33 - 8.83$
        &$8.27 - 8.89$
        &$8.21 - 8.94$
        & $\times$ 
        &$8.53 - 8.62$
        &$8.38 - 8.77$
        \\
        & IO
        &$8.33 - 8.85$
        &$8.26 - 8.92$
        &$8.21 - 8.98$
        & $\times$
        &$8.53 - 8.61$
        &$8.37 - 8.79$
        \\
        \hline
        $\theta_{12}/^{\circ}$
        & NO
        &$34.28 - 34.39$
        &$34.26 - 34.41$
        &$34.25 - 34.42$
        & $\times$ 
        &$35.72 - 35.73$
        &$35.70 - 35.75$
        \\
        & IO
        &$34.27 - 34.39$
        &$34.26 - 34.41$
        &$34.24 - 34.42$
        & $\times$ 
        &$35.72 - 35.73$
        &$35.70 - 35.75$
        \\
        \hline
         $\theta_{23}/^{\circ}$
        & NO
        &$40.3 - 45.6$
        &$39.8 - 49.5$
        &$39.5 - 50.3$
        & $\times$ 
        & $\times$ 
        &$41.5 - 44.7  ~\&~ 48.0 - 49.2$
        \\
        & IO
        &$40.5 - 51.0$
        &$40.1 - 51.4$

        &$39.7 - 51.7$
        & $\times$ 
        & $\times$ 
        &$40.9 - 45.4  ~\&~ 45.6 - 49.2$
        \\
        \hline
         $\delta_{\rm CP}/^{\circ}$
        & NO
        &$248.6 - 272.4$
        &$246.4 - 290.8$
        &$244.7 - 295.2$
        & $\times$ 
        & $\times$ 
        &$226.6 - 239.9  ~\&~ 273.3 - 305.3$
        \\
        & IO
        &$250.0 - 297.7$
        &$247.7 - 300.1$
        &$245.8 - 302.0$
        & $\times$ 
        & $\times$ 
        &$226.2 - 263.9 ~\&~ 266.9 - 312.0$
        \\
        \hline \hline
    \end{tabular}
    \caption{{Allowed ranges for  $\rm{TM}_1$ and $\rm{TM}_2$ mixings obtained using correlations among neutrino oscillation parameters inferred from experimental data.  $\Delta\chi^2{\approx}6.18/9/11.83$ corresponds to 
    $2\sigma$/$2.54\sigma$/$3\sigma$ (2 dof).} 
    }
\label{tab:constr_data}
\end{table}

{Within  allowed regions, the TM$_2$ mixing scheme is most constrained.
Future shrinking of errors of oscillation parameters will affect predictions for discrete flavor models (which can be deduced by comparing $2\sigma$ and $3\sigma$ results for TM$_1$ and TM$_2$ mixing schemes in Fig.3.2 and Table 3.1) and may lead first to the elimination of model TM$_2$.  The outlined here procedure can be applied to the analysis of other neutrino mixing models which predict analytic relations among oscillation parameters.}

{\underline{Predictions for oscillation mixing parameters in discrete flavor models in the context of future neutrino experiments}}

As discussed earlier, a wide class of models with various discrete flavor symmetry groups ($G_f$) exists. With high statistics and their ability to measure the mixing parameters more precisely, the current and future neutrino oscillation experiments provide an excellent testing ground for the flavor symmetry models.  Such studies crucially depend on the breaking pattern of $G_f$ into its residual subgroups for charged lepton sector $G_e$ and neutrino sector $G_\nu$.  For example, in~Ref.~\cite{Blennow:2020snb}, the authors have studied implication of breaking of $G_f \times {CP}$ (with $G_f=A_4, S_4, A_5$) into $G_e > Z_2$, $G_{\nu}=Z_2 \times {CP}$ in the context of ESSnuSB experiment~\cite{ESSnuSB:2013dql}. Such a breaking pattern is usually observed in the semi-direct approach of flavor model building~\cite{King:2019gif}. In this approach the PMNS matrix depends on two free parameters for $G_f = A_4, S_4$, and $A_5$~\cite{Girardi:2016zwz}.
In a similar approach~\cite{Blennow:2020ncm,Petcov:2018snn} considered the breaking pattern into $G_e = Z_k, k > 2$ or $Z_m \times Z_n$, $m, n \geq 2$
and  $G_{\nu} = Z_2 \times CP$ residual symmetry for charged lepton and neutrino sectors, respectively, in the context of ESSnuSB~\cite{Baussan:2013zcy,Wildner:2015yaa}, T2HK~\cite{Abe:2015zbg},    DUNE~\cite{DUNE:2020lwj}, and JUNO~\cite{JUNO:2015sjr} experiments. In this approach, the PMNS matrix is more constrained and depends on a single free angle~\cite{Feruglio:2012cw,Li:2015jxa,DiIura:2015kfa,Ballett:2015wia}.  
In each case, distinct constraints were obtained on the neutrino oscillation parameters $\delta_{\rm CP}$ and  $\theta_{23}$. 
In  Ref.~\cite{Blennow:2020snb} it was demonstrated that out of the 11 (7) one-(two-)parameter models, five (five) are compatible with the present global data at $3 \sigma$.
\begin{figure}[h!]
\includegraphics[width=0.31\textwidth]{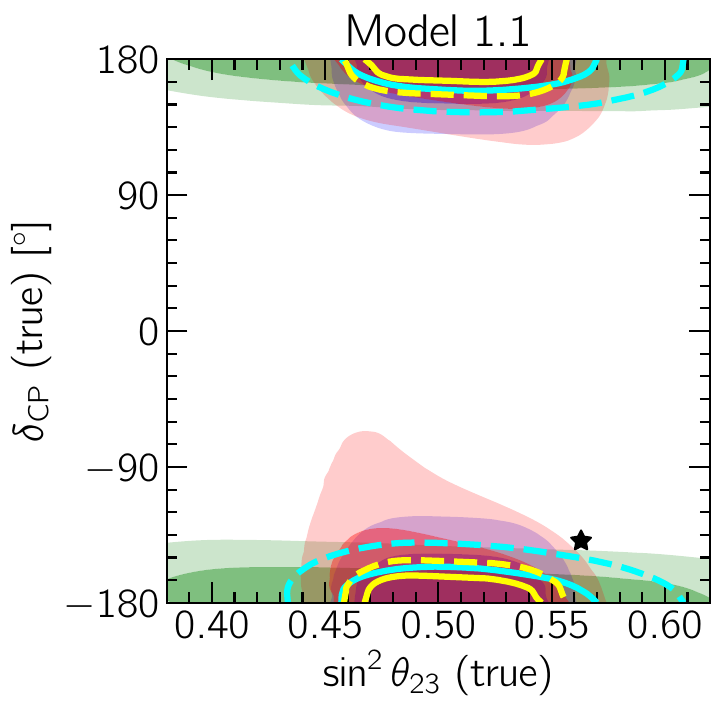}
\hspace{0.02\textwidth}
\includegraphics[width=0.31\textwidth]{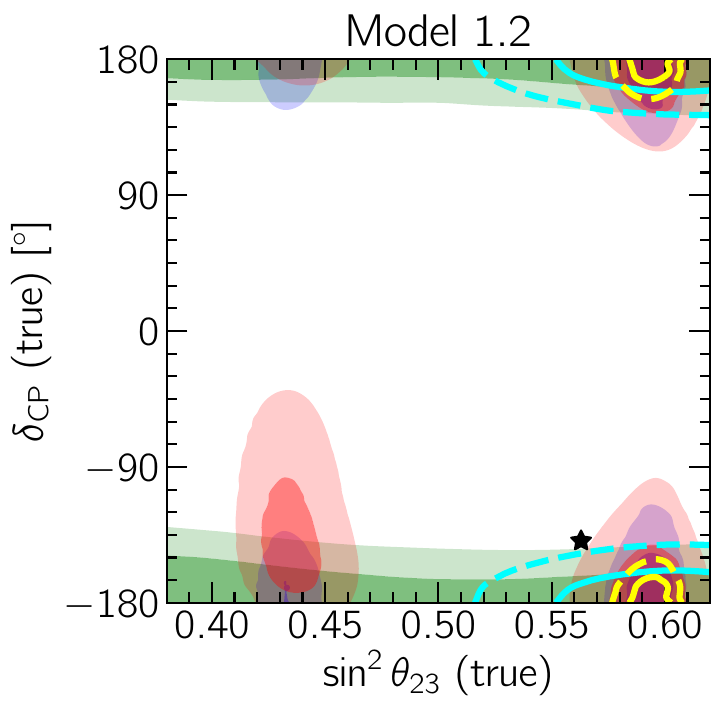}
\hspace{0.02\textwidth}
\includegraphics[width=0.31\textwidth]{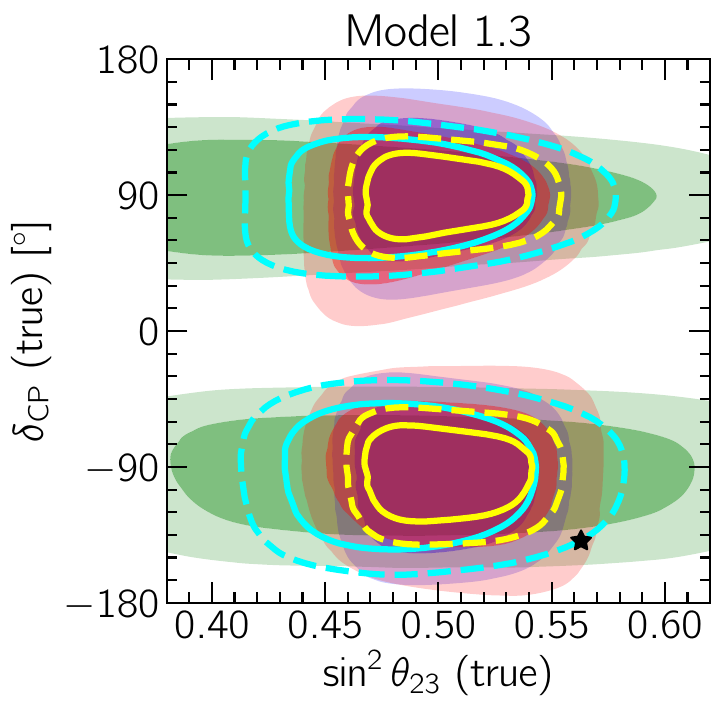}
\\[0.02\textwidth]
\includegraphics[width=0.31\textwidth]{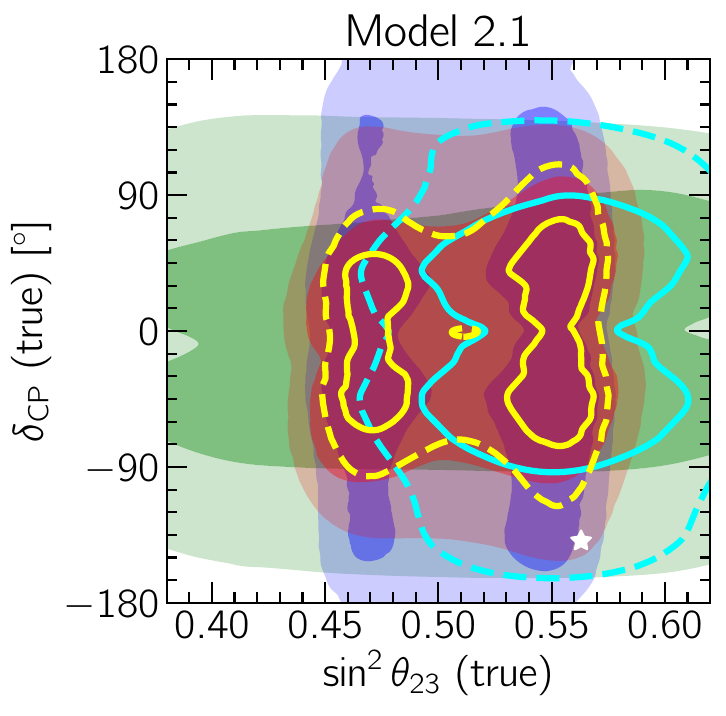}
\hspace{0.02\textwidth}
\includegraphics[width=0.31\textwidth]{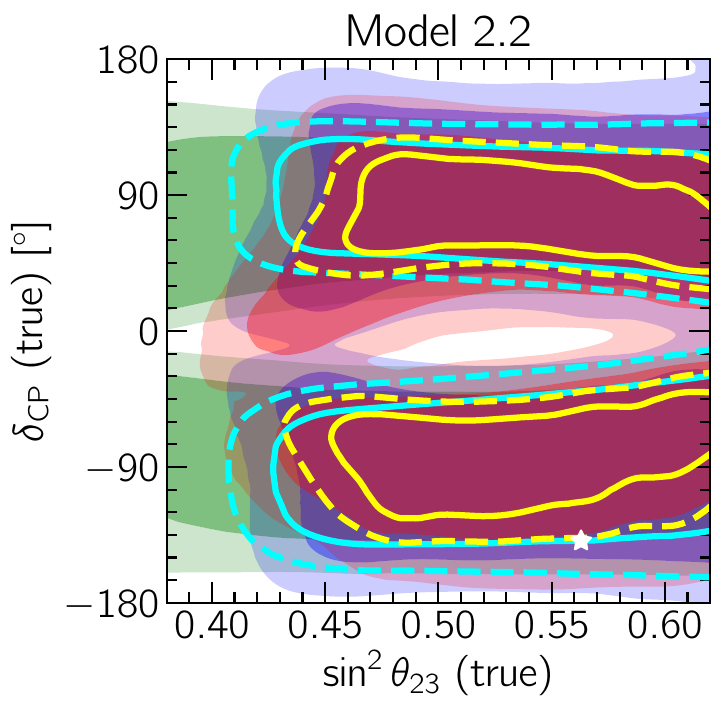}
\hspace{0.02\textwidth}
\includegraphics[width=0.31\textwidth]{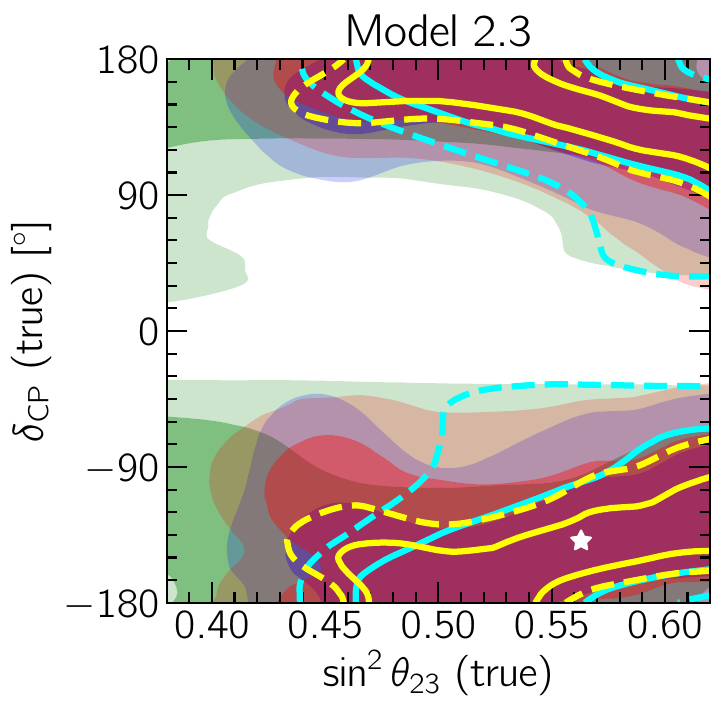}
\caption{Compatibility of one and two-parameter models with any potentially true values of $\sin^2\theta_{23}$ and $\delta_{\rm CP}$ in the context of ESSnuSB,T2HK, DUNE, and their combination.
 The dark (light) green, blue, and red  shaded regions represent $3\sigma$ (5$\sigma$) allowed regions for ESSnuSB,
T2HK, DUNE respectively. Similarly, the continuous (doted) cyan and yellow lines represent 3$\sigma$ (5$\sigma$) allowed regions for  ESSnuSB combined with atmospheric data and ESSnuSB long-baseline experiments. The stars are the best fit values. Figure taken from the arXiv version of ~Ref.~\cite{Blennow:2020ncm}.}
\label{fig:oscexpt2}
\end{figure}
In Fig. \ref{fig:oscexpt2}, we show the compatibility of  one and two-parameter models  with any potentially true values of $\sin^2\theta_{23}$ and $\delta_{\rm CP}$ in the context of ESSnuSB,
T2HK, DUNE, and their combination~\cite{Blennow:2020ncm}. The models are based on $A_5 \times CP$ (Model 1.1), $A _5 \times CP$ (Model 1.2) $S_4 \times CP$ (Model 1.3), $S_4 \times CP$ (Model 1.4), $A _5 \times CP$ (Model 1.5), $A_5$ (Model 2.1), $S_4$ (Model 2.2) and $A_5$ (Model 2.3) discrete groups. A detailed discussion on the specification of each model and their individual compatibility with a combination of experiments can be found in ~Ref.~\cite{Blennow:2020ncm}. In Fig. \ref{fig:oscexpt2}, the dark (light) green, blue, and red  shaded regions represent $3\sigma$ (5$\sigma$) allowed regions for ESSnuSB,
T2HK, DUNE respectively. Similarly, the continuous (doted) cyan and yellow lines represent 3$\sigma$ (5$\sigma$) allowed regions for  ESSnuSB combined with atmospheric data and ESSnuSB long-baseline experiments. The complementarity among these neutrino oscillation experiments offers us some insight into distinguishing various classes of discrete flavor symmetric models. The best fit value for models 1.1, 1.2 falls outside experimentally allowed range of parameters whereas the model 2.3 satisfies all experimental constraints.
As shown in Fig.~\ref{fig:JUNO},  the high-precision measurement of $\sin^2\theta_{12}$ by JUNO will be crucial in discriminating among 
and excluding most of the considered models (see models 1.4 and 1.5 on left and models 1.1, 1.2, 1.4 on right of the   $\sin^2\theta_{12}$   best-fit value. 

\begin{figure*}[h!]
\centering
\includegraphics[width=0.45\textwidth]{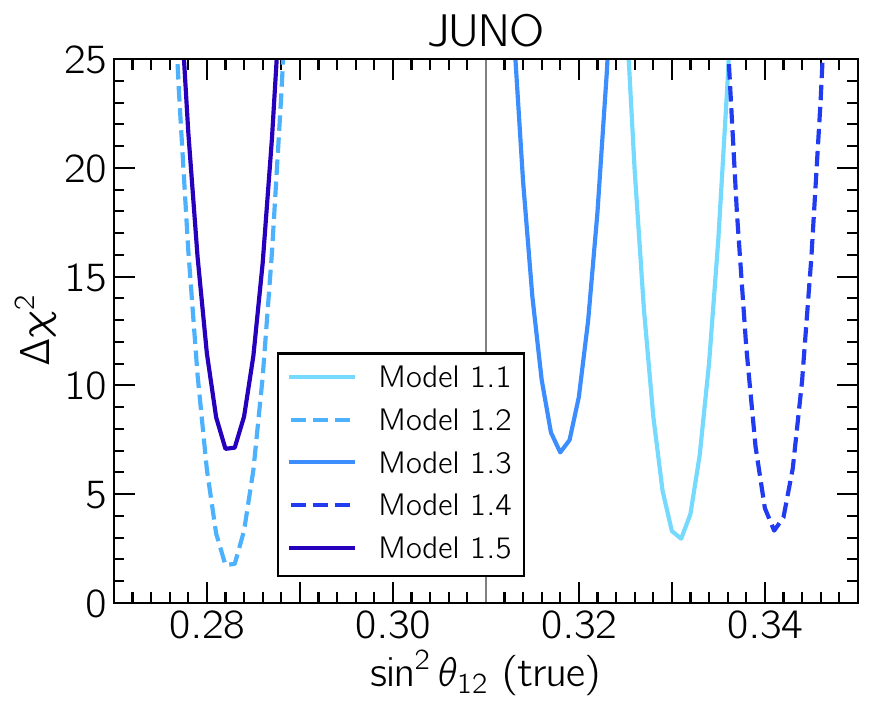}
\hspace{0.05\textwidth}
\caption{Compatibility of one-parameter models with any potentially true value of $\sin^2\theta_{12}$ in the context of JUNO. The {vertical gray line} indicates the best-fit value of $\sin^2\theta_{12}$ from global neutrino oscillation data discussed in ~Ref.~\cite{Blennow:2020ncm}.
Figure taken from the arXiv version of ~Ref.~\cite{Blennow:2020ncm}.}
\label{fig:JUNO}
\end{figure*}

Leveraging DUNE's excellent capability for $\delta_{\rm CP}$ and $\theta_{23}$ measurements, the prospects of generalized CP symmetries with texture zeros~\cite{Nath:2018fvw}  and `bi-large' mixing~\cite{Ding:2019vvi} have also been investigated. 
In another study~\cite{Wang:2018dwk}, the authors have demonstrated an approach to construct operators for neutrino non-standard interactions based on $A_4$ discrete symmetry and its feasibility at DUNE. For studies of consequences of partial $\mu-\tau$ reflection symmetry at DUNE and Hyper-Kamiokande, see ~Refs.~\cite{Chakraborty:2018dew,Nath:2018xkz, Chakraborty:2019rjc}. Guided by the considered discrete flavor symmetry $G_f$, sum rules involving neutrino masses and mixing may also have inherent characteristics~\cite{Gehrlein:2016wlc, Gehrlein:2016fms,Gehrlein:2017ryu,Gehrlein:2022nss} to be confronted with the neutrino experiments mentioned here.  Usually, it is very hard to obtain specific correlations among the neutrino mixing parameters within the modular invariance approach and hence there are very few studies on the viability of these models in the context of neutrino oscillation experiments. Recently, in~Ref.~\cite{Mishra:2023ekx}, the authors have discussed the implication of modular symmetry in neutrino oscillation experiments. In this work,  three different $A_4$ modular symmetric models were considered. The numerical predictions of these three models were tested in the context of  T2HK, DUNE, and JUNO experiments, showing a relative comparison of the models and their compatibility with these experiments.  Furthermore, for a discussion on testing  non-standard neutrino interactions at neutrino oscillation originating  from modular invariance approach (as well as flavor symmetry-based approach), see~Ref.~\cite{Ding:2020yen}.

 \subsection{Neutrinoless Double Beta Decay \label{sec:bb0nu}} 
\begin{figure}[h!]
\includegraphics[width=0.9\textwidth]{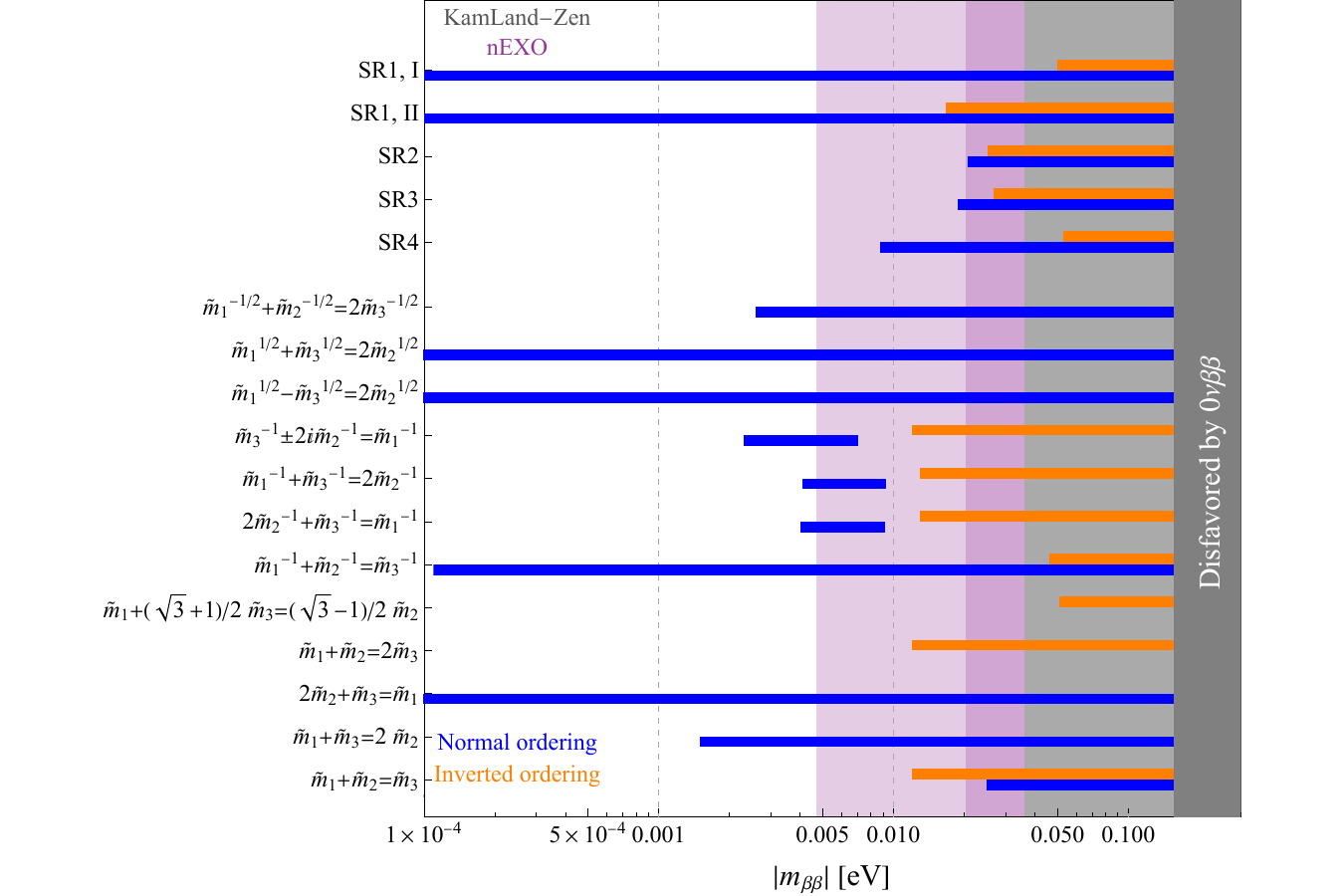}
\caption{A summary  plot of the predictions for $m_{\beta\beta}$ from different mass sum rules for NO and IO. The current experimental bound from KamLand-Zen Ref.~\cite{KamLAND-Zen:2022tow} is shown in gray and the future sensitivity of nEXO Ref.~\cite{nEXO:2021ujk} is in purple. The different shading corresponds to different values of the nuclear matrix elements, leading to the weakest and strongest bounds on $|m_{\beta\beta}|$.
 The figure is updated from  Ref.~\cite{Cirigliano:2022oqy}.  }
\label{fig:sumrule}
\end{figure}
\begin{figure}[h]
	\begin{center}
       \includegraphics[width=.43\textwidth]{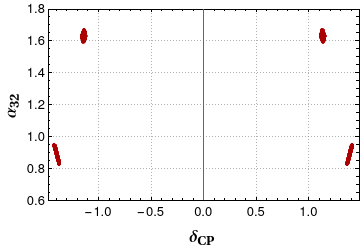}~~
              \includegraphics[width=.45\textwidth]{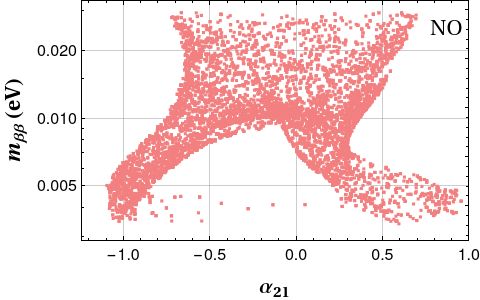}
	\end{center}
\caption{{Correlation between the Dirac CP phase $\delta_{\rm CP}$ and the Majorana phase $\alpha_{32}$ for the TM$_1$ scheme (left panel), see Section \ref{sec:FSdm} for details . Correlation between the Majorana phase $\alpha_{21}$ and $m_{\beta\beta}$ for TM$_2$ mixing scheme, see Section \ref{sec:FLBAU} for details. }}
\label{fig:majorana phase}
\end{figure}

Some flavor models based on discrete symmetries provide predictions for the absolute neutrino mass scale as well as the Majorana phases. Of particular interest are models that predict a correlation between the three complex neutrino mass eigenvalues and the Majorana phases,  $\tilde{m_i} = m_i e^{i \alpha_i}$, hence allow only a certain portion of the parameter space shown in Fig.~\ref{fig:memee} for $0\nu\beta\beta$. This has been extensively studied in the literature in the context of different flavor models, e.g.~TBM~\cite{Hirsch:2008rp}, $\mu-\tau$~\cite{He:2011kn}, $A_4$~\cite{Hirsch:2005mc, Hirsch:2007kh}, $S_4$~\cite{Bazzocchi:2009pv, Bazzocchi:2009da, Ahn:2010nw}, $A_4\times Z_4$~\cite{Boruah:2021qlf}, and $\Delta (3n^2)$~\cite{Hagedorn:2016lva}. 
A comprehensive summary of $0\nu\beta\beta$ predictions from five categories of flavor models, namely, generalized CP, sum rules, charged lepton corrections, texture zeros, and modular symmetries, can be found in  Ref.~\cite{Denton:2023hkx}. As discussed in Ref.~\cite{Cirigliano:2022oqy}, neutrino mass sum rules are present in over sixty flavor models~\cite{Barry:2010yk, King:2013psa, Gehrlein:2016wlc, Gehrlein:2017ryu, Gehrlein:2020jnr}. Fig.~\ref{fig:sumrule} (updated from  Ref.~\cite{Cirigliano:2022oqy}) gives a representation of the predictions for $m_{\beta\beta}$ from different mass sum rules, namely five sum rules connected with modular symmetry models (SR1-I, SR1-II, SR2, SR3, SR4)~\cite{Gehrlein:2020jnr} and twelve sum rules connected with models that predict a correlation between the three complex neutrino mass eigenvalues~\cite{King:2013psa}; see Section \ref{sec:massmodels} for the explanation of the different sum rules.  We can see clearly in Fig.~\ref{fig:sumrule} that the current KamLand-Zen constraint~\cite{KamLAND-Zen:2022tow} has already killed some flavor models, and future experiments like nEXO~\cite{nEXO:2021ujk} will be able to rule out the IO scenario for the remaining models, and also the NO scenario for some models. {In Section \ref{sec:FLBAU}  we will exemplify how predictions for $m_{\beta\beta}$ are constrained when correlated with possible values of cosmological $\eta_B$ parameter (BAU). From Fig.~\ref{fig:0vbb1} it is clear that the application of possible variants of the $\Delta(6n^2)$ group is severely constrained.

Positive signals by neutrinoless double beta decay experiments will essentially establish that neutrinos are Majorana particle. The Majorana phases do not appear in the oscillation probability, hence they can not be constrained from neutrino oscillation data directly~\cite{Bilenky:1980cx}.  However, they can be constrained from low energy neutrino oscillation data indirectly in flavor symmetric models. In the above paragraph, we have already discussed implications of complex mass sum rules in $m_{\beta\beta}$ where Majorana phases are involved. Flavor symmetric models which do not predict any mass sum rule, Majorana phases can still be constrained using neutrino oscillation data. These models felicitate constraints on the  Majorana phases and hence a tight constrained on $m_{\beta\beta}$ can be obtained.

In Section \ref{sec:FSdm}, the TM$_1$ mixing scheme will be discussed in the context of DM, based on the $A_4$ discrete group.  
Here, the single Majorana phase $\alpha_{32}$ can be expressed by TM$_1$ model parameters as given in Eq.~(\ref{eq:majorana phase}) which are further related to the neutrino oscillation mixing angles and $\delta_{\rm CP}$. Thus, using neutrino oscillation data, $\alpha_{32}$ can be severely constrained, see Fig.~\ref{fig:majorana phase} (left panel) and Fig.~\ref{fig:alpha} in Section \ref{sec:FSdm}. In Table~\ref{tab:mass} we make predictions for the light neutrino masses ($m_2,m_3$), their sum ($\sum m_i$) and the effective mass parameter appearing in the neutrinoless double decay ($m_{\beta\beta}$). But the predictions for $m_{\beta\beta}$ fall below the sensitivity of the next generation neutrinoless double beta decay experiments like nEXO, see Fig.~\ref{fig:memee}.
}
\begin{table}[h!]
    \centering
    \begin{tabular}{|c|c|c|c|}
        \hline
        $m_2$ (meV)  & $m_3$ (meV) & $\sum m_i$ (meV)  & $m_{\beta\beta}$ (meV) \\
        \hline
        $8.3-9.0$ & $49.7-51.3$ & $58.0-60.3$ & ${1.61-3.85}$\\
        \hline
    \end{tabular}
    \caption{{Predictions for $m_2$, $m_3$, $\sum m_i$ and $m_{\beta\beta}$ (all in meV) for the FSS TM$_1$ framework.}}
    \label{tab:mass}
\end{table}
{Similarly, for the $A_4$ based TM$_2$ mixing presented in Section \ref{sec:FLBAU}, in Fig.~\ref{fig:majorana phase} (right panel) we have explicitly shown the dependence of $m_{\beta\beta}$ on the Majorana phase $\alpha_{21}$. For correct neutrino masses and mixing the constraint on $\alpha_{21}$, given in Eq.~(\ref{eq:tm2-a21}), within the TM$_2$ framework can also be evaluated as using the allowed parameter space given in the left panel of Fig.~\ref{fig:tm2-abp}. In an identical manner, the other Majorana phase $\alpha_{31}$ can also be constrained. 
\begin{figure}[h]
	\begin{center}
       \includegraphics[width=.35\textwidth]{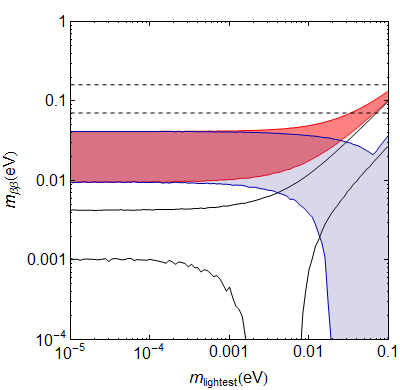}~~
              \includegraphics[width=.35\textwidth]{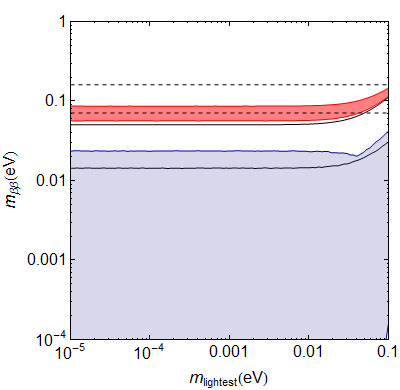}
	\end{center}
\caption{{ $m_{\beta\beta}$ vs lightest neutrino mass for 3+1 neutrino mixing for NO (left) and IO (right). Red regions represent when both of the Majorana phases are zero and blue region represent when both of the Majorana phases are fixed at $90^{\circ}$. Area between the black solid lines represent standard three neutrino allowed region. The dotted lines show the current constraint from KamLAND-Zen. Figure taken from the arXiv version of~\cite{Chakraborty:2019rjc}.  }}
\label{fig:mbbmutau}
\end{figure}

In Section \ref{sec:th13CP}, we mentioned that in case of (partial) $\mu-\tau$ reflection symmetry,  the Majorana phases are fiexd at $0^{\circ}$ and $90^{\circ}$. Such  constrained scenarios lead to tight prediction for $m_{\beta\beta}$. For three neutrino $\mu-\tau$ reflection symmetry, the obvious predictions for $m_{\beta\beta}$ are given in Ref.~\cite{Zhao:2017yvw}. Breaking of $\mu-\tau$ reflection symmetry and its effect in $m_{\beta\beta}$ has also been discussed in ~\cite{Zhao:2017yvw}.  Interestingly, for $\mu-\tau$ reflection symmetry in case of 3+1 neutrino scenario, with  Majorana phases fixd (at $0^{\circ}$, $90^{\circ}$), one can obtain new allowed regions for $m_{\beta\beta}$~\cite{Chakraborty:2019rjc} (when additional sterile Dirac CP violating phases are considered to be zero). In such case, it also disallows a large parameter space for the existing three active neutrino allowed parameter space as shown in Fig.~\ref{fig:mbbmutau}. For a detailed discussion, see~\cite{Chakraborty:2019rjc}. 
}

\subsection{Lepton Flavor and Universality Violation}
Flavor symmetry models can also make distinctive predictions for LFV and LFUV observables. The basic idea is that couplings between flavons and leptons can result in special flavor structures with specific LFV predictions. Thus, discovering LFV signal can provide crucial information to distinguish flavor symmetries and new physics scenarios. See Refs.~\cite{Deppisch:2012vj, Morisi:2012fg} for reviews of the impact of flavor symmetry models on LFV processes. More recent studies of LFV in specific flavor models can be found in Refs.~\cite{Feruglio:2008ht, Kobayashi:2015gwa, Pascoli:2016wlt, Ganguly:2022qxj,Devi:2022scm, Bigaran:2022giz}. 
{For example,  the FSS model~\cite{Ganguly:2023jml} (reproducing  TM$_1$ mixing with $A_4$ discrete flavor symmetry) described in Section \ref{sec:th13CP} also contributes to LFV decays such as $\ell_{\alpha}\to \ell_{\beta}\gamma$ and $\ell_{\alpha}\to 3 \ell_{\beta}$ ($\alpha,\beta = e, \mu,\tau$). Predictions of these LFV decays crucially depend on the VEV alignment of the associated flavons which also dictates the neutrino masses and mixing. {These issues will be discussed in detail in Section \ref{sec:FSdm}.} Owing to the particular flavor structure given in Eqs.~(\ref{eq:seesaw yukawa}) and (\ref{eq:scotoyukuwa}), the scotogenic part within this FSS framework only contributes to the LFV decay $\mu \to e\gamma$ and $\mu \to 3e$ conversion and puts a strong constraint on the allowed parameter space.
\begin{figure}[h!]
\centering
\includegraphics[width=0.4\textwidth]{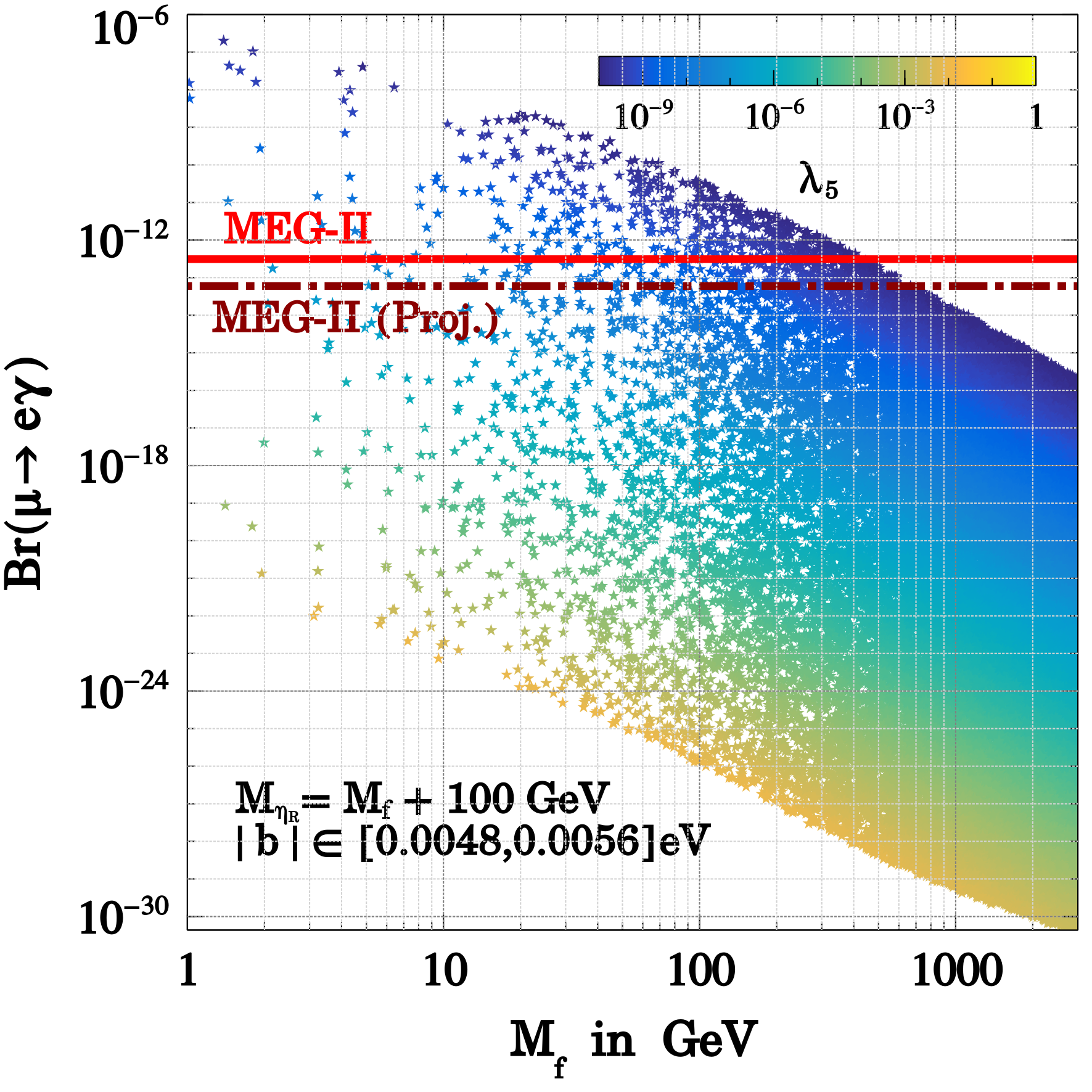}
\includegraphics[width=0.4\textwidth]{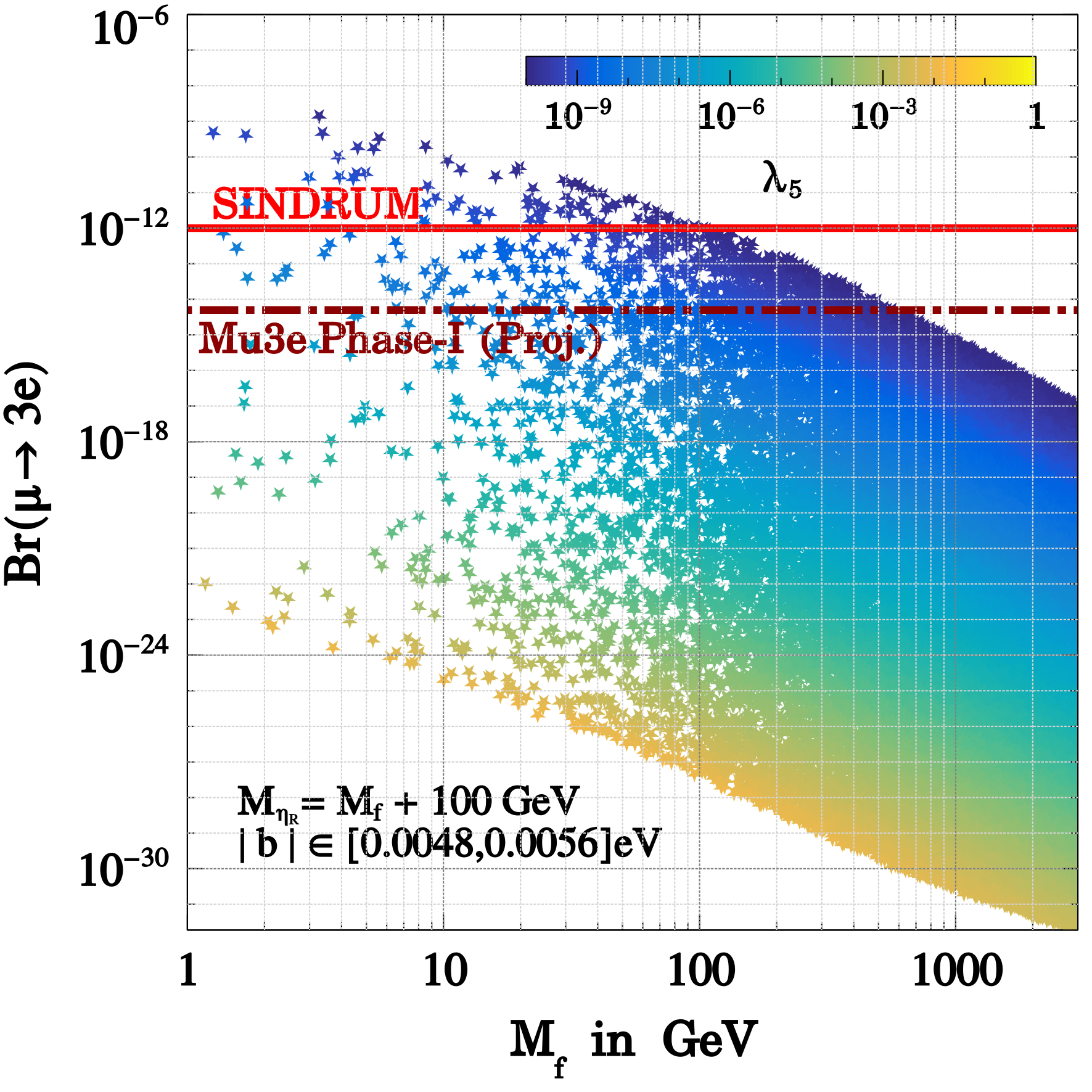}
\caption{Branching ratios for $\mu \to e\gamma$ (left panel) and $\mu \to 3e$ (right panel) against scotogenic fermion   mass $M_f$ satisfying $3\sigma$ allowed range of neutrino oscillation data~\cite{Esteban:2020cvm} as well as correct DM relic density. The horizontal red and magenta lines represent current and future experimental sensitivity. {Masses $M_f, M_{\eta_R}$ and parameters $b, \lambda_5$ are defined in Section \ref{sec:FSdm} where more phenomenological aspects of the FSS model in the context of cosmology will be discussed.} Figure taken from Ref.~\cite{Ganguly:2023jml}.} 
\label{fig:lfvfss}
\end{figure}
In Fig. \ref{fig:lfvfss}, we have shown the prediction for the branching ratios for $\mu \to e\gamma$ (cyan dots, left panel) and $\mu \to 3e$ (blue dots, right panel) against scotogenic fermion ($f$)  mass $M_f$. In this FSS model, $f$ is also a potential DM candidate. In both panels of Fig. \ref{fig:lfvfss}, the dotted regions represent the $3\sigma$ allowed regions which satisfy correct neutrino masses and mixing~\cite{Esteban:2020cvm} as well as correct DM relic density.  The horizontal red lines in both of these panels represent the current sensitivity of MEG~\cite{MEG:2016leq} and SINDRUM~\cite{SINDRUMII:1996fti} experiments and substantially contain the allowed parameter space and restrict DM mass $M_f \gtrsim $ 1750 GeV. More interestingly, this $A_4$ FSS model can be falsified by the future sensitivity (given by horizontal magenta line) of MEG II~\cite{Meucci:2022qbh} and Mu3e Phase-I~\cite{Mu3e:2020gyw} experiment for $\mu \to e\gamma$  and $\mu \to 3e$ decays, respectively.  On the other hand, due to the considered flavor symmetry, the Yukawa couplings $Y_F^{\tau}$,  $Y_N^e$ vanish and hence  $\tau\rightarrow 3e$ processes are strictly disallowed in this model. Thus, with an example of a $A_4$ flavor symmetric scoto-seesaw framework, we find that flavor symmetry can have distinct consequences on lepton flavor violating decays. }  

In some cases, the flavor model predictions for LFV, especially in the tau sector, are expected to be probed in the near future by Belle II. The role of leptonic CP phases (which come out as a prediction in many flavor models) in the LFV observables has been explored recently in Ref.~\cite{Abada:2021zcm}. {Now, considering the mixing of light and heavy RHNs, flavor symmetric models can also constrain the VEV of the flavon fields involved, which helps to realize the desired flavor structure to explain observed leptonic mixing. Following Eqs.~(\ref{ugen1})-(\ref{congr}), the leptonic non-unitarity  can be defined as~\cite{Altarelli:2008yr,Antusch:2008tz,Dev:2009aw,Fernandez-Martinez:2016lgt} 
\begin{eqnarray}
    U\simeq (\mathbb{I}-\eta)U_{\rm PMNS}
\end{eqnarray}
where the non-unitary parameter is defined as $\eta=\frac{1}{2}FF^{\dagger}$ and $F=M_{D}M_R^{-1}$. The present bound on $\eta$ obtained from various non-standard interactions  can be summarized as~\cite{Antusch:2008tz,Fernandez-Martinez:2016lgt, Blennow:2023mqx}
\begin{eqnarray}
   |\eta| \leq \left( \label{eq:nut}
\begin{array}{ccc}
 1.3\times10^{-3}  & 1.2\times10^{-5}  & 1.4\times10^{-3} \\
1.2\times10^{-5}   & 2.2\times10^{-4}  & 6.0\times10^{-4} \\
1.4\times10^{-3}  & 6.0\times10^{-4}  & 2.8\times10^{-3}
\end{array}
\right). 
\end{eqnarray}
In a flavor symmetric model, the mass matrices for the Dirac ($M_D$) and heavy Majorana ($M_R$) neutrinos appearing in $\eta$ can be obtained with the involvement of the flavons. Hence, the constraints on $\eta$ for Eq. (\ref{eq:nut}) can, in principle, constrain the VEV of the associated flavon field. For example, considering an $A_4$ flavor symmetric inverse seesaw scenario in Ref.~\cite{Karmakar:2016cvb}, the authors showed that the flavon VEV ($v_f$) can be constrained as $v_f \leq 6.15 {\lambda}$ TeV, where $\lambda$ is the ratio of the modulus of the coupling constants involved in $M_D$ and $M_R$ respectively. Along with the constrain on $\eta$ given in Eq. (\ref{eq:nut}), the Dirac CP phase $\delta_{\rm CP}$ can also play an instrumental role in obtaining limits on $v_f$; for a detailed discussion see Ref.~\cite{Karmakar:2016cvb}. 
} 

Flavor models can also shed some light on the recent hints for lepton flavor universality violation. In Ref.~\cite{Bigaran:2022kkv}, a comprehensive analysis is given based on the flavor group $G_f = D_{17} \times Z_{17}$ with scalar leptoquarks aiming to explain some anomalies in $B$-physics connected with lepton flavor universality between $\tau$ and $e,\mu$, as well as muon anomalous magnetic moment $g-2$. For other studies of the LFUV in flavor symmetry models, see e.g.~Refs.~\cite{Barbieri:2015yvd, Bordone:2017anc, Fuentes-Martin:2019mun}. 

\section{Flavor Symmetry at Energy Frontier \label{sec:collider}}

Flavor models with extended particle content can lead to interesting collider signals. For instance, in the flavor models with Majorana RHNs, the structure of the complex Dirac Yukawa couplings can be fixed by the flavor (and CP) symmetries. This in turn gives concrete predictions for the LNV/LFV signatures associated with the RHNs, depending on their mass spectrum. In this section, we will illustrate the collider phenomenology in a class of models with residual flavor and CP symmetries. The specific discrete flavor symmetry groups $G_f$ chosen here are the series of groups $\Delta(6\, n^2)$~\cite{Escobar:2008vc}, known to give several interesting neutrino mixing patterns~\cite{King:2014rwa, Hagedorn:2014wha, Ding:2014ora, Ding:2015rwa, Hagedorn:2016lva}. As discussed in Refs.~\cite{Curtin:2018mvb, Chauhan:2021xus}, this framework provides an excellent probe of flavor symmetries in collider experiments. 
For example, it can result in one of the three RHNs to be very long-lived at some special parameter points, termed as points of {\it enhanced residual symmetry} (ERS), which can be searched through {\it long-lived particle} (LLP) searches~\cite{Alimena:2019zri, Alimena:2021mdu}. In comparison, the remaining two RHNs can be probed via prompt/displaced vertex signals at the LHC~\cite{Deppisch:2015qwa, Cai:2017mow} or future hadron collider FCC-hh \cite{FCC:2018byv,FCC:2018vvp}; see discussion in Section~\ref{sec:RHNscolliders}.  

\subsection{Example Group : $\Delta (6n^2)$ \label{sec:delta6pheno}}
 
 The discrete groups $\Delta (6 \, n^2)$, ($n\in Z,\, n \geq 2$) ~\cite{Escobar:2008vc}, can be
characterized by four generators $a$, $b$, $c$ and $d$  along with the identity element $e$, fulfilling the relations
\begin{equation*}
a^3 \, = \, e \; , \;\; c^n \, = \, e \; , \;\; d^n \, = \, e \; , \;\; c \, d \, = \, d \, c\; , \;\; a \, c \, a^{-1} \, = \, c^{-1} d^{-1} \; , \;\; a \, d \, a^{-1} \, = \, c \, ,
\end{equation*}
\begin{equation}
b^2 \, = \, e \; , \;\; (a \, b)^2\, = \, e \; , \;\; b \, c \, b^{-1} \, = \, d^{-1} \; , \;\; b \, d \, b^{-1} \, = \, c^{-1} \, .
\end{equation} 
Upon breaking of the the flavor group $G_f$ at low energies, the neutrino and charged lepton sectors are still invariant under residual flavor and CP symmetry groups. The residual symmetry in the charged lepton sector is chosen to be the diagonal Abelian subgroup of $Z_3$, i.e. $G_\ell=Z_3^{(\mathrm{D})}$, while in the neutrino sector, we choose the residual symmetry to be $G_\nu=Z_2 \times {\rm CP}$. The generators of  $Z_2$ symmetry  $Z (\mathrm{r})$  and CP symmetry  $X (\mathrm{r})$ commute for all representations ${\rm r}$ of $G_f$. 
The transformations related to CP symmetry correspond to the automorphisms of the flavor group. The differences between the residual symmetries $G_\ell$ and $G_\nu$ determine the forms of the lepton mixing matrix, charged lepton mass matrix, the neutrino Yukawa $Y_D$ and the RHN Majorana mass matrix $M_R$. Since, we explicitly choose the charged lepton mass matrix  to be diagonal, the charged lepton sector does not contribute to the lepton mixing. As for the neutrino sector, we assume the Dirac neutrino Yukawa coupling matrix $Y_D$ to be invariant under $G_\nu$ and the Majorana matrix $M_R$ does not break either $G_f$ or CP. The light neutrino masses are obtained by the type-I seesaw formula~\cite{Minkowski:1977sc, Mohapatra:1979ia}:
\begin{align}
    M_\nu =-v^2Y_DM_R^{-1}Y_D^T \, ,
    \label{eq:seesaw}
\end{align}
where $v$ is the SM Higgs VEV. 

As an example, we consider a particular case with generator of $Z_2$ symmetry $Z \, = \, c^{n/2}$, where $c$ is one of the generators of the group $\Delta(6n^2)$ and the corresponding CP transformations reads $X(s)  = \ a \, b \, c^s \, d^{2 s} \, P_{23}$, where $s$ is a parameter that runs from $0$ to $(n-1)$ and $P_{23}$ is the permutation matrix in the 2-3 plane. For this case, the form of $Y_D$ is given by 
\begin{equation}
\label{eq:YDgen}
Y_D \, = \, \Omega^s({3}) \, R_{13} (\theta_L) \, \left(
\begin{array}{ccc}
 y_1 & 0 & 0\\
 0 & y_2 & 0\\
 0 & 0 & y_3
 \end{array}
 \right) \, R_{13} (-\theta_R) \, \Omega^s({3^\prime})^\dagger \; ,
\end{equation}
where the angles $\theta_L$ and $\theta_R$ are free parameters,
with values in the range $[0,\pi)$ and $R_{ij}(\theta)$ denotes rotation by an angle $\theta$ in the $ij$ plane. 
$\Omega^s ({3})$ is a unitary matrix connected with $X ({3}) (s)$, with the following structure:  
 
\begin{equation}
\label{case1Omegain3}
\Omega^s({3}) \, = \, e^{i \phi_s} \, U_{\mathrm{TBM}} \,
\left( \begin{array}{ccc}
 1 & 0 & 0 \\
 0 & e^{-3 \phi_s} & 0\\
 0 & 0 & -1
\end{array}
\right) \, 
\end{equation} 
(where $\phi_s=\pi s/n$), 
whereas the form of the unitary matrix $\Omega^s({3^\prime})$ depends on whether $s$ is even or odd, i.~e.
\begin{align}
 \Omega^{s\; \rm{even}} ({3^\prime}) \, = \, U_{\mathrm{TBM}} \; , \:\:    
 \Omega^{s \; {\rm odd}} ({3^\prime}) \, = \,  U_{\mathrm{TBM}} \, \left(
 \begin{array}{ccc}
 i & 0 & 0\\
 0 & 1 & 0\\
 0 & 0 & i
\end{array} 
 \right)
 \; ,
 \label{Omega3p}
\end{align}
with $U_{\rm TBM}$ given in Eq.~\eqref{eq:TBM}. 
Finally, we can find the form for the PMNS mixing matrix as 
\begin{align}
    U \, = \, \Omega^s({3}) \, R_{13}{(\theta_{\rm L}-\psi)}
    \, K_\nu \, ,
    \label{eq:PMNS1}
\end{align}
where $K_\nu$ is a diagonal matrix with entries equal to $\pm 1$ and $\pm {i}$, making neutrino masses non-negative, and the angle $\psi$ is defined by 
\begin{align}
\tan^2\psi = \frac{m_1+m_3-\sqrt{m_1^2+m_3^2+2m_1m_3\cos(4\theta_R)}}{m_1+m_3+\sqrt{m_1^2+m_3^2+2m_1m_3\cos(4\theta_R)}} \, .
\end{align}
$\theta_L$ in Eq.~\eqref{eq:PMNS1} is determined by reproducing the best-fit values of the measured neutrino mixing angles (cf.~Table~\ref{tab:neutrino_data}).   

As for the RHN Majorana mass matrix $M_R$, since it leaves $G_f$ and CP invariant, its form is s
imply 
\begin{equation}
M_R \, = \, M_N \, \left(
\begin{array}{ccc}
1 & 0 & 0\\
0 & 0 & 1\\
0 & 1 & 0
\end{array}
\right) \, ,
\label{eq:MR}
\end{equation}
with $M_N > 0$ setting the overall mass scale of the RHNs. From Eq.~\eqref{eq:MR}, we see that the three RHNs are exactly degenerate in the flavor symmetry limit. However, if we want to successfully generate the BAU via resonant leptogenesis~\cite{Pilaftsis:2003gt, Dev:2017wwc} using the RHN freeze-out in the early Universe, we need at least two quasi-degenerate RHNs. This can be achieved by introducing a small symmetry breaking term (that can be sourced from higher-dimensional operators)
\begin{equation}
\label{dMRtilde}
\delta M_R = \kappa \, M_N \, \left( \begin{array}{ccc}
2 & 0 & 0\\
0 & 0 & -1\\
0 & -1 & 0
\end{array}
\right)
\end{equation}
with $\kappa\ll 1$. Then the RHN masses acquire a (small) correction
\begin{equation}
\label{eq:RHmassspectrum}
M_1 = M_N \, (1+ 2 \, \kappa) \;\; \mbox{and} \;\; M_2=M_3= M_N \, (1-\kappa) \, ,
\end{equation}
thus making two RHN pairs quasi-degenerate, adequate for resonant leptogenesis (see Section~\ref{subsec:colliderleptogenesis}). {Note that here we have considered general corrections to $M_R$ which are invariant under the residual symmetry $G_\ell$, without specifying any particular breaking mechanism or flavon dynamics. Concrete model realizations where these (higher order) residual symmetry breaking effects are generically present can be found in Refs.~\cite{Feruglio:2019ybq, Ishimori:2010au, King:2013eh}.}

Above we sketched a typical construction connected with CP and discrete flavor symmetries in the lepton sector, leading to the parametrization of the Dirac Yukawa coupling matrix given by Eq.~\eqref{eq:YDgen}. This is a very predictive scenario, with only five real parameters determining the lepton mixing, namely,  three Yukawa couplings $(y_1,y_2,y_3)$ corresponding to the three light neutrino masses, and two rotation angles ($\theta_L$ and $\theta_R$) for lepton mixing. Just two extra parameters ($\kappa$ and $M_N$) are needed for explaining the BAU. For TeV-scale $M_N$, this leads to interesting predictions that can be tested at both energy and intensity frontier experiments. In the following, we will briefly discuss some phenomenological features of this simple scenario. For more details, see  Ref.~\cite{Chauhan:2021xus}.

\subsection{Decay Lengths and Branching Ratios of RHNs \label{sec:RHNscolliders}}

TeV-scale Majorana RHNs give rise to spectacular multilepton signals at the LHC and future colliders~\cite{Deppisch:2015qwa, Cai:2017mow, CMS:2018jxx, ATLAS:2019kpx}. The general problem is to get substantial production rate. 
In the context of minimal type-I seesaw for our TeV-scale RHN scenario, light neutrino masses and mixing require the Yukawa couplings to be significantly suppressed, with values on the order of $10^{-7}$. This, in turn, suppresses the Drell-Yan production of the RHNs at LHC for the smoking-gun signal of same-sign dilepton plus two jets without missing transverse energy~\cite{Keung:1983uu, Datta:1993nm, Han:2006ip, delAguila:2007qnc,Atre:2009rg, Dev:2013wba, Alva:2014gxa, Das:2015toa, Gluza:2015goa, Das:2016hof,Gluza:2016qqv,Das:2017gke}. However, in typical UV-complete RHN models with additional gauge interactions, the production rate can be enhanced without relying on their mixing with the SM neutrinos~\cite{Deppisch:2015qwa}. For example, in $U(1)_{B-L}$ extensions of the SM~\cite{Davidson:1978pm, Marshak:1979fm}, the RHNs can be pair-produced via the $Z'$-mediated process: $pp\to Z'\to N N$~\cite{Buchmuller:1991ce,Basso:2008iv, FileviezPerez:2009hdc, Deppisch:2013cya, 
Kang:2015uoc, Cox:2017eme, Han:2021pun, Das:2022rbl}. Similarly, in the left-right symmetric models based on the $SU(2)_L\times SU(2)_R\times U(1)_{B-L}$ gauge group~\cite{Pati:1974yy, Mohapatra:1974gc,Senjanovic:1975rk}, the RHNs can be produced via the RH current: $pp\to W_R\to N\ell$~\cite{Keung:1983uu, Ferrari:2000sp, Nemevsek:2011hz, Chen:2011hc, Chakrabortty:2012pp, Das:2012ii, Chen:2013foz, Ng:2015hba, Dev:2015kca, Das:2017hmg, Nemevsek:2018bbt, Helo:2018rll, BhupalDev:2019ljp, Li:2022cuq}. Depending on the new gauge boson mass, cross sections up to a few fb are possible at the LHC.

\begin{figure}[t!]
\centering
\includegraphics[width=0.49\textwidth]{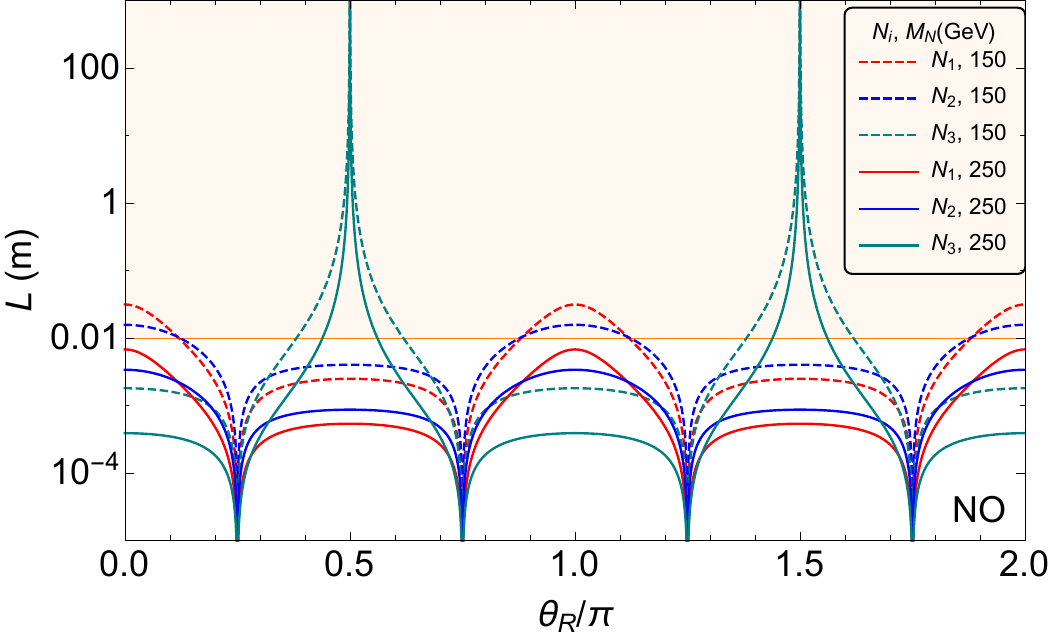}
\includegraphics[width=0.49\textwidth]{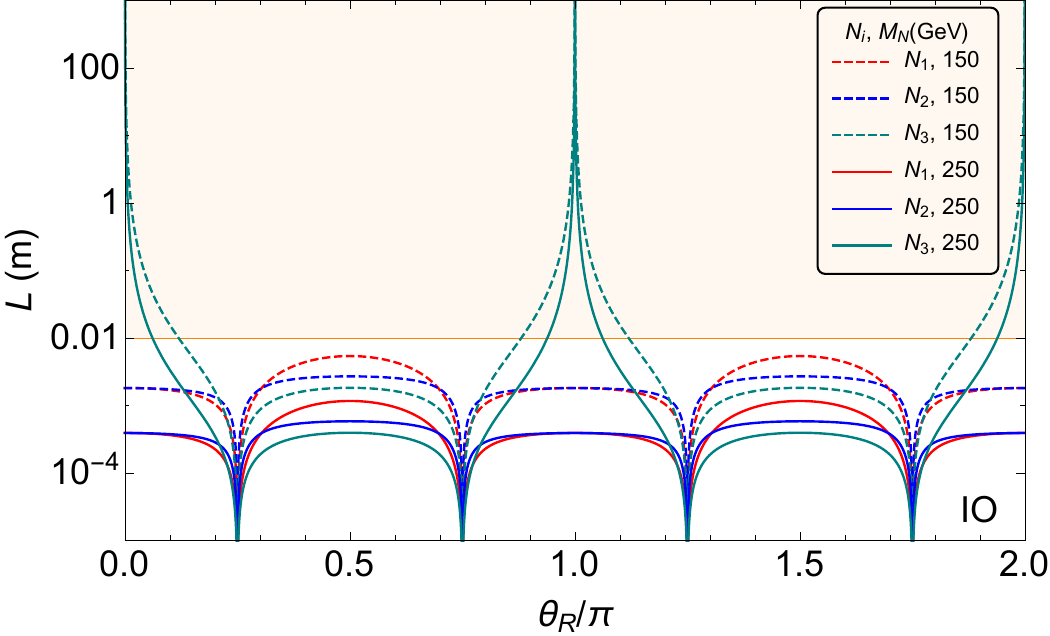}\\
\caption{ Decay lengths for $N_{1,2,3}$ are plotted against $\theta_R$ for different values of the RHN mass scale $M_N$ (with $m_0=0$). The left (right) panel is for NO (IO). The shaded (unshaded) region roughly indicates the displaced/long-lived (prompt) signal regime. Figure adapted from  Ref.~\cite{Chauhan:2021xus} under CC BY 4.0 license.}
\label{fig:dln1}
\end{figure}

After being produced, the RHNs typically decay into SM final states: $N\to W\ell, \, Z\nu, \, h\nu$~\cite{Atre:2009rg}. The total decay width $ \Gamma_i$ of the RHN $N_i$ at the tree-level depends on the Yukawa coupling $Y_D$, and is given by
\begin{equation}
    \Gamma_i \, = \, \frac{(Y_D^\dagger \, Y_D)_{ii}}{8 \, \pi } M_i .
\end{equation}
Thus in flavor symmetry models, the RHN decay lengths depend indirectly on the choice of the generator $Z(\mathbf{r})$ of the $Z_2$ symmetry and the choice of the CP transformation $X(\mathbf{r})$. For the example case considered above, see Eq.~(\ref{eq:YDgen}), the expressions for the decay widths are independent of values of $s$  and only depend on the Yukawa couplings $y_f$ and the angle $\theta_R$:
\begin{subequations}\label{gammaC1}
\begin{alignat}{2}
\Gamma_1 \, & =  \, \frac{M_N}{24 \, \pi} \, \left( 2\, y_1^2 \, \cos^2 \theta_R + y_2^2 + 2\, y_3^2 \, \sin^2 \theta_R \right)  \, , \\
\Gamma_2 \, & =  \,  \frac{M_N}{24 \, \pi} \, \left( y_1^2 \, \cos^2\theta_R + 2 \, y_2^2 + y_3^2\, \sin^2 \theta_R  \right) \, ,  \\
\Gamma_3 \ & =  \, \frac{M_N}{8 \, \pi} \, \left( y_1^2 \, \sin^2 \theta_R + y_3^2 \, \cos^2 \theta_R \right) \, .
\end{alignat}
\end{subequations}
We can convert these decay rates into decay lengths $L_i=\gamma/\Gamma_i$ in the laboratory frame, where $\gamma$ is the boost factor of RHN which can be determined depending on how the RHN is produced. Some numerical results for the decay lengths based on Eqs.~\eqref{gammaC1} are plotted in Fig.~\ref{fig:dln1} as a function of $\theta_R$. Here we have assumed the production of RHNs via $Z^\prime$ with mass $M_{Z^\prime}=4$ TeV in a $U(1)_{B-L}$ model, which implies $\gamma=M_{Z'}/2M_N=8~(13.3)$ for $M_N=250~(150)$ GeV. We find that the decay lengths are connected with neutrino mass ordering, as strong NO arises for $y_1=0$ so that $m_1$ vanishes, $m_2 = y_2^2 \, v^2/M_N$
and $m_3= y_3^2 \, |\cos 2 \, \theta_R| \, v^2/M$, while strong IO arises for $y_3=0$ so that $m_3=0$, $m_1= y_1^2 \, |\cos 2 \, \theta_R| \, v^2/M_N$
and $m_2= y_2^2 \, v^2/M_N$.  
 As can be seen from the decay length expressions~\eqref{gammaC1}, for strong NO and strong IO corresponding to $y_1=0$ and $y_3=0$ respectively (i.e. when $m_0=0$), there are ERS points for $\theta_R \to \pi/2, \, 3 \pi/2$ (NO) or $\theta_R \to 0, \, \pi$ (IO), at which $\Gamma_3\to 0$, i.e.~the RHN $N_3$ becomes long-lived. The larger the ERS is, the smaller the deviation
from points of ERS  will be, i.e.~$\theta_R$ is expected to deviate from $\theta_{R,0}$ by a small amount, $\delta\theta_R=|\theta_R-\theta_{R,0}|$. For $10^{-4}\lesssim \delta\theta_R\lesssim 10^{-2}$, this could lead to displaced vertex signatures from $N_3$ decay that are accessible to future dedicated LLP experiments, such as FASER~\cite{Feng:2022inv} and MATHUSLA~\cite{Curtin:2018mvb}. Most signals from $N_{1,2}$ decays are prompt but can also be slightly displaced depending on the choice of $\theta_R$. The distinction between the two cases (prompt vs displaced) is marked here by $L=1$ cm (horizontal line in Fig.~\ref{fig:dln1}) for the LHC, although the exact value might vary slightly depending on the details of the detector (CMS vs ATLAS).  The points of ERS are of particular relevance for phenomenology since $\theta_L$ deviating
from $\theta_{L,0} =$ $0$ or $\pi$ leads to a non-zero value of the reactor mixing angle $\theta_{13}$.  
ERS points are also relevant for leptogenesis, as discussed in Chapter~\ref{sec:cosmic}. 
 
The underlying Yukawa structure $Y_D$ not only predicts the decay lengths of the RHNs, but also their decay branching ratios (BRs), as the partial decay widths are proportional to $\left|\left(Y_D\right)_{\alpha i}\right|^2$. Considering the decay of long-lived $N_3$ at an LLP detector with $m_0=0$ and $M_N=250$ GeV, we find the following proportion of BRs: 
\begin{equation}
\label{eq:br3}
{\rm BR}(N_3\to e^\pm W^\mp): {\rm BR}(N_3\to \mu^\pm W^\mp): {\rm BR}(N_3\to \tau^\pm W^\mp) \, = \, \left\{\begin{array}{ll}
1: 27.7:18.1 & ({\rm NO}) \\
8.5:1:3.7 & ({\rm IO})
\end{array}\right. ,
\end{equation}
independent of $\theta_R$ and $s$, and almost independent of $M_N$, if $M_N \gg m_W$. Thus, measuring these RHN decay BRs at an LLP detector for at least two charged lepton flavors provides an independent test of the neutrino mass hierarchy at the energy frontier. Decay signals of $N_{1,2}$ at LHC (prompt or displaced vertex) can also be used to test mass hierarchy, but specifically, they depend on $\theta_R$ as well as on the chosen CP symmetry $X (s)$.

\subsection{Lepton Flavor Violation at Colliders}
\label{subsec:SSleptonsLHCRHnu}
The decays of Majorana RHNs into charged leptons lead to LNV as well as LFV signals. Consider the $Z'$-mediated production that leads to the same-sign dilepton final state~\cite{Deppisch:2013cya}: 
\begin{align}
    pp \to Z'\to N_iN_i\to \ell_\alpha^\pm \ell_\beta^\pm +2W^\mp \to \ell_\alpha^\pm \ell_\beta^\pm +4j \, .
    \label{eq:signal}
\end{align}
Since the partial decay widths of RHNs depend on $Y_D$, the LNV signal cross-section is affected by the choice for generator $Z$ of the $Z_2$ symmetry and the choice of the CP transformation $X$, see discussion in Section \ref{sec:delta6pheno}.

\begin{figure}[t!]
\centering
\includegraphics[width=0.49\textwidth]{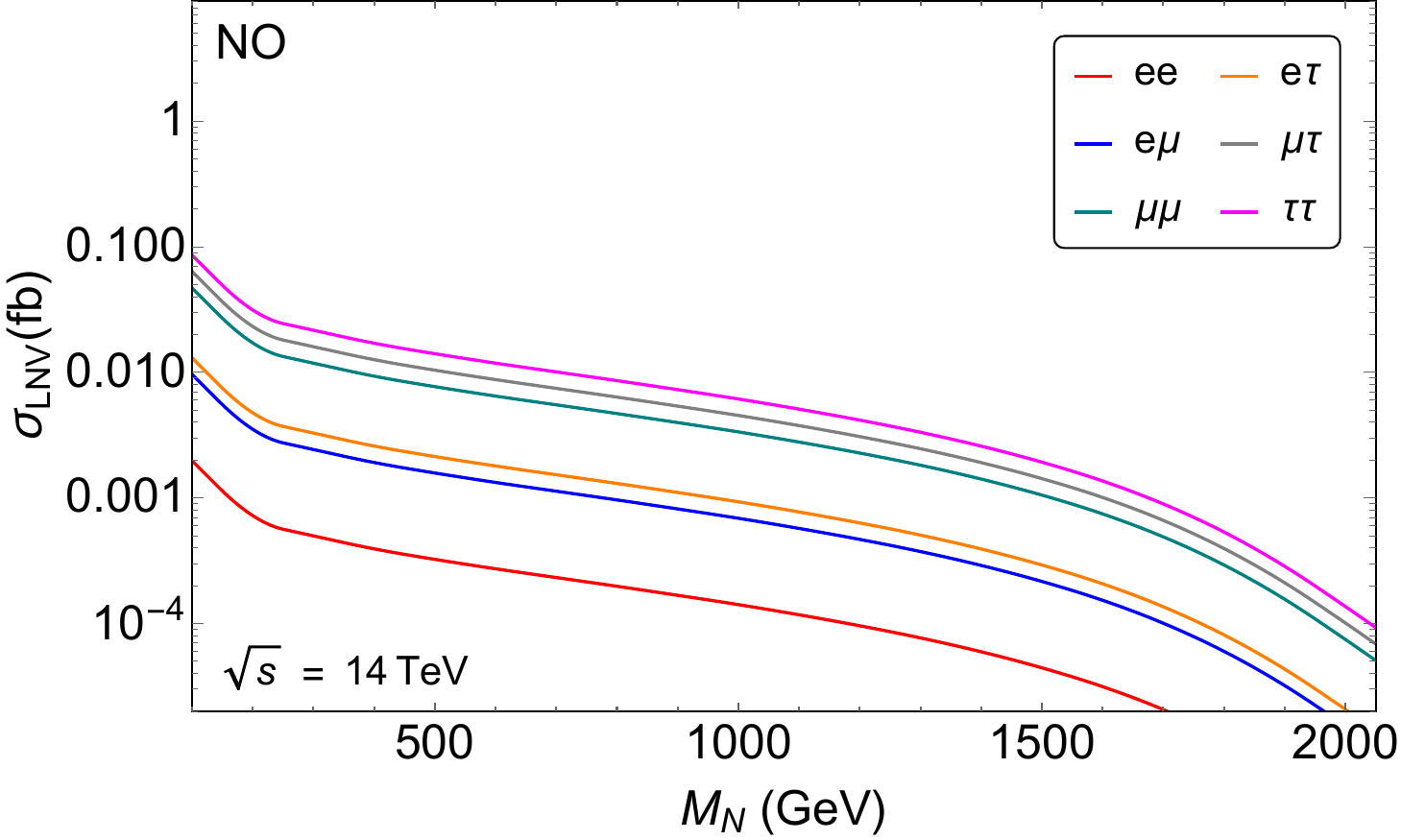}
\includegraphics[width=0.49\textwidth]{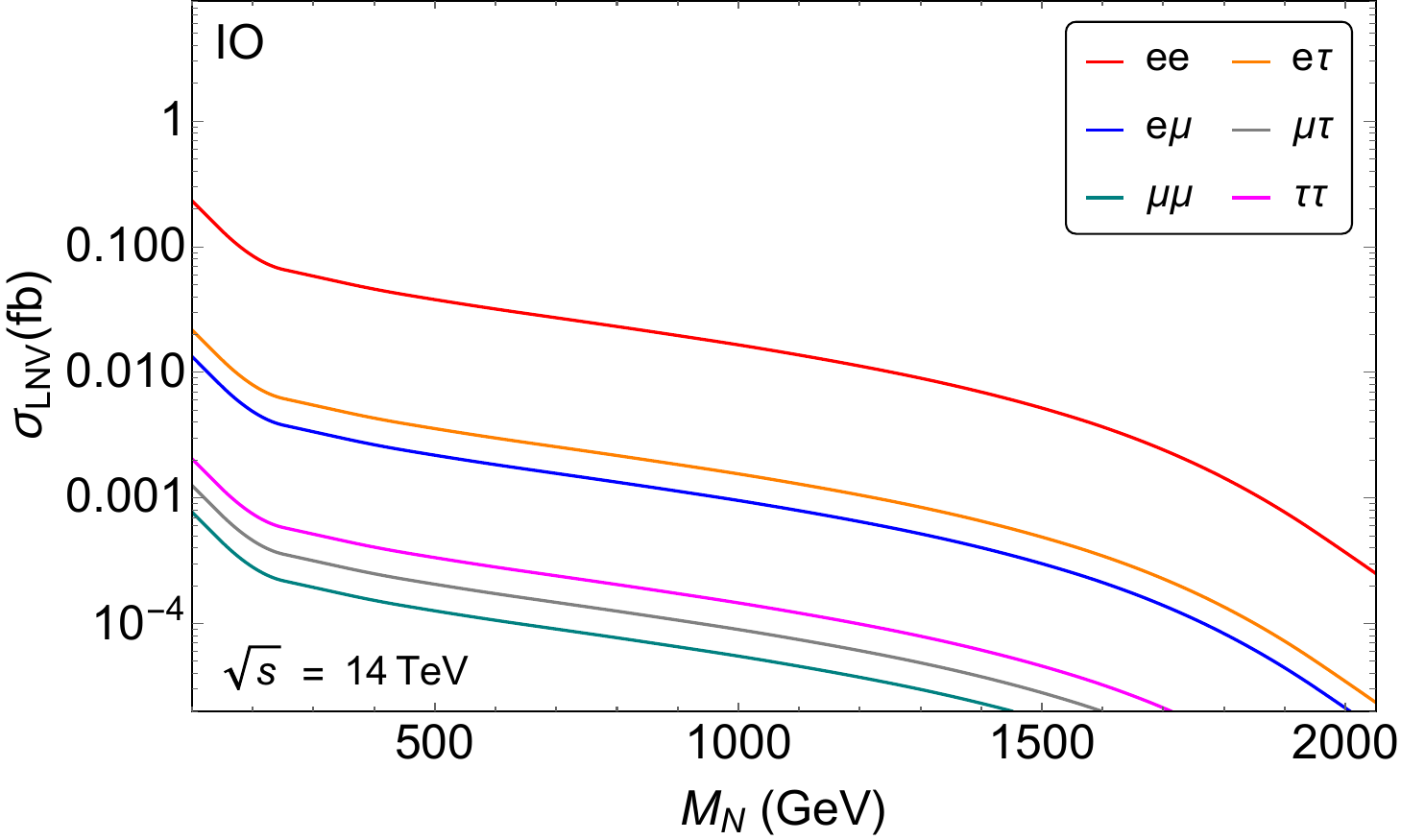}
\caption{Normalized LNV signals as a function of the RHN mass scale $M_N$ at $\sqrt s=14$ TeV LHC for all possible lepton flavor combinations in the strong NO (left) and strong IO (right) limit. Here we have fixed $M_{Z'}= 4$ TeV. Figure taken from the arXiv version of Ref.~\cite{Chauhan:2021xus}.}
\label{fig:collider1}
\end{figure}

We can probe the high-energy CP  phases in the Yukawa coupling matrix at colliders by constructing simple observables out of the same-sign dilepton charge asymmetry. In particular, we can define two observables $\sigma_{\rm LNV}^{\alpha,-}$ (difference) and  $\sigma_{\rm LNV}^{\alpha,+}$ (sum) of the same-sign charged-lepton final states of a given flavor $\alpha$: 
\begin{align}
    \sigma_{\rm LNV}^{\alpha,\pm} \, = \, & \sum_i \sigma_{\rm prod}(pp\to N_iN_i)\left(\left[{\rm BR}(N_i\to \ell_\alpha^-W^+)\right]^2\pm \left[{\rm BR}(N_i\to \ell_\alpha^+W^-)\right]^2 \right)   \times \left[{\rm BR}(W\to jj)\right]^2 \, .
\end{align}
The flavored CP asymmetries $\varepsilon_{i\alpha}$ relevant for leptogenesis turn out to be related to the ratio $\sigma_{\rm LNV}^{\alpha,-}/\sigma_{\rm LNV}^{\alpha,+}$~\cite{Bray:2007ru, Blanchet:2009bu, Dev:2019ljp}. Thus, measuring $\sigma_{\rm LNV}^{\alpha,-}/\sigma_{\rm LNV}^{\alpha,+}$ can help  measure the CP asymmetry, which is predicted by the group theory parameters. The normalized LNV cross sections $\sigma_{\rm LNV}(\ell^\pm_\alpha\ell^\pm_\beta)$ with respect to the new gauge coupling $g'$ for our example case are shown in Fig.~\ref{fig:collider1}. It can be concluded that comparing the LNV final states with different charged-lepton flavor combinations can provide an independent, complementary test of the neutrino mass ordering at the
high-energy frontier. 

\subsection{Correlation between Collider Signals and Leptogenesis}
\label{subsec:colliderleptogenesis}
Leptogenesis~\cite{Fukugita:1986hr} provides an attractive link between two seemingly distinct hints for BSM physics, namely, neutrino masses and mixing, and the  observed BAU, via the seesaw mechanism~\cite{Mohapatra:1979ia}. General details of the leptogenesis mechanism will be discussed in Section~\ref{sec:FLBAU}.   Here, we briefly summarize the collider prospects of testing TeV-scale resonant leptogenesis~\cite{Pilaftsis:2003gt} in the $\Delta(6n^2)$ flavor model discussed in Section~\ref{sec:delta6pheno}, along with an extra $U(1)_{B-L}$ so that TeV-scale RHNs can be produced more efficiently at colliders than in the minimal type-I seesaw. In principle, we could also consider other gauge groups under which the RHNs and the SM are charged, such as the left-right symmetric framework. However, it turns out that in the left-right models, the additional washout effects induced by the right-handed gauge interactions impose a lower bound of $M_{W_R}\gtrsim 20$ TeV to get successful leptogenesis, thereby precluding the possibility of testing it at the LHC~\cite{Frere:2008ct, BhupalDev:2015khe, BhupalDev:2019ljp, Zhang:2020lir}. On the other hand, the corresponding leptogenesis bound in the $U(1)_{B-L}$ model considered here is rather weak due to the double Boltzmann suppression of the $Z'$-induced washout effects~\cite{Blanchet:2009bu, Dev:2017xry, Liu:2021akf}.

 In the presence of the RHNs,  the flavored CP symmetries $\varepsilon_{i\alpha}$ for resonant leptogenesis depend on the structure of $Y_D$ (see Section \ref{sec:FLBAU}), and therefore, can be computed analytically using the form of $Y_D$ given in Eq.~\eqref{eq:YDgen} for the flavor model considered here. 
This CP asymmetry can then be translated into a BAU ($\eta_B$) as described in Section \ref{sec:FLBAU}, and compared with the observed value $\eta_B^{\rm obs}$.  Since the CP asymmetry is a function of $Y_D$, the choice for generator $Z$ of the $Z_2$ symmetry and the choice of the CP transformation $X$ determines the form of $\varepsilon_i$ and in turn, the $\eta_B$. 

\begin{figure}[h!]
\centering
\includegraphics[width=0.49\textwidth]{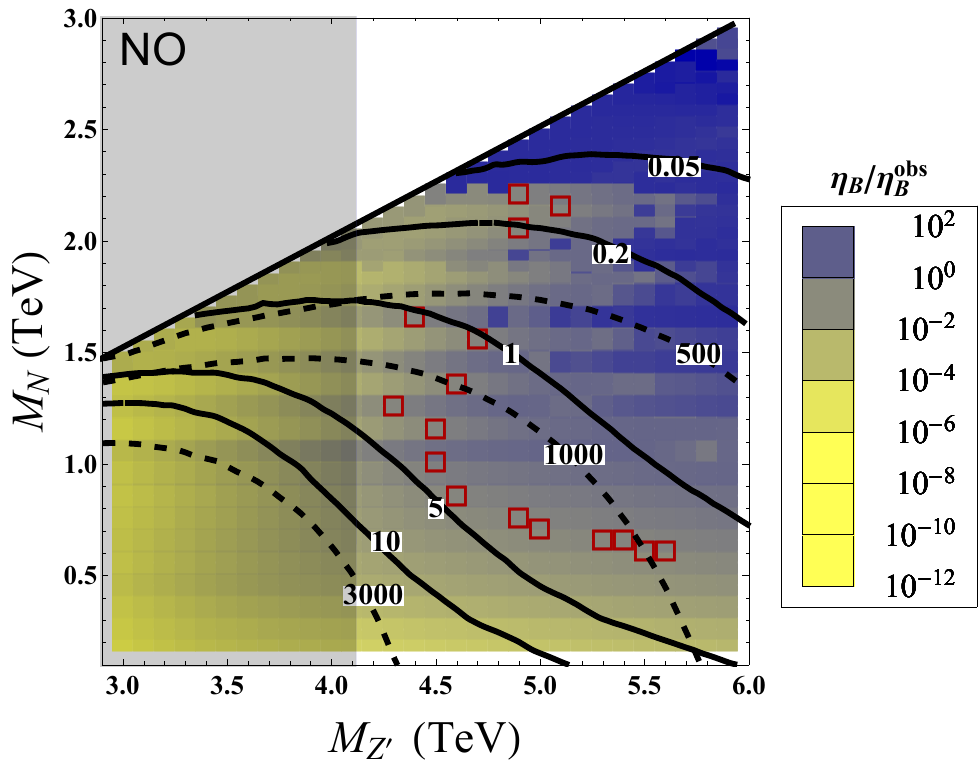}
\includegraphics[width=0.49\textwidth]{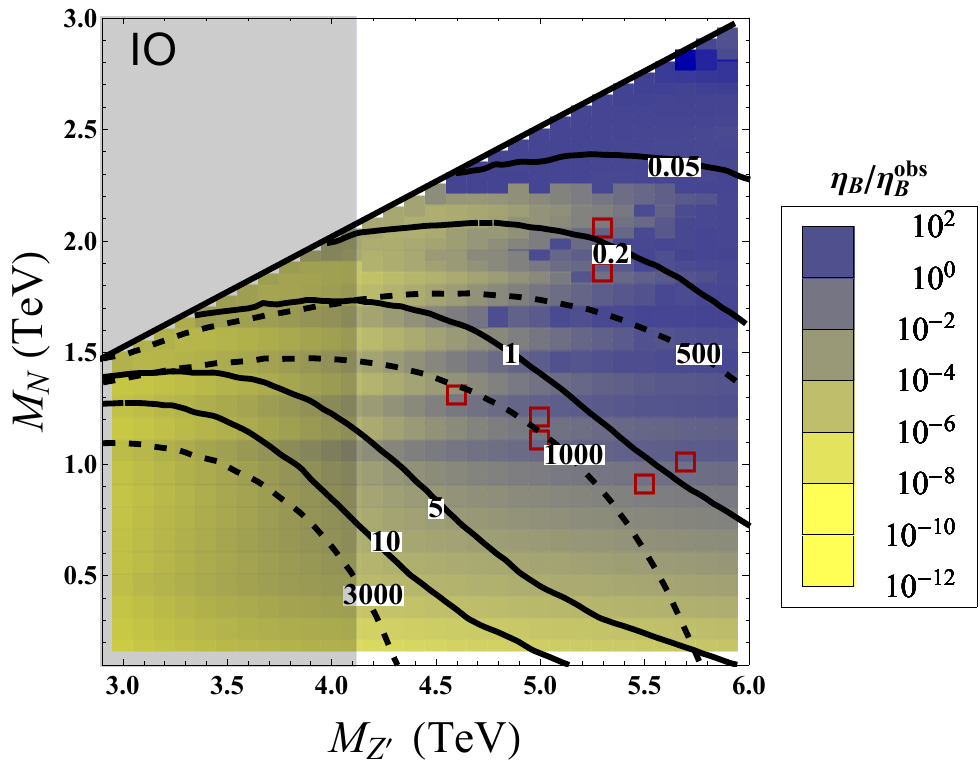}
\caption{Prediction for BAU $\eta_B$ relative to the observed value $\eta_B^{\rm obs}$ in the $(M_{Z'},M_N)$ plane for a fixed $g_{B-L}=0.1$ in a $\Delta(6n^2)$ model for strong NO (IO) in the left (right) panel, with $\theta_R$ set to the respective ERS points. The red boxes correspond to $\eta_B$ within $10\%$ of $\eta_B^{\rm obs}$. The contours show the RHN production cross sections (in ab) at the $\sqrt s=14$ TeV LHC (solid) and at $\sqrt s=100$ TeV FCC-hh (dashed).  The vertical shaded region is the current exclusion from LHC dilepton data. Figure taken from Ref.~\cite{Chauhan:2021xus}.}
\label{fig:etaBZp}
\end{figure}  
\begin{figure}[t!]
\centering
\includegraphics[width=0.75\textwidth]{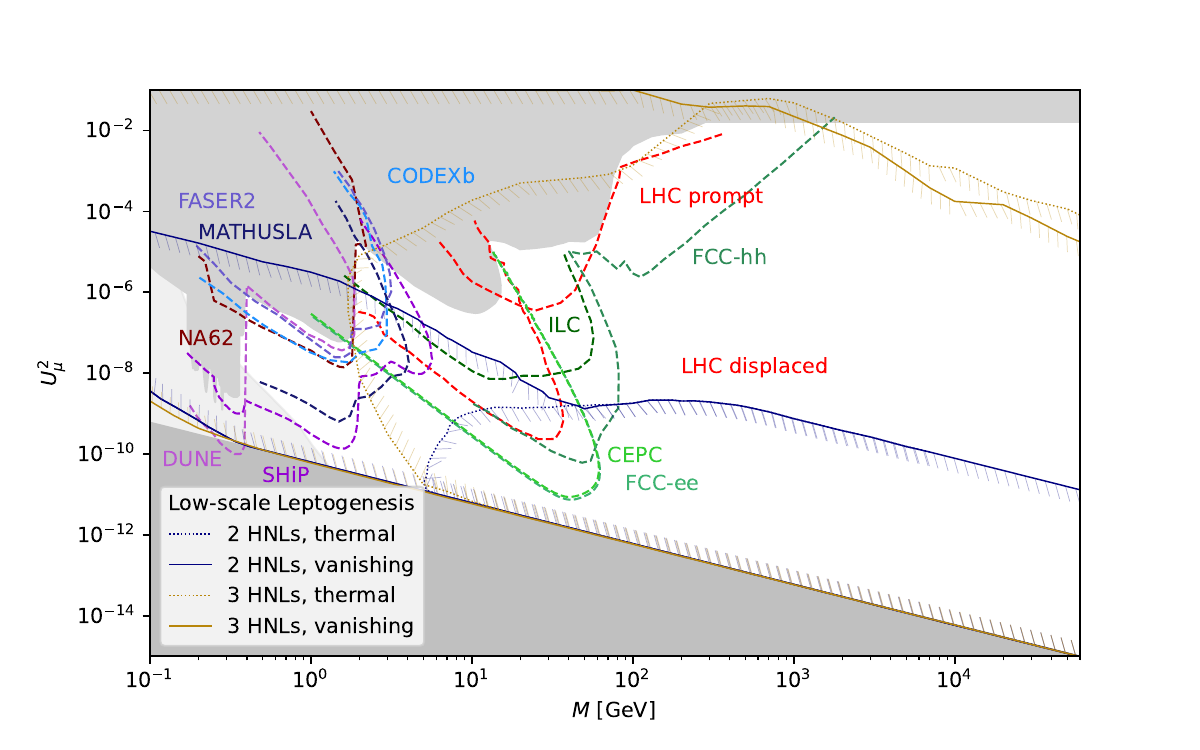}
\caption{Summary of LHC and future collider sensitivities to different low-scale leptogenesis scenarios with two and three RHNs. The upper (lower) shaded regions are excluded by laboratory (seesaw) constraints.  
In the legend, HNLs means `Heavy Neutral Leptons', which are called RHNs in this review. The $x$-axis is the RHN mass scale, and the $y$-axis gives the square of the light-heavy neutrino mixing in the muon flavor, which has the best experimental prospects (compared to electron and tau flavors). Figure taken from the arXiv version of Ref.~\cite{Abdullahi:2022jlv} \label{fig:fcc-lept}.}
\end{figure}
Fig.~\ref{fig:etaBZp} shows the predictions for $\eta_B/\eta_B^{\rm obs}$ in the $(M_{Z'},M_N)$ plane for both NO (left) and IO (right) in a particular case of the $\Delta(6n^2)$ model at ERS points~\cite{Chauhan:2021xus}. The red boxes correspond to $\eta_B$ within $10\%$ of $\eta_B^{\rm obs}$. Now to see whether these points are accessible at colliders, we superimpose the RHN pair-production cross sections $\sigma(pp\to Z'\to N_iN_i)$ in attobarns (ab) for both $\sqrt s=14$ TeV LHC (solid contours) and $\sqrt s=100$ TeV FCC-hh (dashed contours).   Successful leptogenesis for strong NO yields  $\sigma_{\mathrm{prod}}\lesssim 5$ ab at $\sqrt s=14$ TeV LHC, which makes it difficult to get any observable events even with the final target luminosity of 3 ab$^{-1}$ at HL-LHC. The situation worsens for strong IO, where the cross sections are smaller by at least an order of magnitude compared to the strong NO case. On the other hand, a future $100$ TeV hadron collider like FCC-hh~\cite{FCC:2018vvp} can reach a $\sigma_{\mathrm{prod}}$ up to $2000$ ab for the region of successful leptogenesis, which should yield up to 1000 LNV events with 30 ab$^{-1}$ integrated luminosity. In fact, by going to higher $Z'$ masses, the detection prospects at 100 TeV collider can be improved significantly. This is due to relaxed experimental limits on the new gauge coupling $g_{B-L}$ for $M_{Z'}\gtrsim 6$ TeV~\cite{Das:2021esm}.
For instance, at $M_{Z'}=7$ TeV, $g_{B-L}$ can be as large as one. Since $\sigma_{\mathrm{prod}}$ scales as $g_{B-L}^2$ for $M_N<M_{Z'}/2$, apart from a mild suppression due to change in the $Z'$ mass, the cross-section gets enhanced by a factor of 100. The detection prospects will be even better in other $Z'$ model variants like the leptophobic case, where the LHC bound is somewhat weaker.

In general, not working within the context of any particular flavor model, the current status and future prospects of testing low-scale leptogenesis in colliders and other laboratory experiments are summarized in Fig.~\ref{fig:fcc-lept} for the minimal type-I seesaw with either two or three RHNs~\cite{Abdullahi:2022jlv}. It illustrates the fact that LHC and future colliders provide ample opportunity to test low-scale leptogenesis parameter space. One can also note a large enhancement in the allowed mixing-mass space for three RHNs compared to the two RHN scenario. 
The exact picture depends strongly on the leptogenesis scenarios with degenerate or non-degenerate RHNs, freeze-in and freeze-out transitions, thermal initial conditions, LFV and oscillation constraints (e.g. mass of the lightest neutrino $m_0$). For an update, see also the talk in Ref.~\cite{georgisMTTD23}. 
For other works showing the collider/laboratory prospects of testing low-scale leptogenesis, see Refs.~\cite{Chun:2017spz, Klaric:2021cpi,Drewes:2021nqr,Fernandez-Martinez:2022gsu}.
Phenomenological aspects of the high- and low-scale leptogenesis with RHNs in the context of BAU and discrete symmetries  will be further discussed in Chapter \ref{sec:cosmic}.

\subsection{Collider Signals in Other Flavor Models}
In the above discussions involving the group $\Delta (6n^2)$ discussed in Section~\ref{sec:delta6pheno}, it is clear that the analyses of the flavor symmetry models lead to the rich phenomenology connected with intensity and  energy frontiers. Such phenomenological studies of discrete flavor symmetries can be extended to other discrete groups as well. We discuss below a few of these possibilities. 
 
In Ref.~\cite{CarcamoHernandez:2013yiy}, a model has been considered with two Higgs doublets and flavor symmetry $A_{4} \times {Z}_2 \times { Z}^{\prime}_{2}$, which can explain large leptonic mixing angles through a specific alignment of VEVs in the scalar sector. In Ref.~\cite{deMedeirosVarzielas:2015ybd}, some phenomenological consequences of the model for collider physics and the DM problem were further explored. The $Z_2$-even scalar fields of the considered model give two generations of fields that couple completely off-diagonally to the charged leptons. Thus, the model predicts LFV processes $\tau \to \mu \mu e$, $\mu \to e \gamma$ and $e^+e^- \to \tau^+ \mu^-$. As a consequence, LFV processes can also be searched for at hadron colliders (with an expectation of about ten events in each case) through the process $pp \to jj H_2 H_2 \to 2j4\ell$, where $H_2$ denotes either of the neutral components of the second generation $Z_2$-even scalar field. There are also contributions to the diphoton decay width of the Higgs boson. Still, even with couplings to the charged scalars of order unity, the deviations from the SM prediction are beyond the reach of HL-LHC. However, these percent level deviations from SM for scalar masses around TeV could be studied with future lepton colliders. The non-standard sector of this model is rich enough to include DM candidates (scalars and Majorana neutrinos). As shown in the analysis, they can also be probed at hadron colliders, with the possibility to account for the observed relic density for DM with a mass between 47 and 74 GeV or in the interval 600 GeV and 3.6 TeV.

A comprehensive collider study of the parameter space of the flavor symmetry group $A_4$  was performed in Ref.~\cite{Heinrich:2018nip}. At the leading order, breaking the flavor group into residual symmetries $Z_3$($Z_2$) in the charged lepton (neutrino) sector generates the TBM mixing. The required fit to the observed PMNS matrix is achieved through slightly broken residual symmetries induced by a shift in one of the flavon VEVs. A thorough study in constraining the 6-dimensional model parameter space is conducted using the experimental data from $g-2$, MEG, Higgs scalar mixing, Higgs width measurements, and a recast $8$ TeV ATLAS analysis. The most stringent results are obtained from the LFV limit on ${\rm BR}(\mu\to e\gamma)$ set by the MEG experiment, leading to approximately $60\%$ parameter space to be excluded and the recast ATLAS analysis leading to $40\%$ exclusion. 

In another work~\cite{Campos:2014zaa}, the authors consider studies on LFV Higgs decays in the context of 3HDM with $S_4$ symmetry. A remnant $Z_3$ symmetry, which arises due to a specific vacuum alignment, leads to strongly suppressed FCNC. However, this symmetry is slightly broken by perturbations, leading to mixing between the scalars and, hence, to LFV Higgs decays. The original motivation for this work was to explain the $2.4\sigma$ anomaly in $h\rightarrow \mu \tau$ channel reported by CMS in 2015. They find if the extra scalars are light, the contribution to $\ell'\rightarrow \ell\gamma$ can be suppressed while the flavor-violating couplings are still allowed to be large. Due to the $S_4$ symmetry, sizable $h\rightarrow \mu\tau$ leads to enhanced branching fractions also for LFV decays $
h\rightarrow e\mu\, ,e\tau$. Another study  correlating the LFV Higgs and $Z$-boson decays to LFV in the charged lepton sector can be found in  Ref.~\cite{Abada:2022asx}.  

In Ref.~\cite{Cao:2010mp}, collider signatures of $T_7$ flavor symmetry with gauged $U(1)_{B-L}$ are studied for a renormalizable two-parameter neutrino model. \emph{So a mixture of discrete and continuous symmetries is also possible}. Specifically, prospects for $Z'$ production and detection at LHC through decays into neutral Higgs scalars are studied, which subsequently decay into charged leptons with a specific flavor pattern determined from the flavor symmetry group.

As another example, collider signatures of vector-like fermions (VLF) with a non-Abelian flavor symmetry group $Q_6 \times Z_2$ symmetry is studied in Ref.~\cite{Bonilla:2021ize}. This group determines fermion masses as well as mixing. Only the third-generation fermions get their masses directly, while the rest obtain their masses in a see-saw-like mechanism. In this work, genetic algorithms are used to optimize the construction of neural networks that can maximize the statistical significance of a possible discovery (if any) for these VLFs at HL-LHC. While vector-like leptons can only probe masses safely up to $200$ GeV, the prospects for vector-like quarks are better with sensitivity up to $3.8$ TeV.
 
In a unified supersymmetric framework based on  GUT $SU(5)\times A_4$ symmetry~\cite{Belyaev:2018vkl}, studies on muon anomalous magnetic moment
$g-2$ and DM have been performed in the context of LHC data where the right-handed smuon with masses are predicted to be around 100 GeV and with lightest (non-universal) gaugino masses being around 250 GeV.  

\subsection{Higgs in the Diphoton Decay Channel}

The LHC data on the diphoton decay channel of the SM Higgs boson (with mass 125 GeV)~\cite{CMS:2021kom, ATLAS:2022tnm} can also have interesting phenomenological consequences in the context of discrete flavor symmetric scenarios. For example, for the scoto-seesaw FSS TM$_2$ model {which will be discussed in the context of cosmological issues in Section \ref{sec:FSdm},} the partial width of $h\to \gamma\gamma$ receives  additional and significant contributions due to the $\eta^{\pm}$ and $\eta_R$ one-loop effects~\cite{Ganguly:2023jml}.   The signal strength of $h\rightarrow \gamma\gamma$ relative to the SM prediction is given by
\begin{eqnarray}
    R_{\gamma\gamma}&=&\frac{\big[\sigma(gg\to h)\times{\rm Br}(h\to \gamma\gamma)\big]_{\rm Model}}{\big[\sigma(gg\to h)\times{\rm Br}(h\to \gamma\gamma)\big]_{\rm SM}}=\frac{\Gamma^{\rm SM}_{\rm Total}\times \Gamma(h\to \gamma\gamma)_{\rm Model}}{\Gamma^{\rm Model}_{\rm Total}\times \Gamma(h\to \gamma\gamma)_{\rm SM}}. \label{eq:Rgg}
\end{eqnarray}
While computing $R_{\gamma\gamma}$ in our analysis, we have taken the total decay width of the Higgs
boson in the SM as $\Gamma_{\rm Total}^{\rm SM}=4.1$ MeV~\cite{CERNtwiki}.

\begin{figure}[h!]
	\begin{center}
		\includegraphics[width=.4\textwidth]{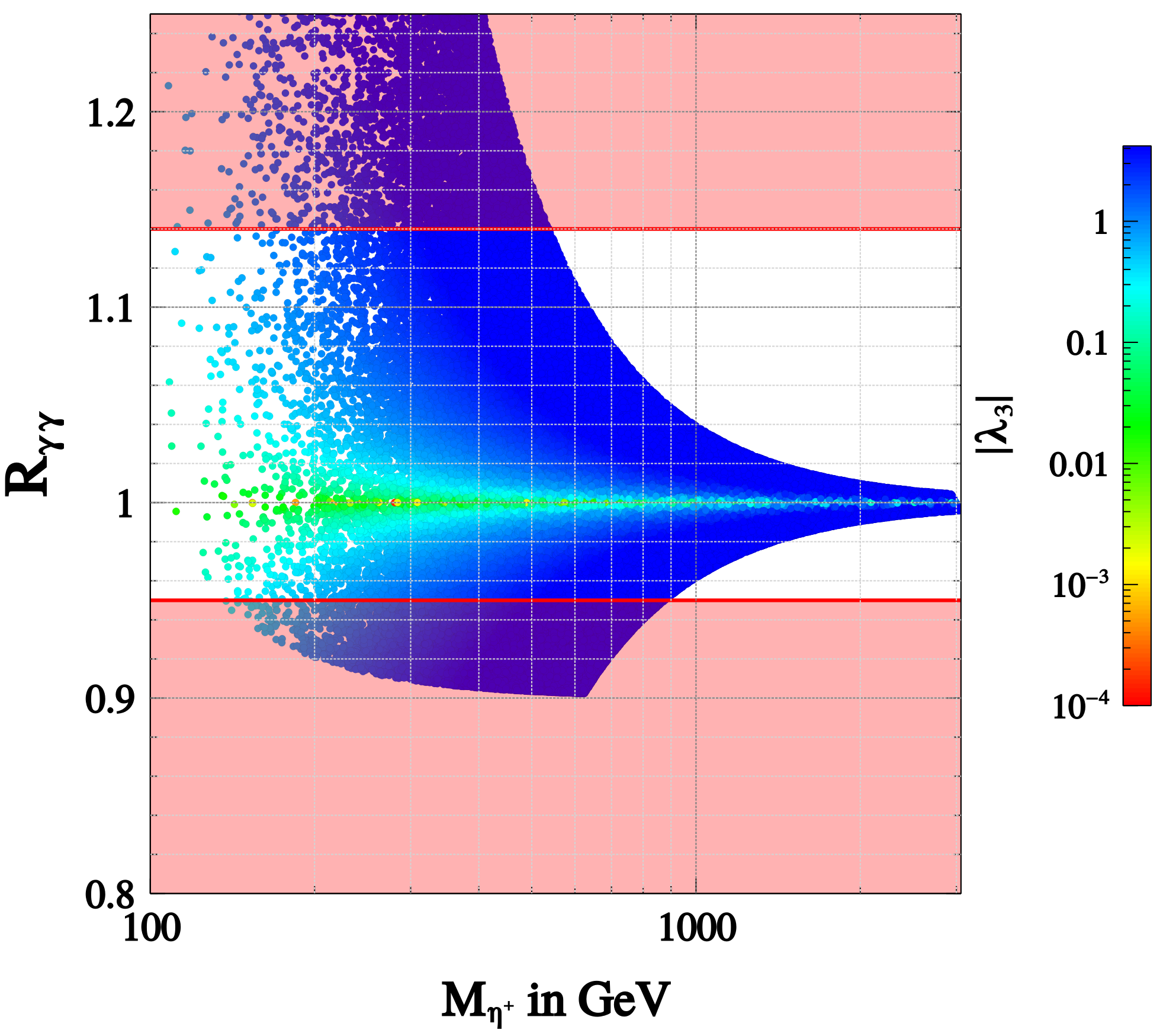}
             \includegraphics[width=.4\textwidth]{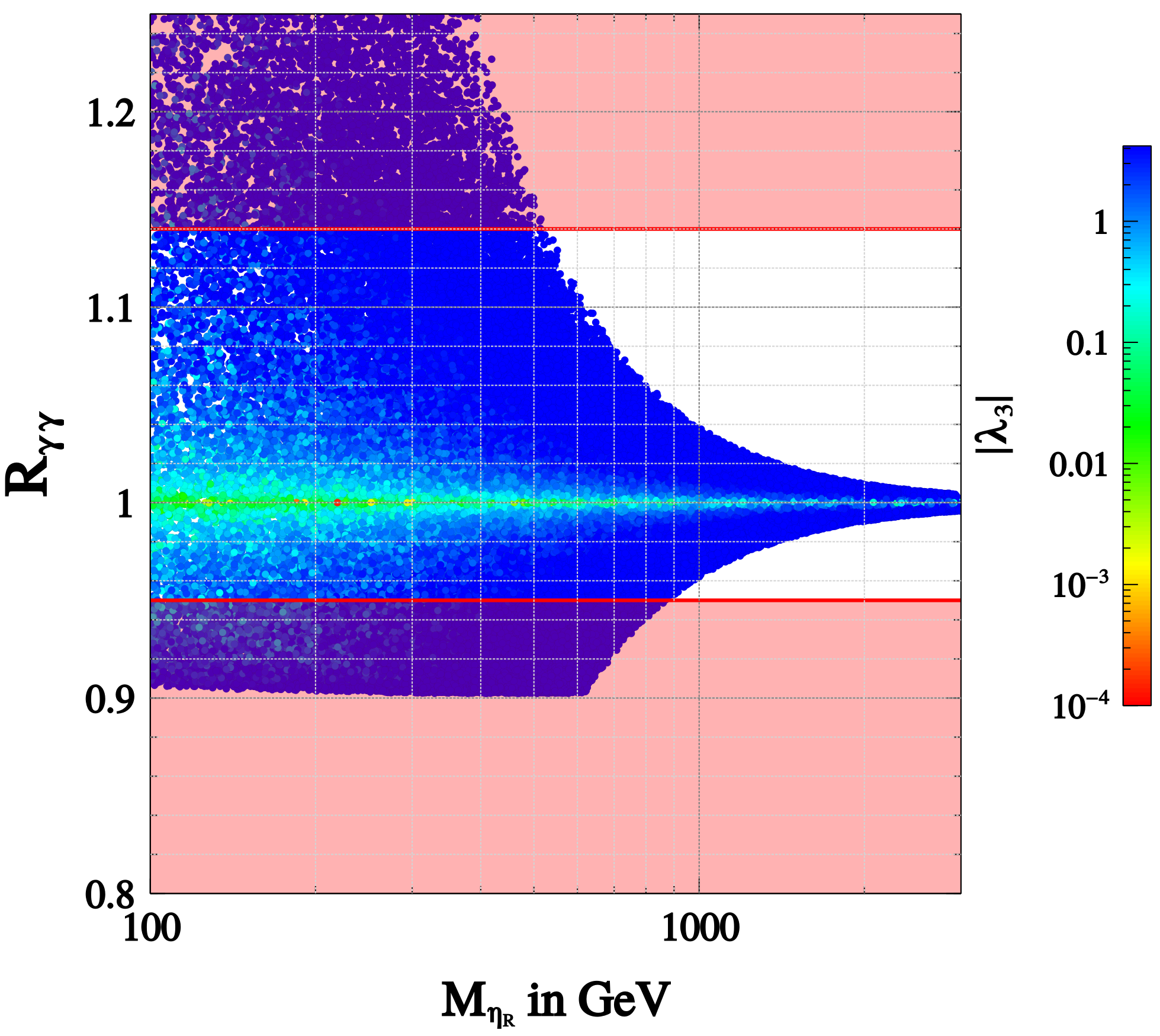}

	\end{center}
\caption{Predictions for the diphoton decay of the Higgs boson within the FSS model~\cite{Ganguly:2023jml}. $R_{\gamma \gamma}$ defined in (\ref{eq:Rgg}) is plotted against  $m_{\eta^{+}}$ (left) and  $m_{\eta_R}$ (right).  The white region is the current experimental allowed range measured by ATLAS~\cite{ATLAS:2022tnm}.  {Masses $M_{\eta^+}, M_{\eta_R}$ and the parameter $\lambda_3$ are defined in Section \ref{sec:FSdm} where more phenomenological aspects of the FSS model in the context of cosmology will be discussed.}}
\label{fig:hgg}
\end{figure}
 
In Fig. \ref{fig:hgg}, we have plotted the prediction for $R_{\gamma\gamma}$ in the context of $A_4$ flavor symmetric scoto-seesaw model~\cite{Ganguly:2023jml} against $m_{\eta^{+}}$ (left panel) and $m_{\eta_R}$  (right panel). 
The horizontal white region ($R_{\gamma\gamma}=1.04^{+0.10}_{-0.09}$) represents  the current allowed region measured by the ATLAS experiment using 139 fb$^{-1}$ of $pp$ collision data at $\sqrt s=13$ TeV~\cite{ATLAS:2022tnm}. Here $\lambda_3$ is a coupling for  $(H^{\dagger}H)(\eta^{\dagger}\eta)$ interaction, $\eta$ being the usual inert scalar doublet appearing in scotogenic models. 
{As we can see, a wide range of scalar boson masses can be probed through the diphoton Higgs decay channel using the present LHC experimental results. With the increasing data collection at LHC and HL-LHC, the precision of $R_{\gamma\gamma}$ will improve, giving prospects for better determination of allowed regions for specific flavor model parameters. Thus, phenomenology-based $R_{\gamma\gamma}$ constraints can be used for further studies and predictions for producing exotic discrete flavor model signals at present and future colliders. 

\section{Flavor Symmetry at Cosmic Frontier}\label{sec:cosmic}
{It is evident from the discussion in Chapter \ref{chapter:theory} that the flavor symmetries may play an important role in understanding the observed pattern of lepton mixings. To explain the origin of tiny neutrino mass, we often work within frameworks (discussed in Section \ref{sec:massmodels}) which also have the potential to explain diverse astrophysical and cosmological phenomenons.}   
In this Chapter, we will discuss some implications of flavor symmetry on various astrophysical and cosmological observables, including DM, BAU and gravitational waves. {Flavor symmetric constrictions can correlate these seemingly uncorrelated sectors with neutrino mixings and make it more predictive. For example, we will consider here frameworks based on the $A_4$  symmetry, the smallest non-Abelian discrete symmetry group with 3-dimensional irreducible representation. Based on this symmetry, we will construct specific models for TM$_1$ and TM$_2$ mixing schemes and discuss their consequences in DM and BAU phenomenology, respectively. }

\subsection{Example Group: $A_4$}\label{sec:a4group}
The $A_4$ discrete symmetry group is a group of even permutations of four
objects. {Interestingly, $A_4$ is a finite, non-Abelian subgroup of $SO(3)$ and $SU(3)$}.  Geometrically, it can be considered as an invariance group of a tetrahedron. It has {$4!/2$} = 12 elements, {which can be divided into four conjugacy classes}. {These twelve elements are: one unit operator, three rotations by $180^{\circ}$, four clockwise
rotations by $120^{\circ}$ and four anti-clockwise rotations by $120^{\circ}$. }
The $A_4$ group has three one-dimensional irreducible representations $1$,$1^{\prime}$ and $1^{\prime\prime}$ and one three dimensional irreducible representation 3. For a detailed discussion on $A_4$ character table and three-dimensional unitary representation of the generators,  see Refs.~\cite{Altarelli:2010gt,Ishimori:2010au}.
Two generators, known as $S$ and $T$ can form all the 12 elements (through multiplications in all possible way) which obey the relation 
\begin{eqnarray}
    S^2=T^3=(ST)^3=1
\end{eqnarray}

{This relation dictates the ‘presentation’ of the group. Therefore, the three 1-dimensional representations are given by
\begin{eqnarray}
    1 \rightarrow (S=1, T=1) \, , \nonumber \\
    1' \rightarrow (S=1, T=\omega) \, , \nonumber \\
    1'' \rightarrow (S=1, T=\omega^2) \, ,
\end{eqnarray}
where $\omega=e^{2i\pi/3}$ is cubic root of unity. Now, the basis for 3-dimensional representation can be written as 
\begin{eqnarray}
	S= \frac{1}{3}\begin{pmatrix}
		-1 & 2 & 2 \\
		2 & -1 & 2 \\
		2 & 2 & -1
	\end{pmatrix} , \quad 
T=\begin{pmatrix}
       1 & 0 & 0\\
        0 & \omega & 0 \\
        0 & 0 & \omega^2
\end{pmatrix}, 
\end{eqnarray}
from which  all 12 matrices of the three dimensional representation of $A_4$ can be
obtained. Alternatively, $S,T$ can be written in another 3-dimensional representation unitary representation, where $S$ is diagonal. The equivalence between these two basis and its effect in the relative phases of the neutrino mass eigen values has been discussed in~\cite{Barry:2010zk}. 
}
The multiplication rules of the singlets and triplets are given by~\cite{Altarelli:2010gt,Ishimori:2010au}
\begin{eqnarray}
	&& 1\otimes 1=1 \, , \nonumber \\ 
	&& 1^{\prime}\otimes1^{\prime\prime}=1 \, ,   \nonumber \\ 
	&& 1^{\prime}\otimes 1^{\prime}=1^{\prime\prime} \, , \nonumber \\
	&& 1^{\prime\prime}\otimes 1^{\prime\prime}=1^{\prime} \, , \nonumber \\
	&& 3\otimes3 =1\oplus1^{\prime}\oplus 1^{\prime\prime}\oplus 3_s\oplus 3_a \, ,
\end{eqnarray}
where the subscripts $``s"$ and $``a"$ denote symmetric and antisymmetric parts respectively. In the $T$ diagonal basis~\cite{Altarelli:2010gt}, writing two triplets as $(x_1,x_2,x_3)$ and $(y_1,y_2,y_3)$ respectively, we can write their products explicitly as 
\begin{eqnarray}
	1&\sim &x_1 y_1+x_2y_3+x_3 y_2, \nonumber\\
	1^{\prime}&\sim &x_3 y_3+x_1y_2+x_2 y_1,  \nonumber\\
	1^{\prime\prime}&\sim &x_2 y_2+x_1y_3+x_3 y_1, \nonumber \\
	3_s&\sim& \frac{1}{3}\begin{pmatrix}
		2 x_1 y_1-x_2 y_3-x_3y_2 \\
		2 x_3 y_3-x_1 y_2-x_2 y_1 \\
		2 x_2 y_2-x_1 y_3-x_3 y_1
	\end{pmatrix} ,\nonumber \\
3_a &\sim & \frac{1}{2}\begin{pmatrix}
       x_2 y_3-x_3 y_2 \\
        x_1 y_2-x_2 y_1 \\
        x_3 y_1-x_1 y_3
\end{pmatrix}.
\end{eqnarray}	
{These multiplication rules will be heavily used to obtain the TM$_1$ and TM$_2$ mixing schemes in the following sections.} 
\subsection{Flavor Symmetry and Dark Matter \label{sec:FSdm}}

There is overwhelming astrophysical and cosmological evidence for the existence of  DM, such as the large-scale structure data, gravitational lensing, and rotation curve of galaxies~\cite{Bertone:2004pz}. However, a laboratory discovery is still awaited. The relic abundance of DM has been measured by WMAP~\cite{WMAP:2012nax}, and more recently by PLANCK~\cite{Planck:2018nkj}, which set it at  $26.8\%$ of the total energy budget of the Universe. However, a broad classification of DM scenarios can satisfy this condition, such as weakly interacting massive particles
(WIMP)~\cite{Steigman:1984ac}, feebly interacting massive particle (FIMP)~\cite{Hall:2009bx}, strongly interacting massive particle (SIMP)~\cite{Hochberg:2014dra}, asymmetric DM (ADM)~\cite{Kaplan:1991ah}, and so on. For reviews of various DM candidates, see e.g. Refs.~\cite{Feng:2010gw, Bernal:2017kxu, Hochberg:2022jfs,  Petraki:2013wwa, Bertone:2018krk}. Over the years, several attempts have been made to connect neutrino physics with DM~\cite{Mohapatra:2002ug, Ma:2006km, Boehm:2006mi, Hambye:2006zn, Ma:2006fn, Allahverdi:2007wt,  Gu:2007ug, Sahu:2008aw,Bhattacharya:2018ljs, Chianese:2020khl,  Chianese:2018dsz,Bhattacharya:2021jli}. In a toy example~\cite{Bhattacharya:2018ljs}, the authors showed a one-to-one correspondence with WIMP DM and type-I seesaw. Earlier, we discussed that discrete flavor symmetric constructions have the potential to explain neutrino masses and mixing as well as can ensure the stability of DM. 

For example~\cite{Bhattacharya:2016lts,Bhattacharya:2016rqj}, with vectorlike singlet ($\chi^0$)-doublet ($\psi$) DM particle spectrum assisted with additional scalars such as $\phi,\eta$ (charged under a global $U(1)$ flavor symmetry) the interaction between DM and neutrino sector can be written as
\begin{equation}
\mathcal{L}_{int}=\left(\frac{\phi}{\Lambda}\right)^n \bar{\psi} \Tilde{H}\chi^0+\frac{(HL^TLH)\phi\eta}{\Lambda^3}.   \label{eq:dmint1}
\end{equation}
The first term in Eq. (\ref{eq:dmint1}), having a Yukawa-like configuration, acts like a Higgs portal coupling of the DM potentially accessible at various ongoing and future direct search and collider experiments.  The second term in Eq. (\ref{eq:dmint1}) plays a crucial role in generating non-zero $\theta_{13}$ as the existing $A_4$ flavons of the theory ensure the TBM mixing. A schematic view is given in Fig.~\ref{fig:dmscm}.
\begin{figure}[h!]
    \centering
    \includegraphics[width=0.5\textwidth]{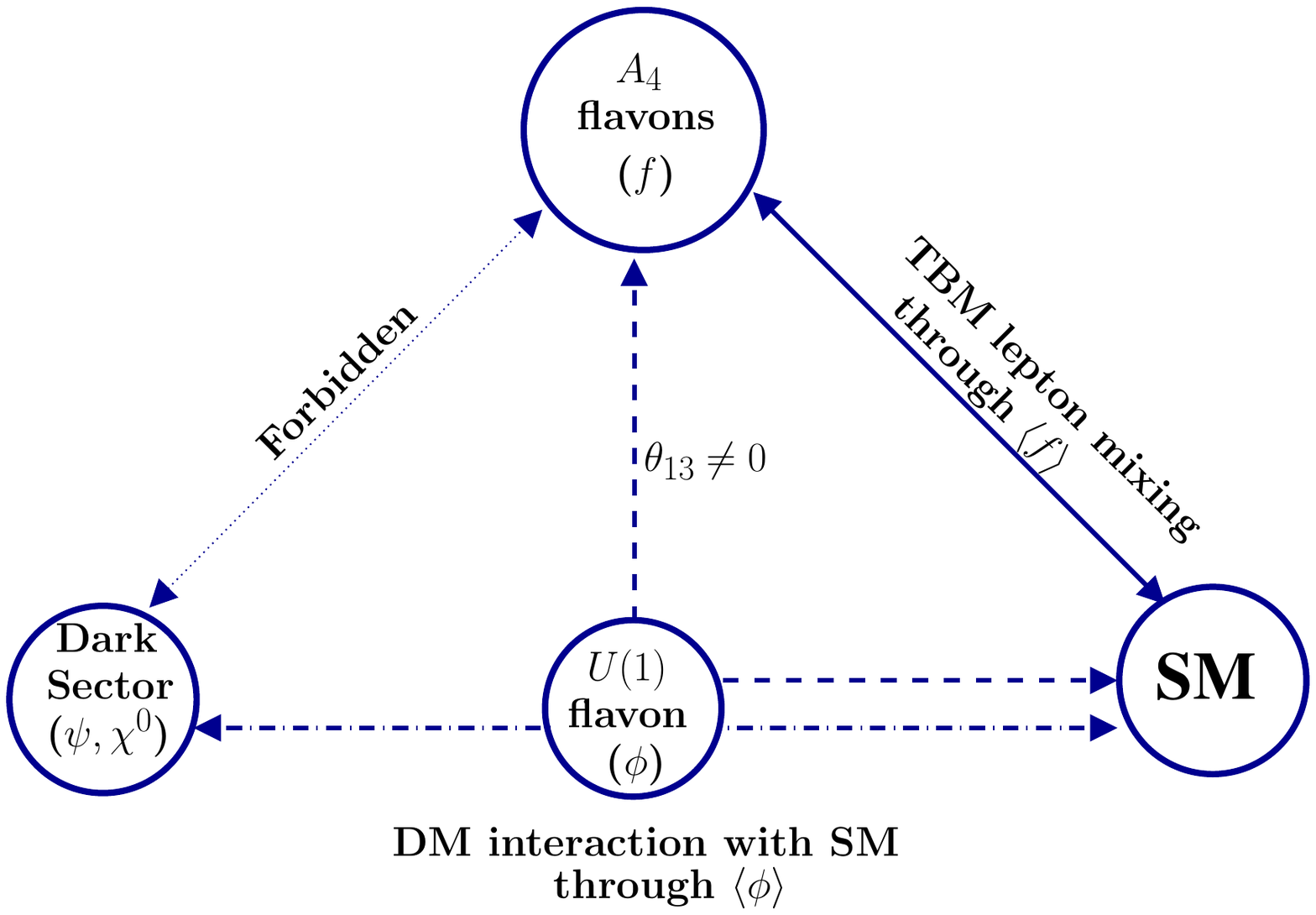}
     \caption{A schematic representation of DM ($\psi, \chi^0$) interaction with SM to generate non-zero $\theta_{13}$ in the presence of the $U(1)$ flavor symmetry. The $A_4$ flavons help in generating base TBM mixing. Figure adapted from the arXiv version of ~Ref.~\cite{Bhattacharya:2016lts}.}
    \label{fig:dmscm}
\end{figure}
Once the $U(1)$ flavor symmetry is broken, it ensures both the stability of DM and generates non-zero $\theta_{13}$. The coupling strength of the DM $\left(\frac{\phi}{\Lambda}\right)^n = \epsilon^n$ is constrained by the correct DM relic abundance, and $\epsilon$ is proportional to the magnitude of $\theta_{13}$. Additionally, the future precise measurement of the leptonic CP phase by T2K and NO$\nu$A experiments will reduce the uncertainty in $n$~\cite{Bhattacharya:2016rqj}. Few other examples of studies of  DM   with discrete flavor symmetry can be found in~Refs.~\cite{Esteves:2010sh, deMedeirosVarzielas:2015ybd, CarcamoHernandez:2013yiy, Hirsch:2010ru, Boucenna:2011tj, Verma:2021koo, Kajiyama:2010sb, Mukherjee:2015axj, Arcadi:2020zai, CentellesChulia:2016fxr,Mukherjee:2017pzq,Gautam:2019pce,deMedeirosVarzielas:2015lmh}. 

Apart from WIMP (or freeze-out) DM paradigm, discrete flavor symmetric constructions can also be extended to the FIMP (or freeze-in) mechanism of DM. In such a scenario, non-thermal DM populating the Universe via freeze-in
mechanism requires tiny dimensionless couplings ($\sim 10^{-12}$). On the other hand, if neutrinos have a tiny Dirac mass, it also requires a coupling of a similar order of magnitude. In~Ref.~\cite{Borah:2018gjk}, the authors have shown that such tiny coupling required in both dark and visible sectors may originate in $A_4$ discrete flavor symmetry. In another effort~\cite{Borah:2019ldn}, it has been shown that the non-zero value of the reactor mixing angle $\theta_{13}$ can originate from Planck scale suppressed operators and can as well realize a super-WIMP DM scenario. The study of DM with non-Abelian discrete symmetries demands more attention to explore the connection between neutrino physics and DM. {In the following,  we now discuss a detailed model for neutrino mixing and its connection with DM phenomenology. } 

\subsubsection*{{TM$_1$ mixing and dark matter phenomenology.} }
Here, we will present an example of the TM$_{1}$ mixing in the context of a hybrid flavor symmetric scoto-seesaw scenario (FSS)~\cite{Ganguly:2022qxj, Rojas:2018wym} where effective neutrino mass is generated via both type-I seesaw and scotogenic contributions.  { The whole framework is embedded within $A_4$ non-Abelian discrete flavor symmetry. Such scoto-seesaw framework was earlier presented to explain the observed hierarchy between the solar and atmospheric neutrino mass scales. From the discussion presented in Chapter \ref{chapter:introduction}, it is evident that the ratio of the solar-to-atmospheric mass-squared difference (r) is found to be
\begin{eqnarray}\label{eq:r}
    r=\Delta m_{21}^2/|\Delta m_{31}^2|\sim 3\times 10^{-2}.
\end{eqnarray}
This may be an indication of the involvement of two different mass scales that might originate from entirely separate mechanisms. In this framework, the SM particle content is extended by incorporating one heavy isosinglet neutral lepton $N_R$ (for the type-I sessaw), a dark fermion $f$ and an inert scalar doublet $\eta$ (for the scotogenic contribution).  Here, the type-I seesaw mechanism gives rise to the larger atmospheric mass scale, whereas the one-loop scotogenic contribution turns out to be the origin of the smaller solar mass scale mediated by a dark fermion, also providing a potential dark matter candidate. Here both $f$ and $\eta$ are considered to be odd under a dark $Z_2$ symmetry, ensuring the stability of dark matter.} The particle content of our model and charge assignment under different symmetries are shown in Table~\ref{table:a4tm1}. The role of each discrete symmetry, particle content and charge assignment in this table are described in detail in Refs.~\cite{Ganguly:2022qxj,Ganguly:2023jml}.  
	 \begin{table}[h!]
		\begin{center}
			\begin{tabular}{ccccccc|cccc}
				\hline
				Fields & $e_R$, $\mu_R$, $\tau_R$ & $L_{\alpha}$ & $H$ & $N_{R}$&  $f$& $\eta$ & $\phi_S$& $\phi_A$ & $\phi_T$ & $\xi$ \\
				\hline
				$A_4$ & $1$ , $1^{\prime\prime}$ , $1^{\prime}$ & $3$ & $1$ & $1$ & $1$ & $1$ & $3$ &$3$& 3 & $1^{\prime\prime}$ \\
				$Z_4$ &  $-1$ & $i$ & $1$ & $1$ & $1$ & $1$  & $i$ & $i$ & $-i$& $1$ \\
				$Z_3$ &  $1$ & $\omega$ &$\omega$ & $1$& 1 & $1$ & $\omega^2$ & $\omega$ & $1$ & $1$ \\
				$Z_2$ & $-1$ & $1$ & $1$ & $1$ & $-1$ & $-1$ & $1$ & $-1$ & $-1$& $-1$ \\
			    \hline
			\end{tabular}
		\caption{ Field contents and transformation under the symmetries of our model. Flavon fields in the second block of the table are introduced to implement the  $A_4$ symmetry. Here $\omega$ is the third root of unity.}
		\label{table:a4tm1}
		\end{center}
	\end{table} 
In this setup, the charged lepton Lagrangian can be written up to the leading order as 
\begin{eqnarray}\label{eq:Lag-cl1}
	 \mathcal{L}_l&=&\frac{y_e}{\Lambda}(\bar{L}\phi_T)_1H e_R + \frac{y_{\mu}}{\Lambda}(\bar{L}\phi_T)_{1'}H \mu_R + \frac{y_{\tau}}{\Lambda}(\bar{L}\phi_T)_{1''}H \tau_R + h.c. \nonumber \\
 &=&{\frac{y_e}{\Lambda} (\bar{L}_1\phi_{T_1}+\bar{L}_2\phi_{T_3}+\bar{L}_3\phi_{T_2})H e_R
 +\frac{y_{\mu}}{\Lambda} (\bar{L}_3\phi_{T_3}+\bar{L}_1\phi_{T_2}+\bar{L}_2\phi_{T_1})H \mu_R} \nonumber\\
 & & {+\frac{y_{\tau}}{\Lambda} (\bar{L}_2\phi_{T_2}+\bar{L}_1\phi_{T_3}+\bar{L}_3\phi_{T_1})H \tau_R} \label{eq:Lag-cl2}
 \end{eqnarray}
where $\Lambda$ is the cut-off scale of the FSS model. $y_e$, $y_{\mu}$ and $y_{\tau}$ are the coupling constants. {In Eq.~(\ref{eq:Lag-cl1}), the terms in the first parenthesis represent products of two $A_4$ triplets forming a one-dimensional representation
which further contract with 1, $1''$ and $1'$ of $A_4$, corresponding to $e_R$, $\mu_R$ and $\tau_R$, respectively. Following multiplication rules given in Section \ref{sec:a4group},
the complete $A_4$ decomposition is written in Eq.~(\ref{eq:Lag-cl2}).  Now, when the flavon $\phi_T$ gets VEV in the direction $\langle \phi_T \rangle=(\langle \phi_{T_1} \rangle+\langle \phi_{T_2} \rangle+\langle \phi_{T_3} \rangle)^T=(v_T,0,0)^T$, the charged lepton Lagrangian can be written as   
\begin{eqnarray}
	 \mathcal{L}_l&=&\frac{y_e}{\Lambda} \bar{L}_1 v_TH e_R
 +\frac{y_{\mu}}{\Lambda} \bar{L}_2v_TH \mu_R 
  +\frac{y_{\tau}}{\Lambda} \bar{L}_3v_T H \tau_R. \label{eq:Lag-cl3}
 \end{eqnarray}
Finally, when the SM Higgs field $h$ also gets non-zero VEV as $\langle h \rangle=v$,   following Eq.~(\ref{eq:Lag-cl3}), the diagonal charged lepton mass matrix can be written as
\begin{eqnarray}\label{eq:total-mass-matrix}
  M_\ell &=& \frac{vv_T}{\Lambda} \begin{pmatrix}
	y_e & 0   & 0 \\
	 0& y_{\mu} & 0 \\
	0 & 0 & y_{\tau}
	\end{pmatrix}. 
\end{eqnarray}
}
Now, the Lagrangian for neutrino mass  contributions in FSS can be written as
\begin{eqnarray}\label{eq:Lagtm1}
	\mathcal{L} = \frac{y_{N}}{\Lambda}(\bar{L}\phi_S)_1\tilde{H} N_{R}+\frac{1}{2}M_{N} \bar{N}_{R}^c N_{R}+ \frac{y_s}{\Lambda^2}(\bar{L}\phi_A)_{1'}\xi i\sigma_2 \eta^* f+ \frac{1}{2}M_f \bar{f}^c f+ h.c.,
\end{eqnarray} 
where $y_N$ and $y_s$ are the coupling constant and $M_N$ is the Majorana mass of the right-handed neutrino $N_R$ while $M_f$ is the mass of the fermion $f$. {Again, following Section~\ref{sec:a4group} 
and Eq.~(\ref{eq:Lagtm1}), the $A_4$ decomposition for the contribution to the neutrino sector can be written as 
\begin{eqnarray}\label{eq:Lagtm1-1}
	\mathcal{L} &=& \frac{y_{N}}{\Lambda}(\bar{L}_1\phi_{S_1}+\bar{L}_2\phi_{S_3}+\bar{L}_3\phi_{S_2})\tilde{H} N_{R} +\frac{1}{2}M_{N} \bar{N}_{R}^c N_{R}\nonumber\\
 & & + \frac{y_s}{\Lambda^2}(\bar{L}_3\phi_{A_3}+\bar{L}_1\phi_{A_2}++\bar{L}_2\phi_{A_1})\xi i\sigma_2 \eta^* f+ \frac{1}{2}M_f \bar{f}^c f+ h.c. \nonumber \\ 
 &=& \frac{y_{N}}{\Lambda}(\bar{L}_2v_{S}-\bar{L}_3v_{S})\tilde{H} N_{R} +\frac{1}{2}M_{N} \bar{N}_{R}^c N_{R} + \frac{y_s}{\Lambda^2}(\bar{L}_1v_{A}+2\bar{L}_2v_{A})\xi i\sigma_2 \eta^* f+ \frac{1}{2}M_f \bar{f}^c f+ h.c.
\end{eqnarray} }
In the above Lagrangian, we put VEVs of $\phi_S$, $\phi_A$ and $\xi$ in directions {$\langle \phi_S \rangle=(\langle \phi_{S_1} \rangle,\langle \phi_{S_2} \rangle,\langle \phi_{S_3} \rangle)^T=(0,-v_S,v_S)^T$, $\langle \phi_A\rangle=(\langle \phi_{A_1} \rangle,\langle \phi_{A_2} \rangle,\langle \phi_{A_3} \rangle)^T=(2  v_A,v_A,0)^T$} and $v_{\xi}$, respectively, getting the appropriate flavor structure. Following Eq. (\ref{eq:Lagtm1}), the   Yukawa coupling for Dirac neutrinos and scotogenic contributions can be written as 
\begin{eqnarray}\label{eq:seesaw yukawa}
   && Y_N=(Y_N^e,Y_N^{\mu},Y_N^{\tau})^T=(0,y_N \frac{v_S}{\Lambda},- y_N \frac{v_S}{\Lambda})^T, \\
    && Y_F=(Y_F^e,Y_F^{\mu},Y_F^{\tau})^T= (y_s \frac{v_{\xi}}{\Lambda}\frac{v_{A}}{\Lambda},y_s \frac{v_{\xi}}{\Lambda}\frac{2 v_{A}}{\Lambda}, 0)^T \equiv  (\kappa,2\kappa,0)^T. \label{eq:scotoyukuwa}
\end{eqnarray}

Finally, with the above Yukawa couplings, the total effective light neutrino mass matrix (with both type-I and scotogenic contributions) is given by 
\begin{eqnarray}
  M_{\nu}&=& -\frac{v^2}{M_N}Y_N^iY_N^j +\mathcal{F}(m_{\eta_R},m_{\eta_I},M_f)M_f Y_f^i Y_f^j \nonumber \\ 
	&=& \begin{pmatrix}
	b & 2b  & 0 \\
	2b & -a+ 4b & a \\
	0 & a & -a
	\end{pmatrix}, 
\end{eqnarray}
where 
\begin{eqnarray}\label{eq:def-a-b}
    a=y_N^2\frac{v^2}{M_N}\frac{v_S^2}{\Lambda^2}, b=y_s^2 \frac{v_{\xi}^2}{\Lambda^2}\frac{v_A^2}{\Lambda^2}\mathcal{F}(m_{\eta_R},m_{\eta_I},M_f) M_f = \kappa^2 \mathcal{F}(m_{\eta_R},m_{\eta_I},M_f)M_f
\end{eqnarray}
 and the loop function $\mathcal{F}$ is written as 
\begin{eqnarray}\label{eq:loopfn}
	\mathcal{F}(m_{\eta_R},m_{\eta_I},M_f)=\frac{1}{32 \pi^2}\Big[\frac{m_{\eta_R}^2 \log\Big(M_f^2/m_{\eta_R}^2\Big)}{M_f^2-m_{\eta_R}^2}-\frac{m_{\eta_I}^2 \log\Big(M_f^2/m_{\eta_I}^2\Big)}{M_f^2-m_{\eta_I}^2}\Big],
\end{eqnarray}
with $m_{\eta_R}$ and $m_{\eta_I}$ being  the masses of the neutral component of $\eta$. The total mass matrix $M_{\nu}$ therefore can be diagonalized by a mixing matrix of TM$_1$ mixing pattern given by 
\begin{eqnarray}\label{eq:unu}
     U&=&  U_{\rm TBM}U_{23}= U_{\rm TBM}\begin{pmatrix}
    1 & 0 & 0 \\
    0 & \cos\theta & \sin\theta e^{-i\psi} \\
    0 & -\sin\theta e^{i\psi} & \cos\theta
    \end{pmatrix}U_M \nonumber \\
    &=&\begin{pmatrix}
    \sqrt{\frac{2}{3}} & \frac{\cos\theta}{\sqrt{3}} & \frac{e^{-i\psi}\sin\theta}{\sqrt{3}} \\
    -\frac{1}{\sqrt{6}} & \frac{\cos\theta}{\sqrt{3}}+\frac{e^{i\psi}\sin\theta}{\sqrt{2}} & -\frac{\cos\theta}{\sqrt{2}}+\frac{e^{-i\psi}\sin\theta}{\sqrt{3}} \\
    -\frac{1}{\sqrt{6}} &\frac{\cos\theta}{\sqrt{3}}-\frac{e^{i\psi}\sin\theta}{\sqrt{2}} & \frac{\cos\theta}{\sqrt{2}}+\frac{e^{-i\psi}\sin\theta}{\sqrt{3}}
    \end{pmatrix}U_M
\end{eqnarray}
where $U_M$ is the Majorana phase matrix defined in Eq.~(\ref{upmns1}). Correlations among the oscillation parameters are given in Eq. (\ref{eq:tm1formulas}). In literature, the TM$_{1}$ mixing has been reproduced using various discrete groups such as $S_4$ in the context of type-I or type-II seesaw scenarios~\cite{Luhn:2013lkn,deMedeirosVarzielas:2012apl,Thapa:2021ehj,Chakraborty:2020gqc}.  

{ We can see that the total light neutrino mass matrix of Eq. (\ref{eq:total-mass-matrix}) contains two parameters $a$ and $b$ associated with the type-I seesaw and scotogenic contributions, which can be complex in general. We can write these parameters as $a=|a|e^{i\phi_a}$ and $b=|b|e^{i\phi_b}$ where $\phi_a$ and $\phi_b$ are the associated phases. For calculational purposes,  we define the parameter $\alpha=|a|/|b|$ and the difference of phases by $\phi_{ab}=\phi_a-\phi_b$. Here, the rotation angle $\theta$ and the phase $\psi$ appearing in Eq. (\ref{eq:unu}) can be expressed in terms of the model parameters as
\begin{eqnarray}\label{eq:theta phi exp}
    \tan\psi=\frac{2\alpha\sin\phi_{ab}}{5-2\alpha\cos\phi_{ab}},\quad \tan2\theta=\frac{2\sqrt{6}}{\cos\psi+2\alpha\cos(\psi+\phi_{ab})}. 
\end{eqnarray}
To obtain the correlation among the mixing angles and phases (assuming diagonal charged lepton mass matrix), we can compare $U_{\nu}=U_{\rm TBM}U_{23}U_M$ of Eq.~(\ref{eq:unu}) with $U_{\rm PMNS}$ given in Eq. (\ref{upmns1}). These correlations can be written as~\cite{deMedeirosVarzielas:2012apl,Luhn:2013lkn} 
\begin{eqnarray}\label{eq:tm1-corel}
   \sin\theta_{13}e^{-i\delta_{\rm CP}}=\frac{e^{-i\psi}\sin\theta}{\sqrt{3}},\quad \sin^2\theta_{12}=1-\frac{2}{3-\sin^2\theta},~~ \sin^2\theta_{23}=\frac{1}{2}\Big(1-\frac{\sqrt{6}\sin2\theta\cos\psi}{3-\sin^2\theta}\Big). 
\end{eqnarray}
These correlations among the three mixing angles are the unique feature of the considered $A_4$ flavor symmetry, giving rise to the  TM$_1$ mixing scheme. More specifically, relations in Eq.~(\ref{eq:tm1-corel}) are general for the TM$_1$ mixing scheme \cite{Luhn:2013lkn,deMedeirosVarzielas:2012apl} where the mixing angles $\theta_{13}$, $\theta_{12}$ are being correlated to each other. The correlation plot among these mixing angles can be found in Ref.~\cite{deMedeirosVarzielas:2012apl} where $\sin^2\theta_{12}$ is restricted to some narrow range corresponding to the $3\sigma$ regions of $\sin^2\theta_{13}$.  Relations in Eq.~(\ref{eq:theta phi exp}) are unique for the considered FSS model. From Eqs.~(\ref{eq:theta phi exp})-(\ref{eq:tm1-corel}), it is clear that the angle  $\theta$ and the associated  phase $\psi$ in $U_{23}$ can be linked with the parameters involved in $M_{\nu}$.   Relations in Eq.~(\ref{eq:tm1-corel}) imply that $\delta_{\rm CP}=\psi$ when $\sin\theta>0$, and $\delta_{\rm CP}=\psi \pm \pi$ for $\sin\theta<0$ which can be written in a compact form as $\tan\delta_{\rm CP}=\tan\psi$. Now, from Eq.~(\ref{eq:total-mass-matrix}), the 
 real and positive mass eigenvalues are calculated as
    \begin{eqnarray}
   m_1&=&0, \label{eq:real mass eigenvalue1} \\
   m_2&=& \frac{|b|}{2}\big[(5-2\alpha\cos\phi_{ab}-P)^2+(Q+2\alpha\sin\phi_{ab})^2\big]^{1/2}, \label{eq:real mass eigenvalue2}\\
   m_3&=&\frac{|b|}{2}\big[(5-2\alpha\cos\phi_{ab}+P)^2+(Q-2\alpha\sin\phi_{ab})^2\big]^{1/2} \label{eq:real mass eigenvalue3}.
    \end{eqnarray}
    where
    \begin{eqnarray}
    &&   P^2=\frac{M\pm \sqrt{M^2+N^2}}{2},\quad Q^2=\frac{-M\pm \sqrt{M^2+N^2}}{2}, \\
    && M=25+4\alpha\cos\phi_{ab}+4\alpha^2\cos2\phi_{ab},\quad N= 4\alpha \sin\phi_{ab}+4\alpha^2\sin2\phi_{ab}.    
    \end{eqnarray}
The phases associated with complex mass eigenvalues ($m^c_i$) are extracted as  $\gamma_i=\phi_b+\phi_i$, $i=2,3$. $i=1$ is excluded here as the lightest mass eigenvalue is zero, and the phase associated with $m_1^c$ is $\gamma_1=0$. Now, $\phi_{2,3}$ in the model can be written as
    \begin{eqnarray}
        \phi_2=\tan^{-1}\Big(\frac{Q+2\alpha\sin\phi_{ab}}{5-2\alpha\cos\phi_{ab}-P}\Big),\quad \phi_3=\tan^{-1}\Big(\frac{Q-2\alpha\sin\phi_{ab}}{5-2\alpha\cos\phi_{ab}+P}\Big)
    \end{eqnarray}
Using the above relations, we can calculate the  Majorana phase $\alpha_{32}$ in $U_{m}$, which can be written as 
\begin{eqnarray}\label{eq:majorana phase}
    \alpha_{32}=\tan^{-1}\Big(\frac{Q-2\alpha\sin\phi_{ab}}{5-2\alpha\cos\phi_{ab}+P}\Big)-\tan^{-1}\Big(\frac{Q+2\alpha\sin\phi_{ab}}{5-2\alpha\cos\phi_{ab}-P}\Big).
\end{eqnarray}
From Eqs.~(\ref{eq:theta phi exp})-(\ref{eq:majorana phase}), we observe that the mixing angles and all the phases depend on parameters $\alpha$ and $\phi_{ab}$ while the light neutrino masses depend on these parameters as well as on $|b|$. Now, we will estimate these model parameters ($\alpha$, $|b|$ and $\phi_{ab}$ ) using neutrino oscillation data on neutrino mixing angles and mass squared differences. With measured values of mixing angles $\theta_{13}$, $\theta_{12}$, $\theta_{23}$ and  mass-squared differences   $\Delta m_{21}^2$,  $|\Delta m_{31}^2|$,  we first estimate  $\alpha$ and  $\phi_{ab}$ using the $3\sigma$ range of neutrino oscillation data~\cite{Esteban:2020cvm}. The allowed ranges for $\alpha$ and $\phi_{ab}$ are plotted in the left panel of Fig \ref{fig:alpha} in the $\alpha-\phi_{ab}$ plane and find the ranges of $\alpha$ to vary between $4.82-5.27$ whereas ranges for $\phi_{ab}$ are $4.72-4.76$ and $5.03-5.06$ radian. 
\begin{figure}[h!]
	\begin{center}
		\includegraphics[width=.39\textwidth]{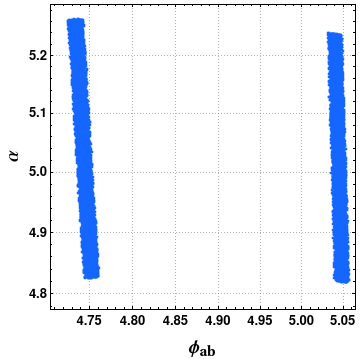}
		\includegraphics[width=.39\textwidth]{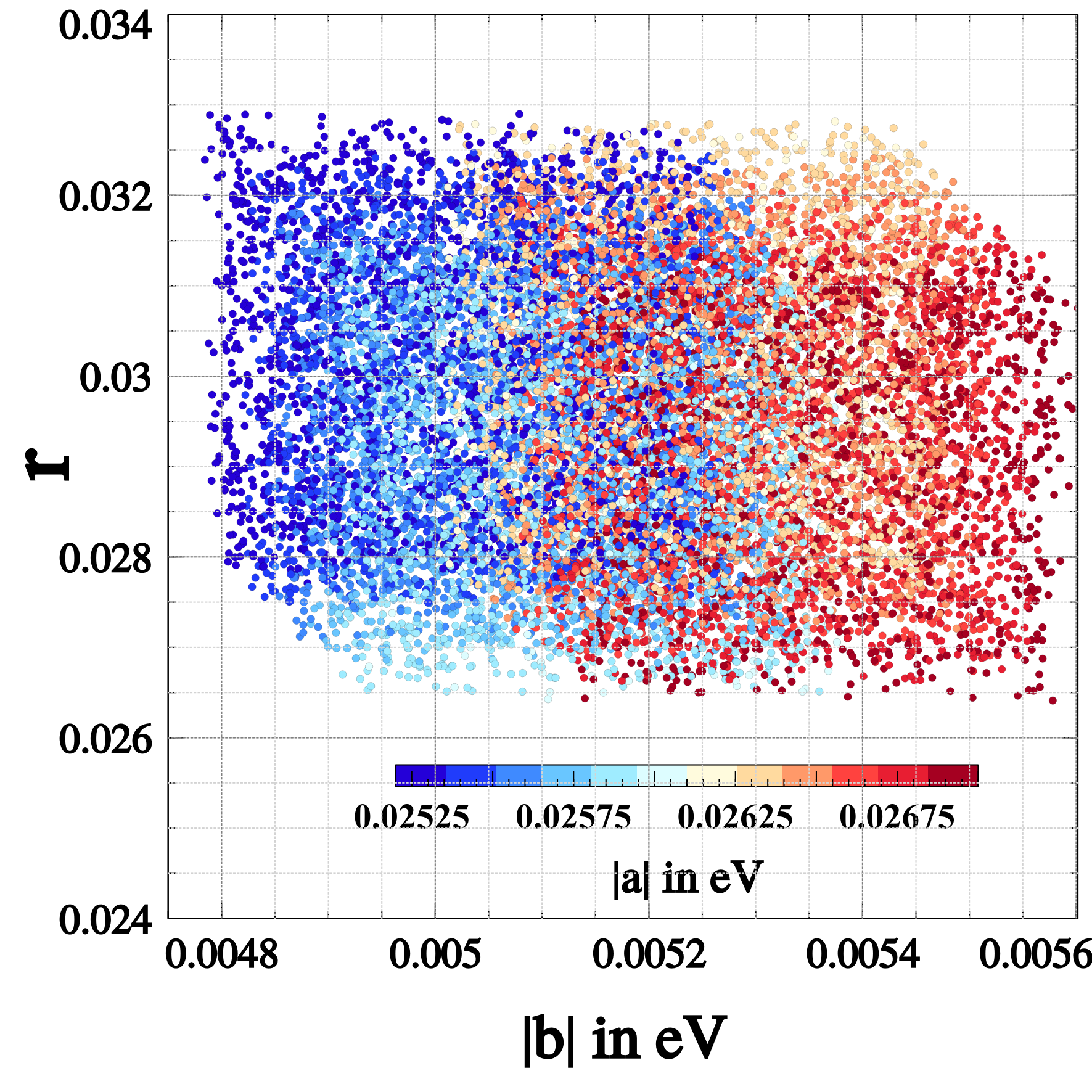}
	\end{center}
\caption{The allowed regions for $\alpha$ and $\phi_{ab}$ (left panel), and $r$ and $|b|$ (right panel) for the $3\sigma$ ranges of neutrino oscillation data~\cite{Esteban:2020cvm}. Figures taken from  Ref.~\cite{Ganguly:2023jml}.}
\label{fig:alpha}
\end{figure}
Subsequently, using Eq.~(\ref{eq:real mass eigenvalue1})-(\ref{eq:real mass eigenvalue3}), and the definition of the mass-squared differences, one can estimate $|a|$, $|b|$ as plotted in the right panel of Fig.~\ref{fig:alpha}. Here we find $|a|\gg |b|$, which explains the hierarchy between the two mass-squared differences. Therefore, the type-I seesaw dominantly contributes to the light neutrino mass.  On the other hand, the small scotogenic contribution is proportional to $|b|$ whose values constrained by neutrino mixing play a crucial role in DM phenomenology.  

Let us now consider the low energy FSS scalar potential which consists of two $SU(2)$ doublet scalars $H$ and $\eta$. The flavons considered here do not mix with these scalars and  the  low energy scalar potential of the model can be written as
\begin{eqnarray}
    V&=&-\mu_{1}^2(H^{\dagger}H)+\mu_{2}^2(\eta^{\dagger}\eta)+\lambda_1 (H^{\dagger}H)^2+\lambda_2 (\eta^{\dagger}\eta)^2+\lambda_3 (H^{\dagger}H)(\eta^{\dagger}\eta)+\lambda_4 (H^{\dagger}\eta)(\eta^{\dagger}H)\nonumber \\
    &&+\frac{\lambda_5}{2}\{(H^{\dagger}\eta)(H^{\dagger}\eta)+h.c.\} \label{eq:scalarpot}
\end{eqnarray}
The doublets in the model can be parameterized as
\begin{eqnarray}
    H=\begin{pmatrix}
        H^+ \\
      v/\sqrt{2}+(h+i\zeta)/\sqrt{2}
    \end{pmatrix},  \quad
    \eta=\begin{pmatrix}
        \eta^+ \\
        (\eta_R+i\eta_I)/\sqrt{2} \label{eq:eta}
    \end{pmatrix}.
\end{eqnarray}
The electroweak gauge symmetry is given by
\begin{eqnarray}
    H=\begin{pmatrix}
        0\\
        v/\sqrt{2}
    \end{pmatrix},\quad 
    \eta=\begin{pmatrix}
        0 \\
        0
    \end{pmatrix}.
\end{eqnarray}
The above symmetry breaking pattern ensures that the $Z_2$ symmetry will remain unbroken and results in two CP even scalars ($h,\eta_R$), one CP odd neutral scalar $\eta_I$ in addition to a pair of charged scalars ($\eta^{\pm}$). Due to the dark $Z_2$ symmetry, there is no mixing between $h$ and $\eta_R$, and $h$ plays the role of the SM Higgs boson. The $Z_2$ symmetry also ensures the stability of the lightest scalar ($\eta_R$ or $\eta_I$) that can act as a dark matter candidate. The masses of all scalars can be written in terms of the following parameters
\begin{eqnarray}
    \{\mu_2,\lambda_1,\lambda_2,\lambda_3,\lambda_4,\lambda_5\}.
\end{eqnarray}
These parameters can be written in terms of physical masses of scalars as
\begin{eqnarray}
 &&   \lambda_1=\frac{m_h^2}{2 v^2},\quad \lambda_3=\frac{2}{v^2}(M_{\eta^\pm}^2-\mu_2^2),\nonumber\\
&& \lambda_4=\frac{M_{\eta_R}^2+M_{\eta_I}^2- 2 M_{\eta^{\pm}}^2}{v^2},\quad \lambda_5=\frac{M_{\eta_R}^2-M_{\eta_I}^2}{v^2}.
\label{eq:scalarpot1}
\end{eqnarray}
We can choose all the $\lambda$s as free parameters or, equivalently the four physical scalar masses, $\lambda_2$ and $\mu_2$, namely
\begin{eqnarray}
   \{ \mu_2^2,m_h, M_{\eta_R}, M_{\eta_I}, M_{\eta^{\pm}},\lambda_2\}.
\end{eqnarray}
The potential mentioned here can satisfy the perturbativity and vacuum stability condition and associated parameters in Eq.~(\ref{eq:scalarpot1}) are important in generating correct DM relic density.  $\lambda_{3,4}$ can be constrained from SM Higgs diphoton signal strength and DM direct search. $\lambda_5$ is relevant in determining the scotogenic Yukawa coupling and hence is constrained from DM relic density, direct search constraints as well as neutrino phenomenology.

In the FSS framework, both type-I seesaw and scotogenic mechanisms are combined to obtain correct neutrino masses and mixing. As explained earlier, these two contributions are crucial to explain the hierarchy of mass scales associated with neutrino oscillation.  Out of three possible DM candidates, we focus on the singlet fermion denoted as $f$, an odd $Z_2$ particle in the scoto-seesaw scenario. We explore various mechanisms that can yield the correct relic density and determine the associated parameter space. Since $f$ is a gauge singlet, its production mechanism is connected to its Yukawa couplings, see Eq.~(\ref{eq:scotoyukuwa}) and Eq.~(\ref{eq:def-a-b}), with SM leptons and the inert doublet $\eta$. DM relic density can be realized through thermal freeze-out or freeze-in mechanisms, depending on the magnitude of Yukawa couplings.  Owing to the flavor symmetry, the magnitude of $|b|$ is such that correct DM relic density through freeze-in mechanism is not allowed. 
\begin{figure}[h]
    \centering
\includegraphics[scale=0.44]{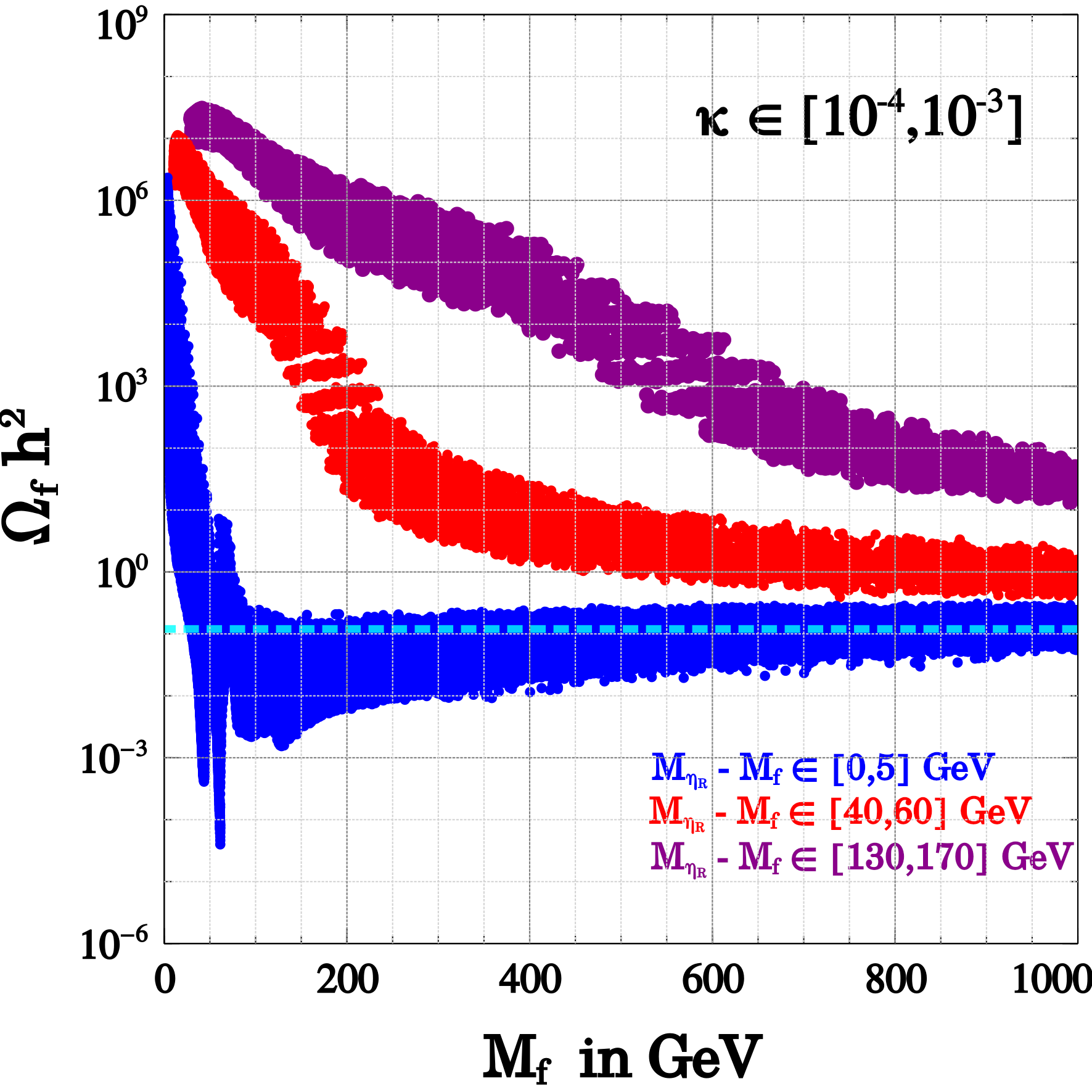}
    \hfil \includegraphics[scale=0.44]{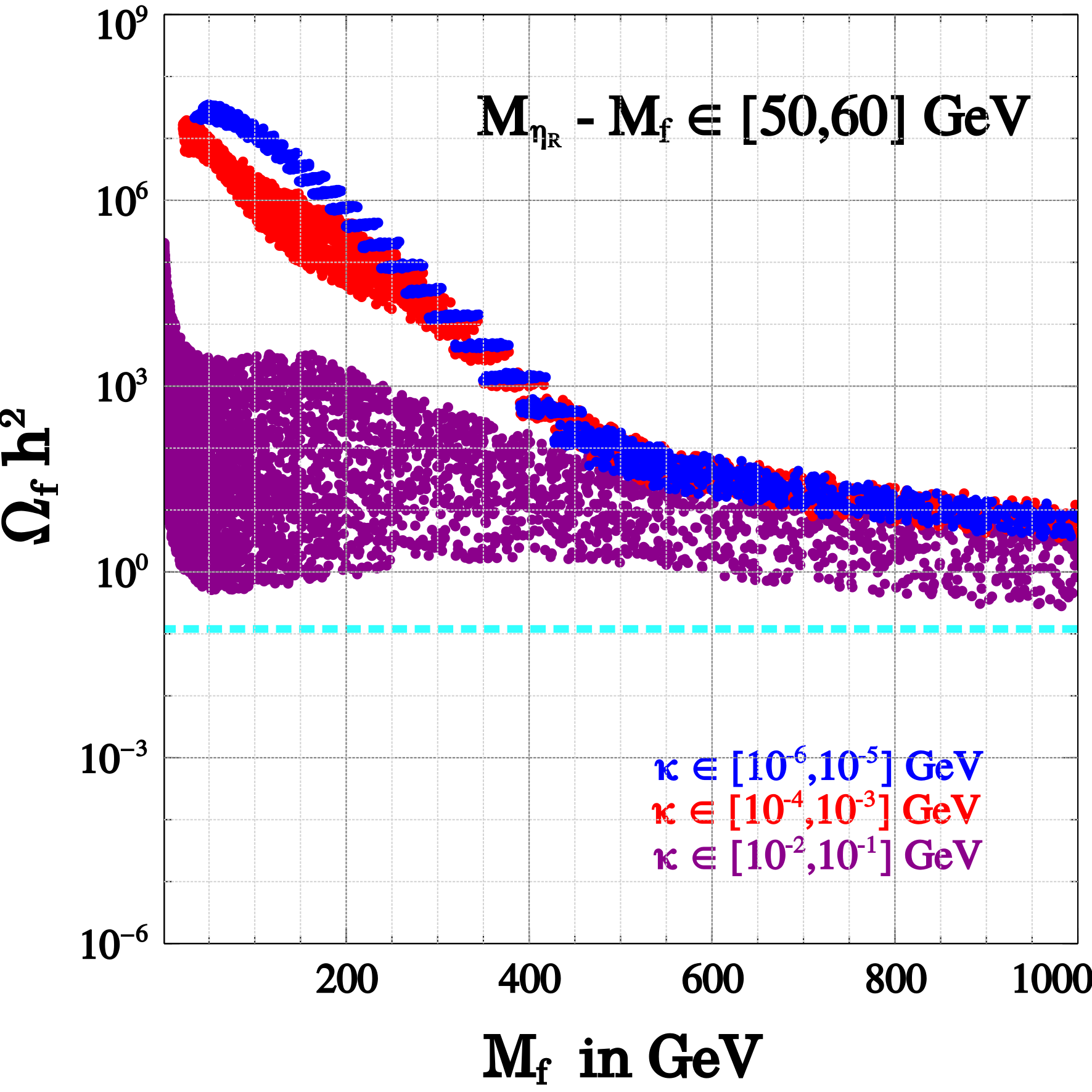}
    \caption{DM relic density as a function of DM mass with the Yukawa couplings (left panel) and mass difference $M_{\eta_R}-M_f$ (right panel) varied randomly as mentioned in the inset of the figure. The horizontal cyan line represents observed relic density~\cite{Planck:2018nkj}. Figures taken from  Ref.~\cite{Ganguly:2023jml}.}
    \label{fig:dmanalysis}
\end{figure}
To discuss the WIMP DM phenomenology in Fig.~\ref{fig:dmanalysis}, we have plotted the relic abundance as a function of DM mass $M_f$ for different values of mass splitting (between the lightest neutral scalar $\eta_R$ and $f$) and Yukawa coupling $\kappa$, see, Eq.~(\ref{eq:scotoyukuwa}). In the left panel of Fig.~\ref{fig:dmanalysis}, the Yukawa coupling $\kappa$ is varied within the range $\kappa \in [10^{-4},10^{-3}]$, while the mass difference between the lightest neutral scalar $\eta_R$ and $f$ (i.e., $M_{\eta_R}-M_f$) is varied in three different ranges such as $[0,5]$, $[40,60]$ and $[130,170]$ GeV respectively. In the right panel of Fig.~\ref{fig:dmanalysis}, the mass difference $(M_{\eta_R}-M_f)$ is varied within a small range of $[50,60]$ GeV for three different ranges of Yukawa couplings. An increase in Yukawa coupling here leads to a decrease in relic density. For a detailed discussion on DM phenomenology within this scoto-seesaw framework, see~\cite{Ganguly:2023jml}.

As mentioned earlier in our FSS framework both type-I seesaw and the scotogenic contribution play an instrumental role in explaining the hierarchy associated with the neutrino mass-squared differences, where scotogenic contribution  $|b|$ is constrained within the range $0.0048-0.0056$ eV.
\begin{figure}[h!]
    \centering
     \includegraphics[scale=0.44]{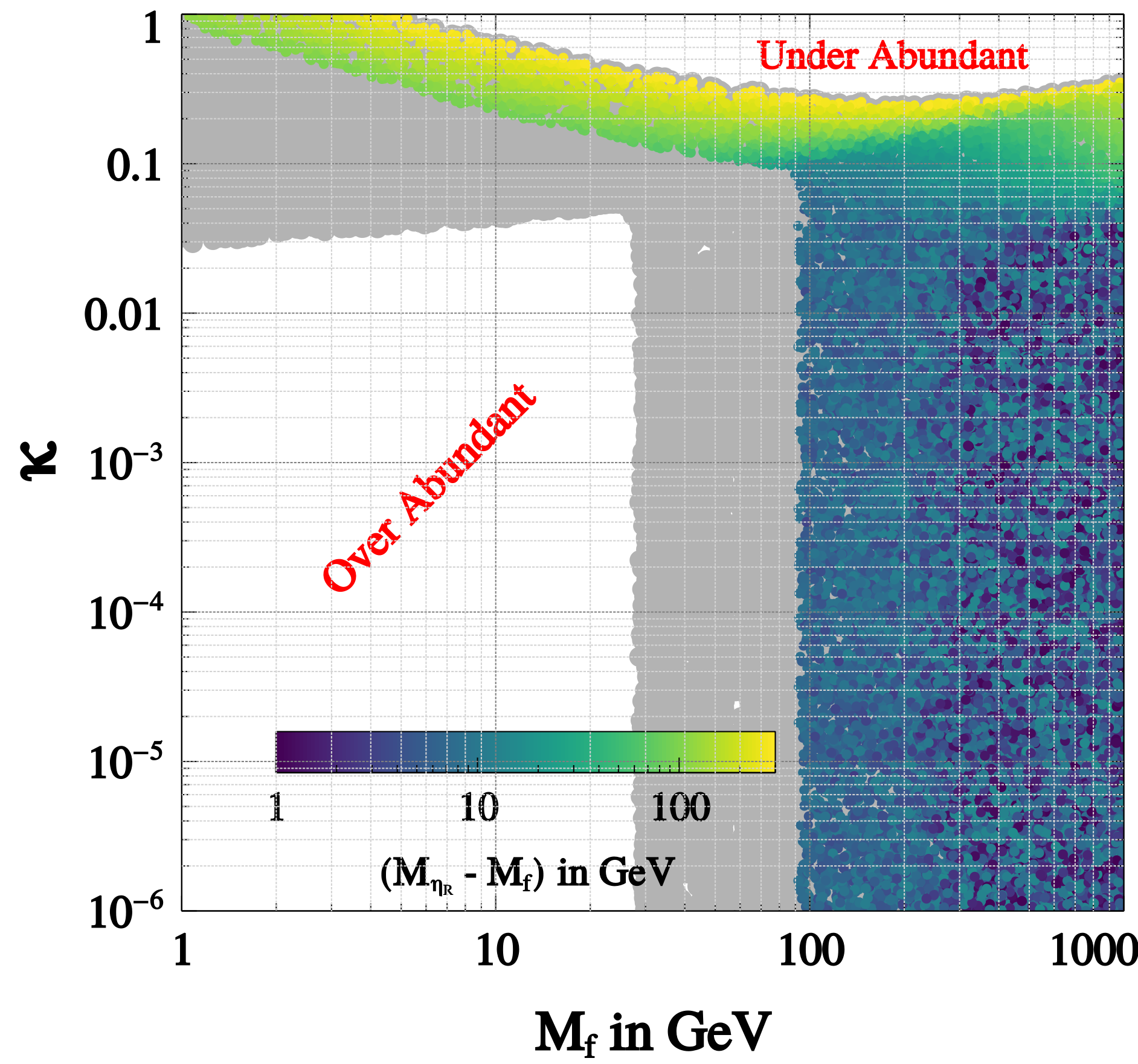}
    \includegraphics[scale=0.44]{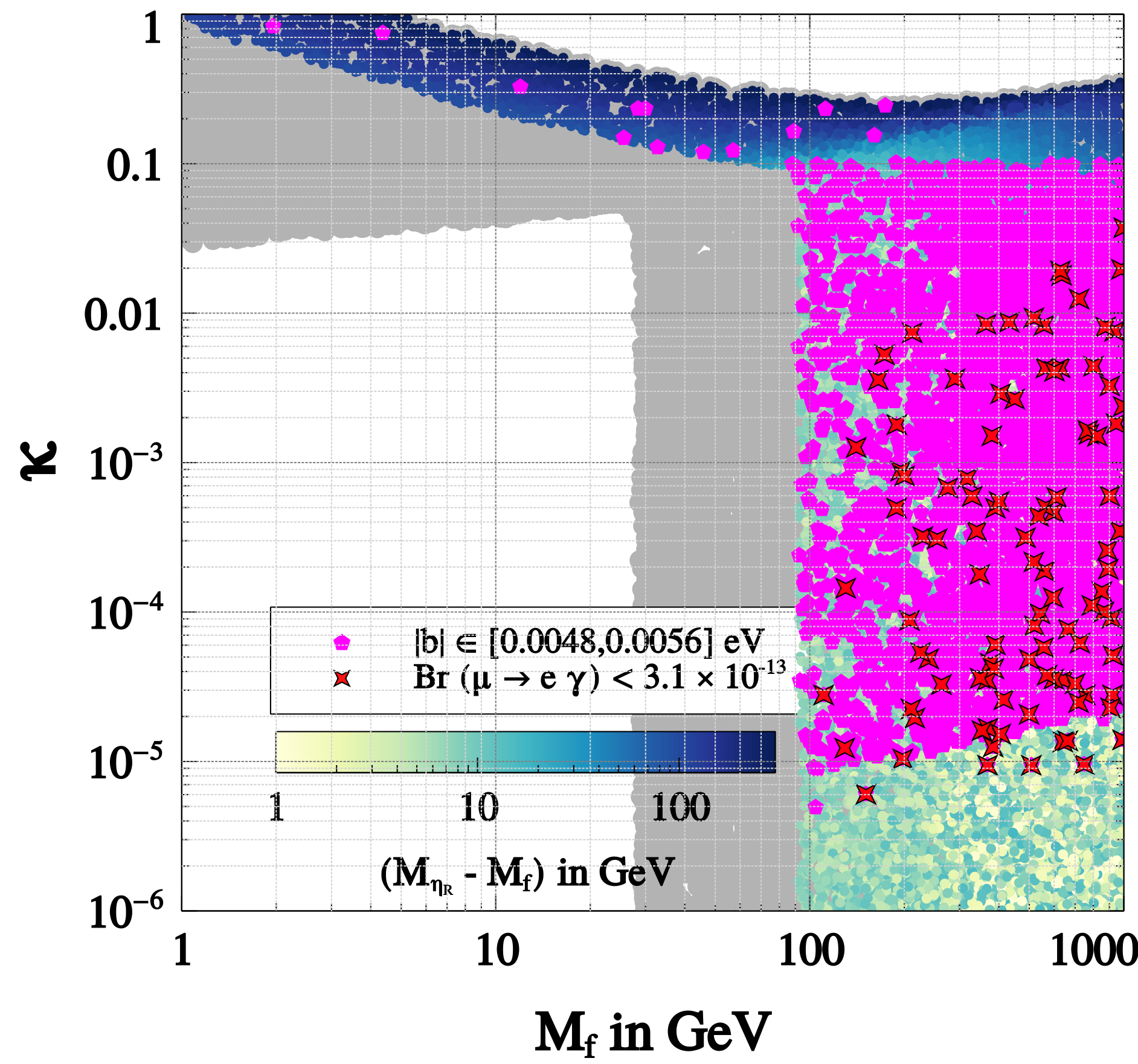}
    \caption{{On left: unconstrained DM predictions. On right: Final parameter space for the two important parameters of the FSS model, namely the Yukawa coupling $\kappa$ and the DM fermion mass $M_f$, after imposing the constraints from DM relic density, direct search, neutrino oscillation data, and LFV {decays}. Figures taken from  Ref.~\cite{Ganguly:2023jml}.}}
    \label{fig:all}
\end{figure}
The estimation of  DM relic density depends on the scotogenic contribution Yukawa coupling $\kappa$ associated with $|b|$ as given in Eq. (\ref{eq:scotoyukuwa}). This dependence is shown in Fig.~\ref{fig:all} (left panel) in the  $M_f - \kappa$ plane. Here the grey-colored points are ruled out by imposing a conservative limit on the doublet scalar mass given by the LEP experiment of about $M_{\eta}\geq 100$ GeV. A bluish color gradient shows allowed regions which satisfy correct relic density and direct search constraints with varying mass difference $M_{\eta_R}-M_f$.
In the right panel of Fig.~\ref{fig:all}, in the same plane,  we have included the allowed parameter space, taking into account neutrino oscillation data and experimental limits on LFV decays~\cite{Ganguly:2023jml}. The magenta-colored points satisfy neutrino oscillation data whereas the red points satisfy LFV decay constraints. In summary, with $\kappa \sim \mathcal{O}(10^{-2}-10^{-5})$ and DM masses between $(100-1000)$ GeV are found to be simultaneously consistent with neutrino oscillation data,  DM relic density, direct search, and LFV decay constraints.
The flavor symmetry construction in the FSS framework allows us to find a common origin for the TM$_1$ mixing scheme and correct DM relic abundance.}

\subsection{Flavor Symmetry and Baryon Asymmetry of the Universe}\label{sec:FLBAU}

Cosmological observations reveal that our Universe possesses a net excess of baryons over antibaryons~\cite{Planck:2018nkj, Planck:2018vyg} which is often termed {\it the baryon asymmetry of the Universe} (BAU). This is measured by the ratio of the excess number density of baryons over antibaryons to photons at present time~\cite{Planck:2018nkj}:
\begin{eqnarray}
    \eta_B^{\rm obs} = \frac{n_B-n_{\bar{B}}}{n_{\gamma}}=(6.12\pm 0.08)\times 10^{-10} \, .
    \label{eq:etaBobs}
\end{eqnarray} 
The mechanism of dynamical generation of the BAU is known as {\it baryogenesis} (see Refs.~\cite{Riotto:1999yt, Cline:2006ts, Bodeker:2020ghk} for reviews) and must satisfy three basic Sakharov conditions, namely, baryon number violation, C and CP violation, and out-of-equilibrium~\cite{Sakharov:1967dj}. Although the SM possesses all three ingredients, it is not sufficient to reproduce the observed BAU~\cite{Kolb:1990vq}. One attractive mechanism to produce the BAU is via {\it leptogenesis}~\cite{Fukugita:1986hr}, where one first produces a lepton asymmetry, and then uses the electroweak sphaleron processes~\cite{Kuzmin:1985mm, Khlebnikov:1988sr, Harvey:1990qw} to convert it to a baryon asymmetry; see Refs.~\cite{Buchmuller:2005eh, Davidson:2008bu, Pilaftsis:2009pk, Fong:2012buy} for reviews. This is particularly appealing, because it relies on the seesaw mechanism which also accounts for the neutrino masses and mixing, thus providing a link between the two seemingly disparate pieces of evidence for BSM physics. 

In Chapter \ref{chapter:theory}, we have briefly reviewed various neutrino mass generation mechanisms, which once augmented with discrete flavor symmetries explain the origin of tiny neutrino masses as well as observed neutrino mixing. Most of these mass-generation mechanisms include additional particles whose involvement also plays a crucial role in leptogenesis. 
The presence of various seesaw realizations of light neutrino
masses in the models with discrete flavor symmetry enables us to study leptogenesis through the decay of associated heavy particles. For example, in type-I~\cite{Minkowski:1977sc,Mohapatra:1979ia, Glashow:1979nm,Yanagida:1979as, GellMann:1980vs, Schechter:1980gr}, II~\cite{Magg:1980ut, Schechter:1980gr, Lazarides:1980nt, Mohapatra:1980yp}, and III~\cite{Foot:1988aq} seesaw mechanisms and variants, fermion singlet RHNs ($N_i$), scalar triplets ($\Delta_i$), and fermionic triplets ($\Sigma_i$) are respectively included.  Each scenario can potentially lead to successful leptogenesis within a wide mass range of these additional particles~\cite{Hambye:2012fh}.  {To elucidate the consequence of flavor symmetry in estimating the observed BAU, we have constructed a detailed framework for the TM$_2$ neutrino mixing and its connection with thermal leptogenesis.   }

\subsubsection*{{Thermal Leptogenesis and TM$_2$ mixing}}
{In what follows, we will study leptogenesis within the framework seesaw mechanisms with  CP-violating decays of right-handed neutrinos in the early Universe.}
{To reproduce the TM$_{2}$ mixing
, we adopt a modified version of the Altarelli-Feruglio (AF) model, {which was initially proposed to explain the TBM mixing}~\cite{Altarelli:2005yx, Karmakar:2014dva}. In this scenario,  light neutrino masses are generated completely via type-I seesaw, and hence three copies of RHNs are included which we consider to be a triplet under $A_4$.  With the involvement of the $A_4$ flavons $\phi_s,\phi_T$ (both triplets) and $\xi$ (singlet) one can obtain the TBM mixing. Now to accommodate nonzero $\theta_{13}$, $\delta_{\rm CP}$, the desired TM$_2$ mixing can be achieved in the involvement of one additional flavon $\xi'$ ($1'$ under $A_4$)  which contributes to the right-handed neutrino mass.  In addition to the $A_4$ discrete symmetry, we also consider a $Z_3$ symmetry which forbids the exchange of $\phi_s$ and $\phi_T$ in the Lagrangian. The complete particle content and their transformations under the symmetries are given in Table~\ref{tab_TM2}.
\begin{table}[h]
\centering \label{tab:tm2fields}
\resizebox{9cm}{!}{%
\begin{tabular}{l|cccccccccc}
\hline
Fields & $e_{R}$, $\mu_{R}$, $\tau_{R}$  & $L$ & $N_R$ & $H$  & $\phi_{S}$ & $\phi_{T}$ &  $\xi$ & $\xi'$\\
\hline
$SU(2)$  & 1  & 2 & 1 & 2  & 1 & 1 & 1&1\\
$A_{4}$ & 1,$1''$, $1'$ &  3 & 3 & 1 &  3 & 3 & 1&$1'$\\
$Z_{3}$ & $\omega$  & $\omega$ & $\omega^{2}$ & 1  & $\omega^{2}$ & 1 & $\omega^{2}$ & $\omega^{2}$\\ 
\hline 
\end{tabular}
}
\caption{{\small Transformation of the fields needed to realize the TM$_2$ mixing. Here $\omega$ is the third root of unity. $\xi'$ is essential to
generate non-zero $\theta_{13}$.}} \label{tab_TM2}
\end{table}
Here we have considered the VEVs for the scalar fields  as $\langle\phi_{S}\rangle=(v_{S},v_{S}, v_{S})^T$,\ $ \langle\phi_{T}\rangle=(v_{T},0,0)^T$,\ 
$ \langle\xi\rangle=v_{\xi}$,\ $\langle\xi'\rangle=v_{\xi'}$ ~\cite{Altarelli:2005yx,Holthausen:2011vd,Karmakar:2014dva}. In Table~\ref{tab_TM2}, $H$ is the $SU(2)$ Higgs doublet (with VEV $v$) {and} singlet under $A_4$. Now, with the symmetries and particle content present in Table~\ref{tab_TM2}, the Lagrangian for the charged leptons can be written as 
\begin{eqnarray}\label{eq:tm2LagCL}
    {\cal{L}}_{CL} &=&\big({y}_{e}(L{\phi_{T}})_1e_{R} +{y}_{\mu}(L{\phi_{T}})_{1'}\mu_{R}+{y}_{\tau}(L{\phi_{T}})_{1''}\tau_{R}\big)\frac{H}{\Lambda}, 
\end{eqnarray}
where $\Lambda$ is the cutoff scale of the theory and   $y_e, y_{\mu}, y_{\tau}$ are
the corresponding coupling constants. Note that each term in the first parentheses 
represents products of two $A_4$ triplets  $L,\phi_T$ which further contracts with $e_{R}$, $\mu_{R}$, $\tau_{R}$, which are charged under $A_4$ as $1,1''$ and $1'$, respectively. Following the prescription given in Eq. (\ref{eq:Lag-cl2}), the charged lepton mass  matrix can be obtained as 
\begin{eqnarray} 
	M_{\ell}&=& \frac{vv_{T}}{\Lambda}\begin{pmatrix}
	y_e & 0  & 0 \\
	0 & y_{\mu} & 0 \\
	0 & 0 & y_{\tau}
	\end{pmatrix}. 
\end{eqnarray}
In the presence of the $A_4$ flavons $\phi_s, \xi, \xi'$ (with VEV $\langle\phi_{S}\rangle=(v_{S},v_{S}, v_{S})^T$, 
$ \langle\xi\rangle=v_{\xi}$, and \ $\langle\xi'\rangle=v_{\xi'}$),  the Lagrangian for the neutrino sector can be written as  
\begin{eqnarray}\label{eq:tm2LagNu}
{\cal{L}}_{\nu}&=&y(\bar LN_{R})\tilde{H}+(x_{A}\xi+x_{B}\phi_{S}+x_{N}\xi')\overline{N^c}_RN_R \nonumber \\
&=&y(L_1N_{R_1}+L_2N_{R_3}+L_3N_{R_2})H+x_{A}(\overline{N^c}_{R_1}N_{R_1}+\overline{N^c}_{R_2}N_{R_3}+\overline{N^c}_{R_3}N_{R_2})\nonumber \\
& & + x_{B}\phi_{S_1}(2\overline{N^c}_{R_1}N_{R_1}-\overline{N^c}_{R_2}N_{R_3}-\overline{N^c}_{R_3}N_{R_2})/3+ x_{B}\phi_{S_1}(2\overline{N^c}_{R_1}N_{R_1}-\overline{N^c}_{R_2}N_{R_3}-\overline{N^c}_{R_3}N_{R_2})/3 \nonumber \\
& & + x_{B}\phi_{S_3}(2\overline{N^c}_{R_3}N_{R_3}-\overline{N^c}_{R_1}N_{R_2}-\overline{N^c}_{R_2}N_{R_1})/3+x_{N}\xi'(\overline{N^c}_{R_2}N_{R_2}+\overline{N^c}_{R_1}N_{R_3}+\overline{N^c}_{R_3}N_{R_1})\label{eq:tm2LagNu2} \, ,
\end{eqnarray}
where  $y$, $x_{A}$, $x_{B}$ are the coupling constants and we have omitted conjugation symbols for simplicity.  After spontaneous 
breaking of electroweak and flavor symmetries, we obtain the Dirac and Majorana mass matrices as 
\begin{eqnarray}
M_{D} = y v Y_{\nu}=y v \begin{pmatrix} 
      1 &0 &0\\
       0 &0 &1\\
       0 &1 &0
\end{pmatrix}, \quad 
M_{R} = \begin{pmatrix}
      a+2b/3 &-b/3 &-b/3\\
       -b/3 &2b/3 &a-b/3\\
       -b/3 &a-b/3 &2b/3
\end{pmatrix}+\begin{pmatrix}
      0 & 0 & d\\
      0 & d & 0\\
      d & 0 & 0
\end{pmatrix}, \label{eq:tm2-matrices}
\end{eqnarray}
where $a=2x_{A}v_{\xi}, b=2x_{B}v_{S}$ and  $d=2x_{N}v_{\xi'}$. {Complex parameters in the neutrino mass matrices can be defined as $a=|a|e^{i\phi_a},b=|b|e^{i\phi_b}$ and $d=|d|e^{i\phi_d}$. 
} The mass matrices for  Dirac and Majorana neutrinos are obtained from the Lagrangian written in Eqs.~(\ref{eq:tm2LagCL}) and (\ref{eq:tm2LagNu2}) following the $A_4$ multiplication rules given in Section~\ref{sec:a4group}. The light neutrino mass matrix can be obtained through the type-I seesaw mechanism using the relation given by Eq.~\eqref{eq:seesaw}. 
 After  diagonalizing $M_{\nu}$ with the tribimaximal mixing matrix $U_{\rm TBM}$ we find, 
\begin{eqnarray}
    M'_{\nu}&=&U^T_{\rm TBM}M_{\nu}U_{\rm TBM} = y^2v^2\begin{pmatrix}
      \frac{2(a+b)-d}{2(a^2-b^2-ad+d^2)} & 0 &\frac{-\sqrt{3}d}{2(a^2-b^2-ad+d^2)}\\
       0&\frac{1}{a+d} &0\\
       \frac{-\sqrt{3}d}{2(a^2-b^2-ad+d^2)} & 0  &\frac{2(b-a)+d}{2(a^2-b^2-ad+d^2)}
\end{pmatrix}\label{eq:tb2mnup}. 
\end{eqnarray}
{The above matrix is diagonal in the limit  $d=0$ (contribution corresponding to $\xi'  $) and {can be diagonalized by the TBM mixing matrix.}} {In this construction, the contribution in the Majorana neutrino mass matrix via $\xi'\Bar{N^c}_RN_R$ plays an instrumental role in generating a deviation from the TBM mixing, accounting for the reactor mixing angle $\theta_{13}$}. Therefore,  the light neutrino mass matrix will no longer be diagonalized by $U_{\rm TBM}$ and a further rotation in the 13-plane can diagonalize  $M'_{\nu}$ given in Eq.~(\ref{eq:tb2mnup}).  So the final diagonalizing matrix for the light neutrino mass matrix is
\begin{eqnarray}
U&=&\frac{M^T_D}{yv}U_{\rm TBM}U^{*}_{13}U_M = \frac{M^T_D}{yv}U_{\rm TBM}.\begin{pmatrix}
      \cos\theta &0 &\sin\theta e^{-i\psi}\\
       0 &1 &0\\
       -\sin\theta e^{i\psi} &0 &\cos\theta
\end{pmatrix}{\rm diag}(e^{i\gamma_1},e^{i\gamma_2},e^{i\gamma_3}), \nonumber \\
&=&\begin{pmatrix}
    \sqrt{\frac{2}{3}}\cos\theta & \frac{1}{\sqrt{3}} & \sqrt{\frac{2}{3}}e^{-i\psi}\sin\theta \\
    -\frac{\cos\theta}{\sqrt{6}} -\frac{e^{i\psi} \sin\theta}{\sqrt{2}} &\frac{1}{\sqrt{3}} &   \frac{\cos\theta}{\sqrt{2}} -\frac{e^{-i\psi} \sin\theta}{\sqrt{6}} \\
     -\frac{\cos\theta}{\sqrt{6}} +\frac{e^{i\psi} \sin\theta}{\sqrt{2}} & \frac{1}{\sqrt{3}} & -\frac{\cos\theta}{\sqrt{2}} -\frac{e^{-i\psi} \sin\theta}{\sqrt{6}}
    \end{pmatrix}{\rm diag}(1,e^{i\alpha_{21}/2},e^{i\alpha_{31}/2}) \,. \label{eq:tm2mixmat}
\end{eqnarray}
The structure of this mixing matrix coincides with the TM$_2$ mixing (see Eq.~(\ref{utm})) obtained in the context of the $A_4$ non-Abelian discrete flavor symmetry. } {Here $\gamma_1,\gamma_2,\gamma_3$ are phases  associated  with real positive mass eigenvalues $m_{1},m_{2},m_{3}$ which can be obtained via type-I seesaw as 
\begin{eqnarray}
    m_i=\frac{(yv)^2}{M_i},~i=1,2,3. 
\end{eqnarray}
\begin{figure}[h!]
	\begin{center}
		\includegraphics[width=.43\textwidth]{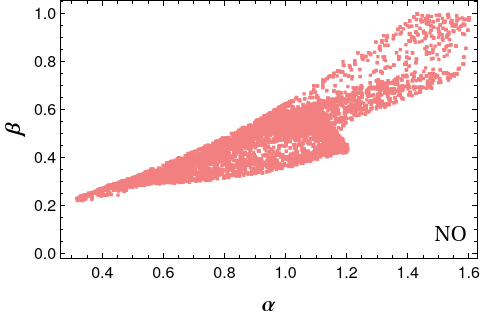}	\includegraphics[width=.43\textwidth]{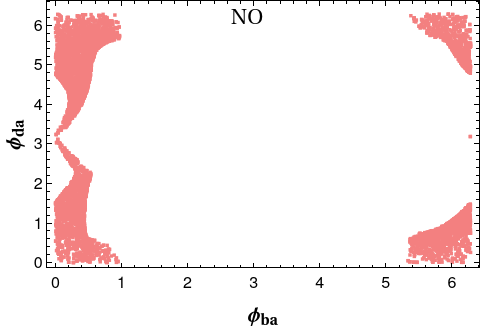}
	\end{center}
\caption{{The allowed regions in $\alpha-\beta$ plane (left panel) and $\phi_{ba}-\phi_{da}$ plane (right panel) for the $3\sigma$ ranges of neutrino oscillation data~\cite{Esteban:2020cvm}.} }
\label{fig:tm2-abp}
\end{figure} 
$M_i$'s are the heavy right-handed neutrino masses. In Eq.~\eqref{eq:tm2mixmat},  $\alpha_{21}=\gamma_1-\gamma_2$, $\alpha_{31}=\gamma_1-\gamma_3$ are the physical Majorana phases. Here the complex light neutrio mass can be written as 
\begin{eqnarray}
    m^c_1=\frac{y^2v^2}{b+\sqrt{a^2+d^2-ad}}, \quad m^c_2=\frac{y^2v^2}{a+d}, \quad m^c_3=\frac{y^2v^2}{b-\sqrt{a^2+d^2-ad}}
\end{eqnarray}
from which the real positive mass eigenvalues for the light neutrinos can be extracted as 
\begin{eqnarray}
    m_1&=&\lambda/\sqrt{(\alpha\cos\phi_{ba}+A)^2+(\alpha\sin\phi_{ba}+B)^2}\label{eq:tm1m1} \, , \\
    m_2&=&\lambda/\sqrt{(1+\beta^2+2\beta\cos\phi_{da})} \, , \\
    m_3&=&\lambda/\sqrt{(\alpha\cos\phi_{ba}-A)^2+(\alpha\sin\phi_{ba}-B)^2} \, , 
\end{eqnarray}
where for compact notation we have defined 
\begin{eqnarray}
    \alpha&=&|b|/|a|, \beta=|d|/|a|, \phi_{ba}=\phi_b - \phi_a, \phi_{da}=\phi_d-\phi_a \, \nonumber \\ 
    P&=&1+\beta^2\cos 2\phi_{da}-\beta\cos \phi_{da},~Q=\beta^2\sin 2\phi_{da}-\beta\sin \phi_{da} \, , \nonumber \\
   A^2&=&(P+\sqrt{P^2+Q^2})/2, B^2=Q^2/4A^2, \lambda=|y^2v^2/a|. 
\end{eqnarray}
The rotation angle ($\theta$) and the associated phase $\psi$ appearing in $U_{13}$ can be linked with the parameters appearing in $M_{\nu}$. Hence, $\theta$ and  $\psi$ can be expressed in terms of $a,b$ and $d$ as
\begin{eqnarray}\label{eq:tm2t2}
    \tan 2\theta&=&\frac{\sqrt{3}\beta\cos\phi_{da}}{(2-\beta\cos\phi_{da})\cos\phi-2\alpha\sin\phi_{ba}\sin\psi}\, , \quad 
    \tan\psi=\frac{\sin\phi_{da}}{\alpha\cos(\phi_{ba}-\phi_{da})}. 
\end{eqnarray}
Comparing the mixing matrix $U$ in Eq.~(\ref{eq:tm2mixmat}) with the $U_{\rm PMNS}$ matrix given in Eq.~(\ref{upmns1}), we
obtain the usual correlations
between the mixing angles and CP violating Dirac phase $\delta_{\rm CP}$ as 
\begin{eqnarray}\label{ang}
\sin\theta_{13}=\sqrt{\frac{2}{3}}\left|\sin\theta\right|, \quad 
\sin^2\theta_{12}=\frac{1}{3(1-\sin^2\theta_{13})}, \quad 
 \sin^2\theta_{23}=\frac{1}{2}
+\frac{1}{\sqrt{2}}\sin\theta_{13}\cos\delta, \quad  \delta_{\rm CP}={\rm arg}[U_{13}] \, ,
\label{ang22}
\end{eqnarray}
up to the order of $\sin^2\theta_{13}$. Here $\sin\theta$ can take positive or negative values depending on the choice of the parameters appearing in Eq.~(\ref{eq:tm2t2}). For both positive and negative values of $\sin\theta$, using Eq.~(\ref{eq:tm2t2}) and (\ref{ang22}), the Dirac CP phase can be written as
\begin{eqnarray}
    \tan\delta_{\rm CP}=\frac{\sin\phi_{da}}{\alpha\cos(\phi_{ba}-\phi_{da})}.\label{eq:tm2dcp} 
\end{eqnarray}
The Majorana phases in the framework can be written as 
\begin{eqnarray}\label{eq:tm2-a21}
    \alpha_{21}&=&\tan^{-1}\left(-\frac{\alpha\sin\phi_{ba}+B}{\alpha\cos\phi_{ba}+A}\right)-\tan^{-1}\left(\frac{-\beta\sin\phi_{da}}{1+\beta\sin\phi_{da}}\right), \\
    \alpha_{31}&=&\tan^{-1}\left(-\frac{\alpha\sin\phi_{ba}+B}{\alpha\cos\phi_{ba}+A}\right)-\tan^{-1}\left(\frac{B-\alpha\sin\phi_{ba}}{A-\alpha\cos\phi_{ba}}\right). \label{eq:tm2-a31}
\end{eqnarray}
As seen from Eq.~(\ref{eq:tm1m1}) to Eq.~(\ref{eq:tm2dcp}), all the mixing angles ($\theta_{12}, \theta_{23}, \theta_{13}$), the Dirac CP phase $\delta_{\rm CP}$ involved in the lepton mixing matrix $U_{\rm PMNS}$ as well as the light neutrino masses ($m_1, m_2, m_3$) are   ultimately determined by the model parameters $\alpha,\beta,\phi_{ba}$ and $\phi_{da}$. 
Hence, using the
$3\sigma$ allowed ranges of the three mixing angles ($\theta_{12}, \theta_{23}, \theta_{13}$) from neutrino oscillation data~\cite{Esteban:2020cvm}
presented in Table \ref{tab:neutrino_data}, we can restrict parameter space for $\alpha,\beta,\phi_{ba}$ and $\phi_{da}$ as plotted in Fig.~\ref{fig:tm2-abp}. 
\begin{figure}[t!]
	\begin{center}
		\includegraphics[width=.43\textwidth]{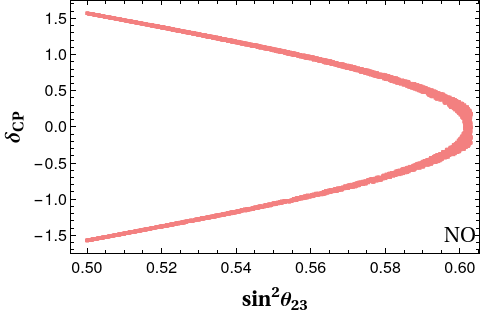}	\includegraphics[width=.43\textwidth]{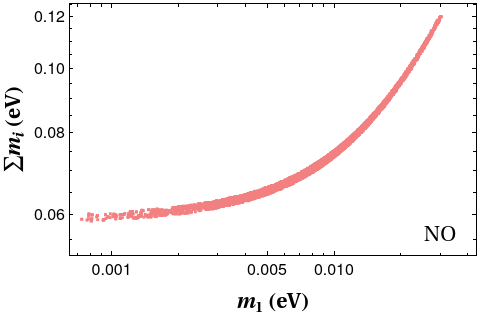}
	\end{center}
\caption{{Predictions for $\sin^2\theta_{23}-\delta_{\rm CP}$ (left panel) and $m_1-\sum m_i$ plane (right panel) for the allowed ranges obtained in Fig.~\ref{fig:tm2-abp}.}  }
\label{fig:tm2-23dcp-mass}
\end{figure} 
For brevity, we will only discuss the results obtained for NO of neutrino masses, however,  it can be easily extended to the IO case. For the 3$\sigma$ range of neutrino oscillation parameters, $\alpha$ and $\beta$ are found to be in the range of $0.3-1.6$ and $0.21-1$ respectively. On the other hand, a full range of ${\phi_{da}}$ is allowed while ${\phi_{ba}}$  is restricted within the range $0-1$ and $5.3-2\pi$ radian. Using these values of the model parameters one can obtain some predictions of the present TM$_2$ scenario, as plotted in Fig.~\ref{fig:tm2-23dcp-mass}. Here the $\sin^2\theta_{23}-\delta_{\rm CP}$ correlation in the left panel shows that only the upper octant of $\theta_{23} (\geq 45^{\circ})$ is allowed. Similarly, the model predictions for the sum of absolute masses of three neutrinos against the lightest mass $m_1$ (for NO) are given in the right panel of Fig.~\ref{fig:tm2-23dcp-mass}. Most interestingly, the Majorana phases $\alpha_{21}$ and $\alpha_{31}$ written in Eq.~(\ref{eq:tm2-a21}) and Eq.~(\ref{eq:tm2-a31}) respectively can also be constrained using low energy neutrino oscillation data. From Eq.~(\ref{2nu0}), it is also clear that estimation of these phases crucially dictates $m_{\beta\beta}$, which can be probed by future double beta decay experiments. 
\begin{figure}[t!]
	\begin{center}
		\includegraphics[width=.43\textwidth]{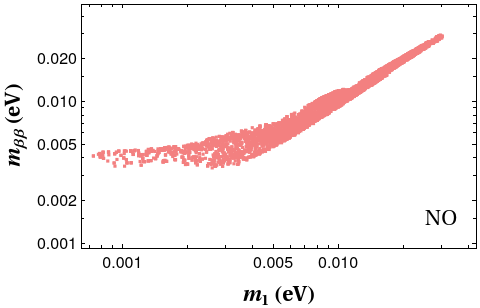}	\includegraphics[width=.43\textwidth]{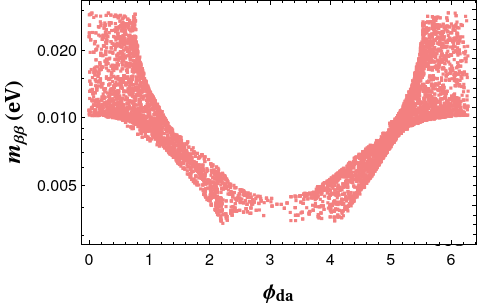}
	\end{center}
\caption{{Predictions for $m_{\beta\beta}$ against the light neutrino mass $m_1$ (left panel) and  the  phase $\phi_{da}$  (right panel) for the allowed ranges obtained in Fig.~\ref{fig:tm2-abp}.}  }
\label{fig:tm2-mbb}
\end{figure} 
In Fig.~\ref{fig:tm2-mbb}, we have plotted the predictions for $m_{\beta\beta}$ against the light neutrino mass $m_1$ (left panel) and the one of the associated phase factor $\phi_{da}$  (right panel) associated in generating non-zero $\theta_{13}$.} 

{Estimation of the Majorana phases as well as the whole flavor symmetric framework also crucially dictates the generation of lepton asymmetry.} For the simple type-I seesaw, lepton asymmetry can be elegantly generated through the out-of-equilibrium decay
of  RHNs in the early Universe. The CP asymmetry parameter $\epsilon_{i\alpha}$ can be evaluated from  the interference between the tree and one-loop level decay amplitudes of the RHN $N_i$ (dropping the subscript $R$) decaying into a lepton doublet $L_\alpha$ with specific flavor $\alpha$ and the Higgs doublet ($H$), i.e.~\cite{Fukugita:1986hr}
\begin{align}
\label{eq:eps1}
    \varepsilon_{i\alpha}=\frac{\Gamma(N_i\to L_\alpha H)-\Gamma(N_i\to \bar{L}_\alpha H^c)}{\Gamma(N_i\to L_\alpha H)+\Gamma(N_i\to \bar{L}_\alpha H^c)} \, ,
\end{align}
where the superscript $c$ stands for the charge conjugation. {Under one-flavor approximation~\cite{Pascoli:2006ie,Blanchet:2006be}, the corresponding asymmetry parameter can be written as~\cite{Covi:1996wh} 
\begin{equation}
 \epsilon_{i}=\frac{1}{8\pi}\sum_{j\ne{i}}\frac{{\rm 
Im}\left[\left(({Y}^{\dagger}_{\nu}{Y}_{\nu})_{ji}\right)^{2}\right]}
{({Y}^{\dagger}_{\nu}{Y}_{\nu})_{ii}}f\left(\frac{m_{i}}{m_{j}}\right),\quad f(x)\equiv{-x}\left(\frac{2}{x^{2}-1}+\log\left(1+\frac{1}{x^{2}}\right)\right)\, , \label{ei}
\end{equation}
in the basis where $M_R$ in Eq.~(\ref{eq:tm2-matrices}) is diagonal and $x=m_{i}/m_{j}$. Therefore, in our set-up, in the diagonal basis of $M_R$, $Y_{\nu}$ in Eq.~(\ref{eq:tm2-matrices})  can be redefined as ${Y}_{\nu}\rightarrow{\rm 
diag}(1,e^{-i\alpha_{21}/2},e^{-i\alpha_{31}/2})U_R^TY_{\nu}$ and 
$U_R=U_{\rm TBM}U_{13}$. In Ref.~\cite{Jenkins:2008rb}, the authors have pointed out that with such structure of the Yukawa matrices, $({Y}^{\dagger}_{\nu}{Y}_{\nu})\propto y^2\mathbf{I}$, hence lepton asymmetry
parameter $ \epsilon_{i}$ in Eq.~(\ref{ei}) vanishes. Therefore in the limit of exact TBM mixing scenario leptogenesis can not be realized~\cite{Jenkins:2008rb}. To get rid of this situation 
with allowed symmetries and particle content mentioned earlier,  one can introduce NLO contributions in the Dirac Yukawa coupling as  
\begin{equation}
\delta Y_{\nu}\propto(\bar L \varphi_T)\tilde{H}N =x_{C}(\bar L \varphi_T)\tilde{H}N/\Lambda+x_{D}(\bar L \varphi_T)\tilde{H}N/\Lambda. 
\end{equation}
Following the $A_4$ multiplication rules given in Section \ref{sec:a4group}, the complete structure of the Dirac Yukawa coupling can replaced by 
 \begin{eqnarray}
Y_{\nu} & = & Y_{{\nu}}+\delta{Y_{\nu}} = y  \begin{pmatrix}
      1 &0 &0\\
       0 &0 &1\\
       0 &1 &0
\end{pmatrix}+\frac{x_{C}v_{T}}{\Lambda}\begin{pmatrix}
              2 & 0 & 0\\
              0 & 0 & -1\\
              0 & -1 & 0
              \end{pmatrix}+\frac{x_{D}v_{T}}{\Lambda}\begin{pmatrix}
              0 & 0 & 0\\
              0 & 0 & -1\\
              0 & 1& 0
              \end{pmatrix}\label{ynu} .
\end{eqnarray}
Such a non-trivial structure of the Yukawa coupling is enough to generate successful leptogenesis without producing any significant effect on the light neutrino masses and mixing obtained via type-I seesaw. Substituting Eq.~(\ref{ynu}) in Eq.~(\ref{ei}), the asymmetry parameter $\epsilon_1$ generated  from the decay of $N_1$ can be written as 
\begin{eqnarray}
\epsilon_{1} = 
\frac{-(v_T/\Lambda)^2}{2\pi}\left[\sin\alpha_{21}
\left(2{\rm Re} (x_{C})^{2}
 \cos^{2}{\theta} + \frac{2{\rm Re}(x_{D})^{2}}{3} \sin^{2}{\theta} + 
\frac{2{\rm Re}(x_{C}){\rm Re}(x_{D})}{\sqrt{3}} \sin{2\theta}\right)
 f\left(\frac{m_{1}}{m_{2}}\right)\right.\nonumber\\
 \left.+\sin\alpha_{31}\left({\rm Re}(x_{C})^{2}\sin^{2}{2\theta} + \frac{{\rm 
Re}(x_{D})^{2}}{3} \cos^{2}{2\theta} +
  \frac{{\rm Re}(x_{C}){\rm Re}(x_{D})}{\sqrt{3}} 
\sin{4\theta}\right)f\left(\frac{m_{1}}{m_{3}}\right)\right].\label{epi0}
\end{eqnarray}
Similarly, expressions for $\epsilon_{2,3}$ can also be obtained from the decay of $N_{2,3}$. With the allowed parameters space obtained from neutrino mixing (in Fig.~\ref{fig:tm2-abp}), we are now in a position to estimate $\epsilon_i$. We have already discussed constraints on the Majorana phases, the rotation angle $\theta$, and light neutrino masses $m_i$.  Finally, with this constraints and  judicious  choices of  $v_{T}/\Lambda, {\rm Re}(x_{C})$ and $ {\rm Re}(x_{D})$  lepton asymmetry of appropriate  magnitude $\mathcal{O}(10^{-6})$ can be generated. This asymmetry further can be converted into BAU with the help of the sphaleron processes, considered in the next subsection.}

\subsubsection*{Resonant Leptogenesis}
In freeze-out leptogenesis via heavy neutrino decay, there are two types of loop-level diagrams involving vertex and wave-function corrections~\cite{Covi:1996wh}. It turns out that the wave-function corrections can be dominant when at least two RHNs have a small mass difference comparable to their widths~\cite{Flanz:1994yx, Pilaftsis:1997dr, Pilaftsis:1997jf}. This is known as the resonant leptogenesis~\cite{Pilaftsis:2003gt, Dev:2017wwc}, which allows the RHN mass scale to be lowered down to the electroweak-scale~\cite{Pilaftsis:2005rv, Deppisch:2010fr, BhupalDev:2014pfm}, without being in conflict with the Davidson-Ibarra bound~\cite{Davidson:2002qv, Buchmuller:2002rq} for thermal leptogenesis.   

In the resonant regime, Eq.~\eqref{eq:eps1} can be written in a compact form as~\cite{Dev:2017wwc}
\begin{align}
    \varepsilon_{i\alpha} \simeq \frac{1}{8\pi \left(Y_D^\dag Y_D\right)_{ii}} \sum_{j\neq i} {\rm Im}\Big[(Y_D^\star)_{\alpha i}(Y_D)_{\alpha j}\Big] {\rm Re}\left[\left(Y_D^\dag Y_D\right)_{ij}\right]{\cal F}_{ij} \, ,
    \label{eq:eps2}
\end{align}
where $Y_D$ is the Yukawa coupling matrix in the basis where the RHN mass matrix is diagonal (this is typically indicated by $\hat{Y}_D$, but we remove the hat for brevity). The resonant enhancement factor is given by 
\begin{align}
    {\cal F}_{ij} \, = \, \frac{M_i M_j (M_i^2-M_j^2)}{(M_i^2-M_j^2)^2+A_{ij}^2} .
\end{align}
Here the $Y_D$ matrices are evaluated in the RHN mass basis, and $A_{ij}$ regulates the behavior of the CP asymmetry in the limit $\Delta M_{ij}\equiv |M_i-M_j|\to 0$. As pointed out in Refs.~\cite{BhupalDev:2014pfm, BhupalDev:2014oar}, in the resonant regime there are two {\it distinct} contributions to the CP asymmetry from RHN mixing and oscillation effects, both of which can be effectively captured by Eq.~\eqref{eq:eps2} but with different regulators:
\begin{align}
    A_{ij}^{\rm mix} \, = \, M_i \Gamma_j \, , \quad A_{ij}^{\rm osc} \, = \, (M_i\Gamma_i+M_j\Gamma_j)\left[\frac{{\rm det}\left({\rm Re}\left(Y_D^\dag Y_D\right)\right)}{\left(Y_D^\dag Y_D\right)_{ii}\left(Y_D^\dag Y_D\right)_{jj}} \right]^{1/2} \, .
    \label{eq:regulator}
\end{align}
The net CP asymmetry is then given by $\varepsilon_{i\alpha}^{\rm tot}=\varepsilon_{i\alpha}^{\rm mix}+\varepsilon_{i\alpha}^{\rm osc}$. This analytic approximation is in good agreement with the full quantum kinetic treatment~\cite{BhupalDev:2014oar, Kartavtsev:2015vto, Klaric:2021cpi} in the strong washout regime. 

Since the regulator part is independent of the lepton flavor $\alpha$, we can sum over $\alpha$ to obtain the total CP asymmetry for a given RHN $N_i$:
    \begin{align}
        \varepsilon_i \ \equiv \ \sum_\alpha \varepsilon_{i\alpha} \, = \, & \frac{1}{8\pi \left(Y_D^\dag Y_D\right)_{ii}} \sum_{j\neq i} {\rm Im}\left[\left(Y_D^\dag Y_D\right)_{ij}\right]  {\rm Re}\left[\left(Y_D^\dag Y_D\right)_{ij}\right]{\cal F}_{ij}  \, .
    \label{eq:eps3}
    \end{align}
Within a semi-analytic Boltzmann approach, the flavor-dependent lepton asymmetry parameter ($\eta_{L_{\alpha}}$) can be written as~\cite{Buchmuller:2004nz,Deppisch:2010fr} 
\begin{eqnarray}
    \eta_{L_{\alpha}} \simeq \frac{3}{2 z_c K^{\rm eff}_{\alpha}} \sum_i \epsilon_{i\alpha} d_i \, ,
\end{eqnarray}
where $z_c=M_N/T_c$, $T_c$ is the critical temperature
below which the electroweak sphalerons freeze-out, $K^{\rm eff}_{\alpha}$ are 
the effective washout factors and $d_i$
are the corresponding dilution factors given in terms of
ratios of thermally-averaged rates for decays and scatterings involving the RHNs~\cite{Deppisch:2010fr, BhupalDev:2014pfm}. The obtained lepton asymmetry then can be converted into the observed BAU via $(B+L)$-violating electroweak sphaleron processes~\cite{Khlebnikov:1988sr, Harvey:1990qw} which gives the final baryon asymmetry 
\begin{eqnarray}
    \eta_B \simeq -0.013 \sum_{\alpha}  \eta_{L_{\alpha}} \, , 
    \label{eq:etaB}
\end{eqnarray} 
where the pre-factor is a product of the sphaleron
conversion rate of 28/79~\cite{Harvey:1990qw} and the entropy dilution
factor of 1/27.3~\cite{BhupalDev:2014pfm}. It was the ratio of Eq.~\eqref{eq:etaB} to the observed value given in Eq.~\eqref{eq:etaBobs} plotted in Fig.~\ref{fig:etaBZp}. 

Let us exemplify the usefulness of Eq.~\eqref{eq:eps2} by applying it to a particular case of the $\Delta(6n^2)$ group discussed in Section~\ref{sec:delta6pheno} with $s$ even. We find that $\varepsilon_{3\alpha}=0$, i.e., the RHN mass eigenstate $N_3$ does not contribute to the CP asymmetry, and in the strong NO and IO limits, 
\begin{align}
\label{eps1NO}
\varepsilon_{1 \alpha}^{\rm NO} \ \approx \ &\frac{y_2 \, y_3}{9} \, \left[-2 \, y_2^2+y_3^2 \, (1- \cos 2 \, \theta_R)\right] \, \sin 3 \, \phi_s  \sin\theta_R \, \sin\theta_{L, \alpha} \, \mathcal{F}_{12} \, , \\
\varepsilon_{1 \alpha}^{\rm IO} \ \approx \ &\frac{y_1 \, y_2}{9} \, \left[-2 \, y_2^2+y_1^2 \, (1+ \cos 2 \, \theta_R)\right] \, \sin 3 \, \phi_s   \cos\theta_R \, \cos\theta_{L, \alpha} \, \mathcal{F}_{12} \, ,
\end{align}
with $\theta_{L, \alpha} = \theta_L + \rho_\alpha \, 4 \pi/3$ and $\rho_e=0$, $\rho_\mu=1$, $\rho_\tau=-1$.
For strong NO (IO) $\varepsilon_{i \alpha}$ becomes very small, if $\theta_R \approx 0 , \, \pi$ ($\theta_R\approx\pi/2,3\pi/2$). In addition, $\mathcal{F}_{ij}$
vanishes for $\cos 2 \, \theta_R=0$.
The CP asymmetries $\varepsilon_{2 \alpha}=-\varepsilon_{1 \alpha}$ with $\mathcal{F}_{12}\leftrightarrow \mathcal{F}_{21}$.
For $s$ odd, similar expressions are obtained with $\sin(3\, \phi_s)\leftrightarrow -\cos(3\, \phi_s)$. 

\begin{figure*}[h!]
\centering
\includegraphics[width=0.48\textwidth]{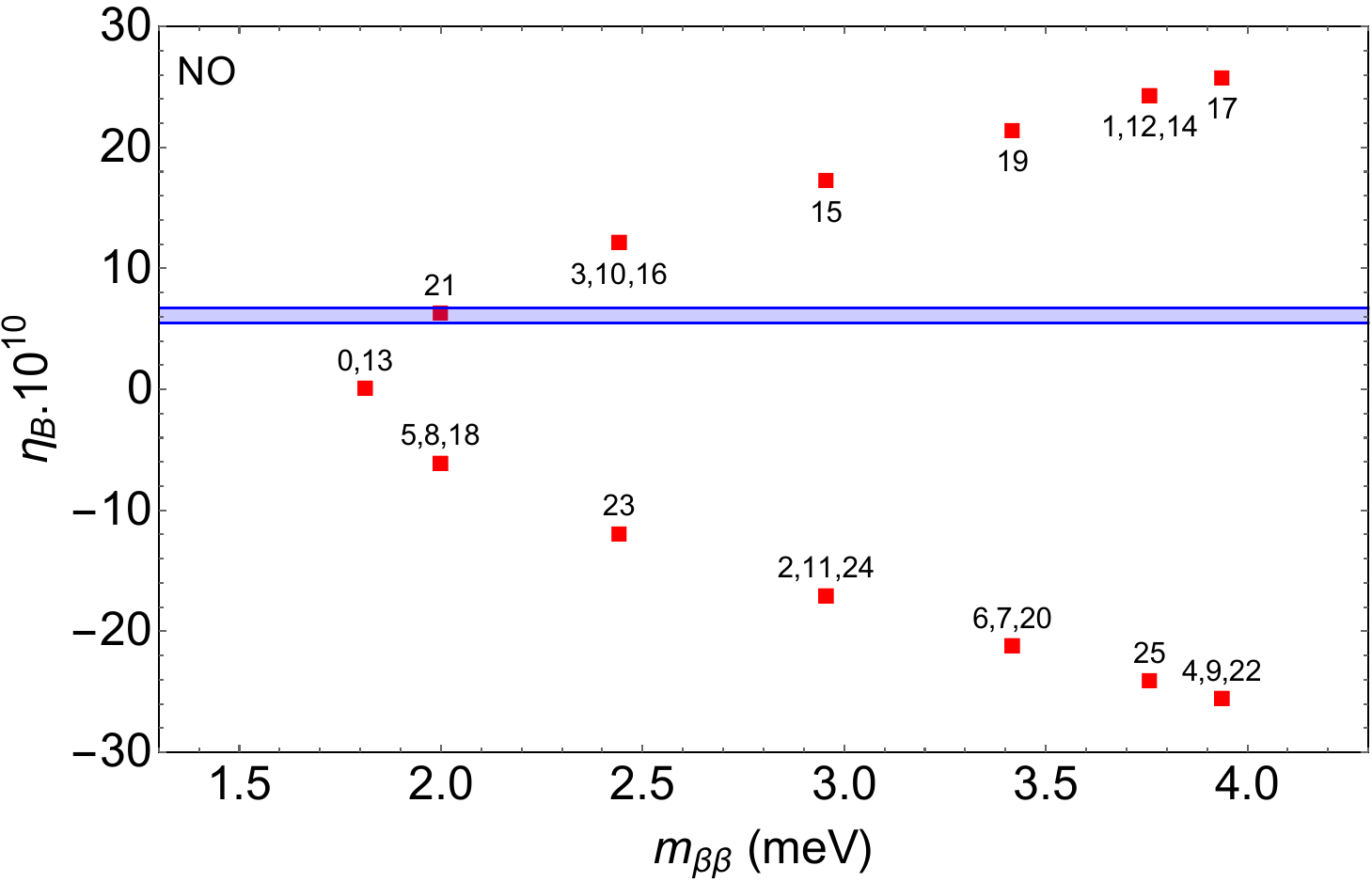}
\hspace{0.1in}
\includegraphics[width=0.48\textwidth]{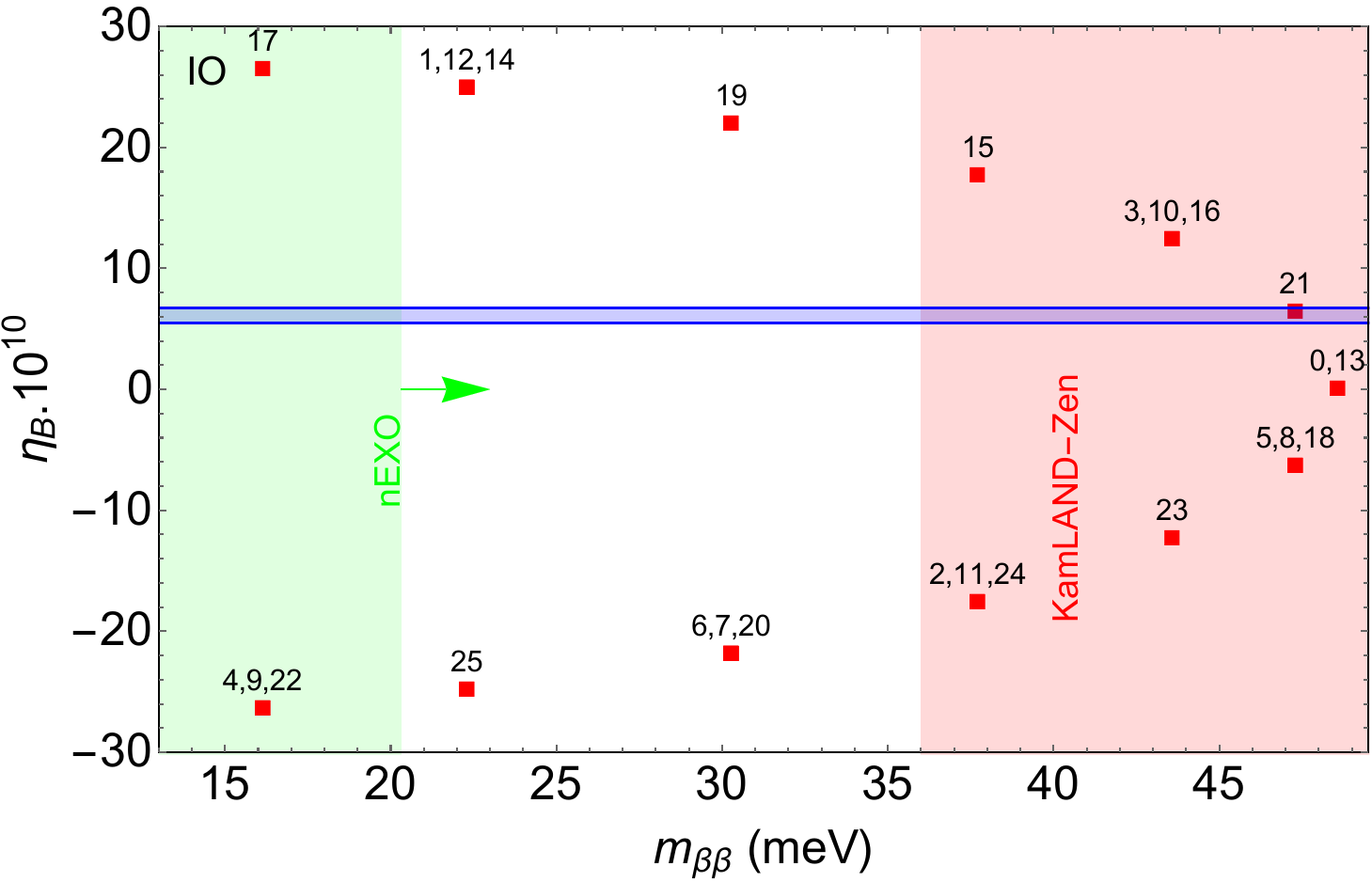}
\caption{Correlation between the predicted BAU $\eta_B$ and the effective neutrino mass $m_{\beta\beta}$ for a $\Delta(6n^2)$ model with $n=26$ and $0\leq s \leq n-1$ (as shown by the numbered points). The blue-shaded horizontal bar corresponds to $\eta_B$ within $10\%$ of $\eta_B^{\rm obs}$~\cite{Planck:2018nkj}. The vertical shaded bands for the IO case indicate the smallest $m_{\beta\beta}$ value (including the NME uncertainties) that is either ruled out by KamLAND-Zen~\cite{KamLAND-Zen:2022tow} (red) or will be accessible to nEXO~\cite{nEXO:2021ujk} (green). Figure taken from Ref.~\cite{Chauhan:2021xus}.}
\label{fig:0vbb1}
\end{figure*}

 These analytic forms of the CP asymmetry enable us to correlate the high-and low-energy CP phases. This is illustrated in Figure~\ref{fig:0vbb1} for the $\Delta(6n^2)$ group with $n=26$, where we plot the predictions for the BAU (which depend on the high-energy CP phases) and for the $0\nu\beta\beta$ observable $m_{\beta\beta}$ (which depends on the low-energy CP phases) corresponding to the $s$ values from 0 to $n-1$. 
The horizontal blue-band corresponds to $\eta_B$ values within 10\% of the observed value. We find that there exist some $s$ values for which the correct BAU can be obtained, while the $m_{\beta\beta}$ predictions for the IO case are either already excluded or will be tested soon in the next-generation $0\nu\beta\beta$ experiments.

\subsubsection*{ARS Mechanism and Unified Picture}
Just like the case with DM relic density which can happen either by freeze-out or freeze-in, the deviation from equilibrium for the RHNs can also happen either via freeze-out when their mass drops below the Hubble temperature), or via freeze-in when the Yukawa couplings are so small that the equilibration rate becomes much lower than the Hubble rate. The resonant leptogenesis mechanism discussed above is an example of the freeze-out scenario, while leptogenesis via oscillations (the so-called ARS mechanism)~\cite{Akhmedov:1998qx} is an example of the freeze-in scenario, which typically happens for GeV-scale RHNs~\cite{Asaka:2005pn, Canetti:2012kh, Shuve:2014zua, Drewes:2017zyw}. These two mechanisms may seem to be quite different, but it was recently shown that both can be described in a unified picture~\cite{Klaric:2020phc, Drewes:2021nqr, 
Klaric:2021cpi}. This can be achieved with matrix-valued quantum kinetic equations for the RHNs and the lepton asymmetries in different SM degrees of freedom. Solving these equations for RHN masses in a wide range from 50 MeV to 70 TeV, Ref.~\cite{Drewes:2022kap} obtained the range of total mixing angle $U^2$ consistent with leptogenesis, as shown in Fig.~\ref{fig:ARS}. Here $\kappa$ refers to the RHN mass splitting parameter, and the contours corrersponding to positive (negative) BAU are shown by solid (dashed) lines. The black line indicates the canonical seesaw line.  The red and blue shaded regions lead to unacceptably large corrections to the light neutrino masses induced by the RHN mass splitting. 

\begin{figure}[h!]
        \centering
        \includegraphics[width = 0.49\textwidth]{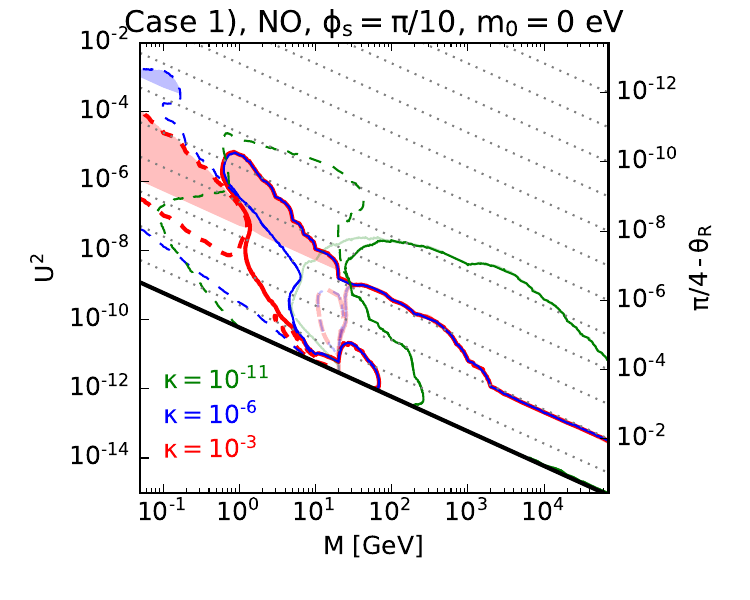}
        \includegraphics[width = 0.49\textwidth]{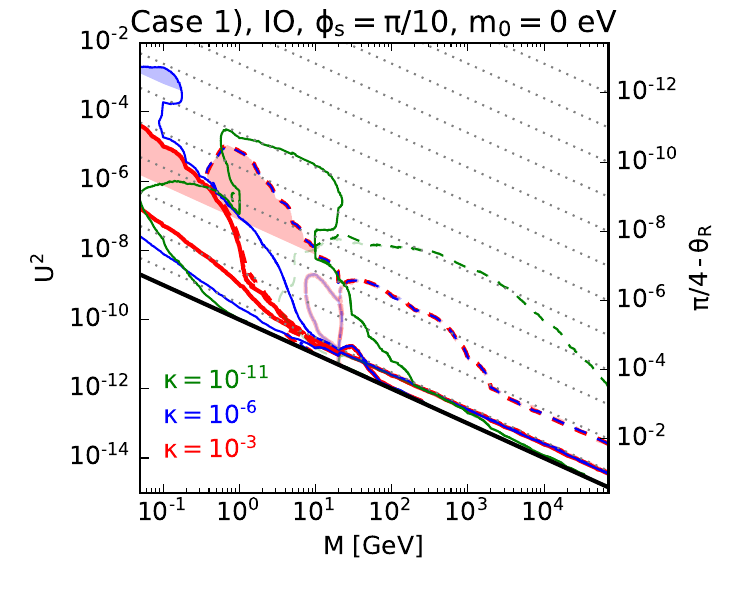}
        \caption{Range of total mixing angle $U^2$ consistent with leptogenesis for a wide range of RHN mass in a $\Delta(6n^2)$ model. Figure taken from the arXiv version of Ref.~\cite{Drewes:2022kap}.}
        \label{fig:ARS}
\end{figure}

\subsubsection*{Other Examples}
As mentioned earlier, in models with discrete flavor symmetry,  structures of the Dirac Yukawa coupling, charged lepton and RHN mass matrices are entirely orchestrated by the associated symmetry.   Hence, the involved discrete symmetry may have a significant impact on neutrino masses, mixing, and Dirac CP phase as well high energy CP violation required for leptogenesis. As TBM mixing was a potential candidate for the lepton mixing matrix, several analyses have been performed to explain TBM mixing and leptogenesis~\cite{Jenkins:2008rb,Hagedorn:2009jy, Branco:2009by}. Interestingly, it was also observed~\cite{Jenkins:2008rb} that in the exact
TBM limit, the CP asymmetry vanishes since  $ {Y}^{\dagger}_{D} {Y}_{D} \propto \mathbb{I}$  (where $ {Y}_{D}$ in the Dirac Yukawa coupling in the basis where RHNs are diagonal). Therefore in the exact TBM scenario, leptogenesis was realized either by introducing higher-order correction in the Dirac Yukawa matrix~\cite{Jenkins:2008rb, Hagedorn:2009jy} or by introducing renormalization group effects~\cite{Branco:2009by}. However, Ref.~\cite{Dev:2015wpa}  derived a no-go theorem for the viability of the minimal radiative resonant leptogenesis.  In the precision era of the neutrino mixing parameters, it is essential to
study the effect of non-zero $\theta_{13}$ and CP phases. In the $A_4$ type-I seesaw scenario~\cite{Karmakar:2014dva}, the authors showed that a minimal (by adding a non-trivial $A_4$ singlet) modification to the existing Altarelli-Feruglio~\cite{Altarelli:2005yx} model may give rise to non-zero $\theta_{13}$ as well as generate the observed BAU. Here, the high-energy CP phases or the Majorana phases appearing in the CP asymmetry also get constrained by the low-energy neutrino oscillation data, which are otherwise insensitive to the oscillation experiments. In a more minimal study~\cite{Datta:2021zzf}, the authors showed that with the presence of only three $A_4$ flavons, both non-zero $\theta_{13}$ and BAU (incorporating renormalization group effects) can be generated. To unify all the sources of CP violation in the theory,  the CP symmetry can be spontaneously
broken by the VEV of a singlet field where its magnitude generates non-zero $\theta_{13}$ and the phase factor becomes directly proportional to the CP asymmetry parameter in general $A_4$ type-I+II scenario~\cite{Karmakar:2015jza}. Recently, it has also been shown that BAU can be successfully generated in a  $S_4$ discrete symmetry framework with TM$_1$ mixing \cite{Chakraborty:2020gqc} taking lepton flavor effects into consideration.  Apart from the example discussed here for $\Delta (6 \, n^2)$, a few other works exploring leptogenesis in similar flavor symmetry models are Refs.~\cite{Hagedorn:2017wjy,Samanta:2018efa,Fong:2021tqj,Drewes:2022kap}.  
For more recent studies  on various discrete flavor symmetry and leptogenesis, see Refs.~\cite{Bjorkeroth:2015tsa, Bjorkeroth:2016lzs,Borah:2017qdu,Das:2019ntw,Gautam:2020wsd,Sarma:2021icl,Boruah:2021ayj}.

\subsection{Flavor Symmetry and Gravitational Waves}

The recent breakthrough in gravitational wave (GW) observations by LIGO~\cite{LIGOScientific:2016aoc} provides us  a new window into the early Universe. A particularly interesting example of early Universe phenomena that can be a stochastic source of GWs is cosmological phase transition~\cite{Witten:1984rs} whose observation may shed light on an array of BSM phenomena, from BAU to GUT physics and inflation~\cite{Caldwell:2022qsj}. As the Majorana mass term in neutrino mass models can be generated by $B-L$ breaking at a high energy scale,  the associated phase transition in the early Universe can produce a possibly observable stochastic
GW background, thus providing a complementary probe of $B-L$ physics like leptogenesis~\cite{Jinno:2016knw, Ellis:2020nnr, Huang:2022vkf, Dasgupta:2022isg}. {In scale-invariant versions of gauged $U(1)$ flavor symmetry models, it is also possible to generate sizable GW which is detectable in future GW experiments such as aLIGO/VIRGO, LISA, DECIGO, and CE, with complementarity signals at colliders and low-energy neutrino experiments~\cite{Dasgupta:2023zrh}; see Fig.~\ref{fig:SNR}.}

\begin{figure}[t!]
    \centering
      \includegraphics[width=0.5\textwidth]{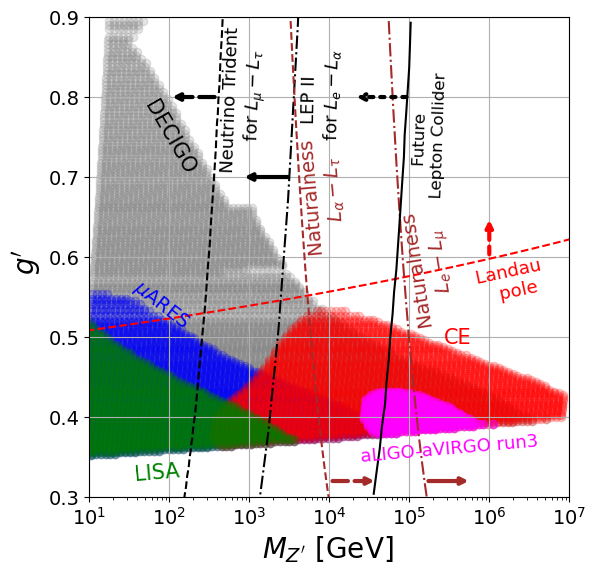}
      \caption{{Sensitivity of the future GW experiments in the mass-coupling plane of scale-invariant $U(1)_{L_\alpha-L_\beta}$ models. The current constraint from non-observation of stochastic GW background in aLIGO-aVIRGO run 3 is shown in purple.  The laboratory constraints from LEP-2 and neutrino trident experiments are shown for comparison, along with the future lepton collider sensitivities. The naturalness and perturbativity (Landau pole) constraints are also shown. Figure taken from Ref.~\cite{Dasgupta:2023zrh}.} }
    \label{fig:SNR}
\end{figure}

As for the discrete flavor symmetry models, a direct test would be the observation of flavons involved in the spontaneous breaking of the discrete symmetry. This is usually assumed to happen at a high scale, which makes it challenging to test experimentally. Whatever the scale of the discrete flavor symmetry spontaneous breaking, it gives rise to degenerate vacua separated by energy barriers leading to a network of cosmic domain walls. This is a serious problem, if the walls are stable, as they could overclose the Universe~\cite{Zeldovich:1974uw, Kibble:1976sj}. Solutions to this domain wall problem have been discussed in the context of non-Abelian discrete symmetries such as $A_4$~\cite{Gelmini:2020bqg} and $S_4$~\cite{Jueid:2023cgp}, which include explicit breaking terms, such that the domain walls collapse in a certain period of time before the symmetry breaking~\cite{Riva:2010jm, Antusch:2013toa, Chigusa:2018hhl,PhysRevD.102.115023}. Such collapse of domain walls in the early Universe could lead to stochastic GWs~\cite{Vilenkin:1981zs, Preskill:1991kd, Gleiser:1998na, Hiramatsu:2013qaa}, thus offering novel probes to test the discrete flavor symmetry models at current and future GW experiments~\cite{Bailes:2021tot}. See also Ref.~\cite{Greljo:2019xan} for other GW imprints of flavor models, such as multipeaked stochastic GW signal from a series of cosmological phase transitions that could be a unique probe of the mechanism behind flavor hierarchies. 

	\section{Summary and Outlook}\label{sec:sum}

The origin of neutrino masses and mixing is a fundamental question in particle physics. In this review, we have discussed flavor model
building strategies mainly based on discrete non-Abelian family symmetries aimed at explaining the patterns of lepton masses and flavor mixing.
We consider fixed patterns (BM, TBM, GR, HG) and more elaborate symmetry groups with unbroken residual symmetries (e.g., $A_4$,  $S_4$,  $A_5$, $T'$,  $\Delta(27)$,  $\Delta(6n^2)$), motivated by the increasingly precise results from neutrino oscillation experiments. In particular, a `large' reactor mixing angle $\theta_{13}$ has been determined, the Dirac CP phase $\delta$ is preferred to be nonzero and the normal mass ordering seems to be favored in the current oscillation data. We discuss the far-reaching implications of these considerations in flavor model building and phenomenology. We also discussed flavor symmetry breaking and mechanisms of mass generation, and flavor symmetry in multi-Higgs doublet models. 

As there are plenty (hundreds) of possible flavor symmetry models, a natural question is: How to falsify or validate any of these models?  Generally, such symmetries are broken at a high scale and beyond our experimental reach. Nonetheless, phenomenological study connected with flavor symmetry effects is rich and   models can be probed effectively in low-energy intensity frontier experiments (like neutrino oscillation experiments, neutrinoless double beta decay, and lepton flavor violation searches), high energy colliders (such as LHC and future lepton/hadron colliders), as well as at the cosmological frontier (via baryogenesis, dark matter and stochastic gravitational wave signals). One of the key tests for such models comes from the seesaw mechanism, namely, the presence of heavy right-handed Majorana neutrinos. Their Yukawa couplings are relevant for collider signals, LFV  and leptogenesis. We give an example of such studies for the $\Delta(6n^2)$ group. Other groups (e.g. $A_4$) and models (e.g. scotogenic) are also discussed in the context of LFV and dark matter effects. 

The concept of flavor symmetry is developing all the time. 
New ideas related to family symmetries also come from modular symmetries or texture zeroes. In the review, we have updated many strict correlations and predictions in models based on TM$_{1}$, TM$_{2}$ mixing,  $\mu-\tau$ reflection symmetries and status of light neutrino mass sum rules in the context of neutrinoless double beta decay.  However, given the plethora of flavor models and the rich phenomenology they offer, we cannot discuss all possibilities here. Our goal in this review was to give a gist of the flurry of activities going on in this field, illustrated with a few example scenarios. With neutrino physics entering the precision era, there is an exciting prospect for more intensive studies of discrete flavor symmetries in the future for both Majorana and Dirac neutrino scenarios.

	
\section*{Acknowledgements}
This work has been supported in part by the Polish National Science Center (NCN) under grant 2020/37/B/ST2/02371, the Freedom of
Research, Mobility of Science, and the Research Excellence Initiative of the University of Silesia in Katowice. The work of GC is supported by the U.S. Department of Energy under the award number DE-SC0020250 and DE-SC0020262. The work of PSBD is supported in part by the U.S. Department of Energy under grant No. DE-SC 0017987.  We thank Julia Gehrlein for discussion on sum rules and for providing Fig.~\ref{fig:sumrule}. We thank Joy Ganguly and Satyabrata  Mahapatra for insights on the FSS model discussed here.  BK thank his collaborators Arunansu Sil, Subhaditya Bhattacharya, Narendra Sahu, Debasish Borah and Srubabati Goswami for fruitful collaboration in the past. For the purpose of
Open Access, the authors have applied a CC-BY public copyright licence to any Author Accepted
Manuscript (AAM) version arising from this submission.

\bibliography{bibliography,biblio2}

\end{document}